\documentstyle[11pt,epsfig]{article}
 \def\3{\ss}
 
 \def\li{\mbox{Li}_2}
 
 \def\eps{\varepsilon}

 \def\yytqb{\gamma\gamma\rightarrow q\bar{q}}

 \def\yqtgq{\gamma q\rightarrow gq}
 \def\ybtgb{\gamma\bar{q}\rightarrow g\bar{q}}
 \def\ygtqb{\gamma g\rightarrow q\bar{q}}

 \def\qptqp{qq'\rightarrow qq'}
 \def\qrtqr{q\bar{q}'\rightarrow q\bar{q}'}
 \def\qbtrp{q\bar{q}\rightarrow \bar{q}'q'}

 \def\qqtqq{qq\rightarrow qq}
 \def\qbtqb{q\bar{q}\rightarrow q\bar{q}}

 \def\qbtgg{q\bar{q}\rightarrow gg}
 \def\qgtqg{qg \rightarrow qg}
 \def\bgtbg{\bar{q}g\rightarrow \bar{q}g}
 \def\ggtqb{gg\rightarrow q\bar{q}}

 \def\ggtgg{gg\rightarrow gg}

 \def\yytqbg{\gamma\gamma\rightarrow q\bar{q}g}

 \def\yqtqqb{\gamma q\rightarrow qq\bar{q}}
 \def\yqtqgg{\gamma q\rightarrow qgg}
 \def\ybtbgg{\gamma \bar{q}\rightarrow \bar{q}gg}
 \def\ygtgqb{\gamma g\rightarrow gq\bar{q}}

 \def\qptqpg{qq'\rightarrow qq'g}
 \def\qrtqrg{q\bar{q}'\rightarrow q\bar{q}'g}
 \def\qbtrpg{q\bar{q}\rightarrow \bar{q}'q'g}
 \def\qgtqpr{qg\rightarrow qq'\bar{q}'}

 \def\qqtqqg{qq\rightarrow qqg}
 \def\qbtqbg{q\bar{q}\rightarrow q\bar{q}g}
 \def\qgtqqb{qg\rightarrow qq\bar{q}}

 \def\qbtggg{q\bar{q}\rightarrow ggg}
 \def\qgtqgg{qg\rightarrow qgg}
 \def\bgtbgg{\bar{q}g\rightarrow\bar{q}gg}
 \def\ggtqbg{gg\rightarrow q\bar{q}g}

 \def\ggtggg{gg\rightarrow ggg}

 \def\lp{\left. }
 \def\rp{\right. }
 \def\lr{\left( }
 \def\rr{\right) }
 \def\le{\left[ }
 \def\re{\right] }
 \def\lg{\left\{ }
 \def\rg{\right\} }
 \def\lb{\left| }
 \def\rb{\right| }

 \def\ts{\frac{-t}{s}}
 \def\tu{\frac{t}{u}}
 \def\us{\frac{-u}{s}}
 \def\ut{\frac{u}{t}}

 \def\sux{\frac{s}{u}}
 
 \def\tux{\frac{t}{u}}
 \def\usx{\frac{u}{s}}
 \def\utx{\frac{u}{t}}

 \def\tus{\frac{tu}{s^2}}
 \def\ust{\frac{us}{t^2}}
 \def\uts{\frac{ut}{s^2}}

 \def\syf{\frac{s}{y_F}}
 \def\tyf{\frac{t}{y_F}}
 \def\uyf{\frac{u}{y_F}}

 \def\sqs{\frac{s}{Q^2}}
 \def\tqs{\frac{t}{Q^2}}
 \def\uqs{\frac{u}{Q^2}}

 \def\zde{\frac{z_a}{1-z_a}}
 \def\edz{\frac{1-z_a}{z_a}}

 \def\ede{\frac{1}{\varepsilon}}
 \def\edze{\frac{1}{2\varepsilon}}
 \def\edes{\frac{1}{\varepsilon^2}}

 \def\edcf{\frac{1}{C_F}}
 \def\ednc{\frac{1}{N_C}}

 \newcommand{\beq}{\begin{equation}}
 \newcommand{\eeq}{\end{equation}}
 \newcommand{\bea}{\begin{eqnarray}}
 \newcommand{\eea}{\end{eqnarray}}
 \newcommand{\nn}{\nonumber \\}
 \renewcommand{\theequation}{\mbox{\arabic{section}.\arabic{equation}}}

\input{feynman.tex}
\input{dina4p.sty}
\title{
  {\baselineskip16pt
    \centerline{\normalsize \tt DESY 97-234 \hfill ISSN 0418-9833}
    \centerline{\normalsize \tt ANL-HEP-PR-97-97 \hfill}
    \centerline{\normalsize \tt hep-ph/9712256 \hfill}
    \centerline{\normalsize \tt December 1997 \hfill}
  }
  \vskip1cm
  {\huge\bf
    Inclusive Jet Production in $\gamma p$ and $\gamma\gamma$ Processes:
    Direct and Resolved Photon Cross Sections in Next-To-Leading Order QCD
  }
  \author{
    {M.\ Klasen$^a$, T.\ Kleinwort$^b$, G.\ Kramer$^c$} \\[5mm]
    {$^a$ Deutsches Elektronen-Synchrotron DESY, Notkestra\ss{}e 85,}\\
    {     D-22607 Hamburg, Germany, e-mail: {\tt klasen@mail.desy.de}}\\[2mm]
    \centerline{and}\\[2mm]
    {     High Energy Physics Division\thanks{Supported by the U.S.\ 
         Department of Energy, Division of High Energy Physics, Contracts
         W-31-109-ENG-38 and DEFG05-86-ER-40272.}, Argonne National Laboratory,}\\
    {     Argonne, IL 60439-4815, U.S.A., email: {\tt klasen@hep.anl.gov}}\\[5mm]
    {$^b$ DESY-IfH Zeuthen, Platanenallee 6,}\\
    {     D-15738 Zeuthen, Germany, e-mail: {\tt tkleinw@ifh.de}} \\[5mm]
    {$^c$ II. Institut f\"ur Theoretische Physik\thanks
          {Supported by Bundesministerium f\"ur Bildung und Wissenschaft,
          Forschung und Technologie, Bonn, Germany under Contract 05\,7HH92P(0)
          and EU Program "Human Capital and Mobility" through Network
          "Physics at High Energy Colliders" under Contract
          CHRX-CT93-0357 (DG12 COMA).} ,
          Universit\"at Hamburg,}\\
    {     Luruper Chaussee 149, D-22761 Hamburg, Germany,}\\
    {     e-mail: {\tt kramer@desy.de}}
    }
  \date{}
}
\begin{document}
\maketitle
%
%\vspace{3cm}
%
\begin{abstract}
\thispagestyle{empty}
The production of jets in low $Q^2$ $ep$ scattering (photoproduction) and in
low $Q^2$ $e^+e^-$ scattering ($\gamma\gamma$ scattering) allows for testing
perturbative QCD and for measuring the proton and photon structure functions.
This requires exact theoretical predictions for one- and two-jet cross
sections. We describe the theoretical formalism, giving sufficient details,
for calculating the direct and resolved processes in $\gamma p$ and
$\gamma\gamma$ reactions in next-to-leading order QCD. We present the complete
analytical results for the Born terms, the virtual, and the real corrections.
To separate singular and regular regions of phase space we use the phase
space slicing method with an invariant mass cut-off. In this way, all soft
and collinear singularities are either canceled or absorbed into the
structure functions. Using a flexible Monte Carlo program, we evaluate
the cross sections numerically and perform various tests and comparisons
with other calculations. We consider the scale dependence of our results
and compare them to data from the experiments H1 and ZEUS at HERA and
OPAL at LEP.
\end{abstract}
\newpage
\setcounter{page}{1}
\section{Introduction}
The analysis of jet production in various high energy processes has
become a major field for testing perturbative QCD. Recently jet
production in $\gamma\gamma$ processes, where the two photons of very small
virtuality are produced in the collision of electrons and positrons,
has come into focus after data have been collected at the TRISTAN
\cite{x1} and LEP \cite{x2} colliders. In addition jet production
in deep inelastic (high $Q^2$) \cite{x3} and low $Q^2$ $ep$ scattering
(equivalent to photoproduction) has been measured at the two HERA
experiments H1 \cite{x4} and ZEUS \cite{x5}.

Jet production in $\gamma\gamma$ and $\gamma p$ collisions has several
similarities. At very small $Q^2$, where $q$ ($Q^2=-q^2$) is the
four-momentum transfer of the electron (positron) producing the virtual
photons in the $\gamma\gamma$ or $\gamma p$ initial state, the emission of
the photon can be described in the Equivalent Photon Approximation.
The spectrum of the virtual photons is approximated by the
Weizs\"acker-Williams (WWA) formula, which depends only on
$y=E_{\gamma}/E_e$, the fraction of the initial electron (positron)
energy $E_e$ transferred to the photon with energy $E_{\gamma}$, and on
$Q_{\max}^2$ (or $\theta_{\max}$), which is the maximal virtuality
(or the maximal electron scattering angle) allowed in the experimental set-up.

Concerning the hard scattering reactions both processes have
so-called direct and resolved components. Thus, in leading order
(LO) QCD the cross section $\sigma(\gamma\gamma\rightarrow \mbox{jets})$
receives contributions from three
distinct parts: (i) the direct contribution (DD), in which the
two photons couple directly to quarks, (ii) the single-resolved
contribution (DR), where one of the photons interacts with the
partonic constituents of the other photon, (iii) the double-resolved
contribution (RR), where both photons are resolved into partonic
constituents before the hard scattering subprocess takes place.
In the DD component (in LO) we have only two high-$p_T$ jets in the
final state and no additional spectator jets. In the DR
contribution one spectator jet originating from low transverse
momentum fragments of one of the photons is present, and in the RR
component we have two such spectator or photon remnant jets.
In the case of $\gamma p$ collisions one of the photons is replaced by
the proton which has no direct interaction. Then we have only the
DR and RR components which are usually referred to as the direct
and the resolved contribution. This means that the DR component
for $\gamma\gamma\rightarrow \mbox{jets}$ has the same structure as the
direct contribution of $\gamma p\rightarrow \mbox{jets}$
and the RR part for $\gamma\gamma\rightarrow \mbox{jets}$
is calculated in the same way as the resolved cross section for
$\gamma p\rightarrow \mbox{jets}$.

Of course, to calculate the resolved cross sections we need a
description of the partonic constituents of the photon. These
parton distributions of the photon are partly perturbative and
non-perturbative quantities. To fix the non-perturbative part one needs
information from experimental measurements to determine the fractional
momentum dependence at a reference scale $M_0^2$. The change with the
factorization scale $M^2 \stackrel{>}{\scriptstyle <} M_0^2$
is obtained from perturbative evolution
equations. Most of the information on the parton distribution functions
(PDF) of the photon comes from photon structure function ($F_2^{\gamma}$)
measurements in deep inelastic $e\gamma$ scattering, where, however, mainly
the quark distribution function can be determined. In $\gamma\gamma$ and
$\gamma p$ high-$p_T$ jet production also the gluon distribution of the photon
enters which can be constrained by these processes.

When we want to proceed to next-to-leading order (NLO) QCD the
following steps must be taken: (i) The hard scattering cross section
for the direct and resolved photon processes are calculated up
to NLO. (ii) NLO constructions for the PDF's of the proton and the
photon are used and are evolved in NLO up to the chosen
factorization scale via the Altarelli-Parisi equations and
convoluted with the NLO hard-scattering cross sections. (iii) To
calculate jet cross sections we must choose a jet definition which
may be either a cluster or a cone algorithm in accordance with the
choice made in the experimental analysis.

There exist several methods for calculating NLO corrections of
jet cross sections in high energy reactions \cite{x6}. As in our
previous work we apply the phase space slicing method with invariant
mass slicing to separate infrared and collinear singular phase space
regions. In this approach the contributions from the singular
regions are calculated analytically with the approximation that
terms of the order of the slicing cut are neglected. In the
non-singular phase space regions the cross section is obtained
numerically. This method allows the application of different
clustering procedures, the use of different variables for describing
the final state together with cuts on these variables as given by
the measurement of the jet cross sections. This method has been
used for the calculation of the NLO DD and DR cross sections
in the case of $\gamma\gamma\rightarrow \mbox{jets}$ \cite{x7} and for
the calculation of the direct cross section in the case of $\gamma p\rightarrow
\mbox{jets}$ \cite{x8}. It is obvious that the NLO calculation of the DR
cross section for the $\gamma\gamma$ case
is the same as of the NLO direct cross section for the $\gamma p$ case.

In the calculation of the $\gamma\gamma$ DD and DR or $\gamma p$
direct cross section
one encounters the photon-quark collinear singularity. This is
subtracted and absorbed into the quark distribution of the photon
in accord with the factorization theorem. This subtraction at the
factorization scale $M^2$ produces an interdependence of the three
components in the $\gamma\gamma$ and the two components in the
$\gamma p$ reaction, so that
a unique separation in DD, DR, and RR (for $\gamma\gamma$) and in direct and
resolved (for $\gamma p$) contributions is valid only in LO. The calculation
of these subtraction terms and of the contributions from the other
singular regions has been presented with sufficient details in
\cite{x7, x8}. The phase space slicing method was applied also for
the calculation of the NLO correction of the $\gamma\gamma$ RR contribution
\cite{x9} and the resolved $\gamma p$ cross section \cite{x10}. Results
for the inclusive two-jet cross sections together with a comparison
to recent experimental data were presented for $\gamma\gamma\rightarrow
\mbox{jets}$ in \cite{x11} and for $\gamma p\rightarrow \mbox{jets}$
in \cite{x12}, respectively. In these
two papers the specific calculations of the corresponding resolved
cross sections in NLO were not explicitly outlined, in particular
it was not shown, how the different infrared and collinear
singularities cancel between virtual and real contributions and
after the subtraction of collinear initial state singularities.
In this work we want to fill this gap. We describe the analytic
calculation of the various resolved terms in the specific singular
regions needed for the NLO corrections. Furthermore we check the
analytical results by comparing with results obtained with other
methods. In order to have a complete presentation we include also
the calculation of the NLO correction to the $\gamma\gamma$ DD cross
section already presented in \cite{x7}. We also include the details of
the DR contributions in the form of the direct $\gamma p$ cross section
taken from \cite{x8}. Details of other material, as for example, various
jet definitions, which were only mentioned in our previous work and which
now become more and more relevant in the analysis of the experimental
data, will also be presented. Since the calculation of the RR (DR) cross
section for $\gamma\gamma\rightarrow\mbox{jets}$ is the same as the
calculation of the $\gamma p\rightarrow\mbox{jets}$ resolved (direct)
cross section we concentrate on the $\gamma p$ case when we present the
details of the calculations. So for the $\gamma\gamma$ case we present
only the DD contribution which has no analogy in the $\gamma p$ case.
We come back to the $\gamma\gamma$ case when we show the
numerical results for specific cases including all three components.

We organize this work in seven main sections:
After this introduction we relate the experimental $ep$ scattering
to photon parton scattering in section 2. We describe the
Weizs\"acker-Williams approximation, discuss the proton and photon
PDF's and explain the experimental and theoretical properties
of various jet definitions. Furthermore, section 2 contains the
master formul{\ae} for one- and two-jet cross sections. These will be
calculated in section 3 in LO and in section 4 in NLO. In both
sections, we calculate the relevant phase space for $2\rightarrow 2$ and
$2\rightarrow 3$
scattering, respectively. The Born matrix elements
are contained in section 3. In section 4 we present the
virtual one-loop matrix elements and the tree-level $2\rightarrow 3$ matrix
elements, which are then integrated over singular regions of phase
space. Next, we demonstrate how all ultraviolet and infrared poles
in the NLO calculation cancel or are removed through renormalization
and factorization into PDF's. A detailed numerical evaluation of
jet cross sections in $\gamma p$ scattering
with the purpose to compare with results of other
work and to make consistency checks is contained in section 5.
We study the
renormalization and factorization scale dependence of one- and
two-jet cross sections including the direct and resolved components.
We also show results for some specific
inclusive one- and two-jet cross sections and compare them with
experimental data, in case they are available, to demonstrate
the usefulness of our methods. Section 6 contains the corresponding
numerical studies and comparisons to data for $\gamma\gamma$ scattering.
The final conclusions are left for section 7.

\setcounter{equation}{0}

\section{Photoproduction of Jets at HERA}

In this section we set up the general framework for theoretical predictions
of the photoproduction of jets in electron-proton scattering. This includes
the separation of the perturbatively calculable parts from the non-perturbative
parts of the cross section and linking the electron-proton to photon-proton
scattering.
This link will be discussed first in section 2.1 and consists in the
Weizs\"acker-Williams or Equivalent Photon Approximation.

The framework of the QCD improved parton model for protons will shortly
be discussed
in section 2.2. This is necessary, since protons are not pointlike but composed
of three valence quarks and sea quarks and gluons. Perturbative QCD is not
applicable at distances comparable to the size of the proton, but only at small
partonic scales due to the asymptotic freedom of QCD. Therefore, the parton
content of the proton is the domain of non-perturbative QCD and has to be
described with universal distribution functions. The scale dependence of these
functions is governed by the Altarelli-Parisi equations.

A similar concept applies for resolved photons, which are discussed in section
2.3. Contrary to direct photons, which are obviously pointlike, resolved
photons can be considered to have a complicated hadronic structure like
protons. However, they have a different valence structure than
protons. Furthermore, a complex relationship between direct and
resolved photons will show up in next-to-leading order of QCD.

Having related the initial state particles electron and proton in the 
experiment to
the photons and partons in perturbative QCD, we can turn our attention to the
final state particles. Section 2.4 describes how the interpretation of partons
as jets changes from leading to next-to-leading order of QCD. Several jet
definition schemes are discussed with respect to their theoretical and
experimental behavior.

The last section 2.5 summarizes the different ingredients of the calculation
and contains the master formul{\ae} for one- and two-jet photoproduction. Also,
the numerous analytical contributions in leading and next-to-leading order of
QCD are organized in a tabular form for transparency. 

\subsection{Photon Spectrum in the Electron}

In photoproduction, one would like to study the hard scattering of real
photon beams off nuclear targets. This has been done in fixed target
experiments, like NA14 at CERN \cite{Aug86} or E683 at Fermilab \cite{Ada94}, 
where real photons are produced e.g.~in pion decay with energies of up
to 400 GeV \cite{Pau92}. If higher energies are required, one must resort
to spacelike, almost real photons radiated
from electrons. This method is employed at the electron-proton collider HERA.
There, electrons of energy $E_e = 26.7$ GeV and recently positrons of
energy $E_e = 27.5$ GeV produce photons with small
virtuality $Q^2$, which then collide with a proton beam of
energy $E_p = 820$ GeV. This corresponds to photon energies of up to
$50$ TeV in fixed target experiments.

On the theoretical side, the calculation of the electron-proton cross
section can be considerably simplified by using the Weizs\"acker-Williams
or Equivalent Photon Approximation. Here, one uses current conservation and
the small photon virtuality to factorize the electron-proton cross section
into a broad-band photon spectrum in the electron and the hard photon-proton
cross section. Already in 1924, Fermi discovered the equivalence between
the perturbation of distant atoms by the field of charged particles flying
by and the one due to incident electromagnetic radiation \cite{Fer24}.
His semi-classical treatment was then extended to high-energy electrodynamics
in 1933-1935 by Weizs\"acker \cite{Wei34} and Williams \cite{Wil34}
independently, who used a Fourier analysis to unravel the
predominance of transverse over longitudinal photons radiated from a
relativistic charged particle. In the fifties, Curtis \cite{Cur56} and
Dalitz and Yennie \cite{Dal57} gave the first field-theoretical derivations
and applied the approximation to meson production in electron-nucleon
collisions. Chen and Zerwas used infinite-momentum-frame techniques for an
extension to photon bremsstrahlung and photon splitting processes
\cite{Che75}. For a recent review of the various approximations see
\cite{x13} and for the application to $\gamma\gamma$ processes see \cite{x14}.

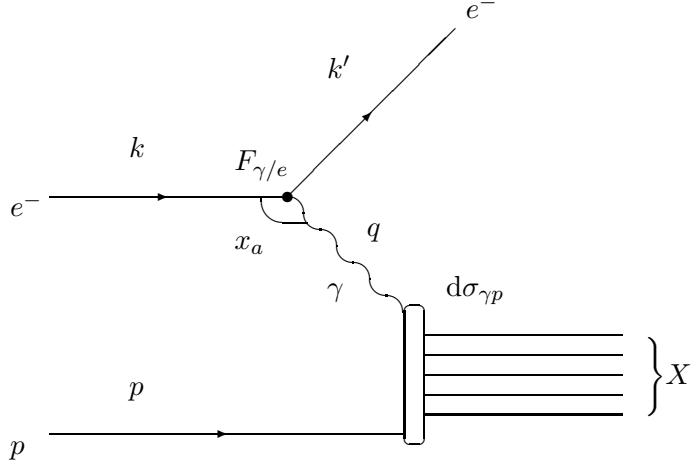
\begin{figure}[htbp]
\begin{center}
\begin{picture}(30000,19500)
\drawline\fermion[\E\REG](3000,10500)[9000]
\drawline\fermion[\NE\REG](12000,10500)[9000]
\drawline\fermion[\E\REG](3000,1500)[13425]
\drawline\fermion[\E\REG](17175,5250)[7500]
\drawline\fermion[\E\REG](17175,4500)[7500]
\drawline\fermion[\E\REG](17175,3750)[7500]
\drawline\fermion[\E\REG](17175,3000)[7500]
\drawline\fermion[\E\REG](17175,2250)[7500]
\drawline\photon[\SE\REG](12000,10500)[7]
\put(12000,10500){\circle*{450}}
\put(12000,10500){\oval(2000,2000)[bl]}
\drawline\fermion[\E\REG](12000,9500)[750]
\put(10000, 8500){$x_a$}
\put(16800,3750){\oval(750,5250)}
\drawarrow[\E\ATTIP](7500,10500)
\drawarrow[\E\ATTIP](9750,1500)
\drawarrow[\NE\ATTIP](15200,13700)
\put(25500,3400){$\Bigg\} X$}
\put(1500,9750){$e^-$}
\put(1500,750){$p$}
\put(13500,6750){$\gamma$}
\put(18750,17250){$e^-$}
\put(6000,12000){$k$}
\put(13500,15000){$k'$}
\put(15000,9000){$q$}
\put(6000,3000){$p$}
\put(10000,11500){$F_{\gamma / e}$}
\put(18000,6750){d$\sigma_{\gamma p}$}
\end{picture}
\end{center}
\caption[Electron-Proton Scattering with Single Photon Exchange]
        {\label{fig9}{\it Electron-proton scattering with single photon
         exchange.}}
\end{figure}

Let us consider the electroproduction process
\beq
 \mbox{electron}(k) + \mbox{proton}(p) \rightarrow \mbox{electron}(k') + X
\eeq
as shown in figure \ref{fig9}, where $k$, $k'$, and $p$ are the four-momenta
of the incoming and outgoing electron and the proton, respectively.
$X$ denotes a generic hadronic system not specified here. $q = k-k'$
is the momentum transfer of the electron to the photon with virtuality
$Q^2 = -q^2 \simeq 0$, and the center-of-mass energy of the process is
$\sqrt{s_H} = \sqrt{(k+p)^2} = 295.9$ GeV ($= 300.3$ GeV for the positron
beam). We restrict ourselves to one-photon
exchange without electroweak $Z^0$ admixture. Two-photon exchange is 
suppressed by an additional order of $\alpha$, and $Z^0$ exchange is suppressed
by the pole in the $Z^0$ mass $1/m_{Z^0}^2$.

The cross section of this process is 
\beq
 \mbox{d}\sigma_{ep} (ep\rightarrow eX)
  = \int\frac{1}{8k.p}\frac{e^2W^{\mu\nu}T_{\mu\nu}}
 {Q^4}\frac{\mbox{d}^3k'}{(2\pi)^32E_e'},
\eeq
where $W^{\mu\nu}$ and $T_{\mu\nu}$ are the usual hadron and lepton tensors.
Exploiting current conservation and the small photon virtuality, we find
\beq
 W^{\mu\nu}T_{\mu\nu} = 4W_1(Q^2=0,q.p) \le Q^2\frac{1+(1-x_a)^2}
 {x_a^2}-2m_e^2\re,
\label{eq1}
\eeq
where
\beq
 x_a = \frac{q.p}{k.p} = 1-\frac{k'.p}{k.p} \in [0,1] 
\eeq
is the fraction of longitudinal momentum carried by the photon.
Next, we have to calculate the phase space for the scattered electron.
Using
\bea
 k' = (E_e',0,E_e'\beta'\sin\theta,E_e'\beta'\cos\theta) &,& 
 k  = (E_e,0,0,E_e\beta),\\
 \beta' = \sqrt{1-\frac{m_e^2}{E_e'^2}} &,&
 \beta = \sqrt{1-\frac{m_e^2}{E_e^2}},
\eea
and integrating over the azimuthal angle, we find
\bea
 \frac{\mbox{d}^3k'}{E_e'} &=& 2\pi\beta'^2E_e'\mbox{d}k'\mbox{d}\cos\theta
                             \nonumber \\
                         &=& \pi\mbox{d}q^2\mbox{d}x_a. \label{eq2}
\eea
Combining eq.~(\ref{eq1}) and (\ref{eq2}) and integrating over the photon
virtuality, we can factorize the electron-proton cross section
into
\beq
 \mbox{d}\sigma_{ep} (ep\rightarrow eX)
 = \int\limits_0^{1}\mbox{d}x_a F_{\gamma/e}(x_a)
 \mbox{d}\sigma_{\gamma p}(\gamma p\rightarrow X),
 \label{eq12}
\eeq
where
\beq
 F_{\gamma/e}(x_a) = \frac{\alpha}{2\pi} \le \frac{1+(1-x_a)^2}{x_a}
 \ln\frac{Q_{\max}^2}{Q_{\min}^2}+2m_e^2x_a\lr\frac{1}{Q_{\min}^2}
 -\frac{1}{Q_{\max}^2}\rr\re
\label{eq3}
\eeq
is the renowned Weizs\"acker-Williams approximation and where
\beq
 \sigma_{\gamma p}(\gamma p\rightarrow X) = 
 -\frac{g_{\mu\nu}W^{\mu\nu}}{8q.p} =
 \frac{W_1(Q^2=0,q.p)}{4q.p}
\eeq
is the photon-proton cross section. We will only use the leading logarithmic
contribution and neglect the second term in eq.~(\ref{eq3}), also calculated
by Frixione et al. \cite{Fri93}.

There exist two fundamentally different experimental situations. At HERA,
the scattered electron is anti-tagged and must disappear into the beam pipe.
For such small scattering angles of the electron, the integration bounds
$Q_{\min}$ and $Q_{\max}$ can be calculated from the equation
\beq
 Q^2 = \frac{m_e^2x_a^2}{1-x_a}+\frac{E_e(1+\beta)(A^2-m_e^2)^2}{4A^3}\theta^2
 +{\cal O} (\theta^4),
\eeq
where $A=E_e(1+\beta)(1-x_a)$.
Using the minimum scattering angle $\theta=0$, we obtain
\beq
 Q_{\min}^2 = \frac{m_e^2x_a^2}{1-x_a},
\eeq
which leads us to the final form of the Weizs\"acker-Williams approximation
used here:
\beq
 F_{\gamma/e}(x_a) = \frac{\alpha}{2\pi}\frac{1+(1-x_a)^2}{x_a}\ln
                         \left(\frac{Q_{\max}^2 (1-x_a)}{m_e^2~x_a^2}\right).
 \label{eq45}
\eeq
At H1 and ZEUS, the maximum virtualities of the photon are given directly as
$Q_{\max}^2 = 0.01~\mbox{GeV}^2$ and $4~\mbox{GeV}^2$, respectively.
If only the maximum scattering angle of the electron is known,
we obtain
\beq
 Q_{\max}^2 = E_e^2(1-x_a)\theta_{\max}^2
\eeq
with $\theta_{\max}$ being of the order of $5^\circ$. This form of $Q_{\max}^2$
will be used for the calculation of the $\gamma\gamma$ cross sections in
accordance with experimental choices of $\theta_{\max}$ at LEP.
Except in equation
(\ref{eq45}), we will assume all particles to be massless in this paper.
Therefore, all results are only valid in the high-energy limit.

When no information about the scattered electron is available, one has
to integrate over the whole phase space thus allowing large transverse
momenta and endangering the factorization property of the cross section.
Then, one only knows that the invariant mass of the produced hadronic
system $X$ has to be bounded from below, e.g. by minimal transverse momenta
or heavy quark mass thresholds, which constrains the maximum virtuality
of the photon only weakly. Possible choices are
\beq
 Q_{\max} = \frac{\sqrt{s_H}}{2},\frac{\sqrt{x_as_H}}{2},E_e, \cdots .
\eeq

At $e^+e^-$ colliders, bremsstrahlung of electrons is not the only source of 
almost real photons. The particles in one bunch experience rapid acceleration
when they enter the electromagnetic field of the opposite bunch producing
beamstrahlung which depends sensitively on the machine parameters.
Even higher luminosities can be achieved by colliding the electron beam at some
distance from the interaction point with a dense laser beam \cite{Aur95}.

\subsection{Parton Distributions in the Proton}

\begin{figure}[htbp]
\begin{center}
\begin{picture}(31000,22500)
\drawline\fermion[\E\REG](3000,13500)[9000]
\drawline\fermion[\E\REG](3000,5250)[9000]
\drawline\fermion[\E\REG](3000,4500)[19500]
\drawline\fermion[\E\REG](3000,3750)[19500]
\drawline\fermion[\E\REG](17200,9750)[5300]
\drawline\fermion[\E\REG](17550,9000)[4950]
\drawline\fermion[\E\REG](17200,8250)[5300]
\drawline\fermion[\NE\REG](12000,5100)[5000]
\drawline\fermion[\NE\REG](12000,13500)[8400]
\drawline\photon[\SE\REG](12000,13500)[6]
\put(12000,13500){\circle*{450}}
\put(16500,9000){\circle{2100}}
\put(12000,4500){\circle*{2100}}
\drawarrow[\E\ATTIP](7500,13500)
\drawarrow[\E\ATTIP](7500,5250)
\drawarrow[\E\ATTIP](7500,4500)
\drawarrow[\E\ATTIP](7500,3750)
\drawarrow[\E\ATTIP](18000,4500)
\drawarrow[\E\ATTIP](18000,3750)
\drawarrow[\E\ATTIP](21000,9750)
\drawarrow[\E\ATTIP](21000,9000)
\drawarrow[\E\ATTIP](21000,8250)
\drawarrow[\NE\ATTIP](15200,16700)
\drawarrow[\NE\ATTIP](13800,6900)
\put(1500,13500){$e^-$}
\put(18000,19500){$e^-$}
\put(6000,15000){$k$}
\put(13500,18000){$k'$}
\put(1500,3750){$p$}
\put(6000,6000){$p$}
\put(10500,6000){$F_{b/p}$}
\put(15000,6000){$p_b=x_b p$}
\put(15000,1500){$p_r=(1-x_b) p$}
\put(12750,9750){$\gamma$}
\put(15000,12000){$q$}
\put(16500,10500){d$\sigma_{\gamma b}$}
\put(24000,8250){$\Bigg\}$ jets}
\put(24000,3750){$\Bigg\}$ remnant}
\end{picture}
\end{center}
\caption[Electron-Proton Scattering in the Parton Model]
        {\label{fig10}{\it Electron-proton scattering in the parton model.}}
\end{figure}
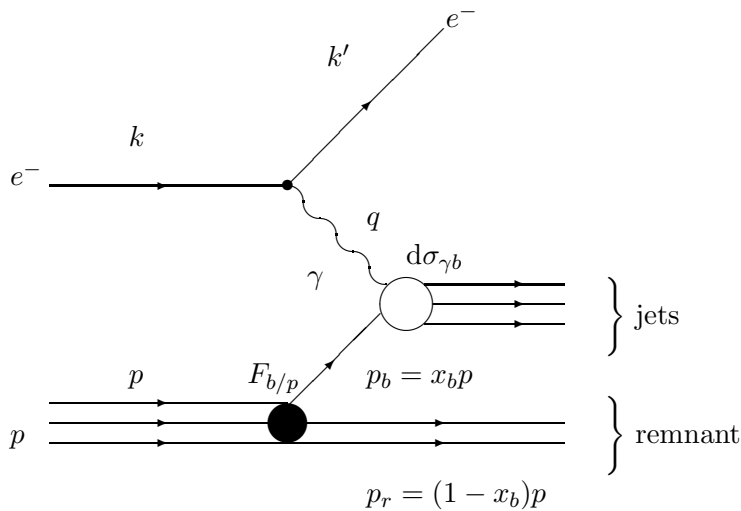

The first evidence that the proton has a substructure came from deep
inelastic electron-proton scattering $ep\rightarrow eX$. Together with
muon- and neutrino-scattering, this process provides us with information
on the distribution of partons (quarks and gluons) in the proton. In the
parton model, scattering off hadrons is interpreted as an incoherent
superposition of scattering off massless and pointlike partons.
This is shown in figure \ref{fig10},
where $p_b$ is the four-momentum of the scattered parton $b$. If the scale
$M_b$, at which the proton is probed, is large enough, the transverse
momentum of the parton will be small and its phase space can be described
by a single variable
\beq
 x_b = \frac{qp_b}{qp} \in [0,1],
\eeq
the longitudinal momentum fraction of the proton carried by the parton.
The other partons in the proton will then not notice the interaction but
form a so-called remnant or spectator jet. Multiple scattering is suppressed
by factors of $1/M_b^2$, and only the leading twist contributes for large
$M_b^2$. The proton remnant carries the four-momentum $p_r$, which
again depends only on $x_b$ and the proton momentum $p$. The proton remnant
will not be counted as a jet in the following.

We can now factorize the photon-proton cross section into
\beq
  \mbox{d}\sigma_{\gamma p} (\gamma p \rightarrow \mbox{jets + remnant}) =
  \sum_b \int\limits_0^1 \mbox{d}x_b F_{b/p} (x_b, M_b^2)
  \mbox{d}\sigma_{\gamma b} (\gamma b \rightarrow \mbox{jets}),
  \label{eq13}
\eeq
where $F_{b/p} (x_b, M_b^2)$ is the probability of finding a parton $b$
(where $b$ may be a quark, an antiquark, or a gluon) within the proton carrying
a fraction $x_b$ of its momentum when probed at the hard scale
$M_b$. The usual choice for deep inelastic scattering, $M_b=Q$, is not
possible for photoproduction, as $Q^2\simeq 0$ here. Instead, one takes
$M_b = \xi E_T$ with $E_T$ being the transverse energy of the outgoing
parton or observed jet and $\xi$ being of order 1.
Contrary to the hard photon-parton scattering cross section
d$\sigma_{\gamma b}$, the parton densities $F_{b/p} (x_b, M_b^2)$
are not calculable in
perturbative QCD and have to be taken from experiment.

In the leading twist approximation, the parton densities extracted from
deep inelastic scattering can be used for any other process with incoming
nucleons like photoproduction or proton-antiproton scattering -- they are
universal. As stated above, they have to be determined by experiment
at some scale $Q^2=Q_0^2$, where one parametrizes the $x$ dependence before
evolving up to any value of $Q^2$ according to the Altarelli-Parisi (AP)
equations \cite{Alt77}.
The input parameters are then determined by a global fit to the data.
The $u$, $d$, and $s$ quark densities are rather well known today from 
muon scattering (BCDMS, NMC, E665) and neutrino scattering (CCFR) experiments,
and the $c$ quark is constrained by open charm production at EMC.
The small-$x$ region, where the gluon and the sea quarks become important,
could, however, not be studied by these experiments.
This is mainly the domain of the HERA experiments H1 and ZEUS
in DIS as well as in photoproduction.

There exist many different sets of parton density functions. Most of them
are easily accessible in the CERN library PDFLIB kept up to date by
Plothow-Besch \cite{Plo95}. 
We will now briefly compare the three parametrizations mainly in use today:
CTEQ \cite{Lai95}, MRS \cite{Mar94}, and GRV \cite{Glu95}.
The first two are very similar to each other, and both take
$Q_0^2 = 4~\mbox{GeV}^2$. 
Very recently, CTEQ and MRS presented new parametrizations (set 4 and set R,
respectively) including new HERA and TEVATRON inclusive jet data
\cite{Lai96,Mar96a}.
The GRV approach is rather different from the two mentioned above.
In order to avoid any free additional parameters, their original version was
based on the assumption that all gluon and sea distributions are generated
dynamically from measured valence quark densities at a very low scale
$Q_0 \simeq {\cal O} (\Lambda )$. The QCD coupling scale $\Lambda$ is fitted
to the data and ranges from 200 to 344 MeV for the parametrizations
considered here and for four quark flavors.

\subsection{Parton Distributions in the Photon}

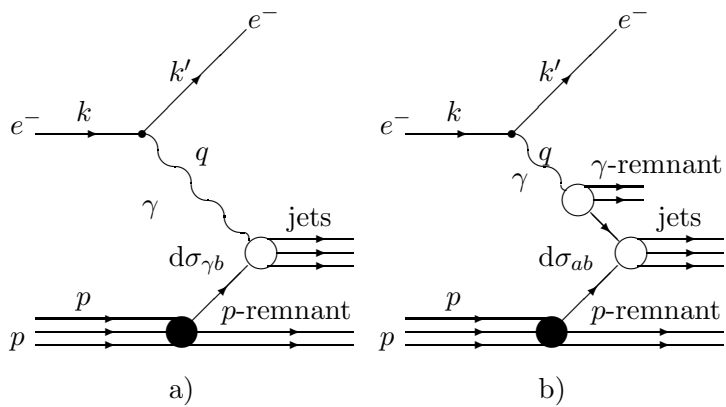
\begin{figure}[htbp]
\begin{center}
\begin{picture}(32000,15000)
\drawline\fermion[\E\REG](1000,10000)[4000]
\drawarrow[\E\ATBASE](3000,10000)
\drawline\fermion[\NE\REG](5000,10000)[5600]
\drawarrow[\NE\ATBASE](7000,12000)
\put(5000,10000){\circle*{300}}
\drawline\photon[\SE\REG](5000,10000)[6]
\put(9500,5500){\circle{1200}}
\put(8750,5950){\oval(600,600)[tr]}
\drawline\fermion[\E\REG](9750,6000)[3250]
\drawarrow[\E\ATTIP](12000,6000)
\drawline\fermion[\E\REG](10100,5500)[2900]
\drawarrow[\E\ATTIP](12000,5500)
\drawline\fermion[\E\REG](9750,5000)[3250]
\drawarrow[\E\ATTIP](12000,5000)
\drawline\fermion[\NE\REG](6750,2750)[3200]
\drawarrow[\NE\ATBASE](8000,4000)
\drawline\fermion[\E\REG](1000,3000)[5500]
\drawarrow[\E\ATTIP](4000,3000)
\drawline\fermion[\E\REG](1000,2500)[12000]
\drawarrow[\E\ATTIP](4000,2500)
\drawarrow[\E\ATTIP](11000,2500)
\drawline\fermion[\E\REG](1000,2000)[12000]
\drawarrow[\E\ATTIP](4000,2000)
\drawarrow[\E\ATTIP](11000,2000)
\put(6500,2500){\circle*{1200}}
\put(0,10000){$e^-$}
\put(0,2000){$p$}
\put(2500,10500){$k$}
\put(6000,12000){$k'$}
\put(9000,14000){$e^-$}
\put(7000,9000){$q$}
\put(5000,7000){$\gamma$}
\put(10500,6500){jets}
\put(2500,3500){$p$}
\put(8000,3000){$p$-remnant}
\put(6000,5000){d$\sigma_{\gamma b}$}
\put(6000,0){a)}
\drawline\fermion[\E\REG](15000,10000)[4000]
\drawarrow[\E\ATBASE](17000,10000)
\drawline\fermion[\NE\REG](19000,10000)[5600]
\drawarrow[\NE\ATBASE](21000,12000)
\put(19000,10000){\circle*{300}}
\drawline\photon[\SE\REG](19000,10000)[3]
\put(21500,7500){\circle{1200}}
\put(21000,8050){\oval(300,300)[bl]}
\drawline\fermion[\SE\REG](22000,7000)[1500]
\drawarrow[\SE\ATBASE](22500,6500)
\drawline\fermion[\E\REG](21750,8000)[2250]
\drawarrow[\E\ATBASE](23000,8000)
\drawline\fermion[\E\REG](22100,7500)[1900]
\drawarrow[\E\ATBASE](23000,7500)
\put(23500,5500){\circle{1200}}
\drawline\fermion[\E\REG](23750,6000)[3250]
\drawarrow[\E\ATTIP](26000,6000)
\drawline\fermion[\E\REG](24100,5500)[2900]
\drawarrow[\E\ATTIP](26000,5500)
\drawline\fermion[\E\REG](23750,5000)[3250]
\drawarrow[\E\ATTIP](26000,5000)
\drawline\fermion[\NE\REG](20750,2750)[3200]
\drawarrow[\NE\ATBASE](22000,4000)
\drawline\fermion[\E\REG](15000,3000)[5500]
\drawarrow[\E\ATTIP](18000,3000)
\drawline\fermion[\E\REG](15000,2500)[12000]
\drawarrow[\E\ATTIP](18000,2500)
\drawarrow[\E\ATTIP](25000,2500)
\drawline\fermion[\E\REG](15000,2000)[12000]
\drawarrow[\E\ATTIP](18000,2000)
\drawarrow[\E\ATTIP](25000,2000)
\put(20500,2500){\circle*{1200}}
\put(14000,10000){$e^-$}
\put(14000,2000){$p$}
\put(16500,10500){$k$}
\put(20000,12000){$k'$}
\put(23000,14000){$e^-$}
\put(20000,9000){$q$}
\put(19000,8000){$\gamma$}
\put(24500,6500){jets}
\put(16500,3500){$p$}
\put(22000,3000){$p$-remnant}
\put(22000,8500){$\gamma$-remnant}
\put(20000,5000){d$\sigma_{ab}$}
\put(20000,0){b)}
\end{picture}
\end{center}
\caption[Generic Diagrams for Direct and Resolved Photoproduction]
        {\label{fig12}{\it Generic diagrams for a) direct and b) resolved
         photoproduction.}}
\end{figure}

Although
the photon is the fundamental gauge boson of quantum electrodynamics (QED),
which is the most accurately tested field theory, many reactions
involving photons are much less well understood.
This is due to the fact that the photon can fluctuate into $q\overline{q}$
pairs which in turn evolve into a complicated hadronic structure.
At HERA, the photon radiated from the electron can thus interact
either directly with a parton in the proton (direct component) or act
as a hadronic source of partons which collide with the partons in the proton
(resolved component, see figure \ref{fig12}). In the latter case,
one does not test the proton structure alone but also the photon 
structure.

Using the factorization theorem, 
the cross section for photon-parton scattering can be written as
\beq
  \mbox{d}\sigma_{\gamma b} (\gamma b \rightarrow \mbox{jets}) =
  \sum_a \int\limits_0^1 \mbox{d}y_a F_{a/\gamma} (y_a, M_a^2)
  \mbox{d}\sigma_{ab} (ab \rightarrow \mbox{jets}).
  \label{eq14}
\eeq
Here, $F_{a/\gamma}(y_a, M_a^2)$ stands for the probability of finding
a parton $a$ with momentum fraction $y_a$ in the photon, which
has to be universal in all processes. The scale $M_a$
is a measure for the hardness of the parton-parton cross section
d$\sigma_{ab}$ calculable in perturbative QCD and is again taken to be of
${\cal O} (E_T)$. The particle $a$ can also be a direct photon. Then,
$F_{\gamma/\gamma}(y_a, M_a^2)$ is simply given by the $\delta$-function
$\delta(1-y_a)$ and does not depend on the hard scale $M_a$.

Before HERA started taking data, information on the hadronic structure
of the photon came almost exclusively from deep inelastic $\gamma^{\ast}\gamma$
scattering at $e^+e^-$ colliders. Similarly to deep inelastic $ep$ scattering,
this is a totally inclusive process well suited to define a photon structure
function. Using $y=Q^2/(x_Bs_H)$, where $Q^2$ denotes the virtuality of the
probing photon $\gamma^{\ast}$, and replacing $F_1$ by the longitudinal
structure function $F_L(x_B,Q^2) = F_2(x_B,Q^2)-2x_BF_1(x_B,Q^2)$,
we can write the deep inelastic scattering (DIS) cross section in the
following form
\beq
 \frac{\mbox{d}^2\sigma}{\mbox{d}x_B\mbox{d}y} =
 \frac{2\pi\alpha^2s_H}{Q^4}\lg\le 1+(1-y)^2 \re F_2^{\gamma}(x_B,Q^2)
 -y^2F_L^{\gamma}(x_B,Q^2)\rg,
 \label{eq9}
\eeq
where in LO $F_2^{\gamma}(x_B,Q^2)$ is related to the singlet quark parton
density in the photon similarly as in deep inelastic $ep$ scattering
\beq
 F_2^{\gamma}(x_B,Q^2) =  \sum_q x_B e_q^2
 (F_{q/\gamma}(x_B,Q^2)+F_{\overline{q}/\gamma}(x_B,Q^2)).
\eeq
Fitting the input parameters to data is not as easy as in the proton case.
First, no momentum sum rule applies for the photonic parton densities as
they are all of LO in QED. This and the subleading nature of the gluonic
process makes a determination of the gluon from $F_2^{\gamma}$ very difficult.
However, a momentum sum rule exists for mesons so that the VMD part of the
gluon is constrained. Second, the cross section for
$F_2^{\gamma}$ is quite small and falls rapidly with increasing $Q^2$
(see eq. (\ref{eq9})) leading to large statistical errors.

Presently, over 20 different parton distributions exist for the photon, and
most of them are available in the PDFLIB \cite{Plo95}.
The VMD input is insufficient to fit the data at
higher $Q^2$ offering two alternatives: GRV \cite{Glu92}, AFG \cite{Aur92},
and SaS \cite{Sch95} follow the same ``dynamical'' philosophy as in the
GRV proton case starting from a simple valence-like input at a low
scale $Q_0^2 = 0.25~...~0.36~\mbox{GeV}^2$. On the other hand, LAC
\cite{Abr91} and GS \cite{Gor92} take a larger value for
$Q_0^2 = 4~...~5.3~\mbox{GeV}^2$,
assume a more complicated ansatz there,
and fit the free parameters to $F_2^{\gamma}$
data. LAC intended to demonstrate the poor constraints
on the gluon assuming a very soft gluon (fits 1 and 2) as well as a very hard
one (fit 3). However, LAC 3 is ruled out now by recent $e^+e^-$ and HERA
data. In their latest parametrization, GS lowered their input scale
to $Q_0^2 = 3~\mbox{GeV}^2$, included all available data on $F_2^{\gamma}$,
and constrained the gluon from jet production at TRISTAN \cite{Gor96}.

The distinction between direct and resolved photon is only meaningful in LO
of perturbation theory. In NLO, collinear singularities arise from the
photon initial state that have to be absorbed into the photon structure
function (cf.~sections 4.2.5 and 4.2.7)
and produce a factorization scheme dependence
as in the proton case. If one requires approximately the same $F_2^{\gamma}$
in LO and NLO, the quark distributions in the $\overline{\mbox{MS}}$-scheme
have quite different shapes in LO and NLO.
This is not the case if the $\mbox{DIS}_{\gamma}$-scheme of GRV is used,
where the direct-photon contribution to $F_2^{\gamma}$ is absorbed into
the photonic quark distributions. This allows for perturbative stability
between LO and NLO results.
Therefore, the separation between direct and resolved process is an artifact
of finite order perturbation theory and depends on the factorization scheme
and scale $M_a$. Experimentally, one tries to get a handle on this by
introducing kinematical cuts, e.g.~on the photon energy fraction taking part
in the hard cross section.

\subsection{Jet Definitions}

Due to the confinement of color charge, neither incoming nor outgoing
partons can be observed directly in experiment, but rather transform into
colorless hadrons. This transformation is a long-distance process and is
not calculable in
perturbative QCD. For the incoming particles, we have already described
how the ignorance of universal parton distributions in hadrons is
parametrized. A similar method can be used for the final state partons,
employing so-called fragmentation functions for the inclusive production
of single hadrons \cite{Bin95}. The transformation of partons into
individual hadrons forming part of the total final state was first studied by
\cite{Fie78}.
Alternatively, one can observe beams of many hadrons going approximately
into the same direction without the need to specify individual hadrons.
The hadrons are then combined into so-called jets by cluster algorithms,
where one starts from an initial cluster in phase space and ends at stable
fixed points for the jet coordinates. These jet definitions should fulfill
a number of important properties. They should \cite{Ell89a}
\begin{itemize}
\item be simple to implement in the experimental analysis,
\item be simple to implement in the theoretical calculation,
\item be defined at any order of perturbation theory,
\item yield a finite cross section at any order of perturbation theory,
\item yield a cross section that is relatively insensitive to hadronization.
\end{itemize}
Although, in principle, hadronization of the final state should be factorizable
from the hard cross section and the initial state, jets do indeed look
quite different in $e^+e^-$ and in, at least partly, hadronic collisions like
$ep$ scattering. This can be attributed to the ``underlying event'' of
remnant jet production from the initial state, which can interfere with the
hard jets in the final state. The main problem here is the determination of
the true jet energy and the subtraction of the remnant pedestal.

In LO QCD, there is a one-to-one correspondence between partons and jets.
This results in a complete insensitivity of theory to the experimentally
used algorithm or to the
resolution parameters. The experimental results depend,
however, on these parameters as well as on detector properties and have to
be corrected stepwise from detector level to hadron level to parton level.
The situation can only be improved by going to NLO in perturbation theory.
Here, the emission of one additional soft or collinear parton is calculable
with correct treatment of the occurring singularities. A hadron jet
can then consist not only of one, but also of two partons. It acquires a
certain substructure, and will depend on the experimental algorithm and
resolution.

Historically, the first resolution criterion was proposed by Sterman and
Weinberg \cite{Ste77}. It adds a particle to a jet if its energy is
smaller than $\varepsilon M$ or its angle with the jet is less
than $2\delta$, which provides a close link to the radiation of secondary
partons. The energy cut handles the soft divergencies and the angular cut the
collinear divergencies. There is some freedom in the choice for $M$.
Usually, one takes the hard scale of the process, e.g.~$Q$ in $e^+e^-$
annihilation.

The PETRA experiments used this $(\epsilon,\delta)$ algorithm and
the so-called JADE cluster algorithm \cite{Bar86}, 
which has the advantage of being invariant.
Two particles are combined into a cluster, if their
invariant mass $s_{ij}=(p_i+p_j)^2$ is smaller than $y M^2$ \cite{Kra84}, 
where $M^2$ is again a typical scale and $y$ is of ${\cal O} (10^{-2})$. In
this way, the soft and collinear divergencies can be described by a single
cut-off $y$. There exist a number of different schemes for
this algorithm regarding the invariant mass of the combination of the two
particles and the combination of the particle four-momenta (JADE-, $E$-,
$E0$-, $P$-, and $P0$-scheme).

\begin{figure}[htbp]
 \begin{center}
  {\unitlength1cm
  \begin{picture}(15,5)
   \epsfig{file=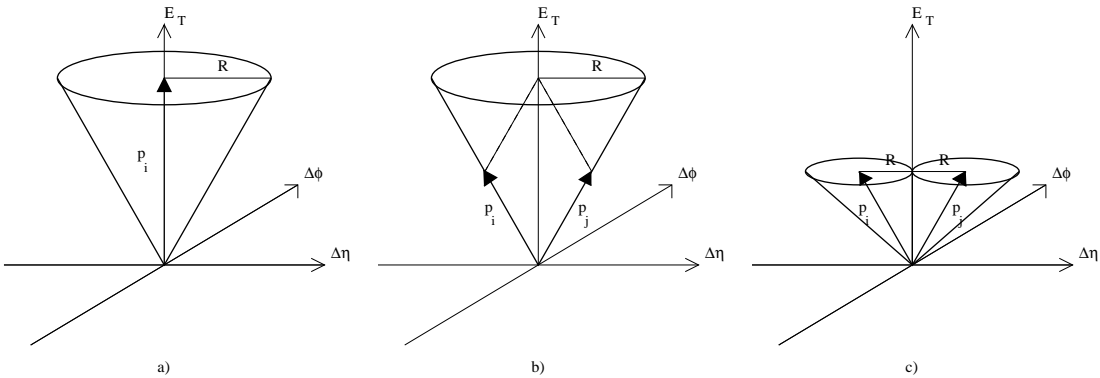,bbllx=44pt,bblly=308pt,bburx=566pt,bbury=486pt,%
           height=5cm,clip=}
  \end{picture}}
  \caption[Jet Cone Definition According to the Snowmass Convention]
          {\label{plot6}{\it
           Jet cone definition according to the Snowmass convention for
           a) a single parton, b) two combined partons with distance $R$
           from the jet axis, and c) two single partons.}}
 \end{center}
\end{figure}

At hadron colliders, cluster algorithms tend to include hadrons from the
remnant jet, which is not present in $e^+e^-$ collisions, into the current
jet. Therefore, algorithms are preferred here which use a cone in
rapidity-azimuth ($\eta-\phi$) space, quite similar to the $\delta$-condition
of Sterman and Weinberg. The (pseudo-)rapidity $\eta=-\ln [\tan(\theta/2)]$
parametrizes the polar angle $\theta$ between the hard jet and the beam axis,
and $\phi$ is the azimuthal angle of the jet around the beam axis. In the case
of cone algorithms, only inclusive cross sections are infrared safe, where an
arbitrary number of particles outside the jet cone can be radiated as long as
they are softer than the observed jets. According to the standardization of the
Snowmass meeting in 1990, calorimeter cells or partons $i$ may have a
distance $R_i$ from the jet center provided that
\beq
 R_i = \sqrt{(\eta_i-\eta_J)^2+(\phi_i-\phi_J)^2} \leq R,
\label{eq10}
\eeq
where $\eta_i$ and $\phi_i$ are the coordinates of the parton or the center
of the calorimeter cell \cite{Hut92}. Typical values for the resolution
parameter $R$ range from 0.7 to 1, where the effects of hadronization and
the underlying event are minimized. The transverse energy of the combined
jet $E_{T_J}$ is calculated from the sum of the particle $E_{T_i}$
\beq
 E_{T_J} = \sum_{R_i\leq R} E_{T_i},
\eeq
and the jet axis is defined by the weighted averages
\bea
 \eta_J  &=& \frac{1}{E_{T_J}}\sum_{R_i\leq R} E_{T_i}\eta_i, \\
 \phi_J  &=& \frac{1}{E_{T_J}}\sum_{R_i\leq R} E_{T_i}\phi_i.
\eea
In perturbative QCD, the final state consists of a limited number of
partons. For a single isolated parton $i$, the partonic and jet parameters
agree ($(E_{T_i},\eta_i,\phi_i) = (E_{T_J},\eta_J,\phi_J)$) as shown
in figure \ref{plot6}a),
% Plot6
whereas two neighboring partons $i$, $j$ will form a combined jet as shown
in figure \ref{plot6}b). Equation (\ref{eq10}) only defines the distances
$R_i$, $R_j$ of each parton from the jet axis so that the
two partons may be separated from each other by
\beq
 R_{ij} = \sqrt{(\eta_i-\eta_j)^2+(\phi_i-\phi_j)^2} \leq
 \frac{E_{T_i}+E_{T_j}}{\max (E_{T_i},E_{T_j})} R.
\eeq
If both partons have equal transverse energy, they may then be separated by
as much as $2R$. As parton $j$ does not lie inside a cone of radius $R$
around parton $i$ and vice versa, one might with some justification also
count the two partons separately as shown in figure \ref{plot6}c). If one
wants to study only the highest-$E_T$ jet, this ``double counting'' must
be excluded.

The selection of the initiating cluster, before a cone is introduced
(``seed-finding''), is not fixed by the Snowmass convention, and different
approaches are possible. The ZEUS collaboration at HERA uses two
different cone algorithms: EUCELL takes the calorimeter cells in a window
in $\eta-\phi$ space as seeds to find a cone with the highest $E_T$. The
cells in this cone are then removed, and the algorithm continues. On the
other hand, PUCELL was adapted from CDF and starts with single calorimeter
cells. It then iterates cones around each of them, until the set of enclosed
cells is stable. In this case it may happen that two stable jets overlap. 
If the overlapping transverse energy amounts to a large fraction of the jets,
they are merged, otherwise the overlapping energy is split. Alternatively, the
overlap could be attributed to the nearest, to the largest, or to both jets. 
The question of overlapping jets cannot be addressed in a next-to-leading
order calculation of photoproduction. There, we only have up to three partons
in the final state, which can form at most one recombined jet and no
overlap.

\begin{figure}[htbp]
 \begin{center}
  {\unitlength1cm
  \begin{picture}(9,5)
   \epsfig{file=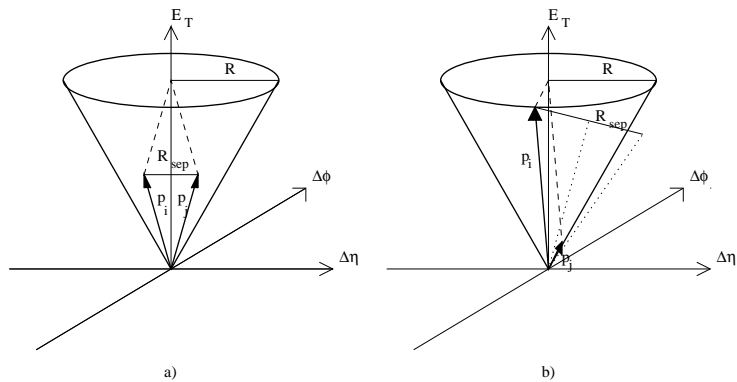,bbllx=60pt,bblly=272pt,bburx=550pt,bbury=524pt,%
           height=5cm,clip=}
  \end{picture}}
  \caption[Jet Cone Definition with an Additional Parameter $R_{\rm sep}$]
          {\label{plot7}{\it
           Jet cone definition with an additional parameter for
           parton-parton separation $R_{\rm sep}$: a) two partons with
           similar or equal transverse energies $E_T$, b) two partons with
           large $E_T$ imbalance.}}
 \end{center}
\end{figure}

Experimentally, jets of type b) in figure \ref{plot6} are
hard to find because of the missing seed in the jet center. This is a problem
in particular with the PUCELL algorithm, which relies on initial clusters and
does
indeed find smaller cross sections than the less affected EUCELL algorithm
\cite{But96}.
One possibility to model this theoretically is to introduce an additional
parameter $R_{\rm sep}$ \cite{y7} to restrict the distance of two partons from
each other:
\beq
 R_{ij} \leq \min\le\frac{E_{T_i}+E_{T_j}}{\max (E_{T_i},E_{T_j})} R,
 R_{\rm sep}\re.
\eeq
$R_{\rm sep} = 2R$ means no restriction.
In figure \ref{plot7}, we can see that for two partons
% Plot7
of similar or equal transverse energies $E_T$, $R_{\rm sep}$ is
the limiting parameter, whereas it is the parton-jet distance $R$ for two
partons with large $E_T$ imbalance. Numerical studies of the $R$ and
$R_{\rm sep}$ dependences will be given in sections 5.3 and 5.5
\cite{But96}.

The JADE algorithm clusters soft particles regardless of how far apart in
angle they are, because only the invariant mass of the clusters
is used. This is improved 
in the $k_T$ or Durham clustering algorithm of Catani et al. \cite{Cat91},
where one defines the closeness of two particles by
\beq 
 d_{ij} = \min (E_{T_i},E_{T_j})^2 R_{ij}^2.
\label{eq11}
\eeq
$R_{ij}$ is again defined in $(\eta-\phi)$ space. For
small opening angles, $R_{ij} \ll 1$, eq.~(\ref{eq11}) reduces to
\beq
 \min (E_{T_i},E_{T_j})^2R_{ij}^2 \simeq \min (E_i,E_j)^2\Delta \theta^2
 \simeq k_T^2,
\eeq
the relative transverse momentum squared of the two particles in the jet.
Similarly, one can define a closeness to the remnant jet particles
$b$, if they are present:
\beq
 d_{ib} = E_{T_i}^2 R_{ib}^2.
\eeq
Particles are clustered into jets as long as they are closer than
\beq
 d_{\{ij,ib\}} \leq d_i^2 = E_{T_i}^2 R^2,
\eeq
where $R$ is an adjustable parameter of ${\cal O} (1)$ analogous to the
cone size parameter in the cone algorithm. Also, one chooses the 
same recombination scheme. Consequently, the $k_T$ algorithm produces
jets that are very similar to those produced by the cone algorithm with
$R = R_{\rm sep}$.

\subsection{Hard Photoproduction Cross Sections}

We have now established the links between the experimentally observed
initial and final states at HERA (electron, proton, and jets) and the
partonic subprocess calculable in perturbative QCD. Formally, we can
combine eqs.~(\ref{eq12}), (\ref{eq13}), and (\ref{eq14}) into
\bea
 && \mbox{d}\sigma_{ep}(ep\rightarrow e~+~\mbox{jets + remnants}) 
 = \label{eq15} \\
 && = \sum_{a,b}\int\limits_0^1\mbox{d}x_aF_{\gamma/e}(x_a)
 \int\limits_0^1\mbox{d}y_aF_{a/\gamma}(y_a,M_a^2)
 \int\limits_0^1\mbox{d}x_bF_{b/p}(x_b,M_b^2)
 \mbox{d}\sigma_{ab}^{(n)}(ab\rightarrow \mbox{jets}), \nonumber
\eea
where $x_a$, $y_a$, and $x_b$ denote the longitudinal momentum fractions
of the photon in the electron, the parton in the photon, and the parton
in the proton, respectively. From now on, we will use the variable $x_a$
as the variable for the {\em parton in the electron} with the consequence
that the incoming partons have momenta $p_a=x_ak$ and $p_b=x_bp$ and
eq.~(\ref{eq15}) becomes
\bea
 && \mbox{d}\sigma_{ep}(ep\rightarrow e~+~\mbox{jets + remnants}) = \\
 && = \sum_{a,b}\int\limits_0^1\mbox{d}x_aF_{\gamma/e}(\frac{x_a}
 {y_a})
 \int\limits_{x_a}^1\frac{\mbox{d}y_a}{y_a}F_{a/\gamma}(y_a,M_a^2)
 \int\limits_0^1\mbox{d}x_bF_{b/p}(x_b,M_b^2)
 \mbox{d}\sigma_{ab}^{(n)}(ab\rightarrow \mbox{jets}). \nonumber
\eea

Next, one has to fix the kinematics for the photoproduction of jets.
All particles are considered to be massless.
In the HERA laboratory system, the positive $z$-axis is taken along
the proton direction such that $k = E_e (1,0,0,-1)$ and $p = E_p (1,0,0,1)$.
For the hard jets, we choose the decomposition of four-momenta into transverse
energies, rapidities, and azimuthal angles $p_i = E_{T_i} (\cosh\eta_i,
\cos\phi_i,\sin\phi_i,\sinh\eta_i)$. The boost from the HERA laboratory system
into the $ep$ center-of-mass system is then simply mediated by a shift in
rapidity $\eta_{\rm boost} = 1/2\ln(E_e/E_p)$. Through energy and momentum
conservation $p_a + p_b = \sum_i p_i$, the final partons are related
kinematically to the invariant scaling variables $x_a$ and $x_b$
\bea
 x_a &=& \frac{1}{2E_e}\sum_i  E_{T_i}e^{-\eta_i} , \label{eq22}\\
 x_b &=& \frac{1}{2E_p}\sum_i  E_{T_i}e^{ \eta_i} . \label{eq23}
\eea

The cross section for the production of an $n$-parton final state from two
initial partons $a$, $b$,
\bea
  \mbox{d}\sigma_{ab}^{(n)} (ab \rightarrow \mbox{jets})
  & = &
  \frac{1}{2s} \overline{|{\cal M}|^2}
  \mbox{dPS}^{(n)},
\eea
depends on the flux factor $1/(2s)$, where $s = x_ax_bs_H$ is the
partonic center-of-mass energy, the $n$-particle phase space
\bea
  \mbox{dPS}^{(n)} & = & \int
  (2\pi )^d \prod_{i=1}^{n} \frac{\mbox{d}^dp_i \delta (p_i^2)}
  {(2\pi )^{d-1}} \delta^d \left( p_a+p_b-\sum_{j=1}^n p_j \right),
  \eea
where $d=4-2\varepsilon$ is the space-time dimension,
and on the squared matrix element $\overline{|{\cal M}|^2}$. The latter is
averaged over initial and summed over final spin and color states.
Since we study almost-real photons with $Q^2\simeq 0$ and treat quarks
and gluons as massless, only transverse polarizations are possible giving
spin average factors of $1/2$ for photons and quarks and
$1/(2(1-\varepsilon))$ for gluons in dimensional regularization.
The quarks form the fundamental representation of the SU(3) color symmetry
group and can therefore carry $N_C = 3$ different colors. The gluons belong
to the adjoint representation of dimension $2N_CC_F=N_C^2-1=8$. This is the
defining equation for
the so-called color factor $C_F$. We therefore average the initial colors
with factors of $1/N_C$ for quarks and $1/(2N_CC_F)$ for gluons.

In leading order QCD, the phase space for two partons $\mbox{dPS}^{(2)}$
in the final state
and the Born matrix elements $T$ have to be
computed as displayed in the first line of table \ref{tab1}.
In next-to-leading order QCD, new contributions arise. Whereas the virtual
corrections $V$ also have a two-particle phase space (second line in table
\ref{tab1}), the real corrections of the final state $F$, of the photon
initial state $I$, and of the proton initial state $J$ require the
inclusion of a third outgoing parton into the phase space $\mbox{dPS}^{(3)}$
(lines three to five in table \ref{tab1}).
We will distinguish between direct and resolved photoproduction in the
subsequent chapters and treat the two contributions in a completely parallel
way.
\begin{table}[htbp]
\begin{center}
\begin{tabular}{|c|c|c|c|}
\hline
 Order &
 Phase Space $\mbox{dPS}^{(n)}$ &
 Direct Matrix Element $\overline{|{\cal M}|^2}$ &
 Resolved Matrix Element $\overline{|{\cal M}|^2}$ \\
\hline
\hline
 LO & & $T_{\gamma b\rightarrow 12}$, sect.~3.2 &
 $T_{ab\rightarrow 12}$, sect.~3.3 \\
\cline{1-1}\cline{3-4}
 & \raisebox{1.5ex}[-1.5ex]{$\mbox{dPS}^{(2)}$, sect.~3.1} &
 $V_{\gamma b\rightarrow 12}$, sect.~4.1.1 & $V_{ab\rightarrow 12}$,
 sect.~4.1.2 \\
\cline{2-4}
 & $\mbox{dPS}^{(3)}$, sect.~4.2.1 & $F_{\gamma b\rightarrow 123}$,
 sect.~4.2.2 & $F_{ab\rightarrow 123}$, sect.~4.2.3 \\
\cline{2-4}
 \raisebox{1.5ex}[-1.5ex]{NLO}
 & & $I_{\gamma b\rightarrow 123}$, sect.~4.2.5 & $I_{ab\rightarrow 123}$,
 sect.~4.2.7 \\
\cline{3-4}
 & \raisebox{1.5ex}[-1.5ex]{$\mbox{dPS}^{(3)}$, sect.~4.2.4} &
 $J_{\gamma b\rightarrow 123}$, sect.~4.2.6 & $J_{ab\rightarrow 123}$,
 sect.~4.2.8 \\
\hline
\end{tabular}
\end{center}
\caption[Overview of Phase Space Parametrizations and Matrix Elements]
        {\label{tab1}{\it Summary of phase space parametrizations and matrix
         elements needed for this NLO calculation of direct and resolved
         photoproduction.}}
\end{table}

The two final state partons produced in a LO process have to balance their
transverse energies
$E_{T_1}=E_{T_2}=E_T$, so that relations (\ref{eq22}) and (\ref{eq23}) simplify
to
\bea
 x_a &=& \frac{E_T}{2E_e} \lr e^{-\eta_1} + e^{-\eta_2} \rr, \label{eq16}\\
 x_b &=& \frac{E_T}{2E_p} \lr e^{ \eta_1} + e^{ \eta_2} \rr. \label{eq17}
\eea
The rapidity $\eta_2$ of the second jet is kinematically fixed by $E_T$,
$\eta_1$, and $x_a$ through the relation
\beq
  \eta_2 = -\ln\lr\frac{2x_aE_e}{E_T}-e^{-\eta_1}\rr.
\label{eq18}
\eeq
In the HERA experiments, $x_a$ is restricted to a fixed interval
$x_{a,\min} \leq x_a \leq x_{a,\max} < 1$. We shall disregard this
constraint and allow $x_a$ to vary in the kinematically allowed range
$x_{a,\min} \leq x_a \leq 1$, where
\beq
  x_{a,\min} = \frac{E_pE_Te^{-\eta_1}}{2E_eE_p-E_eE_Te^{\eta_1}}
\eeq
except when we compare to experimental data. There we shall include the
correct limits on $x_a$ dictated by the experimental analysis.
From eqs.~(\ref{eq16}) and (\ref{eq17}), we can express $x_b$ as a function of
$E_T$, $\eta_1$, and $x_a$:
\beq
  x_b = \frac{x_aE_eE_Te^{\eta_1}}{2x_aE_eE_p-E_pE_Te^{-\eta_1}}.
  \label{eq20}
\eeq
The inclusive two-jet cross section for $ep\rightarrow e~+~\mbox{jet}_1~+~
\mbox{jet}_2~+~X$ is obtained from
\beq
  \frac{\mbox{d}^3\sigma}{\mbox{d}E_T^2\mbox{d}\eta_1\mbox{d}\eta_2}
  = \sum_{a,b} x_a F_{a/e}(x_a,M_a^2) x_b F_{b/p}(x_b,M_b^2)
  \frac{\mbox{d}\sigma}{\mbox{d}t}(ab \rightarrow \mbox{jets}),
  \label{eq19}
\eeq
where 
\beq
 F_{a/e}(x_a,M_a^2) = \int\limits_{x_a}^1\frac{\mbox{d}y_a}{y_a}
 F_{a/\gamma}(y_a,M_a^2)F_{\gamma/e}\lr\frac{x_a}{y_a}\rr
\eeq
defines the parton content in the electron.
d$\sigma$/d$t$ stands for the differential cross section of the partonic
subprocess $ab \rightarrow \mbox{jets}$. The invariants of this process
are $s = (p_a+p_b)^2,~t = (p_a-p_1)^2$, and $u = (p_a-p_2)^2$.
They can be expressed by the final state variables $E_T$, $\eta_1$, and
$\eta_2$ and the initial state momentum fractions $x_a$ and $x_b$:
\bea
  s &=& 4 x_a x_b E_e E_p, \\
  t &=& -2 x_a E_e E_T e^{\eta_1} =  -2 x_b E_p E_T e^{-\eta_2}, \\
  u &=& -2 x_a E_e E_T e^{\eta_2} =  -2 x_b E_p E_T e^{-\eta_1}.
\eea

For the inclusive
one-jet cross section, we must integrate over one of the rapidities
in (\ref{eq19}). We integrate over $\eta_2$ and transform to the variable
$x_a$ using (\ref{eq16}). The result is the cross section for
$ep \rightarrow e+\mbox{jet}+X$, which depends on $E_T$ and $\eta$:
\beq
  \frac{\mbox{d}^2\sigma}{\mbox{d}E_T\mbox{d}\eta}
  = \sum_{a,b} \int_{x_{a,\min}}^1 \mbox{d}x_a x_a 
  F_{a/e}(x_a,M_a^2) x_b F_{b/p}(x_b,M_b^2)
  \frac{4E_eE_T}{2x_aE_e-E_Te^{-\eta}}
  \frac{\mbox{d}\sigma}{\mbox{d}t}(ab\rightarrow \mbox{jet}).
  \label{eq24}
\eeq
Here, $x_b$ is given by (\ref{eq20}) with $\eta_1 = \eta$.
\setcounter{equation}{0}

\section{Leading Order Cross Sections}

In this section, we consider the leading order contributions to the
photoproduction cross section d$\sigma$/d$t$. The leading order direct hard
scattering process is of ${\cal O} (\alpha\alpha_s)$, and the leading order
resolved hard scattering process is
of ${\cal O} (\alpha_s^2)$. Since the latter has to be multiplied with the
photon structure function of ${\cal O} (\alpha/\alpha_s)$, both contributions
are of the same order ${\cal O} (\alpha\alpha_s)$. To this
order, it is necessary to calculate the phase space of two-particle
final states, the tree-level matrix elements for direct photons
$\gamma b\rightarrow 12$, and those for resolved photons $ab\rightarrow 12$.
The two particles produced in the hard scattering then correspond directly
to the two jets that can at most be observed. Either particle can be in
jet one or jet two, so that the cross sections d$\sigma$/d$t$ have to be
considered in their given form and with $(t\leftrightarrow u)$ to give
the complete cross section d$\sigma$/d$t(s,t,u)$+d$\sigma$/d$t(s,u,t)$.
Furthermore, one has to add symmetry factors of $1/n!$ for $n$ identical
particles in the final state. As we will later
go from four to $d=4-2\varepsilon$ dimensions to regularize the
singularities showing up in next-to-leading order, we already give the results
in this section in $d$ dimensions. If one is only interested in leading order
results, one can always set $\varepsilon$ to zero in this section.

Section 3.1 contains the phase space for $2\rightarrow 2$ scattering in
$d$ dimensions. It is calculated in the center-of-mass system of the
incoming or equivalently outgoing particles.
The matrix elements for direct photons are presented in section 3.2. Only
one master diagram for the so-called QCD Compton scattering process $\yqtgq$
contributes. The gluon initiated process is obtained by crossing.
For resolved photons, we have four parton-parton master diagrams. The
corresponding matrix elements are given in section 3.3. In addition, we
show a table from which the matrix elements for all other processes can be
deduced. Section 3.4 contains the direct $\gamma\gamma$ Born matrix element.

\subsection{Phase Space for Two-Particle Final States}

\begin{figure}[htbp]
\begin{center}
\begin{picture}(18000,22500)
\thicklines
\put(9000,10500){\vector(0,1){9000}}
\put(9000,10500){\vector(0,-1){9000}}
\put(9000,10500){\vector(1,1){6000}}
\put(9000,10500){\vector(-1,-1){6000}}
\thinlines
\put(9000,16500){\line(1,0){6000}}
\put(9000,10500){\line(1,0){7500}}
\put(15000,16500){\line(0,-1){6000}}
\put(9750,20250){$p_a$}
\put(9150,12000){$\theta$}
\put(9000,12000){\oval(3000,3000)[tr]}
\put(9750,2250){$p_b$}
\put(15750,17250){$p_1$}
\put(1500,4750){$p_2$}
\end{picture}
\end{center}
\caption[Center-of-Mass System for $2\rightarrow 2$ Scattering]
        {\label{fig15}{\it Center-of-mass system for the scattering of two
         initial into two final partons.}}
\end{figure}
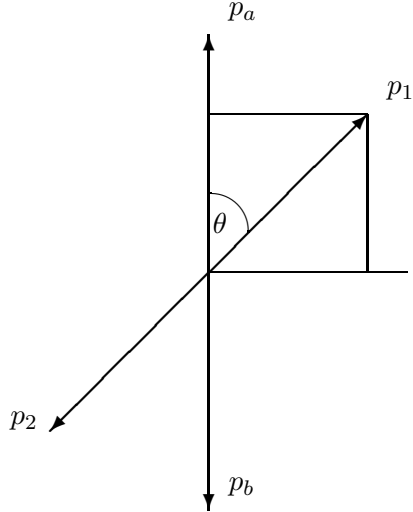

We give here a brief sketch of the phase space calculation for the scattering
of two initial into two final particles in $d=4-2\varepsilon$ dimensions. We
start from the general expression \cite{Byc73}
\beq
  \mbox{dPS}^{(2)}  = \int
  (2\pi )^d \prod_{i=1}^{2} \frac{\mbox{d}^dp_i \delta (p_i^2)}
  {(2\pi )^{d-1}} \delta^d \left( p_a+p_b-\sum_{j=1}^2 p_j \right).
\eeq
The dijet and single-jet cross sections in eqs.~(\ref{eq19}) and (\ref{eq24})
require partonic cross sections differential in the Mandelstam variable $t$.
We therefore insert an additional $\delta$-function with respect to $t$
\beq
  \frac{\mbox{dPS}^{(2)}}{\mbox{d}t}  =  \int
  (2\pi )^d \prod_{i=1}^{2} \frac{\mbox{d}^dp_i \delta (p_i^2)}
  {(2\pi )^{d-1}} \delta^d \left( p_a+p_b-\sum_{j=1}^2 p_j \right)
  \delta (t+2p_ap_1),
\eeq
before we integrate over the $d$-dimensional $\delta$-function leading to
\beq
  \frac{\mbox{dPS}^{(2)}}{\mbox{d}t}  =  \int
  \frac{\mbox{d}^dp_1\delta (p_1^2)}{(2\pi)^{d-2}} \delta
  ((p_a+p_b-p_1)^2) \delta (t+2p_ap_1).
\eeq
Next, we choose the center-of-mass system of the incoming partons $a$, $b$ as
shown in figure \ref{fig15},
where the outgoing partons are of the same energy and can be described by a
single polar angle $\theta$ between $p_1$ and $p_a$. We integrate over
the azimuthal angle $\phi$
\beq
  \frac{\mbox{dPS}^{(2)}}{\mbox{d}t}  =  \int
  \frac{\mbox{d}E_1 E_1^{d-3} \mbox{d}\cos\theta
  (1-\cos^2\theta )^{\frac{d-4}{2}}}
  {2^{d-2}\pi^{\frac{d-2}{2}}\Gamma \lr \frac{d-2}{2} \rr }
  \delta (s-2\sqrt{s}E_1 )\delta (t+\sqrt{s}E_1
  (1-\cos\theta ))
\eeq
and over the remaining $\delta$-functions. The final result is
\cite{Aur87,Arn89,Gra90}
\beq
  \frac{\mbox{dPS}^{(2)}}{\mbox{d}t}
  = \frac{1}{\Gamma(1-\varepsilon)}
  \left( \frac{4\pi s}{tu} \right) ^\varepsilon
  \frac{1}{8\pi s}.
\eeq

\subsection{Born Matrix Elements for Direct Photons}

\begin{figure}[htbp]
\begin{center}
\begin{picture}(11000,6000)
\drawline\photon[\E\REG](1000,5000)[4]
\drawline\fermion[\E\REG](1000,1000)[4000]
\drawarrow[\E\ATTIP](3000,1000)
\put(5500,3000){\oval(1000,5000)}
\drawline\gluon[\E\REG](6000,5000)[4]
\drawline\fermion[\E\REG](6000,1000)[4000]
\drawarrow[\E\ATTIP](8000,1000)
\put(0,1000){$q$}
\put(0,5000){$\gamma$}
\put(10500,1000){$q$}
\put(10500,5000){$g$}
\end{picture}
\end{center}
\caption[Generic $2\rightarrow 2$ Feynman Diagram for Direct Photoproduction]
        {\label{fig1}
         $2\rightarrow 2$ {\it Feynman diagram for direct photoproduction.}}
\end{figure}
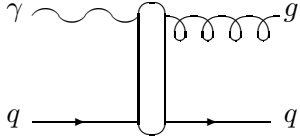

\begin{figure}[htbp]
\begin{center}
\begin{picture}(21000,6000)
\drawline\photon[\E\REG](1000,5000)[4]
\drawline\fermion[\E\REG](1000,1000)[4000]
\drawarrow[\E\ATTIP](3000,1000)
\put(5000,1000){\circle*{300}}
\drawline\fermion[\N\REG](5000,1000)[4000]
\drawarrow[\N\ATTIP](5000,3000)
\put(5000,5000){\circle*{300}}
\drawline\gluon[\E\REG](5000,1000)[4]
\drawline\fermion[\E\REG](5000,5000)[4000]
\drawarrow[\E\ATTIP](7000,5000)
\put(0,1000){$q$}
\put(0,5000){$\gamma$}
\put(9500,1000){$g$}
\put(9500,5000){$q$}
\drawline\photon[\SE\REG](12000,5000)[3]
\drawline\fermion[\NE\REG](12000,1000)[2800]
\drawarrow[\NE\ATTIP](13000,2000)
\put(14000,3000){\circle*{300}}
\drawline\fermion[\E\REG](14000,3000)[4000]
\drawarrow[\E\ATTIP](16000,3000)
\put(18000,3000){\circle*{300}}
\drawline\gluon[\NE\REG](18000,3000)[1]
\drawline\fermion[\SE\REG](18000,3000)[2800]
\drawarrow[\SE\ATTIP](19000,2000)
\put(11000,1000){$q$}
\put(11000,5000){$\gamma$}
\put(20500,1000){$q$}
\put(20500,5000){$g$}
\end{picture}
\end{center}
\caption[Born Diagrams for Direct Photoproduction]
        {\label{fig3}
         {\it Born diagrams for direct photoproduction.}}
\end{figure}
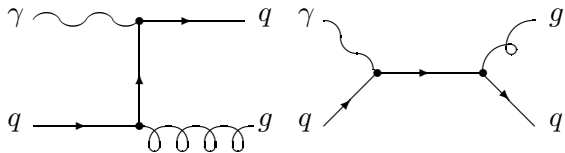

For direct photoproduction of two partons, there is only one generic Feynman
diagram $\yqtgq$ as displayed in figure \ref{fig1}.
Unlike deep inelastic scattering, the leading order process already includes
the production of a hard gluon, which balances the transverse momentum of the
scattered quark. This can either happen in the initial state (left diagram
in figure \ref{fig3}) or in the final state (right diagram in figure
\ref{fig3}) of this so-called ``QCD Compton Scattering''. 
Both diagrams have to be added and squared to give the full matrix element
squared. In $d$ dimensions, the result is
\beq
 |{\cal M}|^2_{\yqtgq}(s,t,u) = e^2e_q^2g^2\mu^{4\varepsilon}T_{\yqtgq}(s,t,u),
 \label{eq25}
\eeq
where the renormalization scale $\mu$ keeps the couplings dimensionless and
where
\beq
 T_{\yqtgq}(s,t,u) = 8N_CC_F(1-\varepsilon)\le(1-\varepsilon) \lr-\frac{u}{s}-
 \frac{s}{u}\rr+2\varepsilon\re.
\eeq
Note that the DIS interference term of the two diagrams is missing as it is
proportional to $Q^2\simeq 0$.
As discussed in section 2.2, the proton not only consists of quarks but also
of gluons giving rise to a ``Boson Gluon Fusion'' process $\ygtqb$ not shown
here. The corresponding matrix element is obtained from eq.~(\ref{eq25}) by
simply crossing the initial quark with the final gluon, or
$(s\leftrightarrow t)$, and multiplying by $(-1)$ for crossing a fermion line.
Similarly, the contribution for incoming anti-quarks $\ybtgb$ can be
calculated by crossing $(s\leftrightarrow u)$.
The LO direct matrix elements have been known for quite some time
\cite{Aur87,Bae89a,Bod92} and are summarized in table \ref{tab2}.
\begin{table}[htbp]
\begin{center}
\begin{tabular}{|c|c|}
\hline
 Process &
 Matrix Element $\overline{|{\cal M}|^2}$ \\
\hline
\hline
 $\yqtgq$ & $[|{\cal M}|^2_{\yqtgq}(s,t,u)]/[4N_C]$ \\
\hline
 $\ybtgb$ & $[|{\cal M}|^2_{\yqtgq}(u,t,s)]/[4N_C]$ \\
\hline
 $\ygtqb$ & $-[|{\cal M}|^2_{\yqtgq}(t,s,u)]/[8(1-\varepsilon)N_CC_F]$ \\
\hline
\end{tabular}
\end{center}
\caption[Squared $2\rightarrow 2$ Matrix Elements for Direct Photoproduction]
        {\label{tab2}{\it Summary of $2\rightarrow 2$ squared matrix elements
         for direct photoproduction.}}
\end{table}

\subsection{Born Matrix Elements for Resolved Photons}

\begin{figure}[htbp]
\begin{center}
\begin{picture}(23000,15000)
\drawline\fermion[\E\REG](1000,14000)[4000]
\drawarrow[\E\ATTIP](3000,14000)
\drawline\fermion[\E\REG](1000,10000)[4000]
\drawarrow[\E\ATTIP](3000,10000)
\put(5500,12000){\oval(1000,5000)}
\drawline\fermion[\E\REG](6000,14000)[4000]
\drawarrow[\E\ATTIP](8000,14000)
\drawline\fermion[\E\REG](6000,10000)[4000]
\drawarrow[\E\ATTIP](8000,10000)
\put(0,10000){$q'$}
\put(0,14000){$q$}
\put(10500,10000){$q'$}
\put(10500,14000){$q$}
\put(5000,8000){a)}
\drawline\fermion[\E\REG](13000,14000)[4000]
\drawarrow[\E\ATTIP](15000,14000)
\drawline\fermion[\E\REG](13000,10000)[4000]
\drawarrow[\E\ATTIP](15000,10000)
\put(17500,12000){\oval(1000,5000)}
\drawline\fermion[\E\REG](18000,14000)[4000]
\drawarrow[\E\ATTIP](20000,14000)
\drawline\fermion[\E\REG](18000,10000)[4000]
\drawarrow[\E\ATTIP](20000,10000)
\put(12000,10000){$q$}
\put(12000,14000){$q$}
\put(22500,10000){$q$}
\put(22500,14000){$q$}
\put(17000,8000){b)}
\drawline\fermion[\E\REG](1000,6000)[4000]
\drawarrow[\E\ATTIP](3000,6000)
\drawline\fermion[\E\REG](1000,2000)[4000]
\drawarrow[\W\ATBASE](3000,2000)
\put(5500,4000){\oval(1000,5000)}
\drawline\gluon[\E\REG](6000,6000)[4]
\drawline\gluon[\E\REG](6000,2000)[4]
\put(0,2000){$\overline{q}$}
\put(0,6000){$q$}
\put(10500,2000){$g$}
\put(10500,6000){$g$}
\put(5000,0){c)}
\drawline\gluon[\E\REG](12700,6000)[4]
\drawline\gluon[\E\REG](12700,2000)[4]
\put(17500,4000){\oval(1000,5000)}
\drawline\gluon[\E\REG](18000,6000)[4]
\drawline\gluon[\E\REG](18000,2000)[4]
\put(12000,2000){$g$}
\put(12000,6000){$g$}
\put(22500,2000){$g$}
\put(22500,6000){$g$}
\put(17000,0){d)}
\end{picture}
\end{center}
\caption[Generic $2\rightarrow 2$ Feynman Diagrams for Resolved
         Photoproduction]
        {\label{fig2}
         $2\rightarrow 2$ {\it Feynman diagrams for resolved photoproduction.}}
\end{figure}
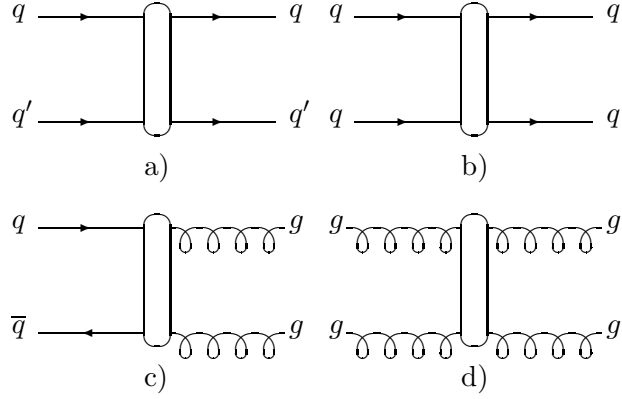

For resolved photoproduction of two partons, we have to calculate the four
generic parton-parton scattering processes $\qptqp$, $\qqtqq$, $\qbtgg$, and
$\ggtgg$. The corresponding Feynman diagrams are displayed in figure
\ref{fig2}.
The first process for the scattering of two non-identical quarks $q$ and $q'$
has already been calculated in 1978 by Cutler and Sivers \cite{Cut78}, shortly
prior to the other processes \cite{Com77}, and was compared to inclusive
jet and hadron production at ISR energies.

\begin{figure}[htbp]
\begin{center}
\begin{picture}(33000,31000)
\drawline\fermion[\E\REG](12000,26000)[4000]
\drawarrow[\E\ATTIP](14000,26000)
\drawline\fermion[\E\REG](12000,30000)[4000]
\drawarrow[\E\ATTIP](14000,30000)
\put(16000,26000){\circle*{300}}
\drawline\gluon[\S\REG](16000,30150)[4]
\put(16000,30000){\circle*{300}}
\drawline\fermion[\E\REG](16000,26000)[4000]
\drawarrow[\E\ATTIP](18000,26000)
\drawline\fermion[\E\REG](16000,30000)[4000]
\drawarrow[\E\ATTIP](18000,30000)
\put(11000,26000){$q'$}
\put(11000,30000){$q$}
\put(20500,26000){$q'$}
\put(20500,30000){$q$}
\put(15500,24000){a)}
\drawline\fermion[\E\REG](6000,22000)[4000]
\drawarrow[\E\ATTIP](8000,22000)
\drawline\fermion[\E\REG](6000,18000)[4000]
\drawarrow[\E\ATTIP](8000,18000)
\put(10000,22000){\circle*{300}}
\drawline\gluon[\S\REG](10000,22150)[4]
\put(10000,18000){\circle*{300}}
\drawline\fermion[\E\REG](10000,22000)[4000]
\drawarrow[\E\ATTIP](12000,22000)
\drawline\fermion[\E\REG](10000,18000)[4000]
\drawarrow[\E\ATTIP](12000,18000)
\put(5000,22000){$q$}
\put(5000,18000){$q$}
\put(14500,22000){$q$}
\put(14500,18000){$q$}
\put(15500,16000){b)}
\drawline\fermion[\E\REG](18000,22000)[4000]
\drawarrow[\E\ATTIP](20000,22000)
\drawline\fermion[\E\REG](18000,18000)[4000]
\drawarrow[\E\ATTIP](20000,18000)
\put(22000,22000){\circle*{300}}
\drawline\gluon[\S\REG](22000,22150)[4]
\put(22000,18000){\circle*{300}}
\drawline\fermion[\SE\REG](22000,22000)[5600]
\drawarrow[\SE\ATTIP](25000,19000)
\drawline\fermion[\NE\REG](22000,18000)[5600]
\drawarrow[\NE\ATTIP](25000,21000)
\put(17000,22000){$q$}
\put(17000,18000){$q$}
\put(26500,22000){$q$}
\put(26500,18000){$q$}
\drawline\fermion[\E\REG](1000,14000)[4000]
\drawarrow[\E\ATTIP](3000,14000)
\drawline\fermion[\E\REG](1000,10000)[4000]
\drawarrow[\W\ATBASE](3000,10000)
\put(5000,14000){\circle*{300}}
\drawline\fermion[\S\REG](5000,14000)[4000]
\drawarrow[\S\ATBASE](5000,12000)
\put(5000,10000){\circle*{300}}
\drawline\gluon[\E\REG](5000,14000)[4]
\drawline\gluon[\E\REG](5000,10000)[4]
\put(0,14000){$q$}
\put(0,10000){$\overline{q}$}
\put(9500,14000){$g$}
\put(9500,10000){$g$}
\put(15500,8000){c)}
\drawline\fermion[\SE\REG](12000,14000)[2800]
\drawarrow[\SE\ATTIP](13000,13000)
\drawline\fermion[\NE\REG](12000,10000)[2800]
\drawarrow[\SW\ATBASE](13000,11000)
\put(14000,12000){\circle*{300}}
\drawline\gluon[\E\REG](13850,12000)[4]
\drawarrow[\S\ATBASE](5000,12000)
\put(18000,12000){\circle*{300}}
\drawline\gluon[\NE\REG](18000,12000)[1]
\drawline\gluon[\SE\REG](18000,12000)[1]
\put(11000,14000){$q$}
\put(11000,10000){$\overline{q}$}
\put(20500,14000){$g$}
\put(20500,10000){$g$}
\drawline\fermion[\E\REG](24000,14000)[4000]
\drawarrow[\E\ATTIP](26000,14000)
\drawline\fermion[\E\REG](24000,10000)[4000]
\drawarrow[\W\ATBASE](26000,10000)
\put(28000,14000){\circle*{300}}
\drawline\fermion[\S\REG](28000,14000)[4000]
\drawarrow[\S\ATBASE](28000,12000)
\put(28000,10000){\circle*{300}}
\drawline\gluon[\SE\REG](28000,14000)[3]
\drawline\gluon[\NE\REG](28000,10000)[3]
\put(23000,14000){$q$}
\put(23000,10000){$\overline{q}$}
\put(32500,14000){$g$}
\put(32500,10000){$g$}
\drawline\gluon[\E\REG](775,6000)[3]
\drawline\gluon[\E\REG](775,2000)[3]
\put(4000,6000){\circle*{300}}
\drawline\gluon[\S\REG](4000,6150)[4]
\put(4000,2000){\circle*{300}}
\drawline\gluon[\E\REG](4000,6000)[3]
\drawline\gluon[\E\REG](4000,2000)[3]
\put(0,6000){$g$}
\put(0,2000){$g$}
\put(7500,6000){$g$}
\put(7500,2000){$g$}
\put(15500,0){d)}
\drawline\gluon[\SE\REG](10000,6000)[1]
\drawline\gluon[\NE\REG](10000,2000)[1]
\put(12000,4000){\circle*{300}}
\drawline\gluon[\E\REG](11925,4000)[1]
\put(13000,4000){\circle*{300}}
\drawline\gluon[\NE\REG](13000,4000)[1]
\drawline\gluon[\SE\REG](13000,4000)[1]
\put(9000,6000){$g$}
\put(9000,2000){$g$}
\put(15500,6000){$g$}
\put(15500,2000){$g$}
\drawline\gluon[\E\REG](17775,6000)[3]
\drawline\gluon[\E\REG](17775,2000)[3]
\put(21000,6000){\circle*{300}}
\drawline\gluon[\S\REG](21000,6150)[4]
\put(21000,2000){\circle*{300}}
\drawline\gluon[\SE\REG](21000,6000)[3]
\drawline\gluon[\NE\REG](21000,2000)[3]
\put(17000,6000){$g$}
\put(17000,2000){$g$}
\put(25500,6000){$g$}
\put(25500,2000){$g$}
\drawline\gluon[\NW\REG](30000,4000)[1]
\drawline\gluon[\SW\REG](30000,4000)[1]
\put(30000,4000){\circle*{300}}
\drawline\gluon[\NE\REG](30000,4000)[1]
\drawline\gluon[\SE\REG](30000,4000)[1]
\put(27000,6000){$g$}
\put(27000,2000){$g$}
\put(32500,6000){$g$}
\put(32500,2000){$g$}
\end{picture}
\end{center}
\caption[Born Diagrams for Resolved Photoproduction]
        {\label{fig4}
         {\it Born diagrams for resolved photoproduction.}}
\end{figure}
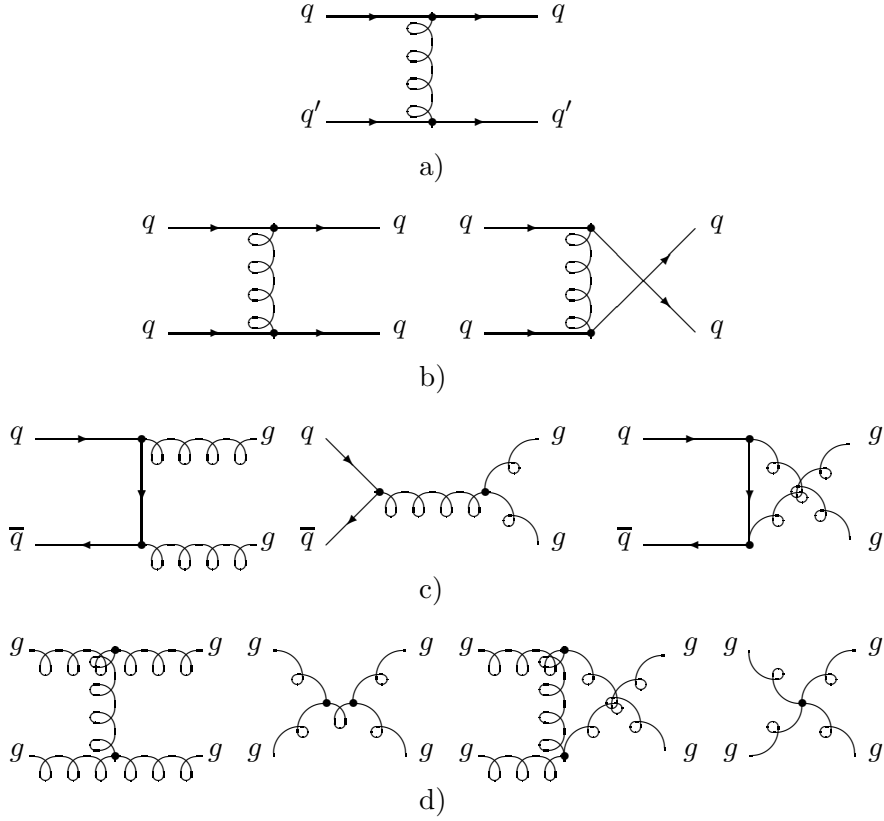
Figure \ref{fig4}a) shows that for $\qptqp$ only the one-gluon exchange between
the different quarks in the $t$-channel contributes. The matrix element
squared is given by
\beq 
 |{\cal M}|^2_{\qptqp}(s,t,u) = g^4\mu^{4\varepsilon}T_{\qptqp}(s,t,u)
 \label{eq46}
\eeq
with
\beq
 T_{\qptqp}(s,t,u) = 4N_CC_F\lr\frac{s^2+u^2}{t^2}-\varepsilon\rr.
\eeq
We will split off the coupling constants for the other Born matrix
elements $|{\cal M}|^2$ as in eq.~(\ref{eq46}).

Equal quark flavors cannot be distinguished in the final state so that both
diagrams in figure \ref{fig4}b) have to be added. In addition to the
squares of the individual diagrams already present in the case of different
quark flavors, the interference term
\beq
 T_{\qqtqq}(s,t,u) = -8C_F(1-\varepsilon)\lr\frac{s^2}{ut}+\varepsilon\rr
\eeq
also contributes to the process $\qqtqq$.

Final gluons are also undistinguishable and can even couple in a non-abelian
way as shown in figure \ref{fig4}c). The resulting matrix element for
the process $\qbtgg$ is
\beq
 T_{\qbtgg}(s,t,u) = 4C_F(1-\varepsilon)\lr\frac{2N_CC_F}{ut}-\frac{2N_C^2}
 {s^2}\rr(t^2+u^2-\varepsilon s^2).
\eeq

The diagrams for the process $\ggtgg$ in figure \ref{fig4}d) give
\beq
 T_{\ggtgg}(s,t,u) = 32N_C^3C_F(1-\varepsilon)^2\lr 3-\frac{ut}{s^2}-
 \frac{us}{t^2}-\frac{st}{u^2}\rr.
\eeq

All other diagrams can be obtained from the above by simple crossing
relations.
The complete result in $d$ dimensions can be found in \cite{Ell86} and is
summarized in table \ref{tab3}.
\begin{table}[htbp]
\begin{center}
\begin{tabular}{|c|c|}
\hline
 Process &
 Matrix Element $\overline{|{\cal M}|^2}$ \\
\hline
\hline
 $\qptqp$ & $[|{\cal M}|^2_{\qptqp}(s,t,u)]/[4N_C^2]$ \\
\hline
 $\qrtqr$ & $[|{\cal M}|^2_{\qptqp}(u,t,s)]/[4N_C^2]$ \\
\hline
 $\qbtrp$ & $[|{\cal M}|^2_{\qptqp}(t,s,u)]/[4N_C^2]$ \\
\hline
 $\qqtqq$ & $[|{\cal M}|^2_{\qptqp}(s,t,u)+|{\cal M}|^2_{\qptqp}(s,u,t)+
              |{\cal M}|^2_{\qqtqq}(s,t,u)]/[4N_C^2]/2!$ \\
\hline
 $\qbtqb$ & $[|{\cal M}|^2_{\qptqp}(u,t,s)+|{\cal M}|^2_{\qptqp}(u,s,t)+
              |{\cal M}|^2_{\qqtqq}(u,t,s)]/[4N_C^2]$ \\
\hline
 $\qbtgg$ & $[|{\cal M}|^2_{\qbtgg}(s,t,u)]/[4N_C^2]/2!$ \\
\hline
 $\qgtqg$ & $-[|{\cal M}|^2_{\qbtgg}(t,s,u)]/[8(1-\varepsilon)N_C^2C_F]$ \\
\hline
 $\bgtbg$ & $-[|{\cal M}|^2_{\qbtgg}(t,u,s)]/[8(1-\varepsilon)N_C^2C_F]$ \\
\hline
 $\ggtqb$ & $[|{\cal M}|^2_{\qbtgg}(s,t,u)]/[16(1-\varepsilon)^2N_C^2C_F^2]$
 \\
\hline
 $\ggtgg$ & $[|{\cal M}|^2_{\ggtgg}(s,t,u)]/[16(1-\varepsilon)^2N_C^2C_F^2]/2!$
 \\
\hline
\end{tabular}
\end{center}
\caption[Squared $2\rightarrow 2$ Matrix Elements for Resolved Photoproduction]
        {\label{tab3}{\it Summary of $2\rightarrow 2$ squared matrix elements
for resolved photoproduction.}}
\end{table}

\subsection{Born Matrix Element for Direct $\gamma\gamma$ Scattering}

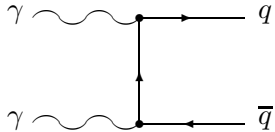
\begin{figure}[htbp]
\begin{center}
\begin{picture}(10500,6000)
\drawline\photon[\E\REG](1000,5000)[4]
\drawline\photon[\E\REG](1000,1000)[4]
\put(5000,1000){\circle*{300}}
\drawline\fermion[\N\REG](5000,1000)[4000]
\drawarrow[\N\ATTIP](5000,3000)
\put(5000,5000){\circle*{300}}
\drawline\fermion[\E\REG](5000,1000)[4000]
\drawarrow[\W\ATBASE](7000,1000)
\drawline\fermion[\E\REG](5000,5000)[4000]
\drawarrow[\E\ATTIP](7000,5000)
\put(0,1000){$\gamma$}
\put(0,5000){$\gamma$}
\put(9500,1000){$\overline{q}$}
\put(9500,5000){$q$}
\end{picture}
\end{center}
\caption{\label{kkkfig1}
         {\it Born diagram for direct $\gamma\gamma$ scattering.}}
\end{figure}

Since the gluon has no electromagnetic charge, direct $\gamma\gamma$
scattering can only proceed through the process $\yytqb$ as shown in figure
\ref{kkkfig1}.
The corresponding matrix element is given by
\beq
 |{\cal M}|^2_{\yytqb}(s,t,u) = e^4e_q^4\mu^{4\varepsilon}T_{\yytqb}(s,t,u),
 \label{kkkeq1}
\eeq
where
\beq
 T_{\yytqb}(s,t,u) = 8N_C(1-\varepsilon)\le(1-\varepsilon) \lr\frac{u}{t}+
 \frac{t}{u}\rr-2\varepsilon\re.
\eeq
It can be obtained from the direct photoproduction matrix element for
$\yqtgq$ in \ref{eq25} through crossing of $s\rightarrow t$,
multiplication with (-1) for the crossing of a fermion line, and
replacing the strong coupling constant $g^2$ by $e^2e_q^2$ and the color
factor $C_F$ by 1.
\setcounter{equation}{0}

\section{Next-To-Leading Order Cross Sections}

As in any calculation of next-to-leading order cross sections, we have to
calculate two types of corrections in photoproduction: virtual and real
corrections. For the direct case, these corrections are of ${\cal O}
(\alpha\alpha_s^2)$, 
and for the resolved case, they are of ${\cal O} (\alpha_s^3)$. Throughout
this section, direct and resolved photon results will
be presented separately, but in a completely parallel way. Whereas the direct
contributions have already been published \cite{Kla93,x8}, the resolved
contributions are presented here for the first time.

The virtual corrections consist of the interference terms of one-loop graphs
with the Born graphs calculated in
section 3 and will be given in section 4.1. We can still use the same phase
space for two final partons as in the last section. The direct matrix elements
are contained in section 4.1.1, those for resolved photons in section 4.1.2.
Again, we display only the relevant master diagrams, from which all
subprocesses can be deduced through crossing. Section 4.1.3 contains the
virtual corrections for direct $\gamma\gamma$ scattering.

The real corrections are derived from the
integration of diagrams with a third parton in the final state over
regions of phase space, where this third parton causes singularities in the
matrix elements. Different methods can be employed here. We choose the phase
space slicing method with an invariant mass cut \cite{Kra84} in section 4.2.
Alternatively, the subtraction method could be used \cite{Ell81,Ell89,x6}.
We calculate the three particle phase space in two different versions for
final state singularities (section 4.2.1) and initial state singularities
(section 4.2.4). These sections are followed by the calculation of the
matrix elements for final state (sections 4.2.2 through 4.2.3) and initial
state corrections (sections 4.2.5 through 4.2.8). The real corrections
for direct $\gamma\gamma$ scattering are presented in section 4.2.9.

Finally, we demonstrate the cancellation of the infrared and collinear
singularities, that
show up in the virtual and real corrections separately, in section 4.3.
The ultraviolet singularities in the virtual corrections are removed by
counter terms in the renormalization procedure. Remaining collinear
singularities in the initial states can be absorbed into the photon and
proton structure functions.

\subsection{Virtual Corrections}

The central blobs in figures \ref{fig1} and \ref{fig2} can also contain loop
corrections to the leading order matrix elements in figures \ref{fig3} and
\ref{fig4}. Up to ${\cal O} (\alpha\alpha_s^2)$, the interference terms
between the direct Born diagrams $T$ and the direct one-loop diagrams
$V$ have to be taken into account according to
\bea
  |{\cal M}|^2({\cal O}(\alpha\alpha_s^2)) &=&
  |\sqrt{\alpha\alpha_s T}+\sqrt{\alpha\alpha_s^3 V}|^2 \nonumber \\
  &=&
  \alpha\alpha_sT+2\alpha\alpha_s^2\sqrt{V^\ast T}+{\cal O}
  (\alpha\alpha_s^3).
\eea
Similarly, the interference terms between the resolved Born diagrams
$T$ and the resolved one-loop diagrams $V$ have to be taken into
account up to ${\cal O} (\alpha_s^3)$.

The inner loop momenta are unconstrained from the outer $2\rightarrow 2$
scattering process. We can therefore still make use of the phase space
calculated in section 3.1, but have to integrate over the inner degrees
of freedom. At the lower and upper integration bounds, we encounter 
infrared (IR) and ultraviolet (UV) divergencies. Using the dimensional
regularization scheme of t'Hooft and Veltman \cite{tHo72}, we integrate in
$d=4-2\varepsilon$ dimensions, thus rendering the integrals finite and keeping
the theory Lorentz and gauge invariant. The IR and UV poles show up as terms
$\propto 1/\varepsilon^2$ and $1/\varepsilon$. As a consequence, the complete
higher order cross sections have to be calculated in $d$ dimensions up to
${\cal O} (\varepsilon^2)$ to ensure that no finite terms are missing.

The ultraviolet divergencies coming from infinite loop momenta can be removed
by renormalizing the fields, couplings, gauge parameters, and masses in the
Lagrangian, since QCD is a renormalizable field theory. This is done by
multiplying the divergent parameters in the Lagrangian with renormalization
constants $Z_i$ and expanding up to the required order in the coupling
constant $g$. The resulting counter terms then render the physical Green's
functions finite. In addition to the $1/\varepsilon$ poles, universal finite
contributions 
\beq
 \frac{1}{\varepsilon}-\gamma_E+\ln (4\pi)
 \label{eq26}
\eeq
are included into the counter terms
according to the modified minimal subtraction ($\overline{\mbox{MS}}$)
scheme \cite{tHo73}, which we employ here.
The coupling constant $g$ is kept dimensionless by multiplying it with
the renormalization scale $\mu$
\beq
 g \rightarrow g\mu^{\varepsilon}.
 \label{eq27}
\eeq
If eq.~(\ref{eq27}) is expanded in powers of $\varepsilon$ and combined with
single poles of the type in eq.~(\ref{eq26}), the renormalization procedure
will
produce an explicit logarithmic dependence on the renormalization scale
$\mu$
\beq
  \frac{1}{\varepsilon} \left( \frac{4\pi\mu^2}{s} \right) ^\varepsilon
  \doteq \frac{1}{\varepsilon} + \ln \frac{4\pi\mu^2}{s},
\eeq
that cancels the first term of the expansion of the running coupling
constant
\beq
  \alpha_s(\mu^2) = \frac{12\pi}{(33-2N_f)\ln \frac{\mu^2}
  {\Lambda^2}} \left( 1-\frac{6(153-19N_f)}{(33-2N_f)^2}
  \frac {\ln (\ln \frac{\mu^2}{\Lambda^2} )}%
{\ln \frac{\mu^2}{\Lambda^2}} \right) + \cdots.
\eeq
$\alpha_s$ is given here in two-loop approximation as appropriate in NLO QCD,
and $N_f$ is the number of flavors in the $q\bar{q}$ loops. 
We choose $\mu = {\cal O} (E_T)$ as for the factorization scales.

Infrared divergencies arise from small loop momenta in the virtual corrections.
They have to cancel eventually against the divergencies from the emission of
real soft and collinear partons according to the Kinoshita-Lee-Nauenberg
theorem \cite{Kin62}. Finally, one obtains finite physical results
in the limit $d\rightarrow 4$. 

\subsubsection{Virtual Corrections for Direct Photons}

\begin{figure}[htbp]
\begin{center}
\begin{picture}(33000,23000)
\drawline\photon[\E\REG](1000,22000)[4]
\drawline\fermion[\E\REG](1000,18000)[4000]
\drawarrow[\E\ATTIP](3000,18000)
\put(5000,22000){\circle*{300}}
\drawline\fermion[\N\REG](5000,18000)[4000]
\drawarrow[\N\ATTIP](5000,20000)
\put(5000,18000){\circle*{300}}
\drawline\gluon[\E\REG](5000,18000)[4]
\drawline\fermion[\E\REG](5000,22000)[4000]
\drawarrow[\E\ATTIP](7000,22000)
\put(0,18000){$q$}
\put(0,22000){$\gamma$}
\put(9500,18000){$g$}
\put(9500,22000){$q$}
\put(15500,16000){a)}
\drawline\gluon[\NE\REG](1200,18000)[1]
\put(1200,18000){\circle*{300}}
\drawline\gluon[\NW\FLIPPED](4800,18000)[1]
\put(4800,18000){\circle*{300}}
\drawline\photon[\E\REG](12000,22000)[4]
\drawline\fermion[\E\REG](12000,18000)[4000]
\drawarrow[\E\ATTIP](14000,18000)
\put(16000,22000){\circle*{300}}
\drawline\fermion[\N\REG](16000,18000)[4000]
\drawarrow[\N\ATTIP](16000,20000)
\put(16000,18000){\circle*{300}}
\drawline\gluon[\W\FLIPPED](17000,18000)[1]
\put(18000,18000){\circle{2000}}
\put(17000,18000){\circle*{300}}
\put(19000,18000){\circle*{300}}
\drawline\gluon[\E\REG](19000,18000)[1]
\drawline\fermion[\E\REG](16000,22000)[4000]
\drawarrow[\E\ATTIP](18000,22000)
\put(11000,18000){$q$}
\put(11000,22000){$\gamma$}
\put(20500,18000){$g$}
\put(20500,22000){$q$}
\drawline\photon[\E\REG](23000,22000)[4]
\drawline\fermion[\E\REG](23000,18000)[4000]
\drawarrow[\E\ATTIP](25000,18000)
\put(27000,22000){\circle*{300}}
\drawline\fermion[\N\REG](27000,18000)[4000]
\drawarrow[\N\ATTIP](27000,20000)
\put(27000,18000){\circle*{300}}
\drawline\gluon[\E\REG](27000,18000)[4]
\drawline\fermion[\E\REG](27000,22000)[4000]
\drawarrow[\E\ATTIP](29000,22000)
\put(22000,18000){$q$}
\put(22000,22000){$\gamma$}
\put(31500,18000){$g$}
\put(31500,22000){$q$}
\drawline\gluon[\SE\FLIPPED](27200,22000)[1]
\put(27200,22000){\circle*{300}}
\drawline\gluon[\SW\REG](30800,22000)[1]
\put(30800,22000){\circle*{300}}
\drawline\photon[\E\REG](1000,14000)[4]
\drawline\fermion[\E\REG](1000,10000)[4000]
\drawarrow[\E\ATTIP](3000,10000)
\put(5000,14000){\circle*{300}}
\drawline\fermion[\N\REG](5000,10000)[4000]
\drawarrow[\N\ATTIP](5000,12000)
\put(5000,10000){\circle*{300}}
\drawline\gluon[\E\REG](5000,10000)[4]
\drawline\fermion[\E\REG](5000,14000)[4000]
\drawarrow[\E\ATTIP](7000,14000)
\put(0,10000){$q$}
\put(0,14000){$\gamma$}
\put(9500,10000){$g$}
\put(9500,14000){$q$}
\put(15500,8000){b)}
\drawline\gluon[\SE\REG](5000,13800)[1]
\put(5000,13800){\circle*{300}}
\drawline\gluon[\NE\FLIPPED](5000,10200)[1]
\put(5000,10200){\circle*{300}}
\drawline\photon[\E\REG](12000,14000)[4]
\drawline\fermion[\E\REG](12000,10000)[4000]
\drawarrow[\E\ATTIP](14000,10000)
\put(16000,14000){\circle*{300}}
\drawline\fermion[\N\REG](16000,10000)[4000]
\drawarrow[\N\ATTIP](16000,12000)
\put(16000,10000){\circle*{300}}
\drawline\gluon[\E\REG](16000,10000)[4]
\drawline\fermion[\E\REG](16000,14000)[4000]
\drawarrow[\E\ATTIP](18000,14000)
\put(11000,10000){$q$}
\put(11000,14000){$\gamma$}
\put(20500,10000){$g$}
\put(20500,14000){$q$}
\drawline\gluon[\S\REG](18000,14150)[4]
\put(18000,14000){\circle*{300}}
\put(18000,10000){\circle*{300}}
\drawline\photon[\E\REG](23000,14000)[4]
\drawline\fermion[\E\REG](23000,10000)[4000]
\drawarrow[\E\ATTIP](25000,10000)
\put(27000,14000){\circle*{300}}
\drawline\fermion[\N\REG](27000,10000)[4000]
\drawarrow[\N\ATTIP](27000,12000)
\put(27000,10000){\circle*{300}}
\drawline\gluon[\E\REG](27000,10000)[4]
\drawline\fermion[\E\REG](27000,14000)[4000]
\drawarrow[\E\ATTIP](29000,14000)
\put(22000,10000){$q$}
\put(22000,14000){$\gamma$}
\put(31500,10000){$g$}
\put(31500,14000){$q$}
\drawline\gluon[\NE\REG](25000,10000)[3]
\put(29000,14000){\circle*{300}}
\put(25000,10000){\circle*{300}}
\drawline\photon[\E\REG](1000,6000)[4]
\drawline\fermion[\E\REG](1000,2000)[4000]
\drawarrow[\E\ATTIP](3000,2000)
\put(5000,6000){\circle*{300}}
\drawline\fermion[\N\REG](5000,2000)[4000]
\drawarrow[\N\ATTIP](5000,4000)
\put(5000,2000){\circle*{300}}
\drawline\gluon[\E\REG](5000,2000)[4]
\drawline\fermion[\E\REG](5000,6000)[4000]
\drawarrow[\E\ATTIP](7000,6000)
\put(0,2000){$q$}
\put(0,6000){$\gamma$}
\put(9500,2000){$g$}
\put(9500,6000){$q$}
\put(15500,0){c)}
\drawline\gluon[\NE\FLIPPED](5000,4000)[1]
\put(5000,4000){\circle*{300}}
\put(7000,6000){\circle*{300}}
\drawline\photon[\E\REG](12000,6000)[4]
\drawline\fermion[\E\REG](12000,2000)[4000]
\drawarrow[\E\ATTIP](14000,2000)
\put(16000,6000){\circle*{300}}
\drawline\fermion[\N\REG](16000,2000)[4000]
\drawarrow[\N\ATTIP](16000,4000)
\put(16000,2000){\circle*{300}}
\drawline\gluon[\E\REG](16000,2000)[4]
\drawline\fermion[\E\REG](16000,6000)[4000]
\drawarrow[\E\ATTIP](18000,6000)
\put(11000,2000){$q$}
\put(11000,6000){$\gamma$}
\put(20500,2000){$g$}
\put(20500,6000){$q$}
\drawline\gluon[\NE\REG](14000,2000)[1]
\put(14000,2000){\circle*{300}}
\put(16000,4000){\circle*{300}}
\drawline\photon[\E\REG](23000,6000)[4]
\drawline\fermion[\E\REG](23000,2000)[4000]
\drawarrow[\E\ATTIP](25000,2000)
\put(27000,6000){\circle*{300}}
\drawline\fermion[\N\REG](27000,2000)[4000]
\drawarrow[\N\ATTIP](27000,4000)
\put(27000,2000){\circle*{300}}
\drawline\gluon[\E\REG](27000,2000)[4]
\drawline\fermion[\E\REG](27000,6000)[4000]
\drawarrow[\E\ATTIP](29000,6000)
\put(22000,2000){$q$}
\put(22000,6000){$\gamma$}
\put(31500,2000){$g$}
\put(31500,6000){$q$}
\drawline\gluon[\SE\REG](27000,4000)[1]
\put(27000,4000){\circle*{300}}
\put(29000,2000){\circle*{300}}
\end{picture}
\end{center}
\caption[Virtual Diagrams for Direct Photoproduction]
        {\label{fig5}{\it
         Virtual diagrams for direct photoproduction. The circle in diagram a)
         denotes a quark, gluon, and ghost loop.}}
\end{figure}
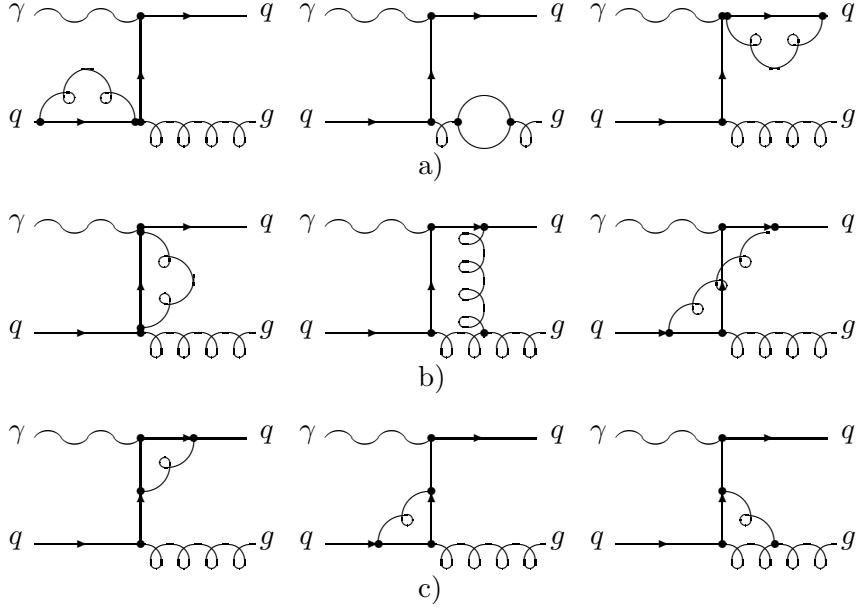

The one-loop corrections to the left diagram of figure \ref{fig3} are
shown in figure \ref{fig5}
and can be classified into a) self-energy diagrams, b) propagator
corrections and box diagrams, and c) vertex corrections.
They contain an additional virtual gluon, which leads to an extra factor
$\alpha_s$. Similar diagrams are obtained for the right diagram of figure
\ref{fig3}. As in leading order, the diagrams for the process $\ygtqb$ can
be obtained from figure \ref{fig5} by crossing the initial quark and the final
gluon or equivalently $(s\leftrightarrow t)$ and multiplying by $(-1)$
for the crossing of a fermion line. However, the virtual corrections can in
general contain logarithms $\ln(x/s-i\eta)$ and dilogarithms $\mbox{Li}_2
(x/s)$, where $x$ denotes different Mandelstam variables before and after
crossing. The $i\eta$-prescription for Feynman propagators then takes care
of additional terms of $\pi^2$, that arise in the quadratic logarithms
with negative argument.

The virtual contributions have been
well known for many years from $e^+ e^- \rightarrow q\bar{q}g$ higher
order QCD calculations \cite{Ell81,Fab82}. For the corresponding
photoproduction cross section, one substitutes $Q^2 = 0$ and performs the
necessary crossings. The result can be found in \cite{Aur87,Bod92,Kla93}.
We have also compared with the results in \cite{Gra90} for deep
inelastic scattering $eq \rightarrow e'gq$ and $e g \rightarrow e' q\bar{q}$,
which can be expressed by the invariants $s$, $t$, and $u$ after setting
$Q^2 = 0$. For completeness and for later use, we write the final result
in the form
\beq
  |{\cal M}|^2_{\gamma b\rightarrow 12} (s,t,u)  = 
  e^2e_q^2g^2 \mu^{4\varepsilon} \frac{\alpha_s}{2\pi}
  \left( \frac{4\pi\mu^2}{s} \right) ^\varepsilon
  \frac{\Gamma(1-\varepsilon)}{\Gamma(1-2\varepsilon)} V_{\gamma b
  \rightarrow 12}(s,t,u)
\eeq
The expression $V_{\yqtgq}(s,t,u)$ is given by
\bea
 V_{\yqtgq}(s,t,u) & = & \le C_F\lr-\frac{2}{\varepsilon^2} +\frac{1}
     {\varepsilon}
     \lr 2\ln \ts - 3 \rr +\frac{2\pi^2}{3} -7 +\ln^2\tu \rr\rp\label{eq28}\\
  && -\frac{N_C}{2}\lr\frac{2}{\varepsilon^2}+\frac{1}{\varepsilon}\lr
     \frac{11}{3}+2\ln\ts-2\ln\us \rr+\frac{\pi^2}{3}+\ln^2\tu
     +\frac{11}{3}\ln\frac{s}{\mu^2}\rr\nonumber\\
  && \lp+\frac{N_f}{3}\lr\ede+\ln\frac{s}{\mu^2}\rr\re T_{\yqtgq}(s,t,u)
     \nonumber\\
  && +8N_CC_F^2\lr-2\ln\us+4\ln\ts-3\frac{s}{u}\ln\us-\lr 2+\frac{u}{s} \rr
    \lr \pi^2+\ln^2\tu \rr\rp\nonumber\\
  && \lp-\lr 2+\frac{s}{u} \rr \ln^2\ts \rr\nonumber\\
  && -4N_C^2C_F\lr 4\ln\ts-2\ln\us-\lr 2+\frac{u}{s} \rr
    \lr \pi^2+\ln^2\tu \rr -\lr 2+\frac{s}{u} \rr \ln^2\ts\rr.\nonumber
\eea
The contributions for incoming anti-quarks and gluons $\ybtgb$ and $\ygtqb$
could be obtained according to table \ref{tab2}, if the imaginary parts were
included above. The result for anti-quarks turns out to be
identical to eq.~(\ref{eq28}), and for the gluon initiated process one
obtains
\bea
  V_{\ygtqb}(s,t,u) & = & \le C_F\lr-\frac{2}{\varepsilon^2}-\frac{3}
     {\varepsilon}
     +\frac{2\pi^2}{3}-7+\ln^2\ts+\ln^2\us\rr\rp\label{eq29}\\
  && -\frac{N_C}{2}\lr\frac{2}{\varepsilon^2}+\frac{1}{\varepsilon}\lr
     \frac{11}{3}-2\ln\ts-2\ln\us \rr+\frac{\pi^2}{3}+\ln^2\frac{tu}{s^2}
     +\frac{11}{3}\ln\frac{s}{\mu^2}\rr\nonumber\\
  && \lp+\frac{N_f}{3}\lr\ede+\ln\frac{s}{\mu^2}\rr\re T_{\ygtqb}(s,t,u)
     \nonumber\\
  && +8N_CC_F^2\lr 2\ln\ts+2\ln\us+3\frac{u}{t}\ln\ts+3\frac{t}{u}\ln\us
     +\lr 2+\frac{u}{t} \rr \ln^2\us\rp\nonumber\\
  && \lp+\lr 2+\frac{t}{u}\rr \ln^2\ts\rr\nonumber\\
  && -4N_C^2C_F\lr 2\ln\ts+2\ln\us +\lr 2+\frac{u}{t}\rr \ln^2\us
     +\lr 2+\frac{t}{u} \rr\ln^2\ts\rr.\nonumber
\eea
All UV poles have been canceled by counter terms through the renormalization
procedure, and
the remaining IR poles are contained in the first three lines of
eqs.~(\ref{eq28}) and (\ref{eq29}). Contrary to some of the finite terms,
they are proportional to the LO Born matrix elements $T_{\yqtgq}$ and
$T_{\ygtqb}$. The explicit dependence of the virtual corrections on the
renormalization scale $\mu$ is contained in the logarithms proportional to
the Born matrix elements and $N_C$ and $N_f$, respectively. All terms of
${\cal O} (\varepsilon)$ and higher have been omitted since they do not
contribute in the physical limit $d\rightarrow 4$.

\subsubsection{Virtual Corrections for Resolved Photons}

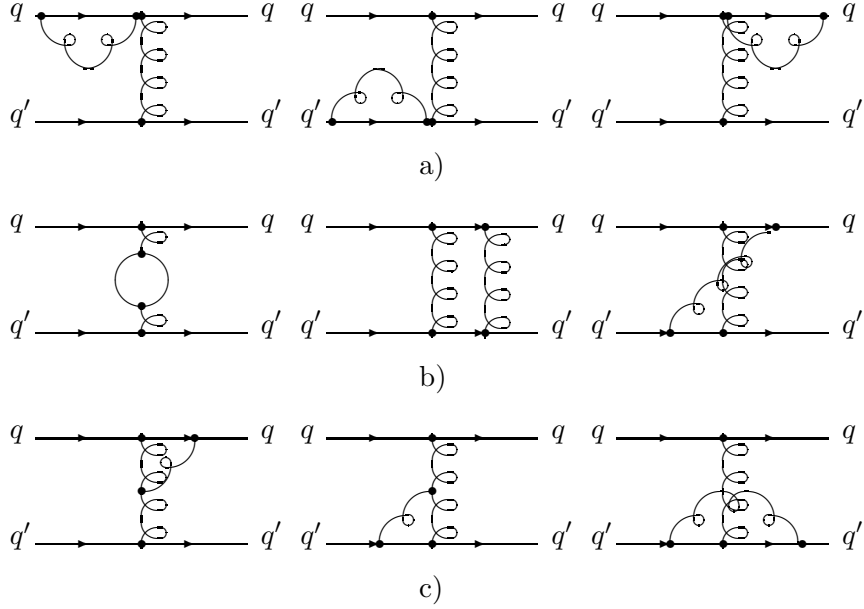
\begin{figure}[htbp]
\begin{center}
\begin{picture}(33000,23000)
\drawline\fermion[\E\REG](1000,22000)[4000]
\drawarrow[\E\ATTIP](3000,22000)
\drawline\fermion[\E\REG](1000,18000)[4000]
\drawarrow[\E\ATTIP](3000,18000)
\put(5000,22000){\circle*{300}}
\drawline\gluon[\N\REG](5000,17800)[4]
\put(5000,18000){\circle*{300}}
\drawline\fermion[\E\REG](5000,18000)[4000]
\drawarrow[\E\ATTIP](7000,18000)
\drawline\fermion[\E\REG](5000,22000)[4000]
\drawarrow[\E\ATTIP](7000,22000)
\put(0,18000){$q'$}
\put(0,22000){$q$}
\put(9500,18000){$q'$}
\put(9500,22000){$q$}
\put(15500,16000){a)}
\drawline\gluon[\SE\FLIPPED](1200,22000)[1]
\put(1200,22000){\circle*{300}}
\drawline\gluon[\SW\REG](4800,22000)[1]
\put(4800,22000){\circle*{300}}
\drawline\fermion[\E\REG](12000,22000)[4000]
\drawarrow[\E\ATTIP](14000,22000)
\drawline\fermion[\E\REG](12000,18000)[4000]
\drawarrow[\E\ATTIP](14000,18000)
\put(16000,22000){\circle*{300}}
\drawline\gluon[\N\REG](16000,17800)[4]
\put(16000,18000){\circle*{300}}
\drawline\fermion[\E\REG](16000,18000)[4000]
\drawarrow[\E\ATTIP](18000,18000)
\drawline\fermion[\E\REG](16000,22000)[4000]
\drawarrow[\E\ATTIP](18000,22000)
\put(11000,18000){$q'$}
\put(11000,22000){$q$}
\put(20500,18000){$q'$}
\put(20500,22000){$q$}
\drawline\gluon[\NE\REG](12200,18000)[1]
\put(12200,18000){\circle*{300}}
\drawline\gluon[\NW\FLIPPED](15800,18000)[1]
\put(15800,18000){\circle*{300}}
\drawline\fermion[\E\REG](23000,22000)[4000]
\drawarrow[\E\ATTIP](25000,22000)
\drawline\fermion[\E\REG](23000,18000)[4000]
\drawarrow[\E\ATTIP](25000,18000)
\put(27000,22000){\circle*{300}}
\drawline\gluon[\N\REG](27000,17800)[4]
\put(27000,18000){\circle*{300}}
\drawline\fermion[\E\REG](27000,18000)[4000]
\drawarrow[\E\ATTIP](29000,18000)
\drawline\fermion[\E\REG](27000,22000)[4000]
\drawarrow[\E\ATTIP](29000,22000)
\put(22000,18000){$q'$}
\put(22000,22000){$q$}
\put(31500,18000){$q'$}
\put(31500,22000){$q$}
\drawline\gluon[\SE\FLIPPED](27200,22000)[1]
\put(27200,22000){\circle*{300}}
\drawline\gluon[\SW\REG](30800,22000)[1]
\put(30800,22000){\circle*{300}}
\drawline\fermion[\E\REG](1000,14000)[4000]
\drawarrow[\E\ATTIP](3000,14000)
\drawline\fermion[\E\REG](1000,10000)[4000]
\drawarrow[\E\ATTIP](3000,10000)
\put(5000,14000){\circle*{300}}
\drawline\gluon[\N\REG](5000,9850)[1]
\put(5000,12000){\circle{2000}}
\put(5000,13000){\circle*{300}}
\put(5000,11000){\circle*{300}}
\drawline\gluon[\N\REG](5000,13000)[1]
\put(5000,10000){\circle*{300}}
\drawline\fermion[\E\REG](5000,10000)[4000]
\drawarrow[\E\ATTIP](7000,10000)
\drawline\fermion[\E\REG](5000,14000)[4000]
\drawarrow[\E\ATTIP](7000,14000)
\put(0,10000){$q'$}
\put(0,14000){$q$}
\put(9500,10000){$q'$}
\put(9500,14000){$q$}
\put(15500,8000){b)}
\drawline\fermion[\E\REG](12000,14000)[4000]
\drawarrow[\E\ATTIP](14000,14000)
\drawline\fermion[\E\REG](12000,10000)[4000]
\drawarrow[\E\ATTIP](14000,10000)
\put(16000,14000){\circle*{300}}
\drawline\gluon[\N\REG](16000,9800)[4]
\put(16000,10000){\circle*{300}}
\drawline\fermion[\E\REG](16000,10000)[4000]
\drawarrow[\E\ATTIP](18000,10000)
\drawline\fermion[\E\REG](16000,14000)[4000]
\drawarrow[\E\ATTIP](18000,14000)
\put(11000,10000){$q'$}
\put(11000,14000){$q$}
\put(20500,10000){$q'$}
\put(20500,14000){$q$}
\drawline\gluon[\S\FLIPPED](18000,14150)[4]
\put(18000,14000){\circle*{300}}
\put(18000,10000){\circle*{300}}
\drawline\fermion[\E\REG](23000,14000)[4000]
\drawarrow[\E\ATTIP](25000,14000)
\drawline\fermion[\E\REG](23000,10000)[4000]
\drawarrow[\E\ATTIP](25000,10000)
\put(27000,14000){\circle*{300}}
\drawline\gluon[\N\REG](27000,9850)[4]
\put(27000,10000){\circle*{300}}
\drawline\fermion[\E\REG](27000,10000)[4000]
\drawarrow[\E\ATTIP](29000,10000)
\drawline\fermion[\E\REG](27000,14000)[4000]
\drawarrow[\E\ATTIP](29000,14000)
\put(22000,10000){$q'$}
\put(22000,14000){$q$}
\put(31500,10000){$q'$}
\put(31500,14000){$q$}
\drawline\gluon[\NE\REG](25000,10000)[3]
\put(25000,10000){\circle*{300}}
\put(29000,14000){\circle*{300}}
\drawline\fermion[\E\REG](1000,6000)[4000]
\drawarrow[\E\ATTIP](3000,6000)
\drawline\fermion[\E\REG](1000,2000)[4000]
\drawarrow[\E\ATTIP](3000,2000)
\put(5000,6000){\circle*{300}}
\drawline\gluon[\N\REG](5000,1800)[4]
\put(5000,2000){\circle*{300}}
\drawline\fermion[\E\REG](5000,2000)[4000]
\drawarrow[\E\ATTIP](7000,2000)
\drawline\fermion[\E\REG](5000,6000)[4000]
\drawarrow[\E\ATTIP](7000,6000)
\put(0,2000){$q'$}
\put(0,6000){$q$}
\put(9500,2000){$q'$}
\put(9500,6000){$q$}
\put(15500,0){c)}
\drawline\gluon[\NE\FLIPPED](5000,4000)[1]
\put(5000,4000){\circle*{300}}
\put(7000,6000){\circle*{300}}
\drawline\fermion[\E\REG](12000,6000)[4000]
\drawarrow[\E\ATTIP](14000,6000)
\drawline\fermion[\E\REG](12000,2000)[4000]
\drawarrow[\E\ATTIP](14000,2000)
\put(16000,6000){\circle*{300}}
\drawline\gluon[\N\REG](16000,1800)[4]
\put(16000,2000){\circle*{300}}
\drawline\fermion[\E\REG](16000,2000)[4000]
\drawarrow[\E\ATTIP](18000,2000)
\drawline\fermion[\E\REG](16000,6000)[4000]
\drawarrow[\E\ATTIP](18000,6000)
\put(11000,2000){$q'$}
\put(11000,6000){$q$}
\put(20500,2000){$q'$}
\put(20500,6000){$q$}
\drawline\gluon[\NE\REG](14000,2000)[1]
\put(14000,2000){\circle*{300}}
\put(16000,4000){\circle*{300}}
\drawline\fermion[\E\REG](23000,6000)[4000]
\drawarrow[\E\ATTIP](25000,6000)
\drawline\fermion[\E\REG](23000,2000)[4000]
\drawarrow[\E\ATTIP](25000,2000)
\put(27000,6000){\circle*{300}}
\drawline\gluon[\N\REG](27000,1800)[4]
\put(27000,2000){\circle*{300}}
\drawline\fermion[\E\REG](27000,2000)[4000]
\drawarrow[\E\ATTIP](29000,2000)
\drawline\fermion[\E\REG](27000,6000)[4000]
\drawarrow[\E\ATTIP](29000,6000)
\put(22000,2000){$q'$}
\put(22000,6000){$q$}
\put(31500,2000){$q'$}
\put(31500,6000){$q$}
\drawloop\gluon[\NE 3](25000,2000)
\put(25000,2000){\circle*{300}}
\put(30000,2000){\circle*{300}}
\end{picture}
\end{center}
\caption[Virtual Diagrams for Resolved Photoproduction]
        {\label{fig6}{\it
         Virtual diagrams for resolved photoproduction. Only the one-loop
         corrections to $\qptqp$ are shown. The circle in diagram b) denotes
         a quark, gluon, and ghost loop.}}
\end{figure}

For the resolved virtual corrections, we show in figure \ref{fig6} only
the one-loop diagrams for $\qptqp$ and
classify them again into a) self-energy diagrams, b) propagator corrections
and box diagrams, and c) vertex corrections. Due to the additional gluon,
these contributions are again one order higher in $\alpha_s$ than the Born
terms as in the direct case. The complete set of diagrams can be found in
\cite{Ell86} as well as the results
\beq
  |{\cal M}|^2_{ab\rightarrow 12} (s,t,u)  = 
  g^4 \mu^{4\varepsilon} \frac{\alpha_s}{2\pi}
  \left( \frac{4\pi\mu^2}{Q^2} \right) ^\varepsilon
  \frac{\Gamma(1-\varepsilon)}{\Gamma(1-2\varepsilon)} V_{ab\rightarrow 12}
  (s,t,u), 
\eeq
where $Q^2$ denotes now an arbitrary scale.
For the diagrams in figure \ref{fig6}, we obtain
\bea
 V_{\qptqp}(s,t,u) & = & \le C_F\lr-\frac{4}{\varepsilon^2}
     -\ede (6+8l(s)-8l(u)-4l(t))\rp\rp\\
  && -\frac{2\pi^2}{3}-16-2l^2(t)+l(t)(6+8l(s)-8l(u))\nonumber\\
  && -2\frac{s^2-u^2}{s^2+u^2}(2\pi^2+(l(t)-l(s))^2+(l(t)-l(u))^2)\nonumber\\
  && \lp+2\frac{s+u}{s^2+u^2}((s+u)(l(u)-l(s))+(u-s)(2l(t)-l(s)-l(u)))\rr
      \nonumber\\
  && +N_C\lr\frac{1}{\varepsilon}\lr 4l(s)-2l(u)-2l(t)\rr
     +\frac{85}{9}+\pi^2+2l(t)(l(t)+l(u)-2l(s))\rp\nonumber\\
  && +\frac{s^2-u^2}{2(s^2+u^2)}(3\pi^2+2(l(t)-l(s))^2+(l(t)-l(u))^2)
     \nonumber\\
  && \lp-\frac{st}{s^2+u^2}(l(t)-l(u))
     +\frac{2ut}{s^2+u^2}(l(t)-l(s))
     +\frac{11}{3}(l(-\mu^2)-l(t))\rr \nonumber \\
  && \lp+\frac{N_f}{2}\lr\frac{4}{3}(l(t)-l(-\mu^2))-\frac{20}{9}\rr\re
     T_{\qptqp}(s,t,u). \nonumber
\eea
The contributions for incoming quarks and anti-quarks of different flavor
can be obtained from table \ref{tab3} as well as the contribution for
identical quark flavors. There one also has to include the interference term
\bea
 V_{\qqtqq}(s,t,u) & = & \le C_F\lr-\frac{4}{\varepsilon^2}
     -\ede (6+4l(s)-4l(t)-4l(u))-\frac{7\pi^2}{6}-16\rp\rp\\
  && \lp-\frac{3}{2}(l(t)+l(u))^2+2l(s)(l(t)+l(u))+2(l(t)+l(u))\rr\nonumber\\
  && +N_C\lr\frac{1}{\varepsilon}(4l(s)-2l(t)-2l(u))+\frac{85}{9}
     +\frac{5}{4}(l(t)+l(u))^2\rp \nonumber\\
  && \lp-l(t)l(u)-2l(s)(l(t)+l(u))
     -\frac{4}{3}(l(t)+l(u))+\frac{5}{4}\pi^2+\frac{11}{3}l(-\mu^2)\rr
     \nonumber \\
  && +\frac{N_f}{2}\lr-\frac{20}{9}+\frac{2}{3}(l(t)+l(u)-2l(-\mu^2))\rr
     \nonumber \\
  && \lp+\frac{1}{N_C}\lr (\pi^2+(l(t)-l(u))^2)\frac{ut}{2s^2}+\frac{u}{s}
     l(t)+\frac{t}{s}l(u)\rr\re T_{\qqtqq}(s,t,u). \nonumber
\eea
For incoming antiquarks, the Mandelstam variables $(s\leftrightarrow u)$ have
to be crossed.
The first lines proportional to the color factors $C_F$ and $N_C$ contain the
IR singular terms in the two processes discussed above. Furthermore,
the complete virtual corrections are proportional to the Born matrix elements.
This is not true for the case $\qbtgg$, where the virtual corrections are
given by
\bea
 V_{\qbtgg}(s,t,u) & = & \le C_F\lr-\frac{2}{\varepsilon^2}-\frac{3}
     {\varepsilon}-7
     -\frac{\pi^2}{3}\rr\rp\\
  && +N_C\lr-\frac{2}{\varepsilon^2}-\frac{11}{3\varepsilon}+\frac{11}{3}
     l(-\mu^2)
     -\frac{\pi^2}{3}\rr\nonumber\\
  && \lp+\frac{N_f}{2}\lr\frac{4}{3\varepsilon}-\frac{4}{3}l(-\mu^2)\rr\re
     T_{\qbtgg}(s,t,u) \nonumber \\
  && +\frac{l(s)}{\varepsilon}\lr\lr 4N_C^3C_F+\frac{4C_F}{N_C}\rr
     \frac{t^2+u^2}
     {ut}-16N_C^2C_F^2\frac{t^2+u^2}{s^2}\rr \nonumber\\
  && +\frac{8N_C^3C_F}{\varepsilon}\lr l(t)\lr\utx-\frac{2u^2}{s^2}\rr+l(u)
     \lr\tux-\frac{2t^2}{s^2}\rr\rr\nonumber\\
  && -\frac{8N_CC_F}{\varepsilon}\lr\utx+\tux\rr (l(t)+l(u))\nonumber\\
  && +f^c(s,t,u)+f^c(s,u,t). \nonumber
\eea
Here, only parts of the IR singular terms are proportional to the Born
matrix element $T_{\qbtgg}$. The finite contributions have been put into the
function
\bea
 f^c(s,t,u) & = & 8N_C^2C_F\le\frac{l(t)l(u)}{N_C}\frac{t^2+u^2}{2tu}\rp\\
  && +l^2(s)\lr\frac{1}{4N_C^3}\frac{s^2}{tu}+\frac{1}{4N_C}\lr\frac{1}{2}
     +\frac{t^2+u^2}{tu}-\frac{t^2+u^2}{s^2}\rr-\frac{N_C}{4}\frac{t^2+u^2}
     {s^2}\rr\nonumber\\
  && +l(s)\lr\lr\frac{5C_F}{4}-\frac{1}{2N_C}-\frac{1}{N_C^3}\rr
     -\lr N_C+\frac{1}{N_C^3}\rr\frac{t^2+u^2}{2tu}
     -\frac{C_F}{2}\frac{t^2+u^2}{s^2}\rr\nonumber\\
  && +\pi^2\lr\frac{1}{8N_C}+\frac{1}{N_C^3}\lr\frac{3(t^2+u^2)}{8tu}
     +\frac{1}{2}\rr+N_C\lr\frac{t^2+u^2}{8tu}-\frac{t^2+u^2}{2s^2}\rr\rr
     \nonumber\\
  && +\lr N_C+\frac{1}{N_C}\rr\lr\frac{1}{8}-\frac{t^2+u^2}{4s^2}\rr
     \nonumber\\
  && +l^2(t)\lr N_C\lr\frac{s}{4t}-\frac{u}{s}-\frac{1}{4}\rr+\frac{1}{N_C}
     \lr\frac{t}{2u}-\frac{u}{4s}\rr+\frac{1}{N_C^3}\lr\frac{u}{4t}
     -\frac{s}{2u}\rr\rr\nonumber\\
  && +l(t)\lr N_C\lr\frac{t^2+u^2}{s^2}+\frac{3t}{4s}-\frac{5u}{4t}
     -\frac{1}{4}\rr
     -\ednc\lr\frac{u}{4s}+\frac{2s}{u}+\frac{s}{2t}\rr
     -\frac{1}{N_C^3}\lr\frac{3s}{4t}+\frac{1}{4}\rr\rr\nonumber\\
  && \lp+l(s)l(t)\lr N_C\lr\frac{t^2+u^2}{s^2}-\frac{u}{2t}\rr
     +\ednc\lr\frac{u}{2s}-\tux\rr+\frac{1}{N_C^3}\lr\sux-\frac{u}{2t}\rr\rr
     \re\nonumber
\eea
Finally, the process $\ggtgg$ gives
\bea
 V_{\ggtgg}(s,t,u) & = & \le N_C\lr-\frac{4}{\varepsilon^2}-\frac{22}
     {3\varepsilon}
     -\frac{67}{9}+\frac{11}{3}l(-\mu^2)+\frac{\pi^2}{3}\rr\rp\\
  && \lp+\frac{N_f}{2}\lr\frac{8}{3\varepsilon}+\frac{20}{9}-\frac{4}{3}
     l(-\mu^2)\rr\re T_{\ggtgg}(s,t,u) \nonumber \\
  && +\frac{32N_C^4C_F}{\varepsilon}l(s)\lr 3-\frac{2tu}{s^2}+\frac{t^4+u^4}
     {t^2u^2}\rr\nonumber\\
  && +\frac{32N_C^4C_F}{\varepsilon}l(t)\lr 3-\frac{2us}{t^2}+\frac{u^4+s^4}
     {u^2s^2}\rr\nonumber\\
  && +\frac{32N_C^4C_F}{\varepsilon}l(u)\lr 3-\frac{2st}{u^2}+\frac{s^4+t^4}
     {s^2t^2}\rr\nonumber\\
  && +8N_C^3C_F(f^d(s,t,u)+f^d(t,u,s)+f^d(u,s,t)), \nonumber
\eea
where, once more, not all IR singularities are proportional to the complete
Born matrix element and the finite terms are given by the function
\bea
 f^d(s,t,u) & = & N_C\le\lr\frac{2(t^2+u^2)}{tu}\rr l^2(s)
     +\lr\frac{4s(t^3+u^3)}{t^2u^2}-6\rr l(t)l(u)\rp\\
  && \lp+\lr\frac{4}{3}\frac{tu}{s^2}-\frac{14}{3}\frac{t^2+u^2}{tu}-14
     -8\lr\frac{t^2}{u^2}+\frac{u^2}{t^2}\rr\rr l(s)-1-\pi^2\re\nonumber\\
  && +\frac{N_f}{2}\le\lr\frac{10}{3}\frac{t^2+u^2}{tu}+\frac{16}{3}
     \frac{tu}{s^2}-2\rr l(s)-\frac{s^2+tu}{tu}l^2(s)\rp\nonumber\\
  && \lp-2\frac{t^2+u^2}{tu}l(t)l(u)+2-\pi^2\re.\nonumber
\eea
All UV singularities have already been absorbed into the Lagrangian, and
terms of ${\cal O} (\varepsilon)$ and higher have been neglected.

As mentioned previously, the crossing of Mandelstam variables according to
table \ref{tab3} may change the signs of the arguments in the logarithms.
This is accounted for in the definition of
\beq
 l(x) = \ln\lb\frac{x}{Q^2}\rb,
\eeq
where $x$ may be any of the variables $s$, $t$, or $u$ and is normalized to
$Q^2=\max(s,t,u)$. Due to the $i\eta$-prescription in the propagators, terms
bilinear in the logarithms may give an extra $\pi^2$:
\bea
 l^2(x) &=& \ln^2\lr \frac{x}{Q^2}\rr-\pi^2,~\mbox{if}~x>0, \\
 l^2(x) &=& \ln^2\lr-\frac{x}{Q^2}\rr,\hspace{0.65cm}\mbox{if}~x<0.
\eea

\subsubsection{Virtual Corrections for Direct $\gamma\gamma$ Scattering}

\begin{figure}[htbp]
\begin{center}
\begin{picture}(33000,15000)
\drawline\photon[\E\REG](1000,14000)[4]
\drawline\photon[\E\REG](1000,10000)[4]
\put(5000,14000){\circle*{300}}
\drawline\fermion[\N\REG](5000,10000)[4000]
\drawarrow[\N\ATTIP](5000,12000)
\put(5000,10000){\circle*{300}}
\drawline\fermion[\E\REG](5000,10000)[4000]
\drawarrow[\W\ATTIP](7000,10000)
\drawline\fermion[\E\REG](5000,14000)[4000]
\drawarrow[\E\ATTIP](7000,14000)
\put(0,10000){$\gamma$}
\put(0,14000){$\gamma$}
\put(9500,10000){$\bar{q}$}
\put(9500,14000){$q$}
\put(15500,8000){a)}
\drawline\gluon[\SE\FLIPPED](5200,14000)[1]
\put(5200,14000){\circle*{300}}
\drawline\gluon[\SW\REG](8800,14000)[1]
\put(8800,14000){\circle*{300}}
\drawline\photon[\E\REG](12000,14000)[4]
\drawline\photon[\E\REG](12000,10000)[4]
\put(16000,14000){\circle*{300}}
\drawline\fermion[\N\REG](16000,10000)[4000]
\drawarrow[\N\ATTIP](16000,12000)
\put(16000,10000){\circle*{300}}
\drawline\fermion[\E\REG](16000,10000)[4000]
\drawarrow[\W\ATTIP](18000,10000)
\drawline\fermion[\E\REG](16000,14000)[4000]
\drawarrow[\E\ATTIP](18000,14000)
\put(11000,10000){$\gamma$}
\put(11000,14000){$\gamma$}
\put(20500,10000){$\bar{q}$}
\put(20500,14000){$q$}
\drawline\gluon[\NE\REG](16200,10000)[1]
\put(16200,10000){\circle*{300}}
\drawline\gluon[\NW\FLIPPED](19800,10000)[1]
\put(19800,10000){\circle*{300}}
\drawline\photon[\E\REG](23000,14000)[4]
\drawline\photon[\E\REG](23000,10000)[4]
\put(27000,14000){\circle*{300}}
\drawline\fermion[\N\REG](27000,10000)[4000]
\drawarrow[\N\ATTIP](27000,12000)
\put(27000,10000){\circle*{300}}
\drawline\fermion[\E\REG](27000,10000)[4000]
\drawarrow[\W\ATTIP](29000,10000)
\drawline\fermion[\E\REG](27000,14000)[4000]
\drawarrow[\E\ATTIP](29000,14000)
\put(22000,10000){$\gamma$}
\put(22000,14000){$\gamma$}
\put(31500,10000){$\bar{q}$}
\put(31500,14000){$q$}
\drawline\gluon[\SE\REG](27000,13800)[1]
\put(27000,13800){\circle*{300}}
\drawline\gluon[\NE\FLIPPED](27000,10200)[1]
\put(27000,10200){\circle*{300}}
\drawline\photon[\E\REG](1000,6000)[4]
\drawline\photon[\E\REG](1000,2000)[4]
\put(5000,6000){\circle*{300}}
\drawline\fermion[\N\REG](5000,2000)[4000]
\drawarrow[\N\ATTIP](5000,4000)
\put(5000,2000){\circle*{300}}
\drawline\fermion[\E\REG](5000,2000)[4000]
\drawarrow[\W\ATTIP](7000,2000)
\drawline\fermion[\E\REG](5000,6000)[4000]
\drawarrow[\E\ATTIP](7000,6000)
\put(0,2000){$\gamma$}
\put(0,6000){$\gamma$}
\put(9500,2000){$\bar{q}$}
\put(9500,6000){$q$}
\put(15500,0){b)}
\drawline\gluon[\NE\FLIPPED](5000,4000)[1]
\put(5000,4000){\circle*{300}}
\put(7000,6000){\circle*{300}}
\drawline\photon[\E\REG](12000,6000)[4]
\drawline\photon[\E\REG](12000,2000)[4]
\put(16000,6000){\circle*{300}}
\drawline\fermion[\N\REG](16000,2000)[4000]
\drawarrow[\N\ATTIP](16000,4000)
\put(16000,2000){\circle*{300}}
\drawline\fermion[\E\REG](16000,2000)[4000]
\drawarrow[\W\ATTIP](18000,2000)
\drawline\fermion[\E\REG](16000,6000)[4000]
\drawarrow[\E\ATTIP](18000,6000)
\put(11000,2000){$\gamma$}
\put(11000,6000){$\gamma$}
\put(20500,2000){$\bar{q}$}
\put(20500,6000){$q$}
\drawline\gluon[\S\REG](18000,6150)[4]
\put(18000,6000){\circle*{300}}
\put(18000,2000){\circle*{300}}
\drawline\photon[\E\REG](23000,6000)[4]
\drawline\photon[\E\REG](23000,2000)[4]
\put(27000,6000){\circle*{300}}
\drawline\fermion[\N\REG](27000,2000)[4000]
\drawarrow[\N\ATTIP](27000,4000)
\put(27000,2000){\circle*{300}}
\drawline\fermion[\E\REG](27000,2000)[4000]
\drawarrow[\W\ATTIP](29000,2000)
\drawline\fermion[\E\REG](27000,6000)[4000]
\drawarrow[\E\ATTIP](29000,6000)
\put(22000,2000){$\gamma$}
\put(22000,6000){$\gamma$}
\put(31500,2000){$\bar{q}$}
\put(31500,6000){$q$}
\drawline\gluon[\SE\REG](27000,4000)[1]
\put(27000,4000){\circle*{300}}
\put(29000,2000){\circle*{300}}
\end{picture}
\end{center}
\caption{\label{kkkfig2}{\it
         Virtual diagrams for direct $\gamma\gamma$ scattering.}}
\end{figure}
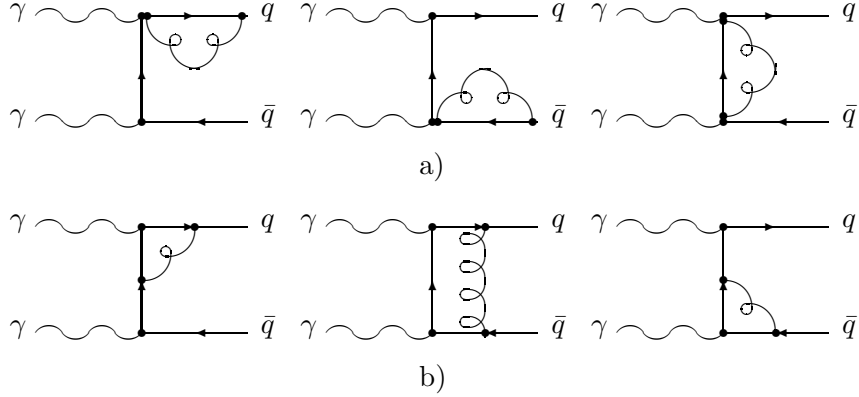

The virtual diagrams contributing to direct $\gamma\gamma$ scattering
are shown in figure \ref{kkkfig2}.
As has already been seen in the Born case, direct $\yytqb$ scattering
is intimately related to $\ygtqb$ scattering. Again, we have to replace
the strong coupling constant by its electromagnetic counterpart resulting
in
\beq
  |{\cal M}|^2_{\gamma\gamma\rightarrow 12} (s,t,u)  = 
  e^4e_q^4 \mu^{4\varepsilon} \frac{\alpha_s}{2\pi}
  \left( \frac{4\pi\mu^2}{s} \right) ^\varepsilon
  \frac{\Gamma(1-\varepsilon)}{\Gamma(1-2\varepsilon)} V_{\gamma \gamma
  \rightarrow 12}(s,t,u).
\eeq
The expression $V_{\yytqb}(s,t,u)$ is given by
\bea
  V_{\yytqb}(s,t,u) & = & C_F\lr-\frac{2}{\varepsilon^2}-\frac{3}
     {\varepsilon}
     +\frac{2\pi^2}{3}-7+\ln^2\ts+\ln^2\us\rr T_{\yytqb}(s,t,u)
     \label{kkkeq2}\\
  && +8N_CC_F\lr 2\ln\ts+2\ln\us+3\frac{u}{t}\ln\ts+3\frac{t}{u}\ln\us
     +\lr 2+\frac{u}{t} \rr \ln^2\us\rp\nonumber\\
  && \lp+\lr 2+\frac{t}{u}\rr \ln^2\ts\rr . \nonumber
\eea
Note that due to the abelian structure of QED, the color class $N_C$, which
is present in eq.~(\ref{eq29}) and arises there from the triple-gluon vertex,
does not appear here.

\subsection{Real Corrections}

For the calculation of the hard scattering cross section in
next-to-leading order, we must include all diagrams with an additional
parton in
the final state. The four-vectors of these subprocesses will be labeled by
$p_ap_b\rightarrow p_1p_2p_3$, where $p_a$ is the momentum of the
incoming photon or parton in the photon and $p_b$ is the momentum of the
incoming parton in the proton. The invariants will be denoted by $s_{ij}=
(p_i+p_j)^2$ and the previously defined Mandelstam variables
$s$, $t$, and $u$. For massless partons, the $2 \rightarrow 3$
contributions can
contain singularities at $s_{ij}=0$. They can be extracted with
the dimensional regularization method and canceled against those associated
with the one-loop contributions.

In order
to achieve this, we go through the following steps. First, we calculate the
phase space for $2\rightarrow 3$ scattering in $d$ dimensions and factorize
it into a regular part corresponding to $2\rightarrow 2$ scattering and a
singular part corresponding to the unresolved two-parton subsystem. Next, the
matrix elements for the $2 \rightarrow 3$ subprocesses are calculated in $d$
dimensions as well. They are squared and averaged/summed over initial/final
state spins and colors and can be used for incoming quarks, antiquarks, and
gluons with the help of crossing.

\begin{figure}[htbp]
\begin{center}
\begin{picture}(32000,7000)
\drawline\photon[\E\REG](1000,6000)[4]
\drawline\fermion[\E\REG](1000,2000)[4000]
\drawarrow[\E\ATTIP](3000,2000)
\put(5000,2000){\circle*{300}}
\drawline\fermion[\N\REG](5000,2000)[4000]
\drawarrow[\N\ATTIP](5000,4000)
\put(5000,6000){\circle*{300}}
\drawline\gluon[\E\REG](5000,2000)[4]
\drawline\gluon[\NE\REG](7000,2000)[1]
\put(7000,2000){\circle*{300}}
\put(5700,1600){X}
\drawline\fermion[\E\REG](5000,6000)[4000]
\drawarrow[\E\ATTIP](7000,6000)
\put(0,2000){$p_b$}
\put(0,6000){$p_a$}
\put(9500,2000){$p_3$}
\put(9500,4000){$p_2$}
\put(9500,6000){$p_1$}
\put(4500,0){a)}
\drawline\photon[\E\REG](12000,6000)[4]
\drawline\fermion[\E\REG](12000,2000)[4000]
\drawarrow[\E\ATTIP](14000,2000)
\put(16000,2000){\circle*{300}}
\drawline\fermion[\N\REG](16000,2000)[4000]
\put(16000,4000){\circle*{300}}
\put(16000,6000){\circle*{300}}
\drawline\gluon[\E\REG](16000,2000)[4]
\drawline\gluon[\E\REG](16000,4000)[4]
\put(15600,4500){X}
\drawline\fermion[\E\REG](16000,6000)[4000]
\drawarrow[\E\ATTIP](18000,6000)
\put(11000,2000){$p_b$}
\put(11000,6000){$p_a$}
\put(20500,2000){$p_3$}
\put(20500,4000){$p_2$}
\put(20500,6000){$p_1$}
\put(15500,0){b)}
\drawline\photon[\E\REG](23000,6000)[4]
\drawline\fermion[\E\REG](23000,2000)[4000]
\drawarrow[\E\ATTIP](25000,2000)
\put(27000,2000){\circle*{300}}
\drawline\fermion[\N\REG](27000,2000)[4000]
\put(27000,4000){\circle*{300}}
\put(27000,6000){\circle*{300}}
\drawline\gluon[\E\REG](27000,2000)[4]
\drawline\gluon[\E\REG](27000,4000)[4]
\put(26600,2500){X}
\drawline\fermion[\E\REG](27000,6000)[4000]
\drawarrow[\E\ATTIP](29000,6000)
\put(22000,2000){$p_b$}
\put(22000,6000){$p_a$}
\put(31500,2000){$p_3$}
\put(31500,4000){$p_2$}
\put(31500,6000){$p_1$}
\put(26500,0){c)}
\end{picture}
\end{center}
\caption[Different Types of Singularities in Three-Body Diagrams]
        {\label{fig17}
         {\it Three-body diagrams with a) final state, b) photon initial state
          and c) proton initial state singularities.}}
\end{figure}
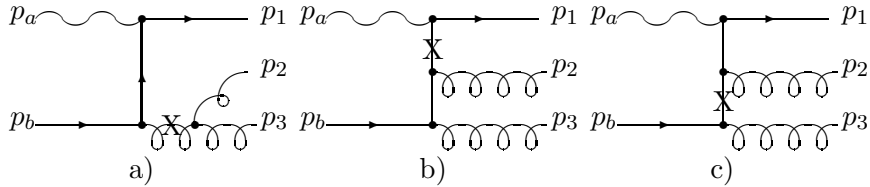

One can distinguish three classes of
singularities in photoproduction $(Q^2\simeq 0)$ depending on which
propagators in the squared matrix elements vanish. Examples for these three
classes are shown in figure \ref{fig17}.
The X marks the propagator leading to the divergence. In the first graph, it
is the invariant $s_{23}$ given by momenta of the final state, which causes
the singularity. Therefore, this divergence
will be called final state singularity. The second graph becomes
singular for $-t_{a1}=s_{a1}=(p_a+p_1)^2=0$, when the photon and the final
quark momentum are parallel. This is the class of photon initial
state singularities. In the third graph, the singularity occurs at
$-t_{b3}=s_{b3}=(p_b+p_3)^2$, where $p_b$ is the initial parton momentum in the proton.
This stands for proton initial state singularity. Since resolved photons behave
like hadrons, they produce similar initial state divergencies.
The first class is familiar from calculations for jet
production in $e^+e^-$ annihilation \cite{Fab82}, the third class
from jet production in deep inelastic $ep$ scattering ($Q^2\neq 0$)
\cite{Gra90}. The second class occurs only for direct photoproduction
\cite{Bod92}.

When squaring the sum of all relevant $2\rightarrow 3$ matrix elements, we
encounter terms, where more than one of the invariants become singular,
e.g.~when one of the gluon momenta $p_3 \rightarrow 0$ so that
$s_{23}=0$ and $-t_{b3}=s_{b3}=0$. These infrared singularities are
disentangled by a partial fractioning decomposition, so that
every term has only one vanishing denominator. This also allows
the separation of the different classes of singularities in figure
\ref{fig17}.

It turns out that the results for direct photoproduction are always
proportional to the LO cross sections involved in the hard scatter
\beq
 F,I,J_{\gamma b\rightarrow 123} = KT_{\gamma b\rightarrow 12},
\eeq
where $F$, $I$, and $J$ denote final state, photon initial state and proton
initial state contributions. For resolved photoproduction, this is no more
true in general but only for the quark-quark scattering subprocesses with
a less complicated color structure than those involving real gluons.
For gluonic processes, only parts of the leading order cross sections can be
factorized
\beq
 F,I,J_{ab\rightarrow 123} = \sum_i K^{(i)}T^{(i)}_{ab\rightarrow 12},
\eeq
where
\beq
 T_{ab\rightarrow 12} = \sum_i T^{(i)}_{ab\rightarrow 12}
\eeq
is again the full Born matrix element.

As the last step, the decomposed matrix elements have to be integrated
up to $s_{ij} \leq y s$. $y$ characterizes the region, where the 
two partons $i$ and $j$ cannot be resolved.
Then, the singular kernels $K$ and $K^{(i)}$ produce terms
$\propto 1/\eps^2$ and $1/\eps$, which will cancel against
those in the virtual diagrams or be absorbed into structure functions,
and finite corrections proportional to $\ln^2 y$, $\ln y$, and $y^0$. For
small values of $y$,
terms of ${\cal O} (y)$ can be neglected. In the following, we shall give
the results for the different classes of singularities separately.

\subsubsection{Phase Space for Three-Particle Final States}

\begin{figure}[htbp]
\begin{center}
\begin{picture}(16750,18250)
\thicklines
\put(9000,7500){\vector(0,1){9000}}
\put(9000,7500){\vector(1,1){6000}}
\put(9000,7500){\vector(-1,1){6000}}
\put(9000,7500){\vector(1,-1){6000}}
\thinlines
\put(9000,13500){\line(1,0){6000}}
\put(9000,7500){\line(1,0){7500}}
\put(15000,13500){\line(0,-1){6000}}
\put(9000,15000){\line(-4,-1){6000}}
\put(9000,7500){\line(-4,-1){6000}}
\put(3000,13500){\line(0,-1){7500}}
\put(9750,17250){$p_a$}
\put(9150,9000){$\chi^\ast$}
\put(8000,9000){$\theta$}
\put(8400,6200){$\phi$}
\put(9000,9050){\oval(3000,4500)[tl]}
\put(9000,9000){\oval(3000,3000)[tr]}
\put(9000,7050){\oval(3000,3000)[b]}
\put(10500,7050){\line(0,1){450}}
\put(15750,14250){$p_2$}
\put(1500,14250){$p_1$}
\put(15000,0){$p_3$}
\end{picture}
\end{center}
\caption[Center-of-Mass System for $2\rightarrow 3$ Scattering]
        {\label{fig16}{\it Center-of-mass system of partons $1$ and $3$
 defining the kinematic variables for the scattering of two initial into three
 final partons. $\chi^\ast$ is defined in the overall center-of-mass system
of partons $a$ and $b$.}}
\end{figure}
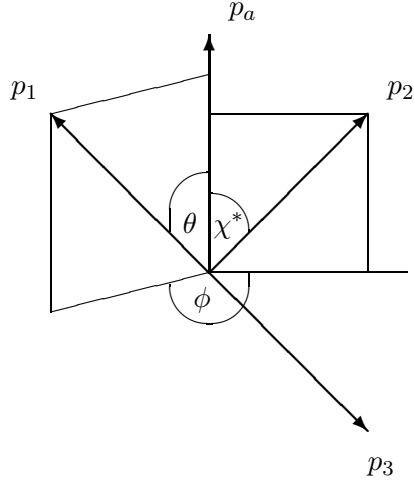

For the real corrections, we have to consider all subprocesses, which
have an additional third parton in the final state attached to the
$2\rightarrow 2$ scattering process. We therefore calculate the phase space
for $2\rightarrow 3$ scattering in this section and choose as the coordinate
system the center-of-mass system of partons $1$ and $3$ as shown in figure
\ref{fig16}.
The angle between $p_a$ and $p_1$ is called $\theta$, the azimuthal angle
between the planes defined by $p_a$ and $p_1$ and $p_a$ and $p_2$ is called
$\phi$, and the angle $\chi^\ast$ between $p_a$ and $p_2$ is defined in
the overall center-of-mass system of the incoming partons $a$ and $b$
denoted by an asterisk $(^\ast)$.
As we have to accommodate a third final particle in the scattering process, the
Mandelstam variables in this section
\bea
 s &=& (p_a+p_b)^2, \\
 t &=& (p_a-p_1-p_3)^2-2p_1p_3, \\
 u &=& (p_a-p_2)^2
\eea
differ slightly from the previously used ones, but still satisfy the relation
$s+t+u=0$ for massless partons. In the limit of soft ($p_3 = 0$)
or collinear ($p_3 \parallel p_1$) particle emission, they can however
be written in a form similar to $2\rightarrow 2$ scattering as
$t=(p_a-\overline{p}_1)^2$ and $u=(p_a-p_2)^2$. Here $\overline{p}_1=p_1+p_3$
represents the four-momentum of the recombined jet, and $p_2$ is the
four-momentum of the second jet.

We start again from the general expression \cite{Byc73}
\beq
  \mbox{dPS}^{(3)}  = \int
  (2\pi )^d \prod_{i=1}^{3} \frac{\mbox{d}^dp_i \delta (p_i^2)}
  {(2\pi )^{d-1}} \delta^d \left( p_a+p_b-\sum_{j=1}^3 p_j \right)
\eeq
and describe the final state singularity with the new variable
\beq
 z' = \frac{p_1p_3}{p_ap_b}.
\eeq
We can then insert two additional $\delta$-functions with respect to $t$ and
$z'$
\beq
  \frac{\mbox{dPS}^{(3)}}{\mbox{d}t\mbox{d}z'}  =  \int
  (2\pi )^d \prod_{i=1}^{3} \frac{\mbox{d}^dp_i \delta (p_i^2)}
  {(2\pi )^{d-1}} \delta^d \left( p_a+p_b-\sum_{j=1}^3 p_j \right)
  \delta (t+s+(p_a-p_2)^2)\delta \lr z'-\frac{2p_1p_3}{s}\rr,
\eeq
before we integrate over the $\delta(p_i^2)$ and the space-like components of
the $d$-dimensional $\delta$-function to eliminate $p_3$. The resulting
expression is
\beq
  \frac{\mbox{dPS}^{(3)}}{\mbox{d}t\mbox{d}z'}  =  \int
  \frac{\mbox{d}^{d-1}p_1\mbox{d}^{d-1}p_2}{(2\pi)^{2d-3}2E_12E_22E_3}
  \delta\lr E_a+E_b-\sum_{j=1}^3E_j\rr\delta(t+s+(p_a-p_2)^2)\delta \lr z'
  -\frac{2p_1p_3}{s}\rr.
\eeq
We now decompose $p_1$ into its energy $E_1$ and angular components
$\theta$ and $\phi$ in the center-of-mass system of partons $1$ and $3$,
and $p_2$ into its components $E_2^{\ast}$, $\chi^{\ast}$, and $\phi_2^{\ast}$
in the overall center-of-mass system with the result
\bea
  \frac{\mbox{dPS}^{(3)}}{\mbox{d}t\mbox{d}z'} &=& \int
  \frac{1}{(2\pi)^{2d-3}8E_3}\frac{2\pi^{\frac{d-3}{2}}}{\Gamma\lr\frac{d-3}{2}
  \rr}E_1^{d-3}\mbox{d}E_1\sin^{d-3}\theta\mbox{d}\theta\sin^{d-4}\phi
  \mbox{d}\phi\frac{\pi^{\frac{d-4}{2}}}{\Gamma\lr\frac{d-2}{2}\rr}
  E_2^{\ast^{d-3}}\mbox{d}E_2^\ast\\
 && \sin^{d-3}\chi^\ast\mbox{d}\chi^\ast\mbox{d}\phi_2^\ast
  \delta\lr E_a+E_b-\sum_{j=1}^3E_j\rr\delta(t+s-2E_a^\ast E_2^\ast
  (1-\cos\chi^\ast))
  \delta \lr z'-\frac{4E_1^2}{s}\rr.\nonumber
\eea
Integrating over the remaining $\delta$-functions and the trivial azimuthal
angle $\phi_2^\ast$ up to $2\pi$, we arrive at
\beq
  \frac{\mbox{dPS}^{(3)}}{\mbox{d}t\mbox{d}z'}  =  \int
  \frac{(16\pi^2)^\eps}{128\pi^3\Gamma(2-2\eps)}z'^{-\eps}
  u^{-\eps}(t+sz')^{-\eps}(b(1-b))^{-\eps}
  \frac{\mbox{d}b}{N_b}\sin^{-2\eps}\phi
  \frac{\mbox{d}\phi}{N_\phi},
\eeq
where we have substituted the polar angle $\theta$ by
\beq
 b=\frac{1}{2}(1-\cos\theta),
\eeq
and $N_b$ and $N_\phi$ are the normalization factors
\bea
  N_b & = & \int_0^1 \mbox{d}b (b(1-b))^{-\eps}
  = \frac{\Gamma ^2 (1-\eps)}{\Gamma (2-2\eps)}, \\
  N_\phi & = &\int_0^\pi \sin^{-2\eps} \phi \mbox{d}\phi
  = \frac{\pi 4^\eps \Gamma (1-2\eps)}{\Gamma^2(1-\eps)}.
  \label{eq30}
\eea
Finally, we can factorize this three particle phase space into 
\beq
 \mbox{dPS}^{(3)} = \mbox{dPS}^{(2)} \mbox{dPS}^{(r)},
\eeq
where
\beq
  \frac{\mbox{dPS}^{(2)}}{\mbox{d}t}
  = \frac{1}{\Gamma(1-\eps)}
  \left( \frac{4\pi s}{tu} \right) ^\eps
  \frac{1}{8\pi s}
\eeq
is the phase space for the two observed jets with momenta $\overline{p}_1$
and $p_2$ already calculated in section 3.1 and
\beq
  \mbox{dPS}^{(r)} =
  \left( \frac{4\pi}{s} \right) ^\eps \frac{\Gamma (1-\eps)}
  {\Gamma (1-2\eps)} \frac{s}{16 \pi ^2} \frac{1}{1-2\eps}
  \mbox{d}\mu_F
\eeq
is the phase space of the unresolved subsystem of partons $1$ and $3$. The
integration measure is
\beq
  \mbox{d}\mu_F =
  \mbox{d}z' z'^{-\eps} \left( 1+\frac{z's}{t} \right)
  ^{-\eps}
  \frac{\mbox{d}b}{N_b} b^{-\eps} (1-b)^{-\eps}
  \frac{\mbox{d}\phi}{N_\phi} \sin^{-2\eps} \phi.
\eeq
The full range of integration in
$\mbox{dPS}^{(r)}$ is given by $z' \in [0,-t/s]$, $b \in [0,1]$, and
$\phi \in [0,\pi ]$. The singular region is defined by the
requirement that partons $p_1$ and $p_3$ are recombined, i.e.~$p_3 = 0$
or $p_3$ parallel to $p_1$, so that $s_{13}=(p_1+p_3)^2=0$.
We integrate over this region up to $s_{13} \leq y s$, which restricts
the range of $z'$ to $0 \leq z' \leq \min\{ -t/s, y \} \equiv y_F$.

\subsubsection{Final State Corrections for Direct Photons}

\begin{figure}[htbp]
\begin{center}
\begin{picture}(23000,6000)
\drawline\photon[\E\REG](1000,5000)[4]
\drawline\fermion[\E\REG](1000,1000)[4000]
\drawarrow[\E\ATTIP](3000,1000)
\put(5500,3000){\oval(1000,5000)}
\drawline\gluon[\E\REG](6000,5000)[4]
\drawline\gluon[\E\REG](6000,3000)[4]
\drawline\fermion[\E\REG](6000,1000)[4000]
\drawarrow[\E\ATTIP](8000,1000)
\put(0,1000){$q$}
\put(0,5000){$\gamma$}
\put(10500,1000){$q$}
\put(10500,5000){$g$}
\put(10500,3000){$g$}
\drawline\photon[\E\REG](13000,5000)[4]
\drawline\fermion[\E\REG](13000,1000)[4000]
\drawarrow[\E\ATTIP](15000,1000)
\put(17500,3000){\oval(1000,5000)}
\drawline\fermion[\E\REG](18000,5000)[4000]
\drawarrow[\E\ATTIP](20000,5000)
\drawline\fermion[\E\REG](18000,3000)[4000]
\drawarrow[\W\ATBASE](20000,3000)
\drawline\fermion[\E\REG](18000,1000)[4000]
\drawarrow[\E\ATTIP](20000,1000)
\put(12000,1000){$q$}
\put(12000,5000){$\gamma$}
\put(22500,1000){$q$}
\put(22500,5000){$q,q'$}
\put(22500,3000){$\overline{q},\overline{q}'$}
\end{picture}
\end{center}
\caption[Generic $2\rightarrow 3$ Feynman Diagrams for Direct Photoproduction]
        {\label{fig7}
         {\it $2\rightarrow 3$ Feynman diagrams for direct photoproduction.}}
\end{figure}
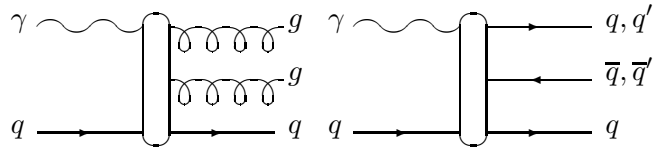

The real corrections to the QCD Compton Scattering process of figure
\ref{fig1} arise from two different mechanisms. Either an additional
gluon can be radiated from the quark or the gluon leading to the left diagram
in figure \ref{fig7},
or a quark-antiquark pair is emitted by a gluon as shown in the
right diagram of figure \ref{fig7}. Both cases lead to an extra factor of
$\alpha_s$, when the matrix elements are squared, so that all real corrections
discussed in the following
are of ${\cal O} (\alpha\alpha_s^2)$. The outgoing quark-antiquark pair can
be of the same or different flavor than the incoming quark. The diagrams
for incoming anti-quarks or gluons can be obtained from figure \ref{fig7}
by crossing a final quark or gluon line with the incoming quark line.
The corresponding matrix elements can then be obtained from table
\ref{tab4}.
\begin{table}[htbp]
\begin{center}
\begin{tabular}{|c|c|}
\hline
 Process &
 Matrix Element $\overline{|{\cal M}|^2}$ \\
\hline
\hline
 $\yqtqgg$ & $[|{\cal M}|^2_{\yqtqgg}(s,t,u)]/[4N_C]/2!$ \\
\hline
 $\ybtbgg$ & $[|{\cal M}|^2_{\yqtqgg}(s,t,u)]/[4N_C]/2!$ \\
\hline
 $\yqtqqb$ & $[|{\cal M}|^2_{\yqtqqb}(s,t,u)]/[4N_C]$ \\
\hline
 $\ygtgqb$ & $[|{\cal M}|^2_{\ygtgqb}(s,t,u)]/[8(1-\eps)N_CC_F]$ \\
\hline
\end{tabular}
\end{center}
\caption[Squared $2\rightarrow 3$ Matrix Elements for Direct Photoproduction]
        {\label{tab4}{\it Summary of $2\rightarrow 3$ squared matrix elements
         for direct photoproduction.}}
\end{table}

All possible topologies and orders of outgoing particles have to be considered
for the full matrix elements of the processes, although they are not shown here
explicitly. The complete result in $d$ dimensions can be found in \cite{Aur87}
in a very compact notation not suitable to isolate the singularities.
We therefore re-calculate the matrix elements with the help of REDUCE
\cite{Hea85}, check their sums with the results in \cite{Aur87},
and keep only those that have singularities in a final state
invariant $s_{ij}=(p_i+p_j)^2$. This can either be a soft or collinear gluon
leading to a quadratic pole (left diagram of figure \ref{fig7}) or a collinear
quark-antiquark pair leading only to a single pole (right diagram of figure
\ref{fig7}). With the help of the singular invariant
\beq
 z' = \frac{p_1p_3}{p_ap_b},
\eeq
we can approximate the nine other invariants describing the $2\rightarrow 3$
scattering process and express them through the $2\rightarrow 2$ Mandelstam
variables $s$, $t$, and $u$ and the variable $b=1/2(1-\cos\theta)$ as defined
in section 4.2.1:
\bea
 p_ap_b &=& \frac{s}{2} \\
 p_ap_1 &=& \frac{s}{2}\frac{-t}{s}b \\
 p_ap_2 &=& \frac{s}{2}\frac{-u}{s} \\
 p_ap_3 &=& \frac{s}{2}\frac{-t}{s}(1-b) \\
 p_bp_1 &=& \frac{s}{2}\frac{-u}{s}b \\
 p_bp_2 &=& \frac{s}{2}\frac{-t}{s} \\
 p_bp_3 &=& \frac{s}{2}\frac{-u}{s}(1-b) \\
 p_1p_2 &=& \frac{s}{2}b \\
 p_1p_3 &=& \frac{s}{2}z' \\
 p_2p_3 &=& \frac{s}{2}(1-b)
\eea
For the subprocess $\yqtqgg$, where a gluon becomes soft or collinear to 
the other outgoing gluon or the quark, the approximated matrix element takes
the form
\bea
  |{\cal M}|^2_{\yqtqgg}(s,t,u) &=& e^2e_q^2g^4\mu^{6\eps}
     \frac{1}{sz'}\le 4C_F\lr(1-b)
     (1-\eps)-2+\frac{-2t/s}{z'-t/s(1-b)}\rr\rp\\
  && \lp-4N_C\lr\frac{-t/s}{z'-t/s(1-b)}-\frac{-u/s}{z'-u/s(1-b)}
     -\frac{2}{z'+(1-b)}+2-b+b^2\rr\re\nn
  && T_{\yqtgq}(s,t,u) \nonumber
\eea
with an identical result for incoming anti-quarks.
The four-fermion diagram only produces a divergence from the collinear
quark-antiquark pair, which can have $N_f$ flavors
\beq
  |{\cal M}|^2_{\yqtqqb}(s,t,u) = e^2e_q^2g^4\mu^{6\eps}
      \frac{1}{sz'}N_f [1-2b(1-b)
     (1+\eps)]T_{\yqtgq}(s,t,u),
\eeq
and the gluon-initiated process $\ygtgqb$ can have a gluon that becomes
soft or collinear to the quark or anti-quark
\bea
  |{\cal M}|^2_{\ygtgqb}(s,t,u) &=& e^2e_q^2g^4\mu^{6\eps}
     \frac{1}{sz'}\le 4C_F\lr(1-b)
     (1-\eps)-2+\frac{2}{z'+(1-b)}\rr\rp\\
  && \lp-2N_C\lr\frac{2}{z'+(1-b)}-\frac{-t/s}{z'-t/s(1-b)}
     -\frac{-u/s}{z'-u/s(1-b)}\rr\re T_{\ygtqb}(s,t,u). \nonumber
\eea

Finally, we integrate the above matrix elements over the singular
phase space dPS$^{(r)}$ given in section 4.2.1.
The necessary integrals are given in appendix A.
The integration over $\phi$ in dPS$^{(r)}$ is trivial, as the matrix elements
do not depend on this variable. This is true for all direct and resolved
photoproduction processes and for all final and initial state contributions,
contrary to the case of deep-inelastic scattering. 
The result is
\beq
  \int\mbox{dPS}^{(r)}
  |{\cal M}|^2_{\gamma b\rightarrow 123} (s,t,u)  = 
  e^2e_q^2g^2 \mu^{4\eps} \frac{\alpha_s}{2\pi}
  \left( \frac{4\pi\mu^2}{s} \right) ^\eps
  \frac{\Gamma(1-\eps)}{\Gamma(1-2\eps)} F_{\gamma b
  \rightarrow 123}(s,t,u).
\eeq
The functions $F_{\gamma b\rightarrow 123}$ are given by
\bea
  F_{\yqtqgg} (s,t,u) &=& \le 2C_F\lr \edes-\edze\lr 2\ln\ts-3\rr\rp\rp\nn
  && \lp-\frac{\pi^2}{3}+\frac{7}{2}-\frac{1}{2}\ln^2\ts+2\ln y_F\ln\ts
     -2\mbox{Li}_2\lr\frac{ y_F s}{t}\rr-\ln^2 y_F-\frac{3}{2}\ln y_F
     \rr\nn
  && -N_C\lr-\frac{2}{\eps^2}-\ede\lr\frac{11}{3}+\ln\ts
     -\ln\us\rr\rp\nn
  && -\frac{1}{2}\ln^2\ts+2\ln y_F\ln\ts-2\mbox{Li}_2\lr\frac{y_Fs}
     {t}\rr-\frac{67}{9}+\frac{11}{3}\ln y_F+\ln^2 y_F\nn
  && \lp\lp+\ln^2\frac{ y_F s}{-u}-\frac{1}{2}\ln^2
     \us+ \frac{2\pi^2}{3}+ 2\mbox{Li}_2\left( \frac{ y_F s}{u} \right)
     \rr\re T_{\yqtgq} (s,t,u), \\
  F_{\yqtqqb} (s,t,u) &=& N_f \le -\frac{1}{3\eps}+\frac{1}{3}\ln y_F
  -\frac{5}{9}\re T_{\yqtgq} (s,t,u),\\
  F_{\ygtgqb} (s,t,u) & = & \le C_F\lr \frac{2}{\eps^2}
     +\frac{3}{\eps}-\frac{2\pi^2}{3}+7-2\ln^2 y_F-3\ln y_F \rr\rp \nn
  && -\frac{N_C}{2}\lr\frac{1}{\eps}\lr\ln\ts+\ln\us\rr
     +\ln^2\frac{ y_F s}{-t} +\ln^2\frac{ y_F s}{-u}-2\ln^2 y_F
     -\frac{1}{2}\ln^2\ts\rp\nn
  && \lp\lp-\frac{1}{2}\ln^2\us+2\mbox{Li}_2\lr\frac{y_Fs}{t}\rr
     +2\mbox{Li}_2\lr\frac{y_Fs}{u}\rr\rr\re T_{\ygtqb} (s,t,u). 
\eea
Terms of ${\cal O} (\eps)$ and of ${\cal O} (y_F)$ have been neglected. The
Spence-functions or dilogarithms Li$_2$ resulting from the calculation are
written down explicitly even though they are also of ${\cal O} (y_F)$. The
complete approximated $2\rightarrow 3$ matrix elements are proportional to the
corresponding Born matrix elements $T_{\gamma b\rightarrow 12}$. This
is also true for the exact matrix elements in four dimensions as calculated
by Berends et al.~\cite{Ber81}, but not for the exact $d$-dimensional
matrix elements \cite{Aur87}.

\subsubsection{Final State Corrections for Resolved Photons}

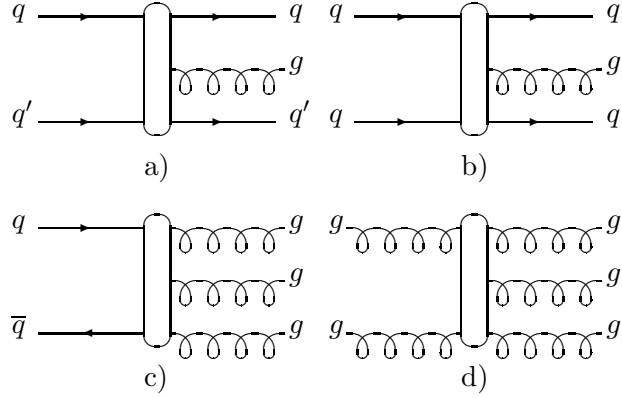
\begin{figure}[htbp]
\begin{center}
\begin{picture}(23000,15000)
\drawline\fermion[\E\REG](1000,14000)[4000]
\drawarrow[\E\ATTIP](3000,14000)
\drawline\fermion[\E\REG](1000,10000)[4000]
\drawarrow[\E\ATTIP](3000,10000)
\put(5500,12000){\oval(1000,5000)}
\drawline\fermion[\E\REG](6000,14000)[4000]
\drawarrow[\E\ATTIP](8000,14000)
\drawline\fermion[\E\REG](6000,10000)[4000]
\drawarrow[\E\ATTIP](8000,10000)
\put(0,10000){$q'$}
\put(0,14000){$q$}
\put(10500,10000){$q'$}
\put(10500,14000){$q$}
\put(5000,8000){a)}
\drawline\gluon[\E\REG](6000,12000)[4]
\put(10500,12000){$g$}
\drawline\fermion[\E\REG](13000,14000)[4000]
\drawarrow[\E\ATTIP](15000,14000)
\drawline\fermion[\E\REG](13000,10000)[4000]
\drawarrow[\E\ATTIP](15000,10000)
\put(17500,12000){\oval(1000,5000)}
\drawline\fermion[\E\REG](18000,14000)[4000]
\drawarrow[\E\ATTIP](20000,14000)
\drawline\fermion[\E\REG](18000,10000)[4000]
\drawarrow[\E\ATTIP](20000,10000)
\put(12000,10000){$q$}
\put(12000,14000){$q$}
\put(22500,10000){$q$}
\put(22500,14000){$q$}
\put(17000,8000){b)}
\drawline\gluon[\E\REG](18000,12000)[4]
\put(22500,12000){$g$}
\drawline\fermion[\E\REG](1000,6000)[4000]
\drawarrow[\E\ATTIP](3000,6000)
\drawline\fermion[\E\REG](1000,2000)[4000]
\drawarrow[\W\ATBASE](3000,2000)
\put(5500,4000){\oval(1000,5000)}
\drawline\gluon[\E\REG](6000,6000)[4]
\drawline\gluon[\E\REG](6000,2000)[4]
\put(0,2000){$\overline{q}$}
\put(0,6000){$q$}
\put(10500,2000){$g$}
\put(10500,6000){$g$}
\put(5000,0){c)}
\drawline\gluon[\E\REG](6000,4000)[4]
\put(10500,4000){$g$}
\drawline\gluon[\E\REG](12700,6000)[4]
\drawline\gluon[\E\REG](12700,2000)[4]
\put(17500,4000){\oval(1000,5000)}
\drawline\gluon[\E\REG](18000,6000)[4]
\drawline\gluon[\E\REG](18000,2000)[4]
\put(12000,2000){$g$}
\put(12000,6000){$g$}
\put(22500,2000){$g$}
\put(22500,6000){$g$}
\put(17000,0){d)}
\drawline\gluon[\E\REG](18000,4000)[4]
\put(22500,4000){$g$}
\end{picture}
\end{center}
\caption[Generic $2\rightarrow 3$ Feynman Diagrams for Resolved
         Photoproduction]
        {\label{fig8}
         {\it $2\rightarrow 3$ Feynman diagrams for resolved photoproduction.}}
\end{figure}

For the four different generic Born diagrams of resolved photoproduction
in figure \ref{fig2}, the real corrections all arise from an additional
gluon in the final state as shown in figure \ref{fig8}.
They are of ${\cal O} (\alpha_s^3)$. All other
diagrams can be obtained from figure \ref{fig8} by crossing one or two
outgoing gluon lines into the initial state, which leads to the gluon
initiated processes with quarks in the final state, or by crossing an incoming
and and outgoing quark line, which leads to the processes with incoming
anti-quarks. Process c) is symmetric under the interchange of the three gluons
and under the interchange of the quark and anti-quark. Process d) is completely
symmetric under the interchange of any of the five gluons. The complete
list of matrix elements is written down in table \ref{tab5}.
\begin{table}[htbp]
\begin{center}
\begin{tabular}{|c|c|}
\hline
 Process &
 Matrix Element $\overline{|{\cal M}|^2}$ \\
\hline
\hline
 $\qptqpg$ & $[|{\cal M}|^2_{\qptqpg}(s,t,u)]/[4N_C^2]$ \\
\hline
 $\qrtqrg$ & $[|{\cal M}|^2_{\qptqpg}(u,t,s)]/[4N_C^2]$ \\
\hline
 $\qbtrpg$ & $[|{\cal M}|^2_{\qptqpg}(t,s,u)
              +|{\cal M}|^2_{\qbtqbg}(s,t,u)]/[4N_C^2]$ \\
\hline
 $\qgtqpr$ & $-[|{\cal M}|^2_{\qbtqbg}(t,s,u)]/[8(1-\eps)N_C^2C_F]$ \\
\hline
 $\qqtqqg$ & $[|{\cal M}|^2_{\qptqpg}(s,t,u)+|{\cal M}|^2_{\qptqpg}(s,u,t)+
               |{\cal M}|^2_{\qqtqqg}(s,t,u)]/[4N_C^2]/2!$ \\
\hline
 $\qbtqbg$ & $[|{\cal M}|^2_{\qptqpg}(u,t,s)+|{\cal M}|^2_{\qptqpg}(u,s,t)+
               |{\cal M}|^2_{\qqtqqg}(u,t,s)+|{\cal M}|^2_{\qbtqbg}(s,t,u)]
 /[4N_C^2]$ \\
\hline
 $\qgtqqb$ & $-[2|{\cal M}|^2_{\qbtqbg}(t,s,u)]/[8(1-\eps)N_C^2C_F]/2!$
 \\
\hline
 $\qbtggg$ & $[|{\cal M}|^2_{\qbtggg}(s,t,u)]/[4N_C^2]/3!$ \\
\hline
 $\qgtqgg$ & $[-|{\cal M}|^2_{\qbtggg}(t,s,u)/3+|{\cal M}|^2_{\qgtqgg}(s,t,u)]
 /[8(1-\eps)N_C^2C_F]/2!$ \\
\hline
 $\bgtbgg$ & $[-|{\cal M}|^2_{\qbtggg}(t,u,s)/3+|{\cal M}|^2_{\qgtqgg}(u,t,s)]
 /[8(1-\eps)N_C^2C_F]/2!$ \\
\hline
 $\ggtqbg$ & $[-|{\cal M}|^2_{\qgtqgg}(t,s,u)+|{\cal M}|^2_{\ggtqbg}(s,t,u)]
 /[16(1-\eps)^2N_C^2C_F^2]$
 \\
\hline
 $\ggtggg$ & $[|{\cal M}|^2_{\ggtggg}(s,t,u)]/[16(1-\eps)^2N_C^2C_F^2]/
 3!$\\
\hline
\end{tabular}
\end{center}
\caption[Final State $2\rightarrow 3$ Matrix Elements for Resolved
         Photoproduction]
        {\label{tab5}{\it Summary of $2\rightarrow 3$ squared matrix elements
         for resolved photoproduction.}}
\end{table}

In four dimensions, these matrix elements were first calculated by Gottschalk
and Sivers \cite{Got80} and by Kunszt and Pietarinen \cite{Kun80}. The
result in $d$ dimensions was given in a compact form by Ellis and Sexton
\cite{Ell86}. We calculate the matrix elements again with the help of
REDUCE \cite{Hea85}, check with the result in \cite{Ell86}, and keep only
terms singular in the final state invariant $z'$. The result can be
expressed through this variable $z'$, $b$, and the Mandelstam variables $s$,
$t$, and $u$. For quark-quark scattering of different flavors in diagram
\ref{fig8}a), we can have a soft or collinear gluon in the final state
\bea
 |{\cal M}|^2_{\qptqpg}(s,t,u) &=& g^6\mu^{6\eps}
   \frac{4}{sz'}\le (2C_F-N_C)\lr \frac{-2}{z'+(1-b)}+\frac{-t/s}{z'-t/s(1-b)}
   +\frac{-u/s}{z'-u/s(1-b)}\rr\rp\nn
  && \lp+C_F\lr(1-b)(1-\eps)-2+\frac{-2u/s}{z'-u/s(1-b)}\rr\re
   T_{\qptqp}(s,t,u).
\eea
For equal flavors, there is additionally the interference contribution
\bea
 |{\cal M}|^2_{\qqtqqg}(s,t,u) &=& g^6\mu^{6\eps}
   \frac{4}{sz'}\le (2C_F-N_C)\lr \frac{-2}{z'+(1-b)}+\frac{-t/s}{z'-t/s(1-b)}
   +\frac{-u/s}{z'-u/s(1-b)}\rr\rp \nn
  && \lp+C_F\lr(1-b)(1-\eps)-2+\frac{2}{z'+(1-b)}\rr\re
   T_{\qqtqq}(s,t,u)
\eea
as the final quark lines in diagram \ref{fig8}b) are undistinguishable and
can be interchanged.
The crossed diagram of quark-antiquark annihilation has an additional
singularity, if the final quark-antiquark pair with $N_f$ possible flavors
becomes collinear
\beq
 |{\cal M}|^2_{\qbtqbg}(s,t,u) = g^6\mu^{6\eps}
   \frac{1}{sz'}N_f\le 1-2b(1-b)(1+\eps)\re
   T_{\qbtgg}(s,t,u).
\eeq
The complete matrix elements given above are proportional to the corresponding
Born matrix elements $T_{ab\rightarrow 12}$.
Quarks and anti-quarks can, of course, also annihilate into three
final gluons as shown in figure \ref{fig8}c), where any pair of two gluons
can produce a soft or collinear singularity:
\bea
 |{\cal M}|^2_{\qbtggg}(s,t,u) &=& g^6\mu^{6\eps}
     \frac{12}{sz'}\le -N_C(b^2-b+2) T_{\qbtgg}(s,t,u)
     \vphantom{\lr\frac{t^2}{tu}\rr}\rp\\
  && +4N_CC_F \nn
  && \lr\frac{-u/s}{z'-u/s(1-b)}(1-\eps)\lr\frac{t^2+u^2
     -\eps s^2}{tu}\rr\lr 2N_CC_F-2N_C^2\frac{tu}{s^2}-N_C^2\frac{u^2}
     {s^2}\rr\rp\nn
  && +\frac{1}{z'+(1-b)}(1-\eps)\lr\frac{t^2+u^2
     -\eps s^2}{tu}\rr\lr N_C^2\frac{t^2+u^2}{s^2}\rr\nn
  && \lp\lp+\frac{-t/s}{z'-t/s(1-b)}(1-\eps)\lr\frac{t^2+u^2
     -\eps s^2}{tu}\rr\lr 2N_CC_F-2N_C^2\frac{tu}{s^2}-N_C^2\frac{t^2}
     {s^2}\rr\rr\re . \nonumber
\eea
Here, only parts of the Born matrix element $T_{\qbtgg}$ can be factorized.
The process $\qgtqgg$ can be obtained by crossing a final gluon with the
incoming anti-quark or $(s\leftrightarrow t)$. In this case, either of the
outgoing gluons can also be radiated from the outgoing quark leading to
\bea
 |{\cal M}|^2_{\qgtqgg}(s,t,u) &=& g^6\mu^{6\eps}
     \frac{4}{sz'}\le C_F\lr(1-b)(1-\eps)-2\rr T_{\qgtqg}(s,t,u)
     \vphantom{\lr\frac{t^2}{tu}\rr}\rp\\
  && -4N_CC_F\nn
  && \lr\frac{-t/s}{z'-t/s(1-b)}(1-\eps)\lr\frac{s^2+u^2
     -\eps t^2}{su}\rr\lr\lr 2N_CC_F-2N_C^2\frac{su}{t^2}\rr\lr -2
     +2\frac{C_F}{N_C}\rr\rp\rp\nn
  && \lp+N_C^2\frac{s^2+u^2}{t^2}\rr\nn
  && +\frac{-u/s}{z'-u/s(1-b)}(1-\eps)\lr\frac{s^2+u^2-\eps t^2}
     {su}\rr\lr 2N_CC_F-2N_C^2\frac{su}{t^2}-N_C^2\frac{u^2}{t^2}\rr\nn
  && \lp\lp+\frac{1}{z'+(1-b)}(1-\eps)\lr\frac{s^2+u^2-\eps t^2}
     {su}\rr\lr 2N_CC_F-2N_C^2\frac{su}{t^2}-N_C^2\frac{s^2}{t^2}\rr\rr\re ,
     \nonumber
\eea
where again only parts of the Born matrix element $T_{\qgtqg}$ can be
factorized.
If also the incoming quark is crossed into the final state, the resulting
quark-antiquark pair in the final state with $N_f$ flavors can produce the
collinear singularity
\beq
 |{\cal M}|^2_{\ggtqbg}(s,t,u) = g^6\mu^{6\eps}
   \frac{1}{sz'}N_f\le 1-2b(1-b)(1+\eps)\re T_{\ggtgg}(s,t,u)
\eeq
proportional to the full Born matrix element.
Finally, the five-gluon process in figure \ref{fig8}d) can have three
different pairs of soft or collinear gluons in the final state with the result
\bea
 |{\cal M}|^2_{\ggtggg}(s,t,u) &=& g^6\mu^{6\eps}
   \frac{12}{sz'}\le -N_C(b^2-b+2) T_{\ggtgg}(s,t,u) \rp\nn
  && +16N_C^4C_F\nn
  && \lr\frac{-t/s}{z'-t/s(1-b)}(1-\eps)^2\lr 3-\frac{2su}
     {t^2}+\frac{s^4+u^4}{s^2u^2}\rr\rp\nn
  && +\frac{-u/s}{z'-u/s(1-b)}(1-\eps)^2\lr 3-\frac{2st}
     {u^2}+\frac{s^4+t^4}{s^2t^2}\rr\nn
  && \lp\lp+\frac{1}{z'+(1-b)}(1-\eps)^2\lr 3-\frac{2tu}
     {s^2}+\frac{t^4+u^4}{t^2u^2}\rr\rr\re . 
\eea
Only parts of the leading-order matrix element $T_{\ggtgg}$ can be
factorized.

Next, the approximated matrix elements have to be integrated over the final
state singularity $z'$ in phase space up to the invariant mass cut $y_F$.
Terms of order ${\cal O} (y_F)$ can be neglected if $y_F$ is chosen
sufficiently small. We use the phase space for the unobserved subsystem
dPS$^{(r)}$ from section 4.2.1 and find
\beq
  \int\mbox{dPS}^{(r)}
  |{\cal M}|^2_{ab\rightarrow 123} (s,t,u)  = 
  g^4 \mu^{4\eps} \frac{\alpha_s}{2\pi}
  \left( \frac{4\pi\mu^2}{Q^2} \right) ^\eps
  \frac{\Gamma(1-\eps)}{\Gamma(1-2\eps)} F_{ab
  \rightarrow 123}(s,t,u)
\eeq
similar to the direct case. The functions $F_{ab\rightarrow 123}$ are given
by
\bea
 F_{\qptqpg}(s,t,u) & = & \le 2(2C_F-N_C)\lr -\frac{1}{2\eps}
     (-2l(s)+l(t)+l(u))-\frac{1}{2}\lr-2l^2\lr\syf\rr\rp\rp\rp\nn
  && \lp+l^2\lr\tyf\rr+l^2\lr\uyf\rr\rr+\frac{1}{4}(-2l^2(s)+l^2(t)+l^2(u))
     +2\mbox{Li}_2\lr-\lb\frac{y_FQ^2}{s}\rb\rr\nn
  && \lp-\mbox{Li}_2\lr-\lb\frac{y_FQ^2}{t}\rb\rr-\mbox{Li}_2\lr-\lb
     \frac{y_FQ^2}{u}\rb\rr\rr \nn
  && +2C_F\lr\frac{1}{\eps^2}+\frac{1}{2\eps}(3-2l(u))
     +\frac{1}{2}l^2(u)+\frac{7}{2}-l^2\lr\uyf\rr-\frac{3}{2}\ln y_F\rp
     \nn 
  && \lp\lp-\frac{\pi^2}{3}-2\mbox{Li}_2\lr-\lb\frac{y_FQ^2}{u}\rb\rr\rr\re
     T_{\qptqp}(s,t,u), \\
 F_{\qqtqqg}(s,t,u) & = & \le 2(2C_F-N_C)\lr -\frac{1}{2\eps}
     (-2l(s)+l(t)+l(u))-\frac{1}{2}\lr-2l^2\lr\syf\rr\rp\rp\rp\nn
  && \lp+l^2\lr\tyf\rr+l^2\lr\uyf\rr\rr+\frac{1}{4}(-2l^2(s)+l^2(t)+l^2(u))
     +2\mbox{Li}_2\lr-\lb\frac{y_FQ^2}{s}\rb\rr\nn
  && \lp-\mbox{Li}_2\lr-\lb\frac{y_FQ^2}{t}\rb\rr-\mbox{Li}_2\lr-\lb
     \frac{y_FQ^2}{u}\rb\rr\rr \nn
  && +2C_F\lr\frac{1}{\eps^2}+\frac{1}{2\eps}(3-2l(s))
     +\frac{1}{2}l^2(s)+\frac{7}{2}-l^2\lr\syf\rr-\frac{3}{2}\ln y_F\rp
     \nn 
  && \lp\lp-\frac{\pi^2}{3}-2\mbox{Li}_2\lr-\lb\frac{y_FQ^2}{s}\rb\rr\rr\re
     T_{\qqtqq}(s,t,u), \\
 F_{\qbtqbg}(s,t,u) & = & N_f\le -\frac{1}{3\eps}+\frac{1}{3}\ln y_F
     -\frac{5}{9}\re T_{\qbtgg}(s,t,u),\\
 F_{\qbtggg}(s,t,u) & = & \le 3N_C\lr\frac{2}{\eps^2}+\frac{11}
     {3\eps}
     -\frac{11}{3}\ln y_F+\frac{67}{9}-\frac{2\pi^2}{3}\rr\re
     T_{\qbtgg}(s,t,u)\nn
  && +6N_CC_F\le\lr-\frac{2}{\eps}l(u)+2l(u)-2l^2\lr\frac{u}{y_F}\rr
     +l^2(u)-4\mbox{Li}_2\lr-\lb\frac{y_FQ^2}{u}\rb\rr\rr\rp\nn
  && \lr N_C^2\lr\tu-\frac{2t^2}{s^2}\rr-\ut-\tu\rr\nn
  && +\lr-\frac{2}{\eps}l(s)+2l(s)-2l^2\lr\frac{s}{y_F}\rr
     +l^2(s)-4\mbox{Li}_2\lr-\lb\frac{y_FQ^2}{s}\rb\rr\rr\nn
  && \lr N_C^2\frac{(t^2+u^2)^2}{uts^2}\rr\nn
  && +\lr-\frac{2}{\eps}l(t)+2l(t)-2l^2\lr\frac{t}{y_F}\rr
     +l^2(t)-4\mbox{Li}_2\lr-\lb\frac{y_FQ^2}{t}\rb\rr\rr\nn
  && \lr N_C^2\lr\ut-\frac{2u^2}{s^2}\rr-\ut-\tu\rr\nn
  && +\lr 2l(u)\lr N_C^2\tu-\frac{s^2}{tu}\rr+2l(s)N_C^2\lr\tu+\ut\rr
     \rp\nn
  && \lp\lp+2l(t)\lr N_C^2\ut-\frac{s^2}{tu}\rr\rr\re , \\
 F_{\qgtqgg}(s,t,u) & = & \le C_F\lr\frac{1}{\eps^2}+\frac{3}
     {2\eps}
     -\frac{3}{2}\ln y_F+\frac{7}{2}-\frac{\pi^2}{3}\rr\re
     T_{\qgtqg}(s,t,u)\nn
  && -N_CC_F\le\lr-\frac{2}{\eps}l(u)+2l(u)-2l^2\lr\frac{u}{y_F}\rr
     +l^2(u)-4\mbox{Li}_2\lr-\lb\frac{y_FQ^2}{u}\rb\rr\rr\rp\nn
  && \lr N_C^2\lr\sux-\frac{2s^2}{t^2}\rr-\usx-\sux\rr\nn
  && +\lr-\frac{2}{\eps}l(s)+2l(s)-2l^2\lr\frac{s}{y_F}\rr
     +l^2(s)-4\mbox{Li}_2\lr-\lb\frac{y_FQ^2}{s}\rb\rr\rr\nn
  && \lr N_C^2\lr\usx-\frac{2u^2}{t^2}\rr-\usx-\sux\rr\nn
  && +\lr-\frac{2}{\eps}l(t)+2l(t)-2l^2\lr\frac{t}{y_F}\rr
     +l^2(t)-4\mbox{Li}_2\lr-\lb\frac{y_FQ^2}{t}\rb\rr\rr\nn
  && \lr 2\frac{s^2+u^2}{t^2}+\frac{1}{N_C^2}\lr\usx+\sux\rr\rr\nn
  && +\lr 2l(u)\lr N_C^2\sux-\frac{t^2}{su}\rr+2l(s)\lr N_C^2\usx-\frac{t^2}
     {su}\rr\rp\nn
  && \lp\lp+2l(t)\lr 2+\frac{1}{N_C^2}\frac{t^2}{su}\rr\rr\re , \\
 F_{\ggtqbg}(s,t,u) & = & N_f\le -\frac{1}{3\eps}+\frac{1}{3}\ln y_F
     -\frac{5}{9}\re T_{\ggtgg}(s,t,u),\\
 F_{\ggtggg}(s,t,u) & = & \le 3N_C\lr\frac{2}{\eps^2}+\frac{11}
     {3\eps}
     -\frac{11}{3}\ln y_F+\frac{67}{9}-\frac{2\pi^2}{3}\rr\re
     T_{\ggtgg}(s,t,u)\nn
  && +48N_C^4C_F\le\lr -\frac{1}{\eps}l(t)+2l(t)-l^2\lr\frac{t}{y_F}\rr
     +\frac{1}{2}l^2(t)-2\mbox{Li}_2\lr-\lb\frac{y_FQ^2}{t}\rb\rr\rr\rp
     \nn
  && \lr 3-\frac{2us}{t^2}+\frac{u^4+s^4}{u^2s^2}\rr\nn
  && +\lr -\frac{1}{\eps}l(u)+2l(u)-l^2\lr\frac{u}{y_F}\rr
     +\frac{1}{2}l^2(u)-2\mbox{Li}_2\lr-\lb\frac{y_FQ^2}{u}\rb\rr\rr\nn
  && \lr 3-\frac{2ts}{u^2}+\frac{t^4+s^4}{t^2s^2}\rr\nn
  && +\lr -\frac{1}{\eps}l(s)+2l(s)-l^2\lr\frac{s}{y_F}\rr
     +\frac{1}{2}l^2(s)-2\mbox{Li}_2\lr-\lb\frac{y_FQ^2}{s}\rb\rr\rr\nn
  && \lp\lr 3-\frac{2tu}{s^2}+\frac{t^4+u^4}{t^2u^2}\rr\re . 
\eea
All terms of ${\cal O} (\eps)$ have been omitted since they do not
contribute in the physical limit $d\rightarrow 4$.

As in the virtual corrections, we account for sign-changing arguments in the
logarithms before and after crossing with the definition of
\beq
 l(x) = \ln\lb\frac{x}{Q^2}\rb,
\eeq
where $x$ is any of the Mandelstam variables $s$, $t$, and $u$ and $Q^2$
is an arbitrary scale chosen to be $Q^2=\max(s,t,u)$. However, there
are no additional terms of $\pi^2$ in the real corrections so that
\beq
 l^2(x) = \ln^2\lb \frac{x}{Q^2}\rb
\eeq
for $x>0$ as well as for $x<0$.

\subsubsection{Phase Space for Three-Particle Final States Revisited}

Having calculated all real corrections coming from final state singularities,
where the singular variable was $s_{13}=(p_1+p_3)^2$, we now turn to the
singularities that arise in the photon and proton initial states in the
variables $-t_{a3}=s_{a3}=(p_a+p_3)^2$ and $-t_{b3}=s_{b3}=(p_b+p_3)^2$. The singular variables
are then defined as
\beq
 z'' = \frac{p_ap_3}{p_ap_b}~\mbox{and}~z''' = \frac{p_bp_3}{p_ap_b},
\eeq
respectively. It is convenient to calculate the three-body phase space in the
same center-of-mass system of the two final state particles {\em now called}
$1$ {\em and} $2$ as in the case of final state corrections (see figure
\ref{fig16}), where now $p_3$ is defined in the overall center-of-mass system
of partons $a$ and $b$. However, we will not choose the angle $\theta$ between
the hard jet and the incoming photon or the related variable $b=1/2(1-\cos
\theta)$ as the second independent variable. Instead it is more convenient to
choose the fraction of the center-of-mass energy going into the hard subprocess
$ab\rightarrow 12$
\beq
 z_a = \frac{p_1p_2}{p_ap_b} \in [X_a,1]~\mbox{and}~
 z_b = \frac{p_1p_2}{p_ap_b} \in [X_b,1],
\eeq
respectively. In this way, the variables $z_a$ and $z_b$ describe the momentum
of the third unobserved particle $p_3=(1-z_a)p_a$ and $p_3=(1-z_b)p_b$. They
are bounded from below through the fractions of the initial electron and proton
energies transferred to the partons with momenta $z_ap_a$ and $z_bp_b$ in the
hard scattering
\beq
 X_a = \frac{p_1p_2}{kp_b}~\mbox{and}~
 X_b = \frac{p_1p_2}{p_ap}.
\eeq

As the three-particle phase space can be calculated for photon and proton
initial state singularities in exactly the same manner, we only consider the
case of the photon initial state in the following. The Mandelstam variables
differ slightly from those used in $2\rightarrow 2$ scattering and final state
corrections in order to accommodate the third unobserved particle radiated from
the initial state
\bea
 s &=& (z_ap_a+p_b)^2 = (p_1+p_2)^2, \\
 t &=& (z_ap_a-p_1)^2, \\
 u &=& (z_ap_a-p_2)^2.
\eea
In the limit of soft ($p_3 = 0$) or collinear ($p_3 \parallel p_a$) particle
emission, they satisfy the relation $s+t+u=sz''\rightarrow 0$ for massless
partons.

The calculation proceeds as follows: We insert three additional
$\delta$-functions with respect to $t$, $z''$, and $z_a$ into the general
expression
\beq
  \mbox{dPS}^{(3)}  = \int
  (2\pi )^d \prod_{i=1}^{3} \frac{\mbox{d}^dp_i \delta (p_i^2)}
  {(2\pi )^{d-1}} \delta^d \left( p_a+p_b-\sum_{j=1}^3 p_j \right)
\eeq
giving
\bea
  \frac{\mbox{dPS}^{(3)}}{\mbox{d}t\mbox{d}z''\mbox{d}z_a}  &=&  \int
    (2\pi )^d \prod_{i=1}^{3} \frac{\mbox{d}^dp_i \delta (p_i^2)}
    {(2\pi )^{d-1}} \delta^d \left( p_a+p_b-\sum_{j=1}^3 p_j \right)\nn
 && \delta (t-(z_ap_a-p_1)^2)\delta\lr z''-\frac{p_ap_3}{p_ap_b}\rr
    \delta\lr z_a-\frac{p_1p_2}{p_ap_b}\rr.
\eea
Next, we integrate over the $\delta(p_i^2)$ and the space-like components of
the $d$-dimensional $\delta$-function to eliminate $p_2$. In the resulting
expression
\bea
  \frac{\mbox{dPS}^{(3)}}{\mbox{d}t\mbox{d}z''\mbox{d}z_a}  &=&  \int
  \frac{\mbox{d}^{d-1}p_1\mbox{d}^{d-1}p_3}{(2\pi)^{2d-3}2E_12E_22E_3}
  \delta\lr E_a+E_b-\sum_{j=1}^3E_j\rr\nn
 && \delta(t-(z_ap_a-p_1)^2)\delta \lr z''
  -\frac{p_ap_3}{p_ap_b}\rr\delta \lr z_a-\frac{p_1p_2}{p_ap_b}\rr,
\eea
we now decompose $p_1$ into its energy and angular components in the
center-of-mass system of partons $1$ and $2$ and $p_3$ in the overall
center-of-mass system
\bea
  \frac{\mbox{dPS}^{(3)}}{\mbox{d}t\mbox{d}z''\mbox{d}z_a} &=& \int
  \frac{1}{(2\pi)^{2d-3}8E_2}\frac{2\pi^{\frac{d-3}{2}}}{\Gamma\lr\frac{d-3}{2}
  \rr}E_1^{d-3}\mbox{d}E_1\sin^{d-3}\theta\mbox{d}\theta\sin^{d-4}\phi
  \mbox{d}\phi\frac{\pi^{\frac{d-4}{2}}}{\Gamma\lr\frac{d-2}{2}\rr}
  E_3^{\ast^{d-3}}\mbox{d}E_3^\ast\nn
 && \sin^{d-3}\chi^\ast\mbox{d}\chi^\ast\mbox{d}\phi_3^\ast
  \delta\lr E_a+E_b-\sum_{j=1}^3E_j\rr\delta(t+2z_aE_aE_1
  (1-\cos\theta))\nn
 && \delta \lr z''-\frac{2z_aE_a^\ast E_3^\ast}{s}(1-\cos\chi^\ast)\rr
  \delta \lr z_a-\frac{4z_aE_1^2}{s}\rr
  .
\eea
Integrating over the remaining $\delta$-functions and the trivial azimuthal
angle $\phi_2^\ast$ up to $2\pi$, we arrive at
\beq
  \frac{\mbox{dPS}^{(3)}}{\mbox{d}t\mbox{d}z''\mbox{d}z_a}  =  \int
  \frac{(16\pi^2)^\eps}{128\pi^3\Gamma^2(1-\eps)}
  z''^{-\eps}(ut)^{-\eps}(1-z'')^{-1+2\eps}(1-z_a-z'')^
  {-\eps}z_a^{-1+\eps}\sin^{-2\eps}\phi\frac{\mbox{d}\phi}
  {N_\phi},
\eeq
where $N_\phi$ is the normalization factor given in eq.~(\ref{eq30}).
Finally, we can factorize this three particle phase space into 
\beq
 \mbox{dPS}^{(3)} = \mbox{dPS}^{(2)} \mbox{dPS}^{(r)},
\eeq
where dPS$^{(2)}$ is the usual phase space for the two observed jets $1$ and
$2$ from section 3.1 and
\beq
  \mbox{dPS}^{(r)} =
  \left( \frac{4\pi}{s} \right) ^\eps \frac{\Gamma (1-\eps)}
  {\Gamma (1-2\eps)} \frac{s}{16 \pi ^2}H(z'')
  \mbox{d}\mu_I
\eeq
is the phase space of the unresolved subsystem of partons $a$ and $3$. The
integration measure is
\beq
  \mbox{d}\mu_I =
  \mbox{d}z'' z''^{-\eps}
  \frac{\mbox{d}z_a}{z_a}\lr\frac{z_a}{1-z_a}\rr^\eps
  \frac{\mbox{d}\phi}{N_\phi} \sin^{-2\eps} \phi
  \frac{\Gamma(1-2\eps)}{\Gamma^2(1-\eps)},
\eeq
and the function
\beq
  H(z'') = (1-z'')^{-1+2\eps}\lr 1-\frac{z''}{1-z_a}\rr^{-\eps}
  = 1+{\cal O} (z'')
\eeq
can be approximated by 1 as it leads only to negligible terms of ${\cal O}
(y)$. The integration of $\mbox{dPS}^{(r)}$ over $z''\in [0,-u/s]$,
$z_a\in[X_a,1]$,
and $\phi\in [0,\pi ]$ is restricted to the singular region of $z''$, when
partons $p_a$ and $p_3$ are recombined, such that
$0 \leq z'' \leq \min\{ -u/s, y \} \equiv y_I$.

\subsubsection{Photon Initial State Corrections for Direct Photons}

Singularities from the initial direct photon arise from diagrams of type b)
in figure \ref{fig17}. The photon here splits up into a quark-antiquark pair,
and one of the two becomes a part of the photon remnant. If the (anti-)quark
is collinear to the incoming photon, a simple $1/\eps$ pole is
produced. Quadratic $1/\eps^2$ poles corresponding to soft {\em and}
collinear singularities do not exist, since there is no direct coupling to a
gluon line. In this specific case, it is therefore not necessary to
perform a partial fractioning decomposition.

The matrix elements for the $2\rightarrow 3$ processes $\yqtqgg$ and $\yqtqqb$
are computed from the generic diagrams in figure \ref{fig7} as in the case
of final state corrections in section 4.2.2. Those for the processes with
incoming anti-quarks and gluons are obtained by simple crossing of the
diagrams or Mandelstam variables according to table \ref{tab4}. We keep
only terms that are singular in the variable
\beq
 z'' = \frac{p_ap_3}{p_ap_b}
\eeq
and express the ten invariants for $2\rightarrow 3$ scattering through
$z''$, the longitudinal momentum fraction $z_a$ transferred from the photon
to the parton entering the hard subprocess, and the variables $s$, $t$, and
$u$ defined in section 4.2.4:
\bea
 p_ap_b &=& \frac{s}{2z_a} \label{eq37}\\
 p_ap_1 &=& \frac{s}{2z_a}\frac{-t}{s} \\
 p_ap_2 &=& \frac{s}{2z_a}\frac{-u}{s} \\
 p_ap_3 &=& \frac{s}{2z_a} z'' \\
 p_bp_1 &=& \frac{s}{2z_a} \frac{-u}{s}z_a \\
 p_bp_2 &=& \frac{s}{2z_a} \frac{-t}{s}z_a \\
 p_bp_3 &=& \frac{s}{2z_a} (1-z_a) \\
 p_1p_2 &=& \frac{s}{2z_a} z_a \\
 p_1p_3 &=& \frac{s}{2z_a} \frac{-t}{s}(1-z_a) \\
 p_2p_3 &=& \frac{s}{2z_a} \frac{-u}{s}(1-z_a) \label{eq38}
\eea
Under these approximations, a singular kernel can be factorized out, which
is universal for all processes of the type $\gamma b\rightarrow 123$ and
describes the splitting of a boson (in this case the photon) into two
fermions (i.e.~the quark-antiquark pair). Consequently, a parton from the
photon scatters now off the parton in the proton and the relevant
parton-parton Born matrix elements show up. In the case of
\bea
  |{\cal M}|^2_{\yqtqgg}(s,t,u) &=& e^2e_q^2g^4\mu^{6\eps}
     \frac{1}{sz''}\le z_a^2+(1-z_a)^2-\eps\re T_{\qbtgg}(s,t,u),
  \label{eq39}
\eea
it is the quark-antiquark annihilation process into two gluons. For
\bea
  |{\cal M}|^2_{\yqtqqb}(s,t,u) &=& e^2e_q^2g^4\mu^{6\eps}
     \frac{1}{sz''}\le z_a^2+(1-z_a)^2-\eps\re T_{\qbtqb}(s,t,u),
\eea
it is the process $\qbtqb$. The result for
\bea
  |{\cal M}|^2_{\ygtgqb}(s,t,u) &=& e^2e_q^2g^4\mu^{6\eps}
     \frac{1}{sz''}\le z_a^2+(1-z_a)^2-\eps\re T_{\qgtqg}(s,t,u) 
  \label{eq40}
\eea
is easily obtained by crossing $(s\leftrightarrow t)$ in the first
process and multiplying by $(-1)$.

As the matrix elements do not depend on the variable $\phi$, the integration
over $\phi$ in dPS$^{(r)}$ is as trivial as before. In addition, the
integration over the simple pole in $z''$ can also be carried out easily.
However, the integration over the momentum fraction $z_a$ involves a
convolution with the photon spectrum in the electron. Since these exist only
in a parametrized form too complicated to integrate, the $z_a$-integration
has to be done numerically and is written down explicitly in the final result
\beq
  \int\mbox{dPS}^{(r)}
  |{\cal M}|^2_{\gamma b\rightarrow 123} (s,t,u)  = 
  \int\limits_{X_a}^1\frac{\mbox{d}z_a}{z_a}
  e^2g^2 \mu^{4\eps} \frac{\alpha_s}{2\pi}
  \left( \frac{4\pi\mu^2}{s} \right) ^\eps
  \frac{\Gamma(1-\eps)}{\Gamma(1-2\eps)} I_{\gamma b
  \rightarrow 123}(s,t,u).
  \label{eq31}
\eeq
The functions $I_{\gamma b \rightarrow 123}$ are given by
\bea
  I_{\yqtqgg} (s,t,u) &=& \le -\ede\frac{1}{2N_C}P_{q\leftarrow \gamma}(z_a)
    +\frac{1}{2N_C}P_{q\leftarrow \gamma}(z_a)\ln\lr y_I\frac{1-z_a}{z_a}\rr
    +\frac{e_q^2}{2}\re T_{\qbtgg} (s,t,u),\hspace{7mm} \\
  I_{\yqtqqb} (s,t,u) &=& \le -\ede\frac{1}{2N_C}P_{q\leftarrow \gamma}(z_a)
    +\frac{1}{2N_C}P_{q\leftarrow \gamma}(z_a)\ln\lr y_I\frac{1-z_a}{z_a}\rr
    +\frac{e_q^2}{2}\re T_{\qbtqb} (s,t,u),\hspace{7mm} \\
  I_{\ygtgqb} (s,t,u) &=& \le -\ede\frac{1}{2N_C}P_{q\leftarrow \gamma}(z_a)
    +\frac{1}{2N_C}P_{q\leftarrow \gamma}(z_a)\ln\lr y_I\frac{1-z_a}{z_a}\rr
    +\frac{e_q^2}{2}\re T_{\qgtqg} (s,t,u).\hspace{7mm}
\eea
Terms of ${\cal O} (\eps)$ and ${\cal O} (y)$ have been omitted as
before.

The collinear $1/\eps$ poles are proportional to the splitting
function
\beq
 P_{q\leftarrow \gamma} (z_a) = 2N_Ce_q^2P_{q\leftarrow g}(z_a)
\eeq
for photons into quarks, where $e_q$ is the fractional charge of the quark
coupling to the photon and
\beq
 P_{q\leftarrow g}(z_a)=\frac{1}{2}\le z_a^2+(1-z_a)^2\re
\eeq
is the Altarelli-Parisi splitting function for gluons into quarks.
This function appears in the evolution equation of the photon structure
function as an inhomogeneous or so-called point-like term (see section 2.3).
Therefore, the photon initial state singularities
can be absorbed into the photon structure function. The necessary
steps are well known \cite{Bod92,Ell80}.
We define the renormalized distribution
function of a parton $a$ in the electron $F_{a/e}(X_a,M_a^2)$ as
\beq
  F_{a/e} (X_a,M_a^2) =
  \int_{X_a}^1 \frac{\mbox{d}z_a}{z_a}
  \left[ \delta_{a\gamma } \delta (1-z_a) + \frac{\alpha}{2\pi}
  R_{q\leftarrow \gamma }(z_a, M_a^2)\right] F_{\gamma /e}
  \lr\frac{X_a}{z_a}\rr
, \eeq
where $R$ has the general form
\beq
  R_{a \leftarrow \gamma } (z_a, M_a^2) =
 -\frac{1}{\eps}P_{q\leftarrow \gamma }(z_a)\frac{\Gamma (1-\eps)}
  {\Gamma (1-2\eps)} \left( \frac{4\pi\mu^2}{M_a^2} \right)
  ^\eps + C_{q\leftarrow \gamma} (z_a)
  \label{eq32}
\eeq
with $C = 0$ in the $\overline{\mbox{MS}}$ scheme.
The renormalized partonic cross section for $\gamma b \rightarrow \mbox{jets}$
is then calculated from
\beq
  \mbox{d}\sigma(\gamma b\rightarrow \mbox{jets}) = \mbox{d}\bar{\sigma}
  (\gamma b\rightarrow \mbox{jets}) - \frac{\alpha}{2\pi} \int
  \mbox{d}z_a R_{q\leftarrow \gamma }(z_a,M_a^2)
  \mbox{d}\sigma (ab\rightarrow \mbox{jets})
  . \label{eq33}
\eeq
d$\bar{\sigma} (\gamma b\rightarrow \mbox{jets})$ is the higher order
cross section before the subtraction procedure, and 
d$\sigma (ab\rightarrow \mbox{jets})$ contains the LO parton-parton
scattering matrix elements $T_{ab\rightarrow 12}(s,t,u)$.
The factor $4\pi\mu^2/M_a^2$ in eq.~(\ref{eq32}) is combined with the factor
$4\pi\mu^2/s$ in eq.~(\ref{eq31}) and leads to $M_a^2$ dependent terms
of the form
\beq
  -\frac{1}{\eps}P_{q\leftarrow \gamma }(z_a)\le
  \lr\frac{4\pi\mu^2}{s}\rr^\eps
  -\lr\frac{4\pi\mu^2}{M_a^2}\rr^\eps\re
  = -P_{q\leftarrow \gamma }(z_a) \ln\lr\frac{M_a^2}{s}\rr,
\eeq
so that the cross section after subtraction in eq.~(\ref{eq33})
will depend on the factorization scale $M_a^2$.

\subsubsection{Proton Initial State Corrections for Direct Photons}

Initial state singularities cannot only show up on the direct photon side,
but also on the proton side. The parton from the proton, which undergoes
the hard scattering, will then radiate a soft or collinear secondary parton.
A quark can, e.g., radiate a gluon as in figure \ref{fig17}c), which will then
not be observed but contributes to the proton remnant. As we can now have
soft gluons, we find not only single but also quadratic poles in $\eps$.
After singling out the matrix elements for the diagrams in figure \ref{fig7}
singular in the variable
\beq
 z''' = \frac{p_bp_3}{p_ap_b},
\eeq
we therefore have to decompose them with partial fractioning using REDUCE
\cite{Hea85}. The invariants are approximated quite similarly to the photon
initial state corrections and are given by
\bea
 p_ap_b &=& \frac{s}{2z_b} \\
 p_ap_1 &=& \frac{s}{2z_b}\frac{-t}{s}z_b \\
 p_ap_2 &=& \frac{s}{2z_b}\frac{-u}{s}z_b \\
 p_ap_3 &=& \frac{s}{2z_b} (1-z_b) \\
 p_bp_1 &=& \frac{s}{2z_b} \frac{-u}{s} \\
 p_bp_2 &=& \frac{s}{2z_b} \frac{-t}{s} \\
 p_bp_3 &=& \frac{s}{2z_b} z''' \\
 p_1p_2 &=& \frac{s}{2z_b} z_b \\
 p_1p_3 &=& \frac{s}{2z_b} \frac{-u}{s}(1-z_b) \\
 p_2p_3 &=& \frac{s}{2z_b} \frac{-t}{s}(1-z_b)
\eea

The result for the first subprocess $\yqtqgg$ turns out to be
\bea
  |{\cal M}|^2_{\yqtqgg}(s,t,u) &=& e^2e_q^2g^4\mu^{6\eps}\frac{1}{sz'''}
     \le 4C_F\lr (1-z_b)(1-\eps)-2+\frac{-2t/s}{z'''-t/s(1-z_b)}\rr\rp\nn
  && \lp -4N_C\lr\frac{-t/s}{z'''-t/s(1-z_b)}-\frac{-u/s}{z'''-u/s(1-z_b)}\rr
     \re T_{\yqtgq}(s,t,u).
\eea
Since the initial quark has to couple to the photon, it cannot vanish in the
beam pipe. The singular kernel only describes the radiation of a gluon from
the quark, factorizing the leading-order QCD Compton process $\yqtgq$. 
According to table \ref{tab4}, we find the same result for incoming
anti-quarks. In the four-fermion process
\beq
  |{\cal M}|^2_{\yqtqqb}(s,t,u) = e^2e_q^2g^4\mu^{6\eps}
     \frac{1}{sz'''}\le \frac{1+(1-z_b)^2}{z_b}-\eps z_b\re T_{\ygtqb}(s,t,u),
\eeq
the initial quark necessarily produces a collinear final state quark. Thus, we
have a non-diagonal transition of the quark-initiated process $\yqtqqb$ into
the gluon-initiated process $\ygtqb$. The other non-diagonal transition
shows up in
\beq
  |{\cal M}|^2_{\ygtgqb,1}(s,t,u) = e^2e_q^2g^4\mu^{6\eps}
     \frac{2}{sz'''}C_F\le (z_b^2+(1-z_b)^2)(1+\eps)-\eps\re T_{\yqtgq}(s,t,u),
\eeq
where the gluon-initiated process $\ygtgqb$ is transformed into the
quark-initiated process $\yqtgq$. The splitting of a gluon into a collinear
quark-antiquark pair is analogous to the splitting of the direct photon in the
last section. However, the gluon also possesses a non-abelian coupling to other
gluons, which can become soft or collinear in
\bea
  |{\cal M}|^2_{\ygtgqb,2}(s,t,u) &=& e^2e_q^2g^4\mu^{6\eps}
     \frac{2}{sz'''}N_C\le 2\lr \frac{1}{z_b}+z_b(1-z_b)-2 \rr\rp\nn
  && \lp+\frac{-t/s}{z'''-t/s(1-z_b)}-\frac{-u/s}{z'''-u/s(1-z_b)}
     \re T_{\ygtqb}(s,t,u)
\eea
and factorizes the photon-gluon fusion process $\ygtqb$.

The complete list of proton initial state corrections given above has to
be integrated over the phase space region, where parton $3$ is an unobserved
part of the proton remnant. The phase space for three-particle final states
is taken from section 4.2.4, where $(z''\leftrightarrow z''')$ and
$(z_a\leftrightarrow z_b)$ have to be interchanged. The relevant
integrals are calculated in appendix B. The result is
\beq
  \int\mbox{dPS}^{(r)}
  |{\cal M}|^2_{\gamma b\rightarrow 123} (s,t,u)  = 
  \int\limits_{X_b}^1\frac{\mbox{d}z_b}{z_b}
  e^2e_q^2g^2 \mu^{4\eps} \frac{\alpha_s}{2\pi}
  \left( \frac{4\pi\mu^2}{s} \right) ^\eps
  \frac{\Gamma(1-\eps)}{\Gamma(1-2\eps)} J_{\gamma b
  \rightarrow 123}(s,t,u),
  \label{eq34}
\eeq
where the functions $J_{\gamma b\rightarrow 123}$ are given by
\bea
  J_{\yqtqgg} (s,t,u) &=& \le 2C_F\lr-\ede\edcf P_{q\leftarrow q}(z_b)
      +\delta (1-z_b) \lr \frac{1}{\eps^2} 
      +\frac{1}{2\eps} \lr 3-2\ln \ts\rr
      +\frac{1}{2}\ln^2\ts+\pi^2\rr\rp\rp \nn
   && +1-z_b+(1-z_b) \ln \lr y_J\frac{1-z_b}{z_b}\rr
      +2R_+\lr\ts\rr-2\ln\lr\ts\lr\frac{1-z_b}{z_b}\rr ^2\rr\nn
   && \lp -2\frac{z_b}{1-z_b}\ln\lr 1+\frac{-t}{y_Js}
      \frac{1-z_b}{z_b}\rr\rr \nn
   && -N_C\lr \delta (1-z_b) \lr \frac{1}{\eps} \ln \ut
      +\frac{1}{2}\ln^2\ts-\frac{1}{2}\ln^2\us\rr
      +2R_+\lr\ts\rr -2R_+\lr\us\rr\rp\nn
   && -2\ln\lr\ts\lr\frac{1-z_b}{z_b}\rr ^2\rr
      +2\ln\lr\us\lr\frac{1-z_b}{z_b}\rr ^2\rr\nn
   && \lp\lp-2\frac{z_b}{1-z_b}\ln\lr 1+\frac{-t}{y_Js}\frac{1-z_b}{z_b}\rr 
      +2\frac{z_b}{1-z_b}\ln\lr 1+\frac{-u}{y_Js}\frac{1-z_b}{z_b}\rr\rr\re\nn
   && T_{\yqtgq}(s,t,u), \\
  J_{\yqtqqb} (s,t,u) &=& \frac{1}{2}\le -\ede\edcf P_{g\leftarrow q}(z_b)
    +\edcf P_{g\leftarrow q}(z_b)\lr\ln\lr y_J\frac{1-z_b}{z_b}\rr+1\rr\rp\nn
    && \lp-2\frac{1-z_b}{z_b}\re T_{\ygtqb} (s,t,u), \\
  J_{\ygtgqb,1} (s,t,u) &=& 2C_F\le -\ede P_{q\leftarrow g}(z_b)
    +P_{q\leftarrow g}(z_b)\lr\ln\lr y_J\frac{1-z_b}{z_b}\rr-1\rr+\frac{1}{2}
    \re\nn
    && T_{\yqtgq} (s,t,u), \\
  J_{\ygtgqb,2} (s,t,u) &=& N_C\le -\ede\frac{1}{N_C}P_{g\leftarrow g}(z_b)
     +\delta (1-z_b)\lr\edes+\ede\ednc \lr\frac{11}{6}N_C
     -\frac{1}{3}N_f\rr-\frac{1}{2\eps}\ln\tus\rp\rp\nn
  && \lp+\frac{1}{4}\ln^2\ts
     +\frac{1}{4}\ln^2\us+\pi^2\rr-2R_+\lr\ts\rr-2R_+\lr\us\rr\nn
  && +2\ln\lr\ts\lr\frac{1-z_b}{z_b}\rr^2\rr
     +2\ln\lr\us\lr\frac{1-z_b}{z_b}\rr^2\rr\nn
  && +2\frac{z_b}{1-z_b}\ln\lr 1+\frac{-t}{y_Js} \frac{1-z_b}{z_b}\rr 
     +2\frac{z_b}{1-z_b}\ln\lr 1+\frac{-u}{y_Js}\frac{1-z_b}{z_b}\rr\nn
  && \lp\lp-4\lr\frac{1-z_b}{z_b}+z_b(1-z_b)\rr\ln\lr y_J\frac{1-z_b}{z_b}\rr
     \re \lr -\frac{N_C}{4}\rr \re T_{\ygtqb} (s,t,u).
\eea
Here, we have used the function
\bea
  R_+(x) & = & \lr \frac{\ln \lr x \lr
             \frac{1-z_b}{z_b}\right)^2\right)}{1-z_b}\right)_+
\eea
as an abbreviation. As the integration over $z_b$ in eq.~(\ref{eq34}) runs
from $X_b$ to 1, the $+$-distributions \cite{Alt78} are defined as
\beq
  D_+[g] = \int_{X_b}^1 \mbox{d}z_b D(z_b) g(z_b) 
          -\int_0^1     \mbox{d}z_b D(z_b) g(1)
, \label{eq44} \eeq
where
\beq
  g(z_b) = \frac{1}{z_b} F_{b'/p}\lr\frac{X_b}{z_b}\rr h(z_b),
\eeq
and $h(z_b)$ is a regular function of $z_b$ \cite{Fur82}. This leads to
additional terms not given here explicitly when eq.~(\ref{eq44}) is transformed
so that both integrals are calculated in the range $[X_b,1]$.

Some of the $J_{\gamma b\rightarrow 123}$ contain infrared singularities
$\propto 1/\eps^2$, which must cancel against the corresponding singularities
in the virtual contributions. The singular parts are decomposed in such a
way that the Altarelli-Parisi kernels in four dimensions
\bea
  P_{q\leftarrow q} (z_b) & = &
    C_F \left[ \frac{1+z_b^2}{(1-z_b)_+} + \frac{3}{2} \delta (1-z_b)
    \right] , \\
  P_{g\leftarrow q} (z_b) & = &
    C_F \left[ \frac{1+(1-z_b)^2}{z_b} \right] , \\
  P_{g\leftarrow g} (z_b) & = &
    2 N_C \left[ \frac{1}{(1-z_b)_+}+\frac{1}{z_b}+z_b(1-z_b)-2 \re
    + \left[ \frac{11}{6}N_C-\frac{1}{3}N_f\right] \delta (1-z_b),\\
  P_{q\leftarrow g} (z_b) & = &
    \frac{1}{2} \left[ z_b^2+(1-z_b)^2 \right]
\eea
proportional to $1/\eps$ are split off. They also appear in the evolution
equations for the parton distribution functions in the proton in section 2.2.
The singular terms proportional to these kernels are absorbed as usual
into the scale dependent structure functions
\beq
  F_{b/p} (X_b,M_b^2)  =
  \int_{X_b}^1 \frac{\mbox{d}z_b}{z_b}
  \left[ \delta_{bb'} \delta (1-z_b) + \frac{\alpha_s}{2\pi}
  R'_{b\leftarrow b'} (z_b, M_b^2) \right] F_{b'/p}
  \lr\frac{X_b}{z_b}\rr
, \eeq
where $F_{b'/p}(X_b/z_b)$ is the LO structure function before
absorption of the collinear singularities and
\beq
  R'_{b \leftarrow b'} (z_b, M_b^2) =
  -\frac{1}{\eps} P_{b\leftarrow b'} (z_b) \frac{\Gamma (1-\eps)}
  {\Gamma (1-2\eps)} \left( \frac{4\pi\mu^2}{M_b^2} \right)
  ^\eps + C'_{b\leftarrow b'} (z_b) \label{eq42}
\eeq
with $C' = 0$ in the $\overline{\mbox{MS}}$ scheme. Then, the renormalized
higher order hard scattering cross section d$\sigma (\gamma b \rightarrow
$jets) is calculated from
\beq
  \mbox{d}\sigma(\gamma b\rightarrow \mbox{jets}) = \mbox{d}\bar{\sigma}
  (\gamma b\rightarrow \mbox{jets}) - \frac{\alpha_s}{2\pi} \int
  \mbox{d}z_b R'_{b\leftarrow b'}(z_b,M_b^2)
  \mbox{d}\sigma (\gamma b'\rightarrow \mbox{jets}) \label{eq43}
. \eeq
d$\bar{\sigma} (\gamma b\rightarrow \mbox{jets})$ is the higher order
cross section before the subtraction procedure, and 
d$\sigma (\gamma b'\rightarrow \mbox{jets})$ contains the lowest order
matrix elements $T_{\yqtgq}(s,t,u)$ and $T_{\ygtqb}(s,t,u)$ in $d$ dimensions.
This well
known factorization prescription \cite{Ell80,Bur80,Arn89}
removes finally all remaining
collinear singularities. It is universal and leads for all processes
to the same definition of structure functions if the choice concerning the
regular function $C'$ in (\ref{eq42}) is kept fixed. Similar to the case of
photon initial state singularities, the higher order cross sections in
(\ref{eq43})
will depend on the factorization scale $M_b$ due to terms of the form
$P_{b\leftarrow b'}(z_b) \ln (M_b^2/s)$.

\subsubsection{Photon Initial State Corrections for Resolved Photons}

We now turn back to the case of resolved photons and consider their initial
state corrections. We start with the singularities on the photonic side of the
hard scattering cross section. For direct photons, there was only one possible
singularity coming from the splitting of the photon into a quark-antiquark
pair as shown in figure \ref{fig17}b). However, resolved photons contribute
to the hard scattering like hadrons through their partonic structure. Therefore
they produce similar poles to the proton initial state singularities as in
figure \ref{fig17}c) or those in section 4.2.6. These poles can be of quadratic
strength due to the radiation of soft and collinear gluons off the quarks in
the photon.

We use the same ${\cal O} (\alpha_s^3)$ matrix elements that were already
calculated for the final state corrections of resolved photons in section
4.2.3. The relevant diagrams are those for parton-parton scattering in figure
\ref{fig8}. As stated before, these diagrams show only the generic types and
have also to be used in their crossed forms for the complete set of diagrams.
All possible processes are listed in table \ref{tab6} together with their
matrix elements, initial spin and color averages, and statistical factors.
Only the matrix elements singular in the variable $z_a''=p_ap_3/p_ap_b$ are
kept, where $p_3$ is the momentum of the parton soft or collinear to the
original parton in the photon with momentum $p_a$. Parton $3$ will then be
a part of the photon remnant. The matrix elements are expressed through
the ten approximated invariants given in eqs.~(\ref{eq37})-(\ref{eq38}) and
decomposed with the help of partial fractioning. It is then possible
to factorize them into singular kernels and parts of the leading-order
parton-parton matrix elements. 
\begin{table}[htbp]
\begin{center}
\begin{tabular}{|c|c|}
\hline
 Process &
 Matrix Element $\overline{|{\cal M}|^2}$ \\
\hline
\hline
 $\qptqpg$ & $[|{\cal M}|^2_{\qptqpg,1}(s,t,u)+|{\cal M}|^2_{\qptqpg,2}(s,t,u)]
             /[4N_C^2]$ \\
\hline
 $\qrtqrg$ & $[|{\cal M}|^2_{\qptqpg,1}(u,t,s)+|{\cal M}|^2_{\qptqpg,2}(u,t,s)]
             /[4N_C^2]$ \\
\hline
 $\qbtrpg$ & $[|{\cal M}|^2_{\qptqpg,1}(t,s,u)]/[4N_C^2]$ \\
\hline
 $\qgtqpr$ & $[ |{\cal M}|^2_{\qgtqpr}(s,t,u)
               +|{\cal M}|^2_{\qgtqpr}(u,t,s)+|{\cal M}|^2_{\qgtqpr}(t,s,u)$\\
           & $ -|{\cal M}|^2_{\qptqpg,2}(t,s,u)]/[8(1-\eps)N_C^2C_F]$ \\
\hline
 $\qqtqqg$ & $[|{\cal M}|^2_{\qptqpg,1}(s,t,u)+|{\cal M}|^2_{\qptqpg,1}(s,u,t)+
               |{\cal M}|^2_{\qqtqqg}(s,t,u)
               $ \\
           & $ +2|{\cal M}|^2_{\qptqpg,2}(s,t,u)]/[4N_C^2]/2!$ \\
\hline
 $\qbtqbg$ & $[|{\cal M}|^2_{\qptqpg,1}(u,t,s)+|{\cal M}|^2_{\qptqpg,1}(u,s,t)+
               |{\cal M}|^2_{\qqtqqg}(u,t,s)
               $ \\
           & $ +|{\cal M}|^2_{\qptqpg,2}(u,t,s)]/[4N_C^2]$ \\
\hline
 $\qgtqqb$ & $[  |{\cal M}|^2_{\qgtqpr}(s,t,u)+|{\cal M}|^2_{\qgtqpr}(s,u,t)
                +|{\cal M}|^2_{\qgtqqb}(s,t,u)$\\
           & $  +|{\cal M}|^2_{\qgtqpr}(u,t,s)+|{\cal M}|^2_{\qgtqpr}(t,u,s)
                +|{\cal M}|^2_{\qgtqqb}(u,t,s)$\\
           & $  +|{\cal M}|^2_{\qgtqpr}(t,s,u)+|{\cal M}|^2_{\qgtqpr}(u,s,t)
                +|{\cal M}|^2_{\qgtqqb}(t,s,u) $ \\
           & $ -2|{\cal M}|^2_{\qptqpg,2}(t,s,u)]/[8(1-\eps)N_C^2C_F]/2!$ \\
\hline
 $\qbtggg$ & $[|{\cal M}|^2_{\qbtggg}(s,t,u)]/[4N_C^2]/3!$ \\
\hline
 $\qgtqgg$ & $[-|{\cal M}|^2_{\qbtggg}(t,s,u)/3+|{\cal M}|^2_{\qgtqgg,1}(s,t,u)
               +|{\cal M}|^2_{\qgtqgg,2}(s,t,u)]/[8(1-\eps)N_C^2C_F]/2!$ \\
\hline
 $\bgtbgg$ & $[-|{\cal M}|^2_{\qbtggg}(t,u,s)/3+|{\cal M}|^2_{\qgtqgg,1}(u,t,s)
               +|{\cal M}|^2_{\qgtqgg,2}(u,t,s)]/[8(1-\eps)N_C^2C_F]/2!$ \\
\hline
 $\ggtqbg$ & $[-|{\cal M}|^2_{\qgtqgg,1}(t,s,u)+|{\cal M}|^2_{\ggtqbg}(s,t,u)]
             /[16(1-\eps)^2N_C^2C_F^2]$
 \\
\hline
 $\ggtggg$ & $[|{\cal M}|^2_{\ggtggg}(s,t,u)]/[16(1-\eps)^2N_C^2C_F^2]/
 3!$\\
\hline
\end{tabular}
\end{center}
\caption[Initial State $2\rightarrow 3$ Matrix Elements for Resolved
         Photoproduction]
        {\label{tab6}{\it Summary of $2\rightarrow 3$ squared matrix elements
         for resolved photoproduction.}}
\end{table}

Let us start with the process $\qptqpg$ in figure \ref{fig8}a). Here, the final
gluon can be soft or collinear to the incoming quark $a$
\bea
  |{\cal M}|^2_{\qptqpg,1}(s,t,u) &=&
     g^6\mu^{6\eps}\frac{2}{sz''}\nn
  && \le (2C_F-N_C)\lr\frac{-2}{z''+(1-z_a)}
     +\frac{-t/s}{z''-t/s(1-z_a)}+\frac{-u/s}{z''-u/s(1-z_a)}\rr\rp\nn
  && \lp +C_F\lr (1-z_a)(1-\eps)-2+\frac{-2u/s}{z''-u/s(1-z_a)}\rr\re
     T_{\qptqp}(s,t,u), \label{eq41}
\eea
so that the original quark will also participate in the Born process $\qptqp$.
However, the quark can also go into the photon remnant. It will then radiate
a gluon that scatters from the quark $b$ with different flavor
\beq
  |{\cal M}|^2_{\qptqpg,2}(s,t,u) =
     g^6\mu^{6\eps}\frac{1}{sz''}\le \frac{1+(1-z_a)^2}{z_a}-\eps z_a
     \re T_{\qgtqg}(s,t,u).
\eeq
For equal flavors in figure \ref{fig8}b), the interference contribution
\bea
  |{\cal M}|^2_{\qqtqqg}(s,t,u) &=&
     g^6\mu^{6\eps}\frac{2}{sz''}\nn
  && \le (2C_F-N_C)\lr\frac{-2}{z''+(1-z_a)}
     +\frac{-t/s}{z''-t/s(1-z_a)}+\frac{-u/s}{z''-u/s(1-z_a)}\rr\rp\nn
  && \lp +C_F\lr (1-z_a)(1-\eps)-2+\frac{2}{z''+(1-z_a)}\rr\re
     T_{\qqtqq}(s,t,u)
\eea
has the same kernel for the color factor $(2C_F-N_C)$, but a crossed version
for the color factor $C_F$, and is proportional to the Born interference
contribution $T_{\qqtqq}$. If the final gluon is crossed into the initial
state for the processes $\qgtqpr$ and $\qgtqqb$ with unlike and like quark
flavors, it can split up into a collinear quark-antiquark pair quite similar
to the splitting of direct photons in section 4.2.5. Consequently, the
singular kernels in
\beq
  |{\cal M}|^2_{\qgtqpr}(s,t,u) =
     g^6\mu^{6\eps}\frac{2}{sz''}C_F\le (z_a^2+(1-z_a)^2)(1+\eps)-\eps\re
     T_{\qptqp}(s,t,u)
\eeq
and
\beq
  |{\cal M}|^2_{\qgtqqb}(s,t,u) =
     g^6\mu^{6\eps}\frac{4}{sz''}C_F\le (z_a^2+(1-z_a)^2)(1+\eps)-\eps\re
     T_{\qqtqq}(s,t,u)
\eeq
are the same as in eqs.~(\ref{eq39})-(\ref{eq40}), and we have the quark-quark
scattering processes $\qptqp$ and $\qqtqq$ on the tree level. The additional
factor of $(1+\eps)$ is due to the averaging of initial gluon spins with
$1/(2(1-\eps))\simeq 1/2(1+\eps)$, whereas the photons were averaged just by
$1/2$. The third process to be considered is taken from figure \ref{fig8}c).
Obviously, only a gluon can go into the photon remnant so that in
\bea
  |{\cal M}|^2_{\qbtggg}(s,t,u) &=& g^6\mu^{6\eps}
     \frac{6}{sz''}\le C_F\lr(1-z_a)(1-\eps)-2\rr T_{\qbtgg}(s,t,u)
     \rp\\
  && +4N_CC_F \nn
  && \lr\frac{-u/s}{z''-u/s(1-z_a)}(1-\eps)\lr\frac{t^2+u^2
     -\eps s^2}{tu}\rr\lr 2N_CC_F-2N_C^2\frac{tu}{s^2}-N_C^2\frac{u^2}
     {s^2}\rr\rp\nn
  && +\frac{1}{z''+(1-z_a)}(1-\eps)\lr\frac{t^2+u^2
     -\eps s^2}{tu}\rr\lr\lr 2N_CC_F-2N_C^2\uts\rr\lr-2+2\frac{C_F}{N_C}
     \rr\rp\nn
  && \lp+N_C^2\frac{t^2+u^2}{s^2}\rr\nn
  && \lp\lp+\frac{-t/s}{z''-t/s(1-z_a)}(1-\eps)\lr\frac{t^2+u^2
     -\eps s^2}{tu}\rr\lr 2N_CC_F-2N_C^2\frac{tu}{s^2}-N_C^2\frac{t^2}
     {s^2}\rr\rr\re \nonumber
\eea
we find the same simple pole for the color factor $C_F$ as in eq.~(\ref{eq41})
and still have quark-antiquark annihilation into gluons in the Born process.
The double poles are, however, more complicated and factorize only parts
of the Born matrix elements. If a final gluon is crossed into the initial
state, a soft or collinear gluon can also be radiated from the initial gluon
\bea
  |{\cal M}|^2_{\qgtqgg,1}(s,t,u) &=& g^6\mu^{6\eps}
     \frac{8}{sz''}\le N_C\lr \frac{1}{z_a}+z_a(1-z_a)-2\rr
     T_{\qgtqg}(s,t,u) \rp\nn
  && -2N_CC_F \nn
  && \lr\frac{-u/s}{z''-u/s(1-z_a)}(1-\eps)\lr\frac{s^2+u^2
     -\eps t^2}{su}\rr\lr 2N_CC_F-2N_C^2\frac{su}{t^2}-N_C^2\frac{u^2}
     {t^2}\rr\rp\nn
  && +\frac{1}{z''+(1-z_a)}(1-\eps)\lr\frac{s^2+u^2
     -\eps t^2}{su}\rr\lr 2N_CC_F-2N_C^2\ust-N_C^2\frac{s^2}{t^2}\rr
     \nn
  && \lp\lp+\frac{-t/s}{z''-t/s(1-z_a)}(1-\eps)\lr\frac{s^2+u^2
     -\eps t^2}{su}\rr\lr N_C^2\frac{s^2+u^2}{t^2}\rr\rr\re,
\eea
leaving a new kernel and a partly factorizable $\qgtqg$ scattering process
behind. Alternatively, the quark can go into the photon remnant
\beq
  |{\cal M}|^2_{\qgtqgg,2}(s,t,u) =
     g^6\mu^{6\eps}\frac{1}{sz''}\le \frac{1+(1-z_a)^2}{z_a}-\eps z_a
     \re T_{\ggtgg}(s,t,u),
\eeq
which leads to a $\ggtgg$ Born process. For collinear quarks, only a single
divergence is possible. The next kernel for $\ggtqbg$ is already known from
$\qgtqpr$ and $\qgtqqb$
\beq
  |{\cal M}|^2_{\ggtqbg}(s,t,u) =
     g^6\mu^{6\eps}\frac{2}{sz''}C_F\le (z_a^2+(1-z_a)^2)(1+\eps)-\eps\re
     T_{\qgtqg}(s,t,u),
\eeq
where a gluon splits into a quark-antiquark pair, but now with a different
leading-order matrix element $\qgtqg$. Finally in
\bea
  |{\cal M}|^2_{\ggtggg}(s,t,u) &=& g^6\mu^{6\eps}
     \frac{12}{sz''}\le N_C\lr\frac{1}{z_a}+z_a(1-z_a)-2\rr
     T_{\ggtgg}(s,t,u) \rp\nn
  && +8N_C^4C_F \nn
  && \lr\frac{-t/s}{z''-t/s(1-z_a)}(1-\eps)^2\lr 3-\frac{2su}
     {t^2}+\frac{s^4+u^4}{s^2u^2}\rr\rp\nn
  && +\frac{-u/s}{z''-u/s(1-z_a)}(1-\eps)^2\lr 3-\frac{2st}
     {u^2}+\frac{s^4+t^4}{s^2t^2}\rr\nn
  && \lp\lp+\frac{1}{z''+(1-z_a)}(1-\eps)^2\lr 3-\frac{2tu}
     {s^2}+\frac{t^4+u^4}{t^2u^2}\rr\rr\re,
\eea
it is clear that the Born process must also be completely gluonic and we
can only have the gluon splitting into two gluons as in $\qgtqgg$. All these
contributions have to be considered several times and also in crossed forms
according to table \ref{tab6}.

What remains to be done is the integration over the phase space of particle
$3$ in a region, where it can be considered a part of the photon remnant.
We do so with the help of the phase space dPS$^{(r)}$ calculated in section
4.2.4 and leave the $z_a$-integration for numerics. Again, the reason is the
analytically not integrable form of the photonic parton densities that have
to be convoluted with the matrix elements. The result is
\beq
  \int\mbox{dPS}^{(r)}
  |{\cal M}|^2_{ab\rightarrow 123} (s,t,u)  = 
  \int\limits_{X_a}^1\frac{\mbox{d}z_a}{z_a}
  g^4 \mu^{4\eps} \frac{\alpha_s}{2\pi}
  \left( \frac{4\pi\mu^2}{Q^2} \right) ^\eps
  \frac{\Gamma(1-\eps)}{\Gamma(1-2\eps)} I_{ab\rightarrow 123}(s,t,u).
\eeq
The integrated matrix elements are put into the functions $I_{ab \rightarrow
123}$:
\bea
 I_{\qptqpg,1}(s,t,u) & = & \le (2C_F-N_C) \lr \delta (1-z_a)
     \lr -\frac{1}{2\eps }(-2l(s)+l(t)+l(u))\rp\rp\rp\nn
  && \lp+\frac{1}{4}(-2l^2(s)+l^2(t)+l^2(u))\rr+\frac{z_a}{(1-z_a)}_+
     (-2l(s)+l(t)+l(u))\nn
  && +2\zde\ln\lr 1+\frac{|s|}{y_IQ^2}\edz\rr-\zde\ln\lr 1+\frac{|t|}{y_IQ^2}
     \edz\rr\nn
  && \lp-\zde\ln\lr 1+\frac{|u|}{y_IQ^2}\edz\rr\rr\nn
  && +C_F\lr-\frac{1}{\eps}\frac{1}{C_F}P_{q\leftarrow q}(z_a)
     +\delta (1-z_a)\lr\frac{1}{\eps^2}+\frac{1}{2\eps}(3-2l(u))
     \rp\rp\nn
  && \lp+\frac{1}{2}l^2(u)+\pi^2\rr+1-z_a+(1-z_a)\ln\lr y_I\edz\rr
     +2R_+\lr\lb\uqs\rb\rr\nn
  && \lp\lp-2l\lr u\lr\edz\rr^2\rr-2\zde\ln\lr 1+\frac{|u|}{y_IQ^2}\edz\rr
     \rr\re \nn
  && T_{\qptqp}(s,t,u), \\
 I_{\qptqpg,2}(s,t,u) & = & \frac{1}{2}\le-\ede\edcf P_{g\leftarrow q}(z_a)
     +\edcf P_{g\leftarrow q}(z_a)\lr\ln\lr
     y_I\edz\rr+1\rr\rp \nn
  && \lp-2\edz\re T_{\qgtqg}(s,t,u), \\
 I_{\qqtqqg}(s,t,u) & = & \le (2C_F-N_C) \lr \delta (1-z_a)
     \lr -\frac{1}{2\eps }(-2l(s)+l(t)+l(u))\rp\rp\rp\nn
  && \lp+\frac{1}{4}(-2l^2(s)+l^2(t)+l^2(u))\rr+\frac{z_a}{(1-z_a)}_+
     (-2l(s)+l(t)+l(u))\nn
  && +2\zde\ln\lr 1+\frac{|s|}{y_IQ^2}\edz\rr-\zde\ln\lr 1+\frac{|t|}{y_IQ^2}
     \edz\rr\nn
  && \lp-\zde\ln\lr 1+\frac{|u|}{y_IQ^2}\edz\rr\rr\nn
  && +C_F\lr-\frac{1}{\eps}\frac{1}{C_F}P_{q\leftarrow q}(z_a)
     +\delta (1-z_a)\lr\frac{1}{\eps^2}+\frac{1}{2\eps}(3-2l(s))
     \rp\rp\nn
  && \lp+\frac{1}{2}l^2(s)+\pi^2\rr+1-z_a+(1-z_a)\ln\lr y_I\edz\rr
     +2R_+\lr\lb\sqs\rb\rr\nn
  && \lp\lp-2l\lr s\lr\edz\rr^2\rr-2\zde\ln\lr 1+\frac{|s|}{y_IQ^2}\edz\rr
     \rr\re \nn
  && T_{\qqtqq}(s,t,u), \\
 I_{\qgtqpr}(s,t,u) & = & 2C_F\le-\ede P_{q\leftarrow g}(z_a)
     +P_{q\leftarrow g}(z_a)\lr\ln\lr
     y_I\edz\rr-1\rr+\frac{1}{2}\re \nn
  && T_{\qptqp}(s,t,u), \\
 I_{\qgtqqb}(s,t,u) & = & 4C_F\le-\ede P_{q\leftarrow g}(z_a)
     +P_{q\leftarrow g}(z_a)\lr\ln\lr
     y_I\edz\rr-1\rr+\frac{1}{2}\re \nn
  && T_{\qqtqq}(s,t,u), \\
 I_{\qbtggg}(s,t,u) & = & \le 3C_F\lr-\ede\edcf P_{q\leftarrow q}(z_a)
     +\delta (1-z_a)
     \lr\edes+\frac{3}{2\eps}+\pi^2\rr\rp\rp\nn
  && \lp\lp +1-z_a+(1-z_a)\ln\lr y_I\edz\rr\rr\re T_{\qbtgg}(s,t,u)\nn
  && +3N_CC_F\le\lr\delta (1-z_a)\lr-\frac{2}{\eps}l(t)+2l(t)+l^2(t)\rr
     +4R_+\lr\lb\tqs\rb\rr\rp\rp\nn
  && \lp -4l\lr t\lr\edz\rr^2\rr-4\zde\ln\lr 1+
     \frac{|t|}{y_IQ^2}\edz\rr\rr\nn
  && \lr N_C^2\lr\ut-\frac{2u^2}{s^2}\rr-\ut-\tu\rr\nn
  && +\lr\delta (1-z_a)\lr-\frac{2}{\eps}l(s)+2l(s)+l^2(s)\rr
     +4R_+\lr\lb\sqs\rb\rr\rp\nn
  && \lp-4l\lr s\lr\edz\rr^2\rr-4\zde\ln\lr 1+
     \frac{|s|}{y_IQ^2}\edz\rr\rr\nn
  && \lr 2\frac{t^2+u^2}{s^2}+\frac{1}{N_C^2}\lr\ut+\tu\rr\rr\nn
  && +\lr\delta (1-z_a)\lr-\frac{2}{\eps}l(u)+2l(u)+l^2(u)\rr
     +4R_+\lr\lb\uqs\rb\rr\rp\nn
  && \lp-4l\lr u\lr\edz\rr^2\rr-4\zde\ln\lr 1+
     \frac{|u|}{y_IQ^2}\edz\rr\rr\nn
  && \lr N_C^2\lr\tu-\frac{2t^2}{s^2}\rr-\ut-\tu\rr\nn
  && +\delta (1-z_a)\lr 2l(t)\lr N_C^2 \ut-\frac{s^2}{tu}\rr
     +2l(s)\lr 2+\frac{1}{N_C^2}\frac{s^2}{tu}\rr\rp\nn 
  && \lp\lp+2l(u)\lr N_C^2 \tu-\frac{s^2}{tu}\rr\rr\re,\\
 I_{\qgtqgg,1}(s,t,u) & = & \le 2N_C\lr-\ede\ednc P_{g\leftarrow g}(z_a)
     \rp\rp\nn
  && +\delta (1-z_a)\lr\edes+\ede\ednc \lr\frac{11}{6}N_C
     -\frac{1}{3}N_f\rr+\pi^2\rr\nn
  && \lp\lp +2\ln\lr y_I\edz\rr\lr\frac{1}{z_a}+z_a(1-z_a)-1\rr
     \rr\re T_{\qgtqg}(s,t,u)\nn
  && -2N_CC_F\le\lr\delta (1-z_a)\lr-\frac{2}{\eps}l(t)+2l(t)+l^2(t)\rr
     +4R_+\lr\lb\tqs\rb\rr\rp\rp\nn
  && \lp -4l\lr t\lr\edz\rr^2\rr-4\zde\ln\lr 1+
     \frac{|t|}{y_IQ^2}\edz\rr\rr\nn
  && \lr N_C^2\frac{(s^2+u^2)^2}{ust^2} \rr\nn
  && +\lr\delta (1-z_a)\lr-\frac{2}{\eps}l(s)+2l(s)+l^2(s)\rr
     +4R_+\lr\lb\sqs\rb\rr\rp\nn
  && \lp-4l\lr s\lr\edz\rr^2\rr-4\zde\ln\lr 1+
     \frac{|s|}{y_IQ^2}\edz\rr\rr\nn
  && \lr N_C^2\lr\usx-\frac{2u^2}{t^2}\rr-\usx-\sux\rr\nn
  && +\lr\delta (1-z_a)\lr-\frac{2}{\eps}l(u)+2l(u)+l^2(u)\rr
     +4R_+\lr\lb\uqs\rb\rr\rp\nn
  && \lp-4l\lr u\lr\edz\rr^2\rr-4\zde\ln\lr 1+
     \frac{|u|}{y_IQ^2}\edz\rr\rr\nn
  && \lr N_C^2\lr\sux-\frac{2s^2}{t^2}\rr-\usx-\sux\rr\nn
  && +\delta (1-z_a)\lr 2l(t) N_C^2 \lr\usx+\sux\rr
     +2l(s)\lr N_C^2 \usx-\frac{t^2}{su}\rr\rp\nn 
  && \lp\lp+2l(u)\lr N_C^2 \sux-\frac{t^2}{su}\rr\rr\re,\\
 I_{\qgtqgg,2}(s,t,u) & = & \frac{1}{2}\le-\ede\edcf P_{g\leftarrow q}(z_a)
     +\edcf P_{g\leftarrow q}(z_a)\lr\ln\lr
     y_I\edz\rr+1\rr\rp \nn
  && \lp-2\edz\re T_{\ggtgg}(s,t,u), \\
 I_{\ggtqbg}(s,t,u) & = & 2C_F\le-\ede P_{q\leftarrow g}(z_a)
     +P_{q\leftarrow g}(z_a)\lr\ln\lr
     y_I\edz\rr-1\rr+\frac{1}{2}\re \nn
  && T_{\qgtqg}(s,t,u), \\
 I_{\ggtggg}(s,t,u) & = & \le 3N_C\lr-\ede\ednc P_{g\leftarrow g}(z_a)\rp\rp
     \nn
  && +\delta (1-z_a)\lr\edes+\ede\ednc \lr\frac{11}{6}N_C
     -\frac{1}{3}N_f\rr+\pi^2\rr\nn
  && \lp\lp +2\ln\lr y_I\edz\rr\lr\frac{1}{z_a}+z_a(1-z_a)-1\rr
     \rr\re T_{\ggtgg}(s,t,u)\nn
  && +12N_C^4C_F\le\lr\delta (1-z_a)\lr-\frac{2}{\eps}l(t)+4l(t)+l^2(t)\rr
     +4R_+\lr\lb\tqs\rb\rr\rp\rp\nn
  && \lp -4l\lr t\lr\edz\rr^2\rr-4\zde\ln\lr 1+
     \frac{|t|}{y_IQ^2}\edz\rr\rr\nn
  && \lr 3-\frac{2us}{t^2}+\frac{u^4+s^4}{u^2s^2}\rr\nn
  && +\lr\delta (1-z_a)\lr-\frac{2}{\eps}l(u)+4l(u)+l^2(u)\rr
     +4R_+\lr\lb\uqs\rb\rr\rp\nn
  && \lp-4l\lr u\lr\edz\rr^2\rr-4\zde\ln\lr 1+
     \frac{|u|}{y_IQ^2}\edz\rr\rr\nn
  && \lr 3-\frac{2ts}{u^2}+\frac{t^4+s^4}{t^2s^2}\rr\nn
  && +\lr \delta (1-z_a)\lr-\frac{2}{\eps}l(s)+4l(s)+l^2(s)\rr
     +4R_+\lr\lb\sqs\rb\rr\rp\nn
  && \lp-4l\lr s\lr\edz\rr^2\rr-4\zde\ln\lr 1+
     \frac{|s|}{y_IQ^2}\edz\rr\rr\nn
  && \lp\lr 3-\frac{2tu}{s^2}+\frac{t^4+u^4}{t^2u^2}\rr\re.
\eea
The absorption of the collinear poles $1/\eps$ proportional to the different
Altarelli-Parisi splitting functions is handled in a completely
analogous way as in section 4.2.6 for proton initial state corrections.
The only difference is that the poles are absorbed into the photon structure
function and not the proton structure function. As always, we omit terms
of higher order in $\eps$ and the invariant mass cut-off $y$.

\subsubsection{Proton Initial State Corrections for Resolved Photons}

Finally, the next-to-leading order ${\cal O} (\alpha_s^3)$ resolved
photoproduction cross section receives also initial state corrections
on the proton side. It is, however, not necessary to calculate these
contributions again. The relevant diagrams in figure \ref{fig8} are the same
as in the last section as is the singularity structure in figure \ref{fig17}c).
This is due to the fact that resolved photons behave like hadrons.

The proton initial state formul{\ae} are obtained from those in section 4.2.7
by interchanging $(z''\leftrightarrow z''')$ and $(z_a\leftrightarrow z_b)$,
so that we consider matrix elements that are singular in the variable
\beq
 z''' = \frac{p_bp_3}{p_ap_b}.
\eeq
The parton $b$ in the proton gives a fraction
\beq
 z_b = \frac{p_1p_2}{p_ap_b}
\eeq
of its momentum to the $2\rightarrow 2$ hard scattering process and the
rest of $(1-z_b)$ to particle $3$ in the proton remnant. The list of
approximated invariants is the same as in section 4.2.6 for proton initial
state corrections of direct photons, and the list of contributing matrix
elements is the same as in table \ref{tab6}.

The integration over the singular region of phase space gives
\beq
  \int\mbox{dPS}^{(r)}
  |{\cal M}|^2_{ab\rightarrow 123} (s,t,u)  = 
  \int\limits_{X_b}^1\frac{\mbox{d}z_b}{z_b}
  g^4 \mu^{4\eps} \frac{\alpha_s}{2\pi}
  \left( \frac{4\pi\mu^2}{Q^2} \right) ^\eps
  \frac{\Gamma(1-\eps)}{\Gamma(1-2\eps)} J_{ab\rightarrow 123}(s,t,u),
\eeq
where the $z_b$-integration is done numerically to allow for inclusion
of the proton structure function. The functions 
\beq
 J_{ab\rightarrow 123}(s,t,u)=I_{ab\rightarrow 123}(s,t,u)
\eeq
are identical to those from the last chapter. For the IR singularities and
finite contributions proportional to the $\delta(1-z_b)$-function, the
integration can of course be carried out trivially. Thus, these singularities
cancel against those from the virtual corrections for resolved
photoproduction. The collinear singularities proportional to the
Altarelli-Parisi splitting functions are now absorbed into the proton
structure functions and not the photon structure functions as before.

\subsubsection{Real Corrections for Direct $\gamma\gamma$ Scattering}

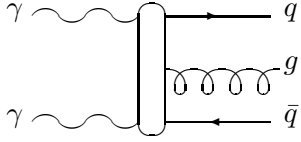
\begin{figure}[htbp]
\begin{center}
\begin{picture}(11000,6000)
\drawline\photon[\E\REG](1000,5000)[4]
\drawline\photon[\E\REG](1000,1000)[4]
\put(5500,3000){\oval(1000,5000)}
\drawline\fermion[\E\REG](6000,5000)[4000]
\drawarrow[\E\ATTIP](8000,5000)
\drawline\gluon[\E\REG](6000,3000)[4]
\drawline\fermion[\E\REG](6000,1000)[4000]
\drawarrow[\W\ATBASE](8000,1000)
\put(0,1000){$\gamma$}
\put(0,5000){$\gamma$}
\put(10500,1000){$\bar{q}$}
\put(10500,5000){$q$}
\put(10500,3000){$g$}
\end{picture}
\end{center}
\caption{\label{kkkfig3}
         {\it $2\rightarrow 3$ Feynman diagram for direct $\gamma\gamma$
          scattering.}}
\end{figure}

Real corrections to direct $\gamma\gamma$ scattering arise through the
radiation of a gluon off one of the quark lines in the underlying Born
process $\yytqb$. This can be inferred from figure \ref{kkkfig3}.
The calculation of the final and initial state singular parts proceeds along
the same lines as for direct and resolved photoproduction.

Let us first consider the final state singularity. There, the 
approximated matrix element
\beq
  |{\cal M}|^2_{\yytqbg}(s,t,u) = e^4e_q^4g^2\mu^{6\eps}
     \frac{1}{sz'} 4C_F\lr(1-b)
     (1-\eps)-2+\frac{2}{z'+(1-b)}\rr T_{\yytqb}(s,t,u)
\eeq
is integrated over the singular phase space to give
\beq
  \int\mbox{dPS}^{(r)}
  |{\cal M}|^2_{\gamma\gamma\rightarrow 123} (s,t,u)  = 
  e^4e_q^4 \mu^{4\eps} \frac{\alpha_s}{2\pi}
  \left( \frac{4\pi\mu^2}{s} \right) ^\eps
  \frac{\Gamma(1-\eps)}{\Gamma(1-2\eps)} F_{\gamma\gamma
  \rightarrow 123}(s,t,u)
\eeq
with
\beq
  F_{\yytqbg} (s,t,u)  =  C_F\lr \frac{2}{\eps^2}
     +\frac{3}{\eps}-\frac{2\pi^2}{3}+7-2\ln^2 y_F-3\ln y_F \rr
     T_{\yytqb} (s,t,u). 
\eeq

Turning to the initial state, we find the approximated matrix element
\beq
  |{\cal M}|^2_{\yytqbg}(s,t,u) = e^4e_q^4g^2\mu^{6\eps}
     \frac{1}{sz''}\le z_a^2+(1-z_a)^2-\eps\re T_{\yqtgq}(s,t,u).
\eeq
After integration, this yields
\beq
  \int\mbox{dPS}^{(r)}
  |{\cal M}|^2_{\gamma \gamma\rightarrow 123} (s,t,u)  = 
  \int\limits_{X_a}^1\frac{\mbox{d}z_a}{z_a}
  e^4e_q^2 \mu^{4\eps} \frac{\alpha_s}{2\pi}
  \left( \frac{4\pi\mu^2}{s} \right) ^\eps
  \frac{\Gamma(1-\eps)}{\Gamma(1-2\eps)} I_{\gamma \gamma
  \rightarrow 123}(s,t,u)
\eeq
with
\beq
  I_{\yytqbg} (s,t,u) = \le -\ede\frac{1}{2N_C}P_{q\leftarrow \gamma}(z_a)
    +\frac{1}{2N_C}P_{q\leftarrow \gamma}(z_a)\ln\lr y_I\frac{1-z_a}{z_a}\rr
    +\frac{e_q^2}{2}\re T_{\yqtgq} (s,t,u).\hspace{7mm}
\eeq
The pole is proportional to the Altarelli-Parisi splitting function and is
absorbed into the photon parton density.

\subsection{Finite Next-To-Leading Order Cross Sections}

We conclude this section with a summary of all singularities that appeared
in the next-to-leading order cross section of jet photoproduction. There were
three types of singularities:
\begin{itemize}
\item UV singularities in the virtual corrections
\item IR singularities in the virtual and real corrections
\item Collinear singularities in the initial state real corrections
\end{itemize}
All of them were regularized dimensionally by going from four to $d=4-2\eps$
dimensions, where the regulator $\eps$ had a positive sign for the ultraviolet
(UV) and a negative sign for the infrared (IR) divergencies.

The ultraviolet divergencies were encountered in the calculation of the
virtual diagrams in section 4.1. The diagrams had an additional inner 
``virtual'' particle and were classified into self-energy diagrams, propagator
corrections, box diagrams, and vertex corrections. The inner loop momenta
could not be observed and had to be integrated up to infinity. The
resulting UV singularities could be removed by renormalizing the fields,
couplings, gauge parameters, and masses in the Lagrangian through
multiplicative renormalization constants $Z_i$, which show up in perturbation
theory as counter terms order by order in the strong coupling $\alpha_s$.
The counter terms were not given explicitly, so that all matrix element
formul{\ae} in section 4.1 are already UV-divergence free and all fields,
couplings etc.~have to be considered physical and renormalized. As an example,
the counter term for the QCD Compton graph $\yqtgq$ has the form
\bea
  |{\cal M}|^2_{\yqtgq,CT}(s,t,u) &=& e^2e_q^2g^2\mu^{4\eps}\frac{\alpha_s}
     {2\pi}\lr\frac{4\pi\mu^2}{s}\rr^\eps\frac{\Gamma(1-\eps)}{\Gamma(1-2\eps)}
     \nn
  && \lr\frac{1}{\eps}+\ln\frac{s}{\mu^2}\rr\lr\frac{1}{3}N_f-\frac{11}{6}N_C
     \rr T_{\yqtgq}(s,t,u)
\eea
in the $\overline{\mbox{MS}}$ scheme and leads to a logarithmic dependence of
the cross section on the renormalization scale $\mu$.

The second type of singularities, IR divergencies, were produced at the lower
end of the loop integration in the virtual corrections and through soft or
collinear real particle emission in the real corrections. They were written
down explicitly throughout the last sections and have to cancel according
to the Kinoshita-Lee-Nauenberg theorem \cite{Kin62}. We demonstrate
this cancellation separately for direct and resolved photoproduction in
tables \ref{tab7} and \ref{tab8} and for direct $\gamma\gamma$ scattering in
table \ref{kkktab1}.

\begin{table}[htbp]
\begin{center}
\begin{tabular}{|c|c|c|c|}
\hline
 Process  & Color Factor & NLO Correction & Singular Parts of Matrix Elements\\
\hline
\hline
 $\yqtgq$ & $C_F$        & Virtual Corr.  & $\le-\frac{2}{\eps^2}-\frac{1}
  {\eps}(3-2l(t))\re T_{\yqtgq}(s,t,u) $\\
          &              & Final State    & $\le+\frac{1}{\eps^2}+\frac{1}
  {2\eps}(3-2l(t))\re T_{\yqtgq}(s,t,u) $\\
          &              & Initial State  & $\le+\frac{1}{\eps^2}+\frac{1}
  {2\eps}(3-2l(t))\re T_{\yqtgq}(s,t,u) $\\
\cline{2-4}
          & $N_C$        & Virtual Corr.  & $\le-\frac{1}{\eps^2}-\frac{1}
  {2\eps}\lr\frac{11}{3}-2l(s)+2l(t)-2l(u)\rr\re T_{\yqtgq}(s,t,u) $\\
          &              & Final State    & $\le+\frac{1}{\eps^2}+\frac{1}
  {2\eps}\lr\frac{11}{3}-~~l(s)+~~l(t)-~~l(u)\rr\re T_{\yqtgq}(s,t,u) $\\
          &              & Initial State  & $\le\hspace{8.5mm}     +\frac{1}
  {2\eps}\lr\hspace{6.5mm}-~~l(s)+~~l(t)-~~l(u)\rr\re T_{\yqtgq}(s,t,u) $\\
\cline{2-4}
          & $N_f$        & Virtual Corr.  & $+\frac{1}{3\eps} T_{\yqtgq}(s,t,u)
  $\\
          &              & Final State    & $-\frac{1}{3\eps} T_{\yqtgq}(s,t,u)
  $\\
\hline
\end{tabular}
\end{center}
\caption[Cancellation of IR Singularities for Direct Photoproduction]
        {\label{tab7}{\it Cancellation of IR singularities from virtual, final
 state, and initial state NLO corrections for the direct partonic subprocesses
 and different color factors.}}
\end{table}

There is only one generic $2\rightarrow 2$ diagram $\yqtgq$ in direct
photoproduction as shown in figure \ref{fig1}, from which the photon-gluon
fusion process can be deduced with the help of crossing relations. Therefore
we show only this process in table \ref{tab7}, but for the three contributing
color factors $C_F$, $N_C$, and $N_f$. The real corrections for the first two
classes come from the process $\yqtqgg$ and are equally divided among final
and initial state singularities. The last class, however, occurs only
in the splitting of the final gluon into an additional quark-antiquark pair
with $N_f$ flavors. This is the reason why no initial state singularity is
present here.

\begin{table}[htbp]
\begin{center}
\begin{tabular}{|c|c|c|c|}
\hline
 Process  & Color Factor & NLO Correction & Singular Parts of Matrix Elements\\
\hline
\hline
 $\qptqp$ & $C_F$        & Virtual Corr.  & $\le-\frac{4}{\eps^2}-\frac{1}
  {\eps}(6+8l(s)-8l(u)-4l(t))\re T_{\qptqp}(s,t,u) $\\
          &              & Final State    & $\le+\frac{2}{\eps^2}+\frac{1}
  {\eps}(3+4l(s)-4l(u)-2l(t))\re T_{\qptqp}(s,t,u) $\\
          &              & Initial State  & $\le+\frac{2}{\eps^2}+\frac{1}
  {\eps}(3+4l(s)-4l(u)-2l(t))\re T_{\qptqp}(s,t,u) $\\
\cline{2-4}
          & $N_C$        & Virtual Corr.  & $\le+\frac{1}{\eps}(4l(s)-2l(u)
  -2l(t))\re T_{\qptqp}(s,t,u) $\\
          &              & Final State    & $\le-\frac{1}{\eps}(2l(s)-~l(u)
  -~l(t))\re T_{\qptqp}(s,t,u) $\\
          &              & Initial State  & $\le-\frac{1}{\eps}(2l(s)-~l(u)
  -~l(t))\re T_{\qptqp}(s,t,u) $\\
\hline
 $\qqtqq$ & $C_F$        & Virtual Corr.  & $\le-\frac{4}{\eps^2}-\frac{1}
  {\eps}(6+4l(s)-4l(t)-4l(u))\re T_{\qqtqq}(s,t,u) $\\
          &              & Final State    & $\le+\frac{2}{\eps^2}+\frac{1}
  {\eps}(3+2l(s)-2l(t)-2l(u))\re T_{\qqtqq}(s,t,u) $\\
          &              & Initial State  & $\le+\frac{2}{\eps^2}+\frac{1}
  {\eps}(3+2l(s)-2l(t)-2l(u))\re T_{\qqtqq}(s,t,u) $\\
\cline{2-4}
          & $N_C$        & Virtual Corr.  & $\le+\frac{2}{\eps}(2l(s)-l(t)
  -l(u))\re T_{\qqtqq}(s,t,u) $\\
          &              & Final State    & $\le-\frac{1}{\eps}(2l(s)-l(t)
  -l(u))\re T_{\qqtqq}(s,t,u) $\\
          &              & Initial State  & $\le-\frac{1}{\eps}(2l(s)-l(t)
  -l(u))\re T_{\qqtqq}(s,t,u) $\\
\hline
 $\qbtgg$ & $C_F$        & Virtual Corr.  & $\le-\frac{2}{\eps^2}-\frac{3}
  {\eps}\re T_{\qbtgg}(s,t,u) $\\
          &              & Initial State  & $\le+\frac{2}{\eps^2}+\frac{3}
  {\eps}\re T_{\qbtgg}(s,t,u) $\\
\cline{2-4}
          & $N_C$        & Virtual Corr.  & $\le-\frac{2}{\eps^2}-\frac{11}
  {3\eps}\re T_{\qbtgg}(s,t,u) $\\
          &              & Final State    & $\le+\frac{2}{\eps^2}+\frac{11}
  {3\eps}\re T_{\qbtgg}(s,t,u) $\\
\cline{2-4}
          & $N_f$        & Virtual Corr.  & $+\frac{2}{3\eps} T_{\qbtgg}(s,t,u)
  $\\
          &              & Final State    & $-\frac{1}{3\eps} T_{\qbtgg}(s,t,u)
  $\\
          &              & Initial State  & $-\frac{1}{3\eps} T_{\qbtgg}(s,t,u)
  $\\
\cline{2-4}
          & $1$          & Virtual Corr.  & $+\frac{1}{\eps}l(s)\lr\lr 4N_C^3
  C_F+\frac{4C_F}{N_C}\rr\frac{t^2+u^2}{tu}-16N_C^2C_F^2\frac{t^2+u^2}{s^2}\rr
  $ \\
          &              & Final State    & $-\frac{1}{2\eps}l(s)\lr\lr 4N_C^3
  C_F+\frac{4C_F}{N_C}\rr\frac{t^2+u^2}{tu}-16N_C^2C_F^2\frac{t^2+u^2}{s^2}\rr
  $ \\
          &              & Initial State  & $-\frac{1}{2\eps}l(s)\lr\lr 4N_C^3
  C_F+\frac{4C_F}{N_C}\rr\frac{t^2+u^2}{tu}-16N_C^2C_F^2\frac{t^2+u^2}{s^2}\rr
  $ \\
\cline{2-4}
          & $8N_C^3C_F$  & Virtual Corr.  & $+\frac{1}{\eps}\lr l(t)\lr\utx
  -\frac{2u^2}{s^2}\rr+l(u)\lr\tux-\frac{2t^2}{s^2}\rr\rr $\\
          &              & Final State    & $-\frac{1}{2\eps}\lr l(t)\lr\utx
  -\frac{2u^2}{s^2}\rr+l(u)\lr\tux-\frac{2t^2}{s^2}\rr\rr $\\
          &              & Initial State  & $-\frac{1}{2\eps}\lr l(t)\lr\utx
  -\frac{2u^2}{s^2}\rr+l(u)\lr\tux-\frac{2t^2}{s^2}\rr\rr $\\
\cline{2-4}
          & $8N_CC_F$    & Virtual Corr.  & $-\frac{1}{\eps}\lr\utx+\tux\rr
  (l(t)+l(u)) $\\
          &              & Final State    & $+\frac{1}{2\eps}\lr\utx+\tux\rr
  (l(t)+l(u)) $\\
          &              & Initial State  & $+\frac{1}{2\eps}\lr\utx+\tux\rr
  (l(t)+l(u)) $\\
\hline
 $\ggtgg$ & $N_C$        & Virtual Corr.  & $\le-\frac{4}{\eps^2}-\frac{22}
  {3\eps}\re T_{\ggtgg}(s,t,u) $\\
          &              & Final State    & $\le+\frac{2}{\eps^2}+\frac{11}
  {3\eps}\re T_{\ggtgg}(s,t,u) $\\
          &              & Initial State  & $\le+\frac{2}{\eps^2}+\frac{11}
  {3\eps}\re T_{\ggtgg}(s,t,u) $\\
\cline{2-4}
          & $N_f$        & Virtual Corr.  & $+\frac{4}{3\eps}T_{\ggtgg}(s,t,u)
  $\\
          &              & Final State    & $-\frac{2}{3\eps}T_{\ggtgg}(s,t,u)
  $\\
          &              & Initial State  & $-\frac{2}{3\eps}T_{\ggtgg}(s,t,u)
  $\\
\cline{2-4}
          & $32N_C^4C_F$ & Virtual Corr.  & $+\frac{1}{\eps} \lr l(s)\lr 3
  -2\tus+\frac{t^4+u^4}{t^2u^2}\rr + \mbox{cycl.~perm.} \rr $\\
          &              & Final State    & $-\frac{1}{2\eps}\lr l(s)\lr 3
  -2\tus+\frac{t^4+u^4}{t^2u^2}\rr + \mbox{cycl.~perm.} \rr $\\
          &              & Initial State  & $-\frac{1}{2\eps}\lr l(s)\lr 3
  -2\tus+\frac{t^4+u^4}{t^2u^2}\rr + \mbox{cycl.~perm.} \rr $\\
\hline
\end{tabular}
\end{center}
\caption[Cancellation of IR Singularities for Resolved Photoproduction]
        {\label{tab8}{\it Cancellation of IR singularities from virtual, final
state, and initial state NLO corrections for the resolved partonic subprocesses
and different color factors.}}
\end{table}

For resolved photoproduction, we have the four generic processes in figure
\ref{fig2}. They are presented in table \ref{tab8} and divided further into
color factor classes. All other processes can be obtained through crossing.
The real corrections for quark-quark scattering with different and like
flavors arise simply from the emission of an additional gluon and factorize
the complete Born matrix element. Final and initial state corrections
contribute at equal parts. For processes involving more gluons, the situation
is more complex. In $\qbtgg$, an additional gluon leads to different color
factors depending on whether it is radiated in the initial state $(C_F)$ or
the final state $(N_C)$. A final gluon can also split up into $N_f$ flavors
accounting for half of the real corrections in the class $N_f$. The other
half comes from the process $\qgtqgg$, where the initial gluon splits up into
a quark-antiquark pair with $N_f$ flavors. So far, the factorization property
of $\qbtgg$ holds. However, the logarithmic contributions to the emission of
a third initial or final gluon in the color classes $1$, $8N_C^3C_F$, and
$8N_CC_F$ only factorize parts of the leading order cross section. In the
completely gluonic process $\ggtgg$ it does not matter where the third gluon
is radiated (color class $N_C$). Still, a final gluon can split up into
$N_f$ flavors or an initial gluon can have come from $N_f$ different quarks.
Finally, the logarithmic contributions proportional to $32N_C^4C_F$ only
factorize parts of the Born cross section, but are symmetric under cyclic
permutations of the Mandelstam variables as a completely symmetric process
must be.

\begin{table}[htbp]
\begin{center}
\begin{tabular}{|c|c|c|c|}
\hline
 Process  & Color Factor & NLO Correction & Singular Parts of Matrix Elements\\
\hline
\hline
 $\yytqb$ & $C_F$        & Virtual Corr.  & $\le-\frac{2}{\eps^2}-\frac{1}
  {\eps}(3-2l(t))\re T_{\yytqb}(s,t,u) $\\
          &              & Final State    & $\le+\frac{2}{\eps^2}+\frac{1}
  {\eps}(3-2l(t))\re T_{\yytqb}(s,t,u) $\\
\hline
\end{tabular}
\end{center}
\caption{\label{kkktab1}
 {\it Cancellation of IR singularities from virtual and final
 state NLO corrections for direct $\gamma\gamma$
 scattering.}}
\end{table}

For direct $\gamma\gamma$ scattering, there is only one Born matrix element
$\yytqb$ and only one color class $C_F$. The virtual singularities are
canceled by the final state singularities alone as shown in table
\ref{kkktab1}.

The third and last class of singularities are those in the initial state from
collinear real particle emission. Although they are generally classified as
infrared singularities as well, they are not included in the tables \ref{tab7},
\ref{tab8}, and \ref{kkktab1}
above. These single poles proportional to the Altarelli-Parisi
splitting functions $P_{q\leftarrow q}$(z), $P_{g\leftarrow q}(z)$, $P_{q
\leftarrow g}(z)$, and $P_{g\leftarrow g(z)}$ do not cancel against similar
poles from virtual corrections. They are absorbed into the renormalized photon
and proton structure functions according to the $\overline{\mbox{MS}}$ scheme
and leave behind a logarithmic dependence of the hard cross section on the
factorization scales $M_a$ and $M_b$.

The finite next-to-leading order cross section
for the photoproduction of two jets was given in section 2.5 as
\beq
  \frac{\mbox{d}^3\sigma}{\mbox{d}E_T^2\mbox{d}\eta_1\mbox{d}\eta_2}
  = \sum_b x_a F_{a/e}(x_a,M_a^2) x_b F_{b/p}(x_b,M_b^2)
  \frac{\mbox{d}\sigma}{\mbox{d}t}(ab \rightarrow p_1p_2)
\eeq
with the partonic cross section
\beq
  \frac{\mbox{d}\sigma}{\mbox{d}t} (ab \rightarrow p_1p_2) =
  \frac{1}{2s} \overline{|{\cal M}|^2}\frac{\mbox{dPS}^{(2)}}{\mbox{d}t}.
\eeq
We can now return to the physical four dimensions letting $\eps\rightarrow 0$
everywhere in the calculation.
The strong coupling constant $\alpha_s$ and the parton densities in the
electron $F_{a/e}(x_a,M_a^2)$ and in the proton $F_{b/p}(x_b,M_b^2)$ are
renormalized and defined in the $\overline{\mbox{MS}}$ scheme.
\setcounter{equation}{0}

\section{Numerical Results for Photoproduction}

In this section we present numerical results first
for single and dijet NLO cross
sections in complete photoproduction, as they were defined in section 2.5. We
use the analytical results that have been calculated in leading order in
section 3 and in next-to-leading order in section 4. All UV, IR, and collinear
initial state singularities canceled or could be removed through a
renormalization procedure, leading to finite results in four dimensions.
Equivalent numerical results for single and dijet NLO cross sections in
complete $\gamma\gamma$ collision are presented in the next chapter.

The cross section formul{\ae} are implemented in a  FORTRAN computer
program, which consists of four main parts as explained in table \ref{tab9}.
\begin{table}[htbp]
\begin{center}
\begin{tabular}{|c|c|c|}
\hline
 Part No. & Number of Jets & Contributions \\
\hline
\hline
 1        & 2              & Analytical contributions in LO and NLO \\
\hline
 2        & 2              & Numerical contributions in jet cone 1 \\
\hline
 3        & 2              & Numerical contributions in jet cone 2 \\
\hline
 4        & 3              & Numerical contributions outside jet cones \\
\hline
\end{tabular}
\end{center}
\caption[Organization of the FORTRAN Program]
        {\label{tab9}{\it Organization of the  FORTRAN program into
         four main parts.}}
\end{table}
The first part contains the analytical results of sections 3 and 4, and the
numerical parts are needed for an implementation of the Snowmass jet definition
(see section 5.1). Obviously, the third part is only needed for two-jet cross
sections. For one-jet cross sections, this region of phase space is included
in part four.

The main task of the computer program is to integrate over different
parameters. For example, total or partly differential cross sections are
integrated over $E_T$, $\eta_1$ and/or $\eta_2$. Furthermore, the momentum
fractions of the initial photon in the electron $x_a$, the parton in the photon
$y_a$, and the parton in the proton $x_b$ have to be integrated. Additional
momentum fractions $z_a$ and $z_b$ appear in the initial state corrections.
Finally, the numerical contributions have to be integrated over the phase
space of the third parton $E_{T_3}$, $\eta_3$, and $\phi_3$. The two
$\delta$-functions for the momentum fractions of the partons going into the
hard $2\rightarrow 2$ scattering $X_a$ and $X_b$,
\beq
 \delta \lr X_a -\frac{1}{2E_e}\sum_i E_{T_i}e^{-\eta_i} \rr~\mbox{and}~
 \delta \lr X_b -\frac{1}{2E_p}\sum_i E_{T_i}e^{~\eta_i} \rr,
\eeq
reduce the total number of integrations by two. The non-trivial task of
computing up to seven-dimensional integrals is solved with the Monte Carlo
routine VEGAS written by G.P. Lepage \cite{Lep78}, which adapts the spacing
of the integration bins to the size of the integrand in the bin.

The input parameters for our predictions will be kept constant throughout
this chapter, if not stated otherwise. All cross sections are for HERA
conditions, where electrons of energy $E_e=26.7$~GeV collide with protons
of energy $E_p=820$~GeV. Positive rapidities $\eta$ correspond to the proton
direction. We use the Weizs\"acker-Williams approximation of eq.~(\ref{eq45})
with $Q_{\rm max}^2 = 4~\mbox{GeV}^2$ as in the ZEUS experiment, but
do not restrict the range of longitudinal photon momentum $x_a$ in the
electron. For the parton densities in the proton, we choose the
next-to-leading order parametrization  CTEQ3 in the $\overline{\mbox{MS}}$
scheme, which already includes HERA deep inelastic scattering data.
The corresponding $\Lambda$ value of
$\Lambda^{(4)}=239$~MeV is also used in the two-loop calculation of the strong
coupling constant $\alpha_s$ with four flavors. We do not use the one-loop
approximation for leading order calculations, because the effects of the
next-to-leading order hard scattering contributions are then isolated. The
parton densities in the photon are taken from the NLO fit of GRV and
transformed from the DIS$_{\gamma}$ into the $\overline{\mbox{MS}}$ scheme. The
renormalization and factorization scales are equal to $E_T$. We use the
Snowmass jet definition with $R=1$, no jet double counting, and no
$R_{\rm sep}$
parameter. This makes the cross sections independent of the phase space
slicing parameter $y$, which is fixed at $y=10^{-3}$. This value is
sufficiently small to justify the omission of ${\cal O} (y)$ terms in the
analytical calculation (see section 5.1).

In section 5.1, we check our analytical results with respect to $y$-cut
independence and with two different existing programs for direct and resolved
one-jet production. Section 5.2 contains studies of the dependence of the
cross sections on renormalization and factorization scales and on various
cancellation mechanisms. Theoretical predictions for one- and two-jet
cross sections are presented in sections 5.3 and 5.4, before we conclude
this section with a comparison of our calculation to data from the H1 and ZEUS
collaborations in section 5.5.

\subsection{Check of the Analytical Results}

The first check of our analytical results for the photoproduction of jets in
next-to-leading order of QCD has already been
given in section 4.3. There it has been
shown that all infrared divergencies cancel in a consistent way between the
virtual and the real corrections, giving a finite cross section in four
dimensions. We have therefore missed none of the singular terms. This
cancellation mechanism could only work since we integrated the $2\rightarrow 3$
matrix elements over regions, where two final state particles or an initial
and a final state particle had an invariant mass $s_{ij}=(p_i+p_j)^2$ smaller
than a fraction $y_{F,I,J}$ of the center-of-mass energy $s$. Naturally,
this leads to a dependence of the cross sections on these phase space slicing
parameters $y_{F,I,J}$. We choose $y_{\rm cut}=y_F=y_I=y_J$
in the following.

One possibility to deal with the dependence of the cross section on the
invariant mass cut-off $y_{\rm cut}$ is to use it as a definition for the
experimentally observed jets. As we noted already in section 2.4, the
experimental jet definition has only a correspondence in theory beyond the
leading order. Furthermore, we mentioned a special kind of algorithm, i.e.~the
JADE cluster algorithm, which uses exactly the invariant mass criterion to
cluster hadrons into jets. For $e^+e^-$-experiments, one can then identify
the theoretical (partonic) with the experimental (hadronic) cut-off
\beq
 y_{\rm cut}^{\rm theory} = y_{\rm cut}^{\rm experiment}.
\eeq
In this case, the corresponding cross sections are for {\em exclusive} dijet
production.

The numerical value of $y_{\rm cut}$ is constrained in a two-fold manner.
First, the real corrections contain simple and quadratic logarithmic terms $\ln
y_{\rm cut},~\ln^2 y_{\rm cut}$, coming from the single and quadratic $1/\eps$
poles, which force the cross section to become negative for $y_{\rm cut} <
10^{-2}$. Therefore one should choose $y_{\rm cut} \geq 10^{-2}$.
Second, we have always assumed the singular region to be small in the real
corrections. This means we have omitted terms like $y_{\rm cut} \ln y_{\rm
cut},~y_{\rm cut},$ ... This forces us to take $y_{\rm cut} \ll 1$. The
typical value for the JADE algorithm is therefore given by $y_{\rm cut} \simeq
10^{-2}$.

The omitted terms of ${\cal O} (y_{\rm cut})$ can be calculated numerically to
improve the precision of the calculation. This makes it necessary to include
the full and un-approximated $2\rightarrow 3$ matrix elements in the FORTRAN
program, which are then integrated numerically in the region between the
analytical cut-off $y_{\rm cut}^{\rm theory}$ and the experimental cut-off
$y_{\rm cut}^{\rm experiment}$. It is important to note that the full
$2\rightarrow 3$ matrix elements have to be integrated {\em after partial
fractioning}. Regions, where a single invariant is small but is not a pole
in the matrix element, cannot be neglected. The theoretical invariant mass
cut is then reduced to a purely technical variable, on which the physical
prediction will no more depend. If one integrates over the phase space outside
the experimental cut-off as well, one arrives at {\em inclusive} dijet cross
sections. In this case, the full $2\rightarrow 3$ matrix elements contribute as
a leading-order process through the production of three jets, where the third
jet is unobserved.

% Plot9
\begin{figure}[p]
 \begin{center}
  {\unitlength1cm
  \begin{picture}(12,8)
   \epsfig{file=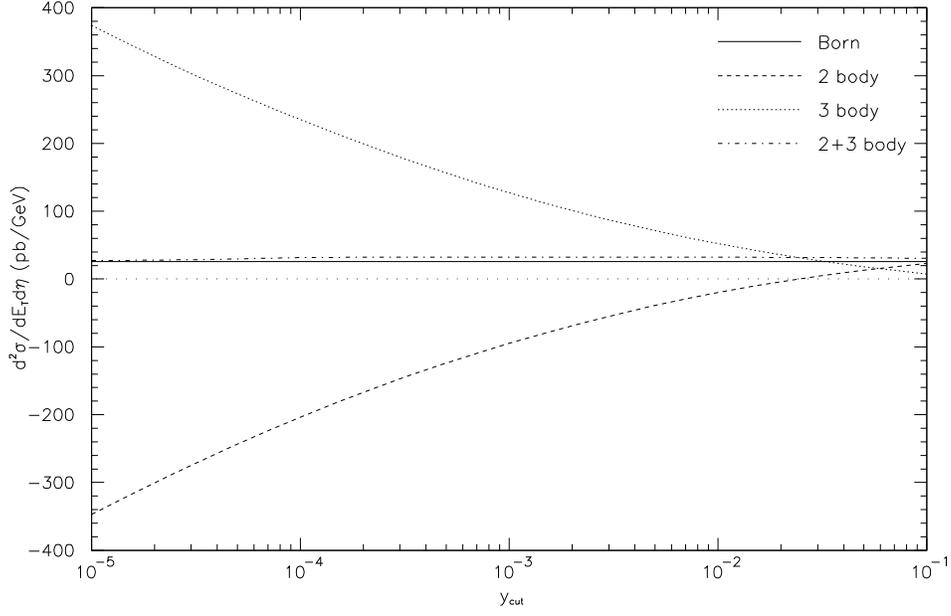,bbllx=520pt,bblly=95pt,bburx=105pt,bbury=710pt,%
           height=12cm,clip=,angle=270}
  \end{picture}}
  \caption[$y_{\rm cut}$-Independence for Direct Photoproduction]
          {\label{plot9}{\it Inclusive single-jet cross section
           d$^2\sigma/$d$E_T$d$\eta$ for direct photons at $E_T=20$~GeV and
           $\eta=1$, as a function of $y_{\rm cut}$. The analytical (dashed)
           and numerical (dotted) contributions have a quadratic logarithmic
           dependence on $y_{\rm cut}$, which cancels in the full
           next-to-leading order result (dot-dashed curve).}}
 \end{center}
\end{figure}

% Plot10
\begin{figure}[p]
 \begin{center}
  {\unitlength1cm
  \begin{picture}(12,8)
   \epsfig{file=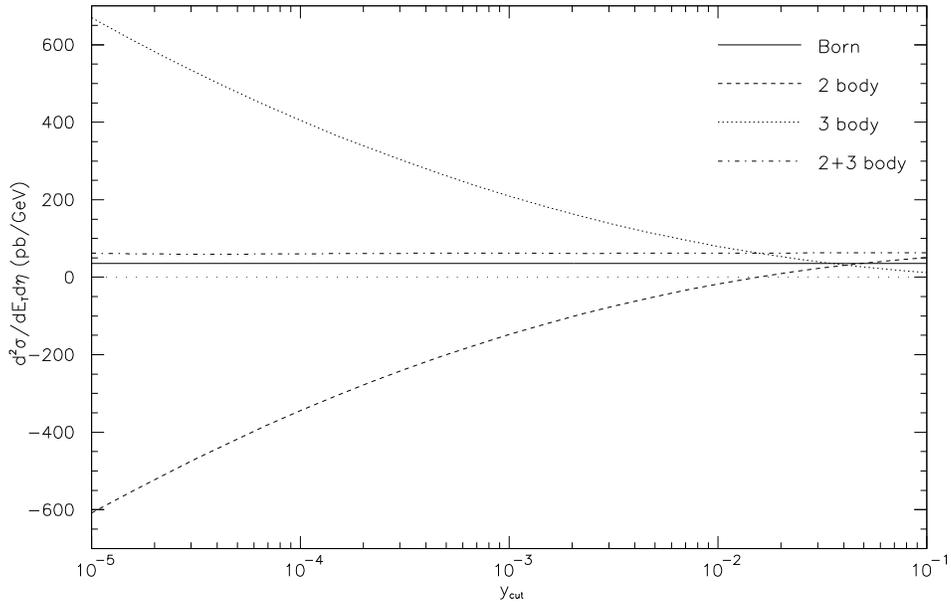,bbllx=520pt,bblly=95pt,bburx=105pt,bbury=710pt,%
           height=12cm,clip=,angle=270}
  \end{picture}}
  \caption[$y_{\rm cut}$-Independence for Resolved Photoproduction]
          {\label{plot10}{\it Inclusive single-jet cross section
           d$^2\sigma/$d$E_T$d$\eta$ for resolved photons at $E_T=20$~GeV and
           $\eta=1$, as a function of $y_{\rm cut}$. Again the sum of two-body
           (dashed) and three-body (dotted) curves is
           $y_{\rm cut}$-independent (dot-dashed curve) like the leading
           order (full) curve.}}
 \end{center}
\end{figure}

At HERA, we have to deal with a partly hadronic process, for which cluster
algorithms are not so well suited. We will therefore follow the two experiments
H1 and ZEUS and use the Snowmass cone algorithm, where hadrons $i$ are combined
into a single jet, if they lie inside a cone in azimuth-rapidity space from the
jet center. The experimental cone size $R$ was already defined in section 2.4
\beq
 R_i = \sqrt{(\eta_i-\eta_J)^2+(\phi_i-\phi_J)^2} \leq R
\eeq
and will be chosen to be $R=1$ in the following. As $R$ takes the role
of $y_{\rm cut}^{\rm experiment}$, we expect the sum of the analytical two-body
and numerical three-body cross sections to be independent of $y_{\rm cut}^{\rm
theory}$. This is the second decisive test of our analytical results.

In figure \ref{plot9} we plot the inclusive single-jet cross section
d$^2\sigma/$d$E_T$d$\eta$ for direct photons at $E_T=20$~GeV and $\eta=1$ as
a function of $y_{\rm cut}$. The leading order prediction (full curve) is
trivially independent of $y_{\rm cut}$. The analytical two-body contributions
(dashed curve) exhibit a quadratic logarithmic dependence on $y_{\rm cut}$ and
turn negative below $y_{\rm cut}= 2.3\cdot 10^{-2}$. We add the numerical
contributions inside and outside the jet cone into the three-body contributions
(dotted curve), which are also quadratically logarithmically dependent on
$y_{\rm cut}$, but positive. The sum of two- and three-body contributions is
the full inclusive next-to-leading order result (dot-dashed curve) and is
independent of $y_{\rm cut}$. This proves that the $y_{\rm cut}$-dependent
finite terms in our analytical results are correct. Towards very small values
of $y_{\rm cut}\simeq 10^{-5}$, the dot-dashed curve drops slightly. This
is due to the limited accuracy in the numerical integration of the three-body
contributions and can be remedied with increased computer power.

Figure \ref{plot10} is the analogue to figure \ref{plot9} for
resolved photoproduction. The two-body curve turns negative at a slightly
smaller value of $y_{\rm cut}= 1.5\cdot 10^{-2}$. As the number of
contributing partonic subprocesses is much larger than in the direct case,
more computer time was needed for similar statistical accuracy.

The ultimate test for our analytical and numerical results consists in the
comparison of our prediction with existing predictions, that were obtained
by B\"odeker \cite{Bod92} (direct case) and by Salesch \cite{Sal93}
(resolved case) with different methods. This can only be done for the
single-jet
inclusive cross section d$^2\sigma/$d$E_T$d$\eta$, since dijet cross sections
are not available. B\"odeker's results are obtained with the subtraction
method of Ellis et al.~\cite{Ell89} to cancel soft and collinear singularities.
Salesch calculated finite cone corrections in addition to the results by
Aversa et al. \cite{Ave88}, who used the so-called small cone approximation.
Here, a small jet cone radius $\delta$ is used to slice the phase space and
isolate the final state divergencies. The dependence on $\delta$ cancels, when
the finite cone corrections are added. We take exactly the same parameters in
our calculation and in
the reference calculations. Furthermore, we choose $y_{\rm cut} =
10^{-3}$ to ensure that we are not sensitive to the omission of terms of
${\cal O} (y_{\rm cut})$.

\begin{figure}[p]
 \begin{center}
  {\unitlength1cm
  \begin{picture}(12,8)
   \epsfig{file=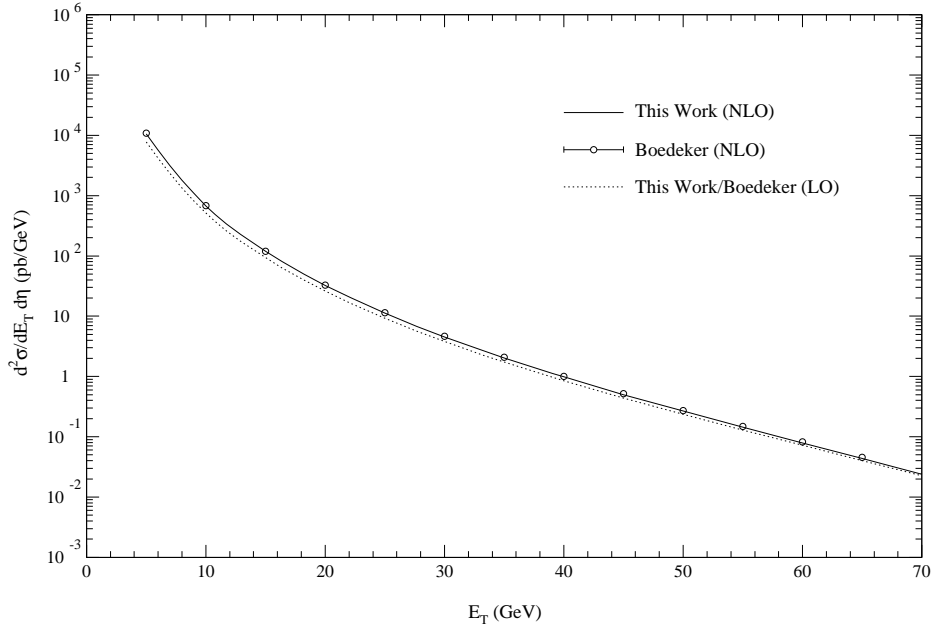,bbllx=520pt,bblly=95pt,bburx=105pt,bbury=710pt,%
           height=12cm,clip=,angle=270}
  \end{picture}}
  \caption[Comparison of $E_T$-Dependence with B\"odeker for Direct Photons]
          {\label{plot11}{\it Inclusive single-jet cross section
           d$^2\sigma/$d$E_T$d$\eta$ for direct photons at
           $\eta=1$, as a function of $E_T$. The subtraction method results
           (open circles) agree with our phase space slicing prediction (full
           curve).}}
 \end{center}
\end{figure}

\begin{figure}[p]
 \begin{center}
  {\unitlength1cm
  \begin{picture}(12,8)
   \epsfig{file=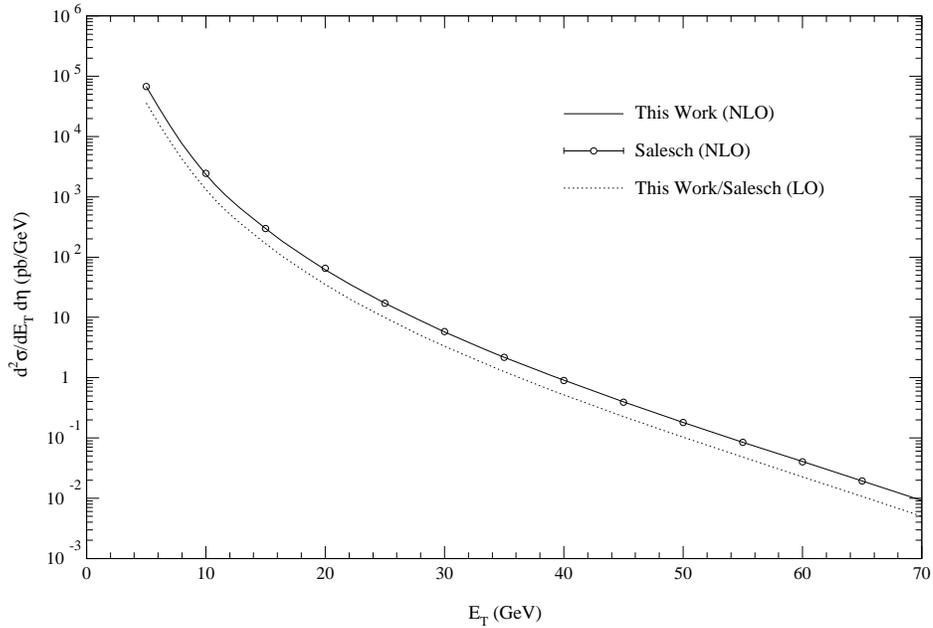,bbllx=520pt,bblly=95pt,bburx=105pt,bbury=710pt,%
           height=12cm,clip=,angle=270}
  \end{picture}}
  \caption[Comparison of $E_T$-Dependence with Salesch for Resolved Photons]
          {\label{plot12}{\it Inclusive single-jet cross section
           d$^2\sigma/$d$E_T$d$\eta$ for resolved photons at
           $\eta=1$, as a function of $E_T$. The small cone approximation
           and finite cone corrections (open circles) agree with our
           invariant mass cut method (full curve). The leading order (dotted)
           curve is shown for comparison.}}
 \end{center}
\end{figure}

The $E_T$ dependence is checked in figure \ref{plot11}
with B\"odeker for direct photons and in figure \ref{plot12} with Salesch
for resolved photons at $\eta = 1$. Our prediction (full curve) agrees
with the older calculations (open circles) at a level better than 1\%. The
numerical statistics are so good that the error bars in B\"odeker's and
Salesch's programs are not seen. For comparison, we include the leading
order dotted curves, where the predictions are identical.

% Plot11

% Plot12

\begin{figure}[p]
 \begin{center}
  {\unitlength1cm
  \begin{picture}(12,8)
   \epsfig{file=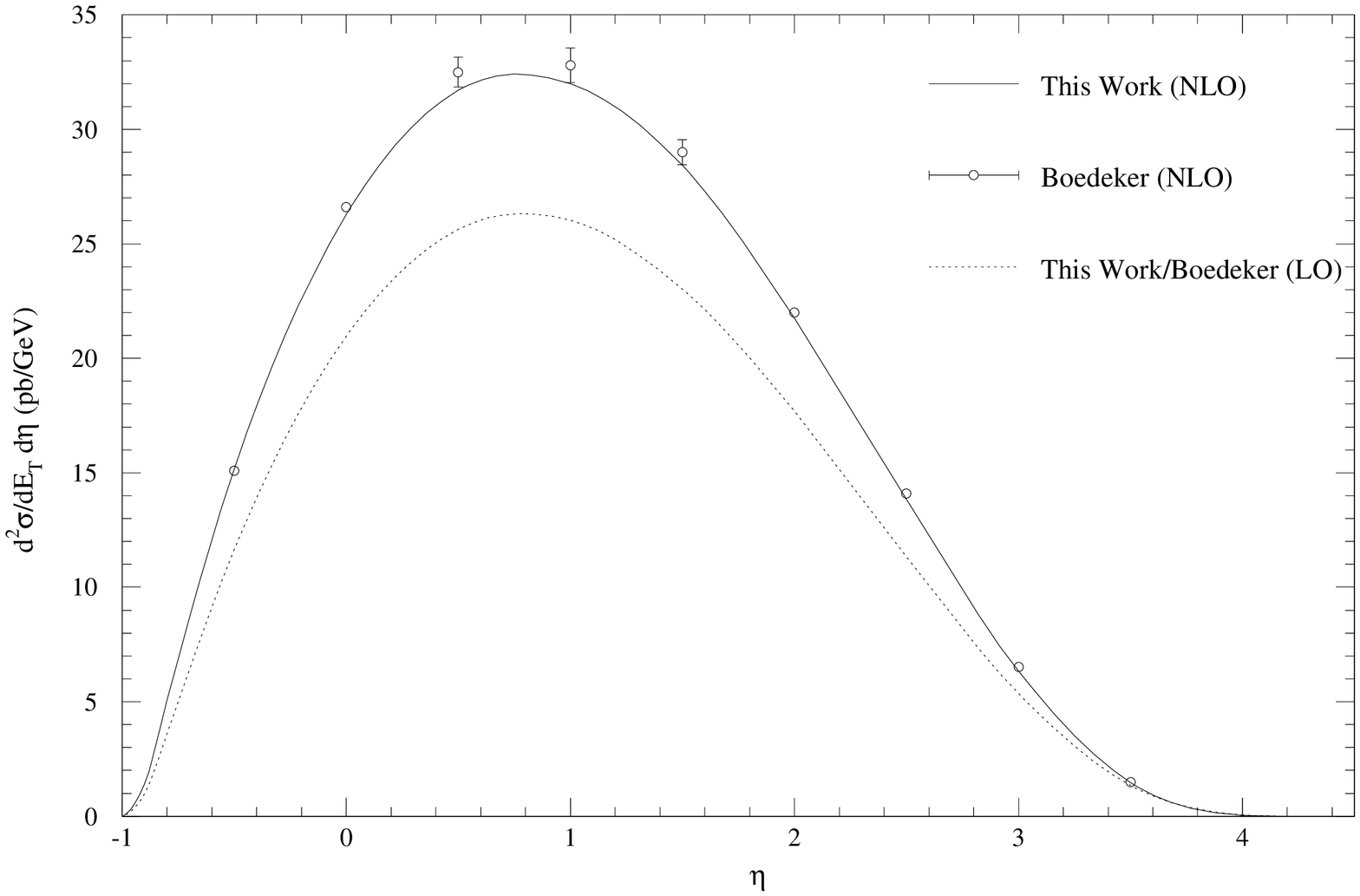,bbllx=520pt,bblly=95pt,bburx=105pt,bbury=710pt,%
           height=12cm,clip=,angle=270}
  \end{picture}}
  \caption[Comparison of $\eta$-Dependence with B\"odeker for Direct Photons]
          {\label{plot13}{\it Inclusive single-jet cross section
           d$^2\sigma/$d$E_T$d$\eta$ for direct photons at
           $E_T=20$~GeV, as a function of $\eta$. Our predictions (full curve)
           agree with B\"odeker's predictions (open circles).}}
 \end{center}
\end{figure}

\begin{figure}[p]
 \begin{center}
  {\unitlength1cm
  \begin{picture}(12,8)
   \epsfig{file=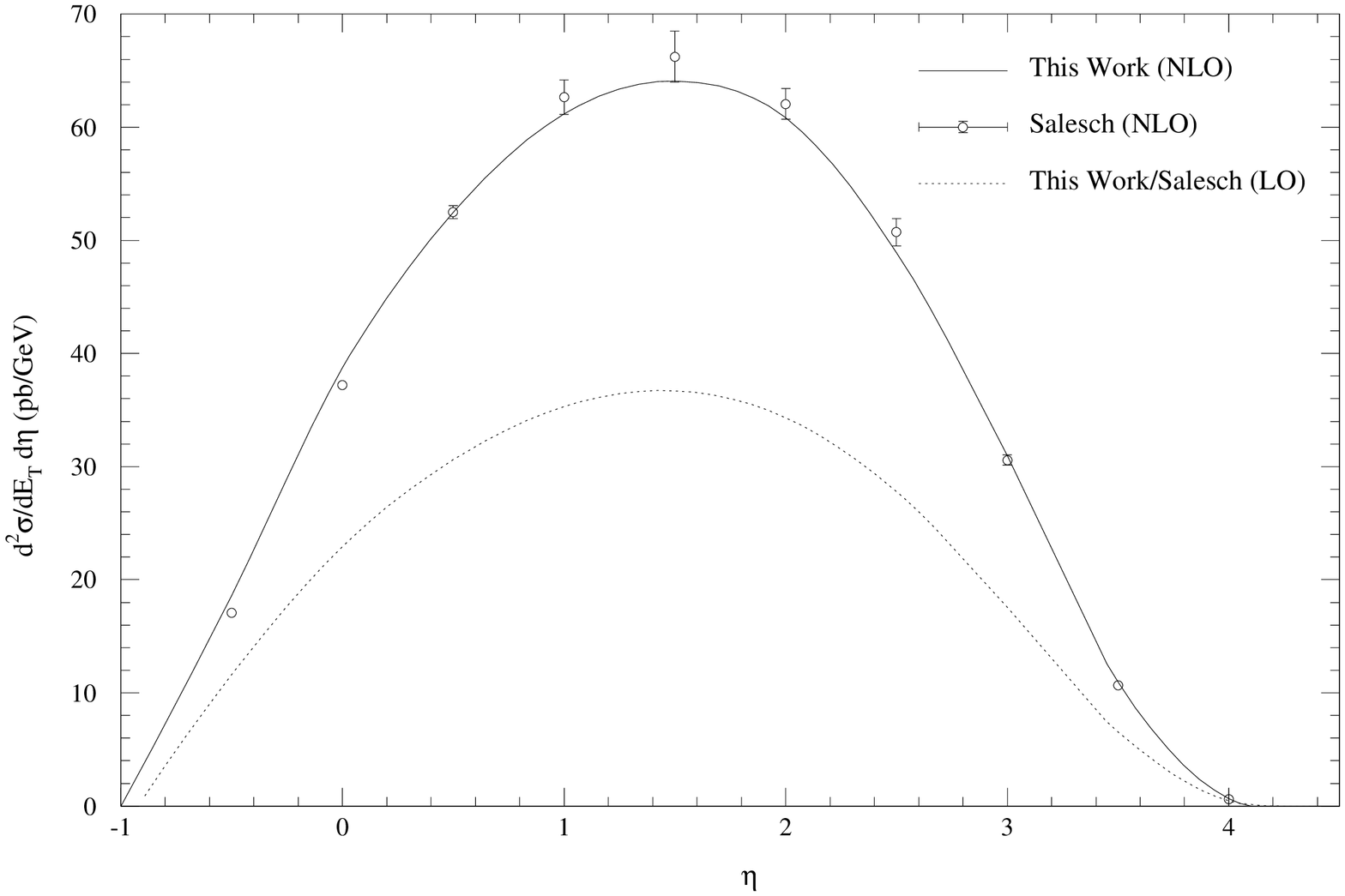,bbllx=520pt,bblly=95pt,bburx=105pt,bbury=710pt,%
           height=12cm,clip=,angle=270}
  \end{picture}}
  \caption[Comparison of $\eta$-Dependence with Salesch for Resolved Photons]
          {\label{plot14}{\it Inclusive single-jet cross section
           d$^2\sigma/$d$E_T$d$\eta$ for resolved photons at
           $E_T=20$~GeV, as a function of $\eta$. Good agreement is seen
           for Salesch's (open circles) and our calculation (full curve).
           The leading order (dotted) curve is shown for comparison.}}
 \end{center}
\end{figure}

We compare the same inclusive single-jet cross section as a rapidity
distribution for $E_T = 20$~GeV. We choose now a linear scale for the ordinate
in figures \ref{plot13} for direct and \ref{plot14} for resolved
photoproduction. This makes it possible to see the error bars in the
reference calculations (open circles) which are at the same level as the
agreement between the different programs. Errors of $\sim 1\%$ present in
our program are not shown. With more computer time, the agreement would
certainly be even better.

From these comparisons, we conclude that also the cut-independent finite
terms in our analytical results and in our numerical FORTRAN program are
correct. Even more, it was possible to check the transverse energy, rapidity,
and other distributions not shown here for every single direct and resolved
partonic subprocess. We found perfect agreement everywhere.

% Plot13

% Plot14

It is clear that the results presented in this work can also be applied to the
calculation of inclusive jet cross sections in $p\overline{p}$ collisions
by just replacing the photon parton distributions with those of the antiproton
and considering only the resolved case. We could have extended our tests
further by comparing with published results for the $p\overline{p}$ case.
Such results have been given in \cite{Ell89} using the subtraction method
and in \cite{y30} based on the phase-space slicing method. In connection
with work on the large transverse momentum production of single inclusive
jets in $p\overline{p}$ collisions \cite{y31}, in which the program based on
\cite{Sal93} was used, we compared with results from \cite{Ell89} and
found good agreement. A comparison with the two-jet results in \cite{y30}
has not been attempted.

After completion of this work, two new complete calculations of jet production
in low $Q^2$ $ep$ collisions have been presented. In \cite{y33}, the subtraction
method for canceling infrared and collinear singularities was applied to
obtain various inclusive one- and two-jet cross sections. These authors
compared inclusive single-jet and two-jet cross sections with our published
results \cite{x12} and found good agreement \cite{y33}. Another work
\cite{y32} uses also the phase-space slicing method, but with two
separate parameters for the infrared and the collinear singular regions
as in their earlier work \cite{Bae89a}. Their results compare to our results
in \cite{x12}. A direct thorough comparison of the results obtained with
the same input has not been done yet.

\subsection{Renormalization- and Factorization-Scale Dependences}

From the numerical checks in section 5.1, we can now be sure to have a reliable
computer program for the calculation of jet cross sections in photoproduction.
We will use this program here to study the dependence of these cross sections
on the three relevant scales:
\begin{itemize}
\item the renormalization scale $\mu$
\item the factorization scale for the photon $M_a$
\item the factorization scale for the proton $M_b$
\end{itemize}
We expect a reduced dependence in next-to-leading order for all three scales
with respect to the leading order. As stated in sections 4.1, 2.2, and 2.3, we
take $\mu$, $M_a$, and $M_b$ to be of ${\cal O} (E_T)$. Therefore, we will
always plot the single-jet inclusive cross section d$^2\sigma$/d$E_T$d$\eta$
as a function of scale$/E_T$ in the following. Direct and resolved
photoproduction will be presented separately at a fixed transverse jet energy
of $E_T=20$~GeV and at a rapidity of $\eta=1$, where the cross sections are at
their maximum.

\begin{figure}[p]
 \begin{center}
  {\unitlength1cm
  \begin{picture}(12,8)
   \epsfig{file=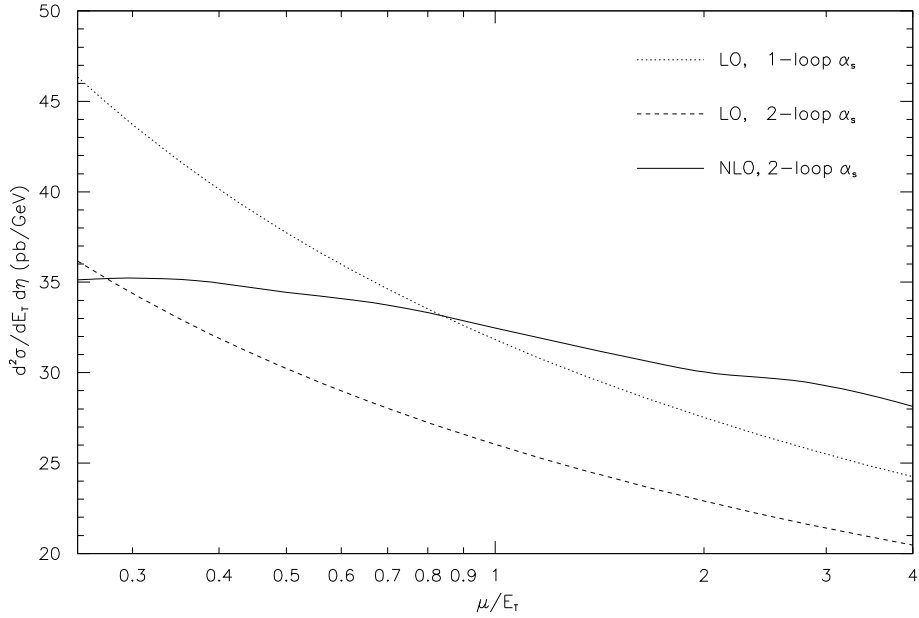,bbllx=520pt,bblly=95pt,bburx=105pt,bbury=710pt,%
           height=12cm,clip=,angle=270}
  \end{picture}}
  \caption[Renormalization Scale Dependence for Direct Photoproduction]
          {\label{plot15}{\it Inclusive single-jet cross section
           d$^2\sigma/$d$E_T$d$\eta$ for direct photons at $E_T=20$~GeV and
           $\eta=1$, as a function of $\mu/E_T$. From $\mu=1/4E_T$ to $\mu=4
           E_T$, the leading order predictions drop by one half. The two-loop
           curve (dashed) is 20\% smaller than the one-loop curve (dotted).
           The next-to-leading order (full) curve only varies by 20\%.}}
 \end{center}
\end{figure}

\begin{figure}[p]
 \begin{center}
  {\unitlength1cm
  \begin{picture}(12,8)
   \epsfig{file=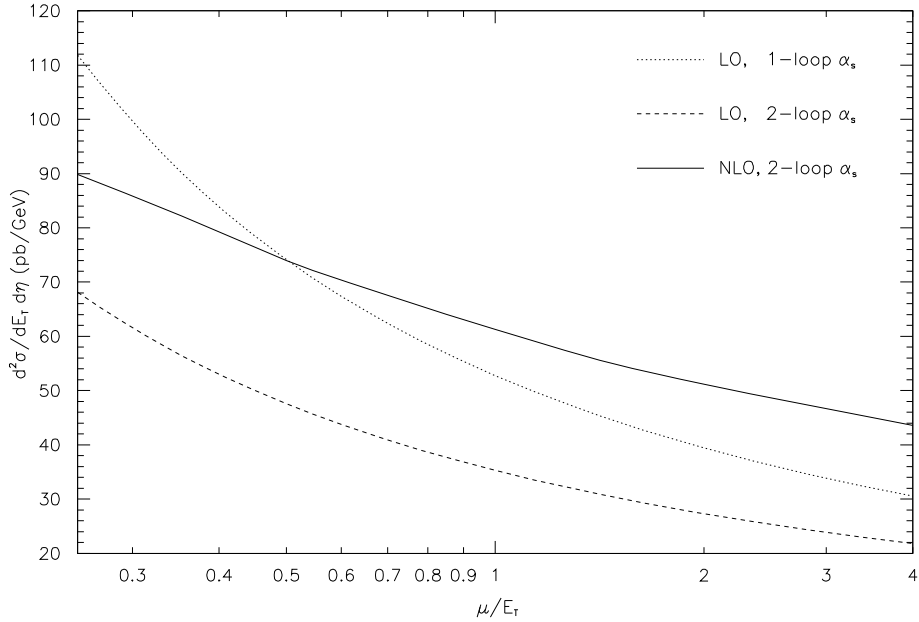,bbllx=520pt,bblly=95pt,bburx=105pt,bbury=710pt,%
           height=12cm,clip=,angle=270}
  \end{picture}}
  \caption[Renormalization Scale Dependence for Resolved Photoproduction]
          {\label{plot16}{\it Inclusive single-jet cross section
           d$^2\sigma/$d$E_T$d$\eta$ for resolved photons at $E_T=20$~GeV and
           $\eta=1$, as a function of $\mu/E_T$. From $\mu=1/4E_T$ to $\mu=4
           E_T$, the leading order predictions drop by a factor of four. The
           two-loop curve (dashed) is 40\% smaller than the one-loop curve
           (dotted). The next-to-leading order (full) curve only varies by a
           factor of two.}}
 \end{center}
\end{figure}

We start with the dependence on the renormalization scale $\mu$. Leading
order ${\cal O} (\alpha\alpha_s)$ cross sections depend on $\mu$ only through
the running of the strong coupling constant $\alpha_s(\mu^2)$, which has to be
implemented in the one-loop approximation
\beq
  \alpha_s(\mu^2) = \frac{12\pi}{(33-2N_f)\ln \frac{\mu^2}{\Lambda^2}}
\eeq
for consistency. The results are shown as dotted curves in figures \ref{plot15}
and \ref{plot16}. For direct photons, the curve drops by one half when
going from $\mu=1/4E_T$ to $\mu=4E_T$. The resolved photon hard cross section
is one order higher in $\alpha_s$ $({\cal O} (\alpha_s^2))$ and therefore
drops even by a factor of four. This strong $1/\ln(\mu^2)$-behavior is
slightly weakened in the two-loop approximation
\beq
  \alpha_s(\mu^2) = \frac{12\pi}{(33-2N_f)\ln \frac{\mu^2}
  {\Lambda^2}} \left( 1-\frac{6(153-19N_f)}{(33-2N_f)^2}
  \frac {\ln (\ln \frac{\mu^2}{\Lambda^2} )}%
{\ln \frac{\mu^2}{\Lambda^2}} \right),
\eeq
where one adds further logarithmic terms in $\mu^2$. These curves are shown
in dashed form in figures \ref{plot15} and \ref{plot16}. However, going from
one-loop to two-loop $\alpha_s$ affects not so much the shape but the overall
normalization. The cross sections are reduced by 20\% in the direct and by
almost 40\% in the resolved case. Although not physically relevant, the dashed
curves can be compared to the full curves and enable us to see solely the
effect of the explicit logarithmic terms
\beq
  \frac{1}{\varepsilon} \left( \frac{4\pi\mu^2}{s} \right) ^\varepsilon
  \doteq \frac{1}{\varepsilon} + \ln \frac{4\pi\mu^2}{s},
\eeq
that appear in the one-loop corrections to the hard cross section in section
4.1. They cancel partly the $\alpha_s$-dependence, so that the full curves
in figures \ref{plot15} and \ref{plot16} are much flatter compared to the
leading order curves with one- and two-loop $\alpha_s$. The direct curve
varies by only 20\% and exhibits a maximum near $\mu/E_T\simeq 1/2$. According
to the principle of minimal sensitivity \cite{Ste81}, $\mu = 1/2 E_T$
would then be the optimal scale. The dependence of the resolved
photoproduction cross section is also weakened and amounts to a factor of two
in NLO. It meets the one-loop LO result at $\mu \simeq 1/2 E_T$. This scale
is sometimes preferred since the NLO corrections are small in its vicinity.

% Plot15

% Plot16

\begin{figure}[p]
 \begin{center}
  {\unitlength1cm
  \begin{picture}(12,8)
   \epsfig{file=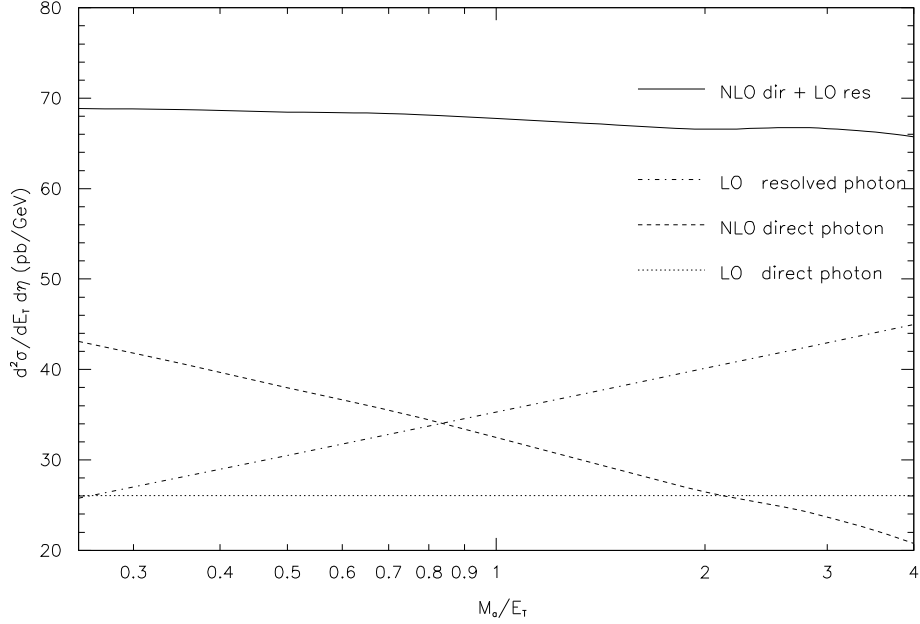,bbllx=520pt,bblly=95pt,bburx=105pt,bbury=710pt,%
           height=12cm,clip=,angle=270}
  \end{picture}}
  \caption[Photon Factorization Scale Dependence for Direct and Complete
           Photoproduction]
          {\label{plot17}{\it Inclusive single-jet cross section
           d$^2\sigma/$d$E_T$d$\eta$ for direct photons at $E_T=20$~GeV and
           $\eta=1$, as a function of $M_a/E_T$. The LO direct curve (dotted)
           is independent of $M_a$ as is the sum (full curve) of NLO direct
           (dashed) and LO resolved (dot-dashed) contributions.}}
 \end{center}
\end{figure}

\begin{figure}[p]
 \begin{center}
  {\unitlength1cm
  \begin{picture}(12,8)
   \epsfig{file=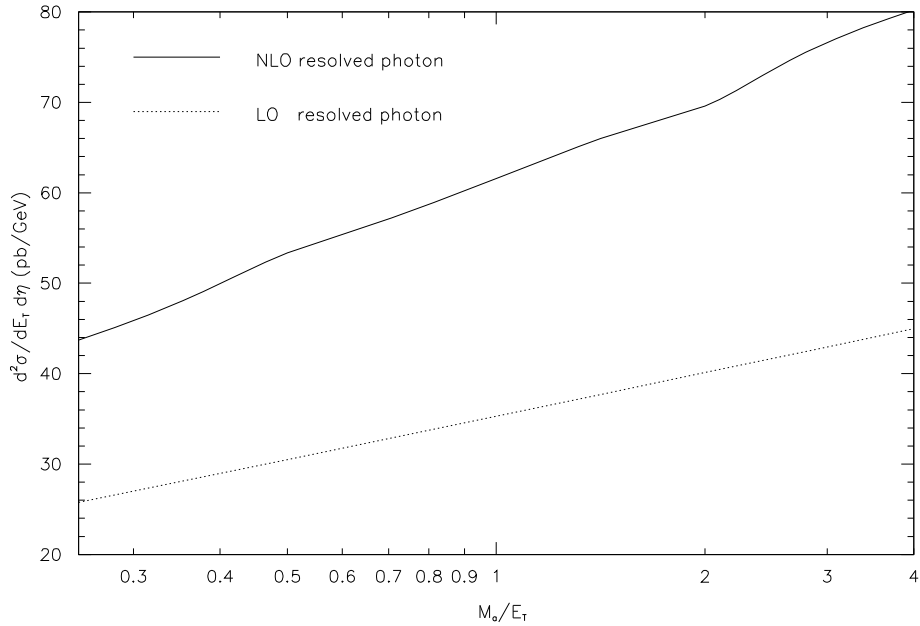,bbllx=520pt,bblly=95pt,bburx=105pt,bbury=710pt,%
           height=12cm,clip=,angle=270}
  \end{picture}}
  \caption[Photon Factorization Scale Dependence for Resolved Photoproduction]
          {\label{plot18}{\it Inclusive single-jet cross section
           d$^2\sigma/$d$E_T$d$\eta$ for resolved photons at $E_T=20$~GeV and
           $\eta=1$, as a function of $M_a/E_T$. The logarithmic behavior of
           the resolved part is slightly steeper in NLO (full curve) than
           in LO (dotted curve).}}
 \end{center}
\end{figure}

The next scale that we will study is the factorization scale in the photon
$M_a$. It separates the soft parton content in the photon given by the parton
distribution function $F_{a/\gamma}(y_a,M_a^2)$ from the hard partonic cross
section d$\sigma_{ab} (ab\rightarrow\mbox{jets})$. As discussed in section 2.3,
direct and resolved processes are only separable in leading order. The direct
parton distribution is then simply a $\delta$-function $\delta(1-y_a)$ and does
not depend on any scale. This can be seen from the dotted curve in figure
\ref{plot17}. The resolved part depends, however, strongly on the scale $M_a$
as could already be inferred from $e^+e^-$-data on the $F_2^{\gamma}$ structure
function. We expect the $M_a$-dependence to be dominated
by the asymptotic pointlike contribution to $F_2^{\gamma}$
\beq
 F_2^{\gamma}(x_B,M_a^2) = \alpha\le\frac{1}{\alpha_s(M_a^2)}a(x_B)+b(x_B)\re
\eeq
that behaves like $\ln M_a^2$.
Clearly we see this behavior in the leading order dot-dashed curve in figure
\ref{plot17}, which is identical to the dotted curve in figure \ref{plot18}
and is obtained with the GRV photon distributions. As the GRV distributions
are fitted in the DIS$_{\gamma}$-scheme, we transform them into the
$\overline{\mbox{MS}}$ scheme by adding the correct finite terms independent
of $M_a$. The renormalization scheme dependence was studied by B\"odeker, 
Kramer, and Salesch \cite{Bod94} and will not be studied here again.
The main result is that the dependence
disappears in next-to-leading order, because the transformation of
the resolved part is compensated through scheme dependent terms in the initial
state singularities of the direct photon. The renormalization of the
singularities in section 4.2.5
\beq
  F_{a/\gamma} (y_a,M_a^2) =
  \int_{y_a}^1 \frac{\mbox{d}z_a}{z_a}
  \left[ \delta_{a\gamma } \delta (1-z_a) + \frac{\alpha}{2\pi}
  R_{q\leftarrow \gamma }(z_a, M_a^2)\right] F_{\gamma /\gamma}
  \lr\frac{y_a}{z_a}\rr
\eeq
gave also rise to scale dependent terms
\beq
  R_{a \leftarrow \gamma } (z_a, M_a^2) =
 -\frac{1}{\eps}P_{q\leftarrow \gamma }(z_a)\frac{\Gamma (1-\eps)}
  {\Gamma (1-2\eps)} \left( \frac{4\pi\mu^2}{M_a^2} \right)
  ^\eps
\eeq
through the $1/\eps$ pole
\beq
  -\frac{1}{\eps}P_{q\leftarrow \gamma }(z_a)\le
  \lr\frac{4\pi\mu^2}{s}\rr^\eps
  -\lr\frac{4\pi\mu^2}{M_a^2}\rr^\eps\re
  = -P_{q\leftarrow \gamma }(z_a) \ln\lr\frac{M_a^2}{s}\rr.
\eeq
We can see this negative logarithmic behavior $(-\ln M_a^2)$
in the dashed curve in
figure \ref{plot17} for next-to-leading order direct photoproduction. It has
the opposite sign than the LO resolved contribution, so that the sum of both
will eventually be almost independent of $M_a$ (full curve in figure
\ref{plot17}). The next-to-leading order resolved contribution has a similar,
but slightly steeper shape than the leading order and is shown as a full curve
in figure \ref{plot18}. Therefore, the dependence of the complete NLO
calculation on the photon factorization scale is slightly stronger than in the
full line of figure \ref{plot17}. The difference will be compensated in
next-to-next-to-leading order (NNLO) of direct photoproduction, which is
beyond the scope of our calculation. These findings agree with the
earlier analysis of the $M_a$ dependence of the inclusive single-jet
cross sections in \cite{Bod94} and constitute a nice check of our
formalism.

% Plot17

% Plot18

\begin{figure}[h]
 \begin{center}
  {\unitlength1cm
  \begin{picture}(12,7.5)
   \epsfig{file=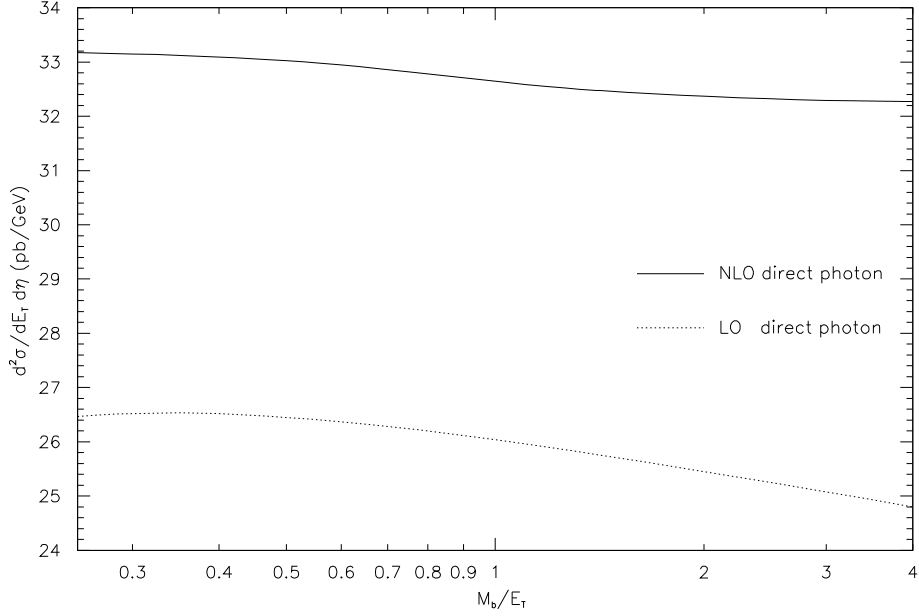,bbllx=520pt,bblly=95pt,bburx=105pt,bbury=710pt,%
           height=12cm,clip=,angle=270}
  \end{picture}}
  \caption[Proton Factorization Scale Dependence for Direct Photoproduction]
          {\label{plot19}{\it Inclusive single-jet cross section
           d$^2\sigma/$d$E_T$d$\eta$ for direct photons at $E_T=20$~GeV and
           $\eta=1$, as a function of $M_b/E_T$. The weak behavior of the
           LO (dotted) curve is even more reduced in NLO (full curve).}}
 \end{center}
\end{figure}

\begin{figure}[h]
 \begin{center}
  {\unitlength1cm
  \begin{picture}(12,8)
   \epsfig{file=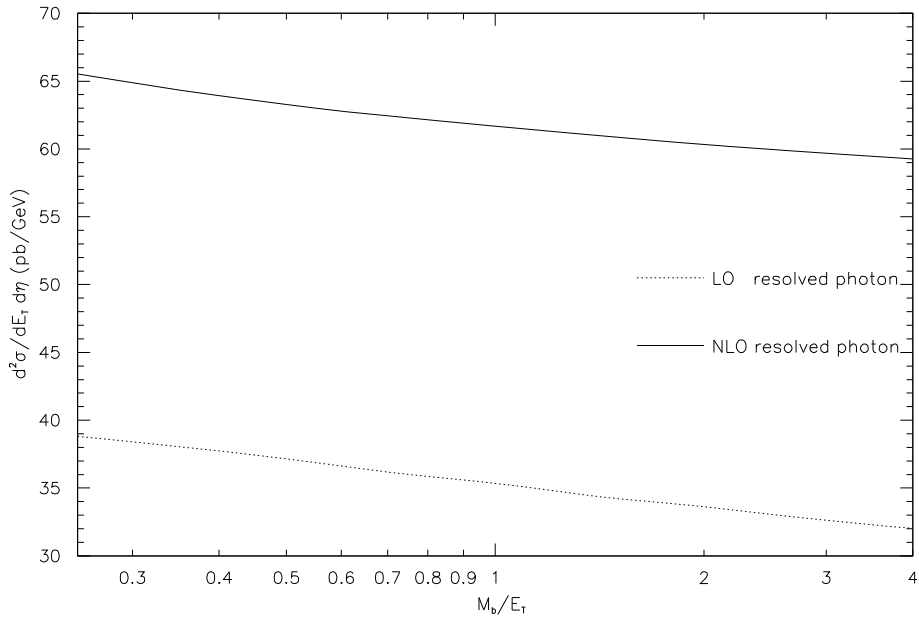,bbllx=520pt,bblly=95pt,bburx=105pt,bbury=710pt,%
           height=12cm,clip=,angle=270}
  \end{picture}}
  \caption[Proton Factorization Scale Dependence for Resolved Photoproduction]
          {\label{plot20}{\it Inclusive single-jet cross section
           d$^2\sigma/$d$E_T$d$\eta$ for resolved photons at $E_T=20$~GeV and
           $\eta=1$, as a function of $M_b/E_T$. The approximately logarithmic
           dependence in LO (dotted curve) is reduced in NLO (full curve).}}
 \end{center}
\end{figure}

\begin{figure}[h]
 \begin{center}
  {\unitlength1cm
  \begin{picture}(12,8)
   \epsfig{file=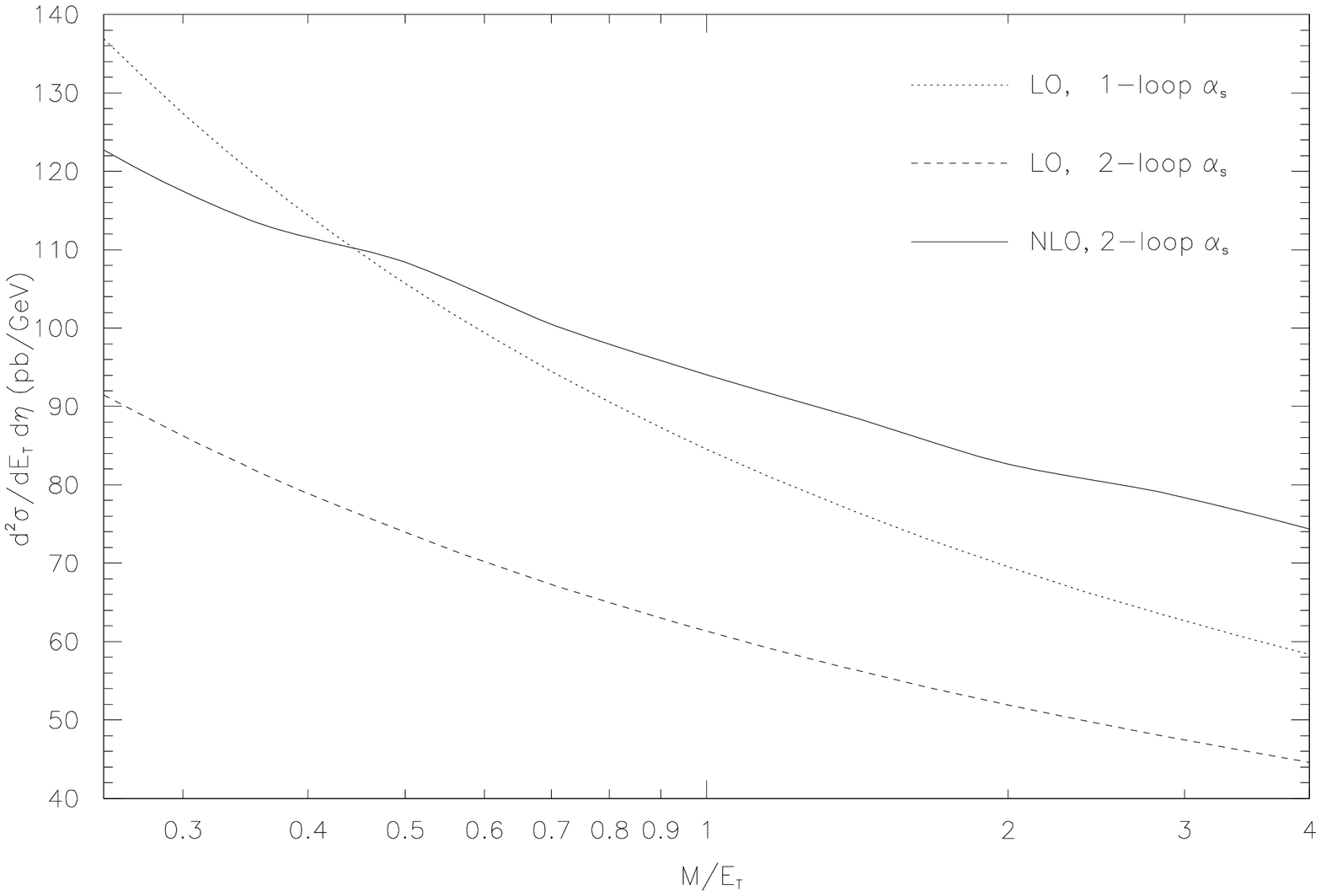,bbllx=520pt,bblly=95pt,bburx=105pt,bbury=710pt,%
           height=12cm,clip=,angle=270}
  \end{picture}}
  \caption[Total Scale Dependence for Complete Photoproduction]
          {\label{plot49}{\it Complete inclusive single-jet cross section
           d$^2\sigma/$d$E_T$d$\eta$ at $E_T=20$~GeV and $\eta=1$, as a
           function of $M/E_T$ for $M=\mu=M_a=M_b$. The strong dependence in
           LO (dotted and dashed curves) is reduced in NLO (full curve).}}
 \end{center}
\end{figure}

For the proton, we have a similar scale $M_b$ as in the photon case that
separates the soft and the hard part of the hadronic and partonic cross
sections. Of course, the proton is not point-like and does not have a direct
component. The scale dependence of the proton distribution functions
$F_{b/p}(x_b,M_b^2)$ is governed by the AP equations. If we solve them
iteratively, we can assume that in
\beq
 \frac{\mbox{d}F_{b/p}(x_b,M_b^2)}{\mbox{d}\ln M_b^2} =
 \frac{\alpha_s^{(0)}}{2\pi}\int\limits_{x_b}^1\frac{\mbox{d}z_b}{z_b}
 P_{b\leftarrow b'}\lr\frac{x_b}{z_b}\rr F_{b/p}^{(0)}(z_b)
\eeq
the starting distributions $F_{b/p}^{(0)}(z_b)$ and the coupling constant
$\alpha_s^{(0)}$ are scale independent. Then, the integral over $\mbox{d}
\ln M_b^2$ can be carried out leading to a simple logarithmic behavior
\beq
 F_{b/p}(x_b,M_b^2) =
 \frac{\alpha_s^{(0)}}{2\pi}\int\limits_{x_b}^1\frac{\mbox{d}z_b}{z_b}
 P_{b\leftarrow b'}\lr\frac{x_b}{z_b}\rr F_{b/p}^{(0)}(z_b)
 \ln \lr \frac{M_b^2}{s}\rr.
\eeq
For the LO resolved photon contribution, this approximation is obviously good
enough as can be seen from the dotted curve in figure \ref{plot20}. The cross
section drops by 20\% when going from $M_b=1/4E_T$ to $M_b=4E_T$. Of course,
the full solution of the AP equations is not so simple, and the LO direct
photon contribution in figure \ref{plot19} deviates slightly from a simply
logarithmic graph. The variation of the cross section here amounts only to
6.5\%. In next-to-leading order, initial state singularities arise as in the
photon case and are absorbed into the proton structure function
\beq
  F_{b/p} (x_b,M_b^2)  =
  \int_{x_b}^1 \frac{\mbox{d}z_b}{z_b}
  \left[ \delta_{bb'} \delta (1-z_b) + \frac{\alpha_s}{2\pi}
  R'_{b\leftarrow b'} (z_b, M_b^2) \right] F_{b'/p}
  \lr\frac{x_b}{z_b}\rr.
\eeq
Again, finite terms accompany the pole in the $\overline{\mbox{MS}}$ scheme
\beq
  R'_{b \leftarrow b'} (z_b, M_b^2) =
  -\frac{1}{\eps} P_{b\leftarrow b'} (z_b) \frac{\Gamma (1-\eps)}
  {\Gamma (1-2\eps)} \left( \frac{4\pi\mu^2}{M_b^2} \right)
  ^\eps,
\eeq
which depend on $M_b^2$ through
\beq
  -\frac{1}{\eps}P_{b\leftarrow b'}(z_b)\le
  \lr\frac{4\pi\mu^2}{s}\rr^\eps
  -\lr\frac{4\pi\mu^2}{M_b^2}\rr^\eps\re
  = -P_{b\leftarrow b'}(z_b) \ln\lr\frac{M_b^2}{s}\rr.
\eeq
These logarithms cancel the leading logarithmic behavior of the parton
distribution functions in the proton. Consequently, the full next-to-leading
order curves in figures \ref{plot19} and \ref{plot20} depend only very weakly
on the renormalization scale $M_b$ in the proton.
Finally, we consider the total scale dependence for complete photoproduction
in figure \ref{plot49}. The dependence of the inclusive single-jet cross
section on $M=\mu=M_a=M_b$ is dominated by the renormalization scale
dependence in the one-loop (dotted curve) and two-loop (dashed curve)
approximation for $\alpha_s$. Since resolved photoproduction is more important
than direct photoproduction at $E_T=20$~GeV, figure \ref{plot49} resembles
figure \ref{plot16} more than figure \ref{plot15}. The total scale dependence
is reduced by almost a factor of two in next-to-leading order (full curve).

% Plot19

% Plot20

% Plot49

\subsection{One-Jet Cross Sections}

In the last section, we have studied the dependence of the inclusive
single-jet cross section on the various scales, which cannot be predicted
from theory. It was shown that the next-to-leading order cross section is much
less dependent on the renormalization and factorization scales than the
leading order cross section, due to cancellation mechanisms. We continue
to study one-jet cross sections here, but shift the emphasis towards
experimentally relevant distributions. The two main experimental observables in
the double differential cross section
\beq
  \frac{\mbox{d}^2\sigma}{\mbox{d}E_T\mbox{d}\eta}
  = \sum_b \int_{x_{a,\min}}^1 \mbox{d}x_a x_a 
  F_{a/e}(x_a,M_a^2) x_b F_{b/p}(x_b,M_b^2)
  \frac{4E_eE_T}{2x_aE_e-E_Te^{-\eta}}
  \frac{\mbox{d}\sigma}{\mbox{d}t}(ab\rightarrow p_1p_2)
\eeq
are the hardness or transverse energy $E_T$ of the observed jet and its
orientation in the detector, i.e.~the angle $\theta$ it forms with the
beam axis. Instead of this angle, we use the pseudorapidity $\eta = -\ln
[\tan(\theta/2)]$ as usual. As the electron is only weakly deflected, it
stays in the beam pipe like the proton remnant. Therefore, the jets are
homogeneously distributed in the azimuthal angle $\phi$, and we have
integrated over $\phi$ in the theoretical prediction. The single-jet
cross sections for direct photoproduction have already been published 
\cite{x8}, those for resolved photoproduction are shown here for the
first time with our phase space slicing method. Using a different method
similar results have been published in \cite{Kra94,Sal93,y28,y29}.

\begin{figure}[p]
 \begin{center}
  {\unitlength1cm
  \begin{picture}(12,8)
   \epsfig{file=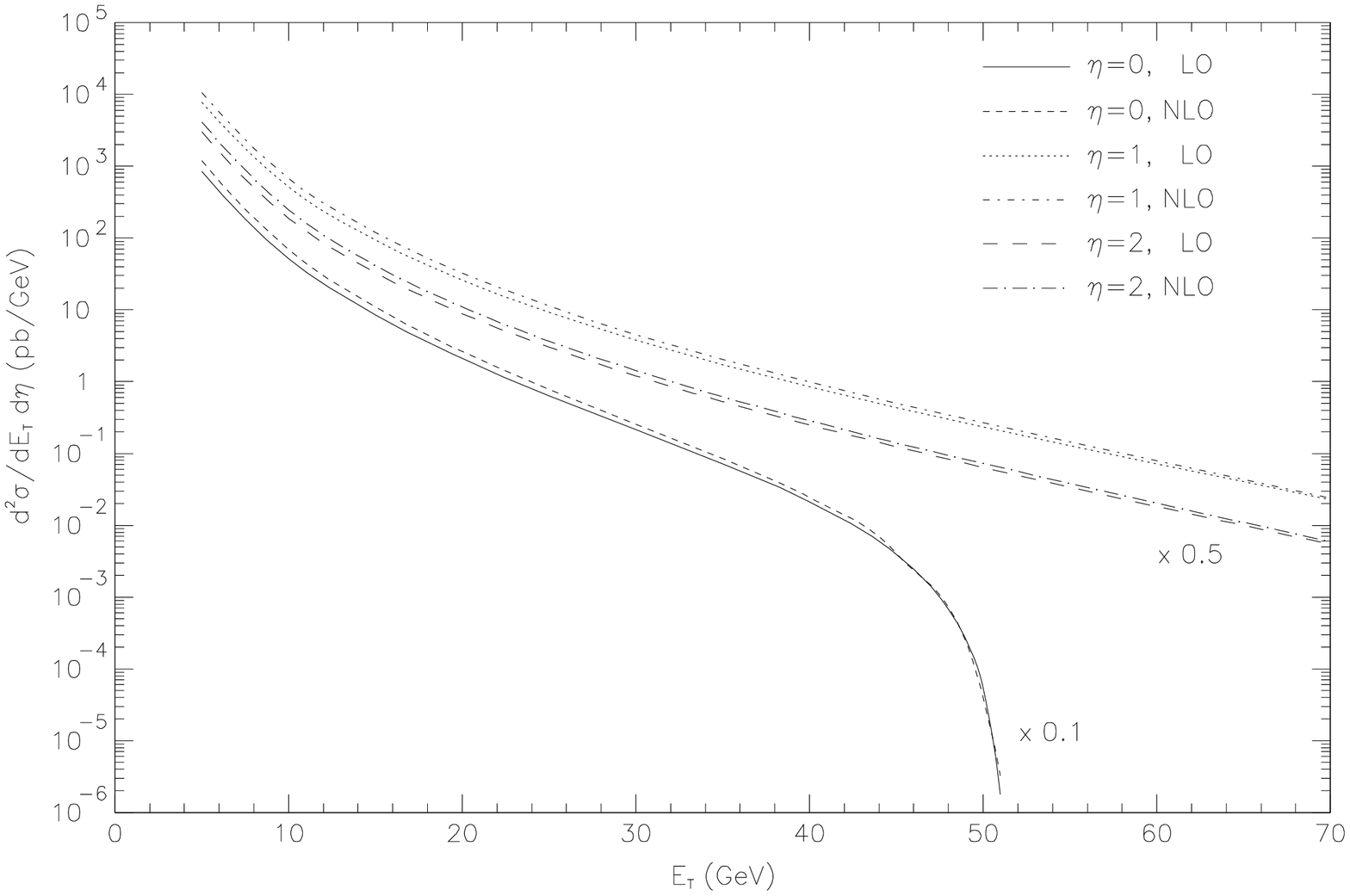,bbllx=520pt,bblly=95pt,bburx=105pt,bbury=710pt,%
           height=12cm,clip=,angle=270}
  \end{picture}}
  \caption[$E_T$-Dependence of Single-Jet Cross Section for Direct Photons]
          {\label{plot21}{\it Inclusive single-jet cross section
           $\mbox{d}^2\sigma/\mbox{d}E_T\mbox{d}\eta$ for direct photons
           as a function of $E_T$ for various rapidities $\eta = 0, 1, 2$
           in LO and NLO. The cross section for $\eta = 0~(\eta = 2)$ is
           multiplied by a factor of $0.1~(0.5)$.}}
 \end{center}
\end{figure}

We start with distributions in the jet transverse energy $E_T$, which is 
at the same time a measure for the resolution of partons within the photon and
the proton. In figure \ref{plot21}, we plot the direct photon contribution
for transverse energies between 5 and 70~GeV. In this interval,
the cross section drops
by almost six orders of magnitude. The next-to-leading order curves are
always above the leading order predictions. To separate the curves for three
different rapidities, we multiplied those for $\eta=0$ and $\eta=2$ by factors
of 0.1 and
0.5, respectively. At $\eta=0$, the phase space restricts the accessible
range in $E_T$ to $E_T < 50$~GeV. The so-called $k$-factor, i.e.~the ratio of
NLO to LO, drops from 1.8 at 5~GeV to 1.2 at 70~GeV for $\eta=1$, so that
the higher order corrections are more important at small scales, where the
strong coupling $\alpha_s$ is large. Had we calculated the leading order with
one-loop $\alpha_s$, the correction would, however, not be so large.

\begin{figure}[p]
 \begin{center}
  {\unitlength1cm
  \begin{picture}(12,8)
   \epsfig{file=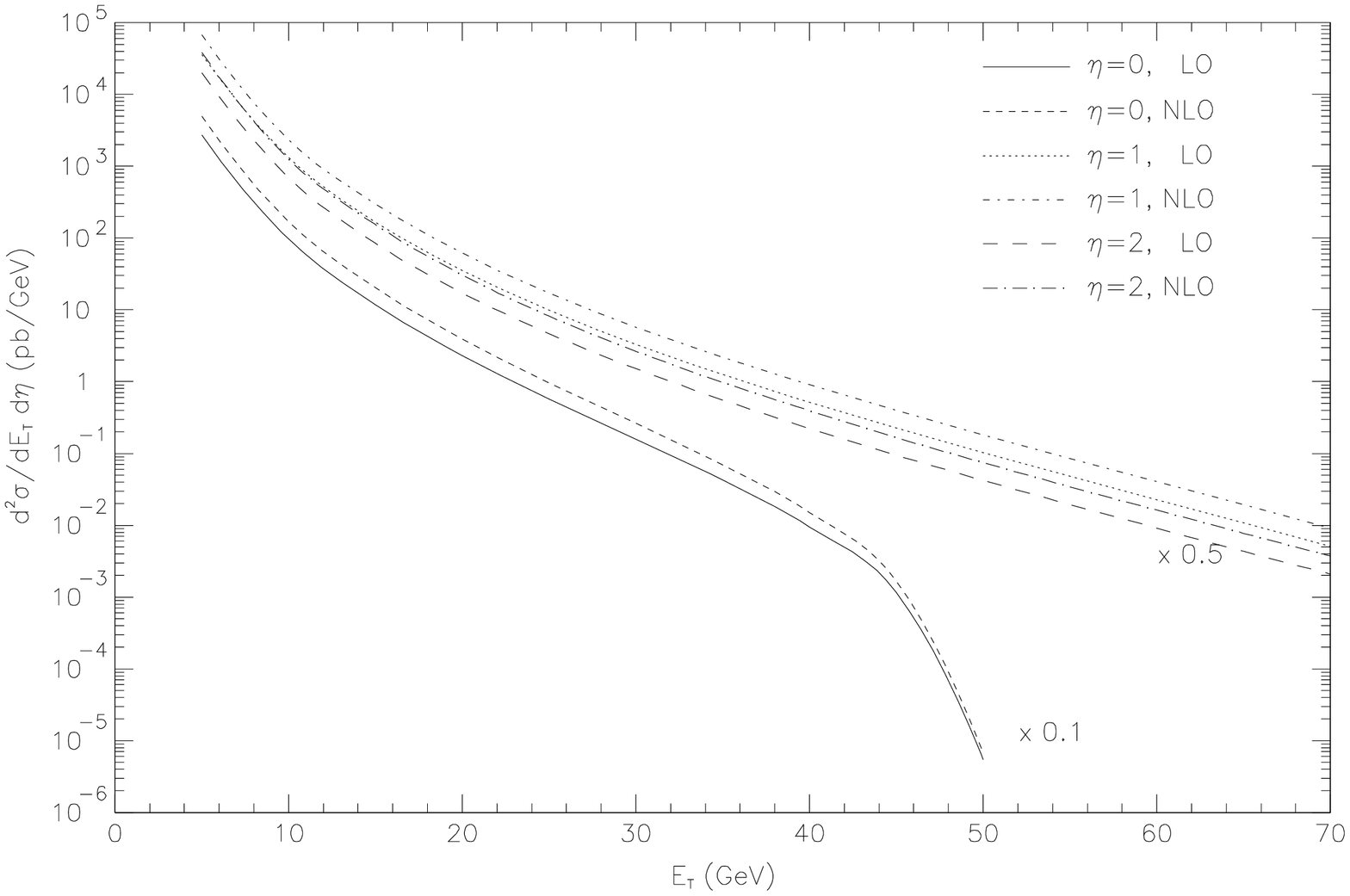,bbllx=520pt,bblly=95pt,bburx=105pt,bbury=710pt,%
           height=12cm,clip=,angle=270}
  \end{picture}}
  \caption[$E_T$-Dependence of Single-Jet Cross Section for Resolved Photons]
          {\label{plot22}{\it Inclusive single-jet cross section
           $\mbox{d}^2\sigma/\mbox{d}E_T\mbox{d}\eta$ for resolved photons
           as a function of $E_T$ for various rapidities $\eta = 0, 1, 2$
           in LO and NLO. The cross section for $\eta = 0~(\eta = 2)$ is
           multiplied by a factor of $0.1~(0.5)$.}}
 \end{center}
\end{figure}

The corresponding curves for resolved photoproduction are shown in figure
\ref{plot22}. Here, the curves for $\eta=1$ and $\eta=2$ lie closer to
each other in spite of the rescaling for $\eta=2$. This already hints at a
broader maximum in the rapidity distribution compared to the direct case
(see below). The NLO and LO curves lie further apart due to the large
$k$-factor of about 1.8 over the whole $E_T$-range. This can be understood
from the non-perturbative nature of resolved photons, where higher order
corrections are more important at all scales due to the many channels
involved.

% Plot21

% Plot22

\begin{figure}[p]
 \begin{center}
  {\unitlength1cm
  \begin{picture}(12,8)
   \epsfig{file=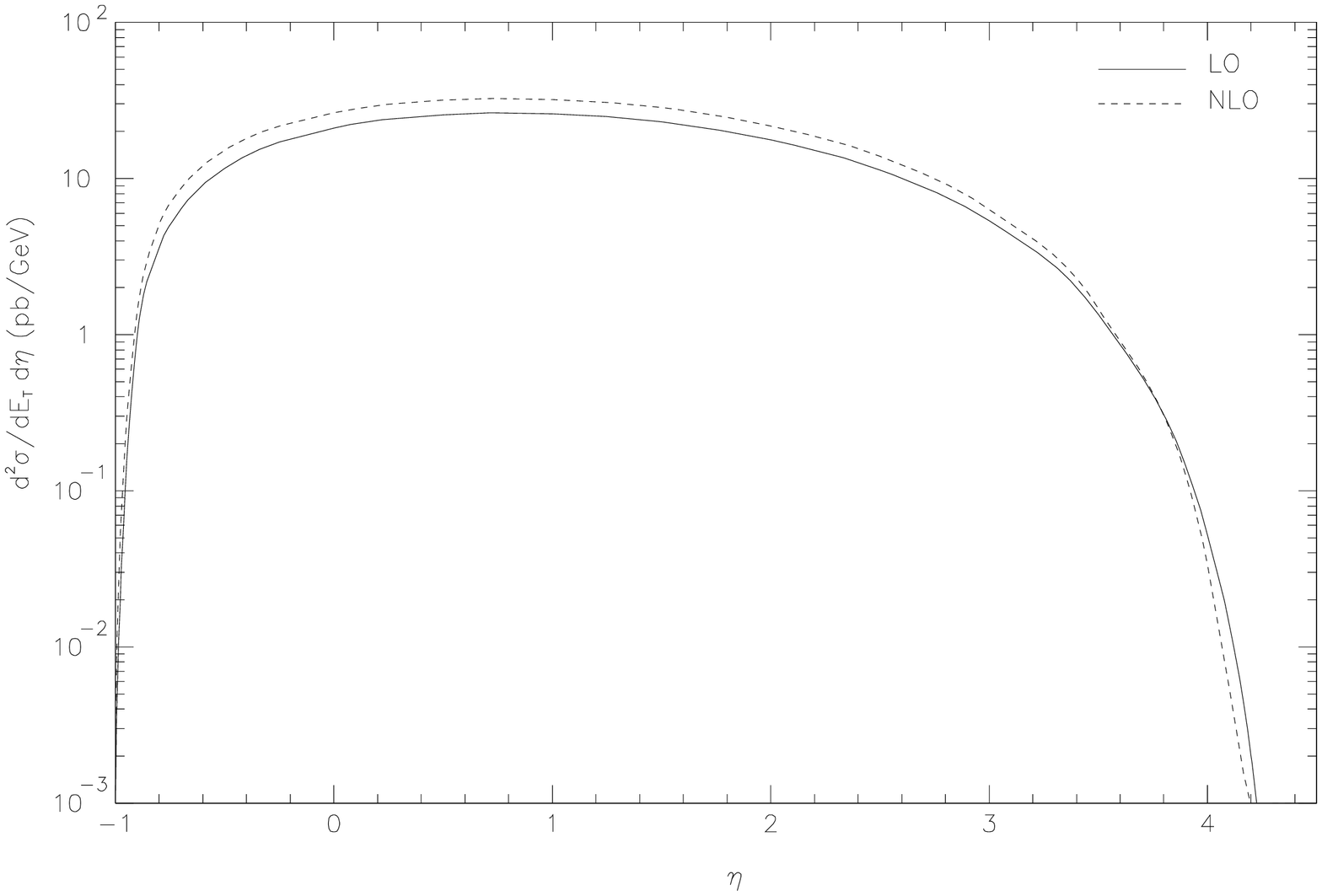,bbllx=520pt,bblly=95pt,bburx=105pt,bbury=710pt,%
           height=12cm,clip=,angle=270}
  \end{picture}}
  \caption[$\eta$-Dependence of Single-Jet Cross Section for Direct Photons]
          {\label{plot23}{\it Inclusive single-jet cross section
           $\mbox{d}^2\sigma/\mbox{d}E_T\mbox{d}\eta$ for direct photons
           as a function of $\eta$ for $E_T = 20$~GeV in LO and NLO.}}
 \end{center}
\end{figure}

The rapidity distribution for direct photons is presented in figure
\ref{plot23} at a transverse energy of $E_T=20$~GeV. In the available phase
space of $\eta\in[-1,4]$, the cross section exhibits a rather broad maximum and
steep edges. The $k$-factor is 1.25 in the central region.

\begin{figure}[p]
 \begin{center}
  {\unitlength1cm
  \begin{picture}(12,8)
   \epsfig{file=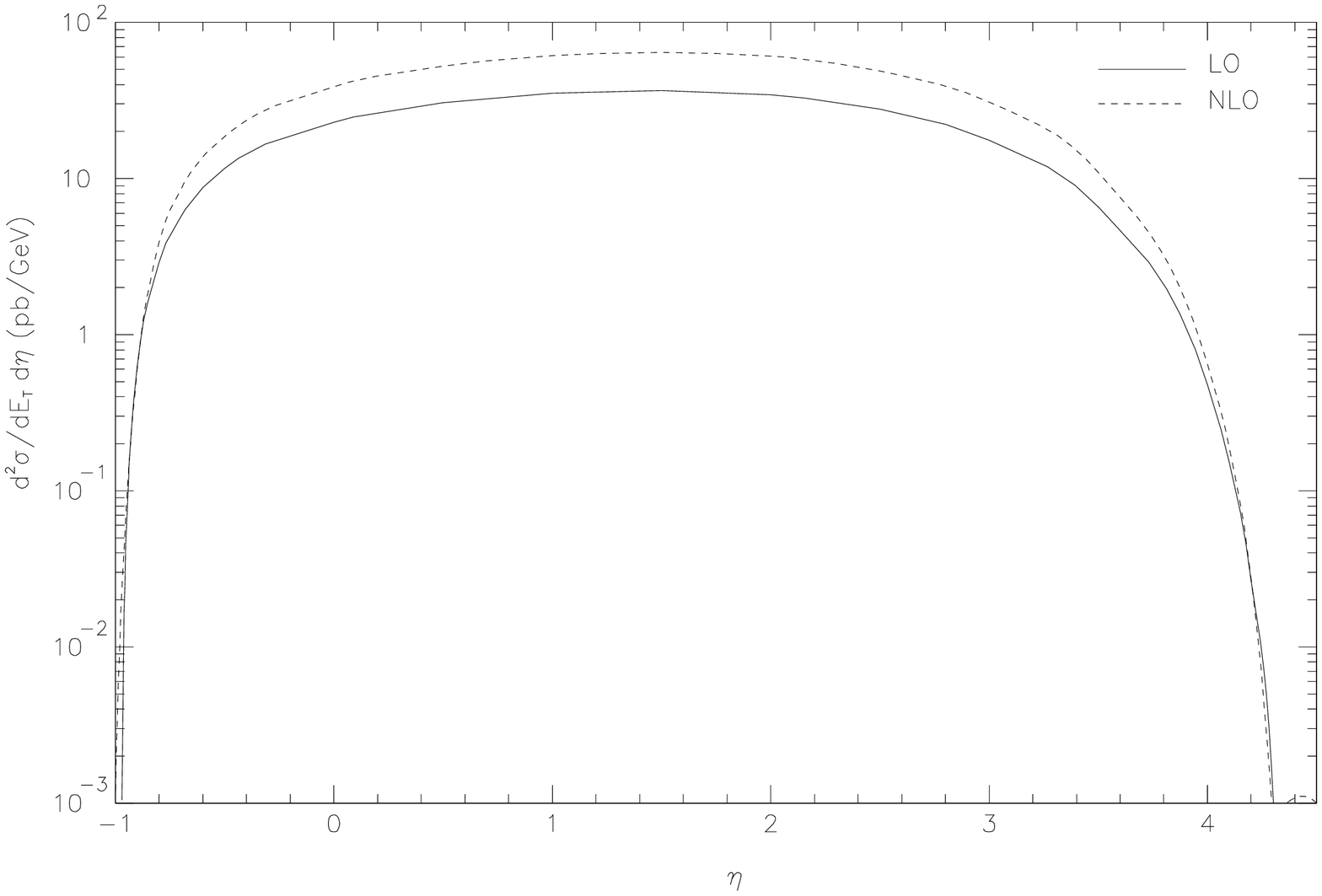,bbllx=520pt,bblly=95pt,bburx=105pt,bbury=710pt,%
           height=12cm,clip=,angle=270}
  \end{picture}}
  \caption[$\eta$-Dependence of Single-Jet Cross Section for Resolved Photons]
          {\label{plot24}{\it Inclusive single-jet cross section
           $\mbox{d}^2\sigma/\mbox{d}E_T\mbox{d}\eta$ for resolved photons
           as a function of $\eta$ for $E_T = 20$~GeV in LO and NLO.}}
 \end{center}
\end{figure}

We have already mentioned that the corresponding distribution for resolved
photons is expected to have an even broader maximum. This proves to be true
in figure \ref{plot24}, where the maximum is also slightly shifted in the 
proton direction from $\eta = 1$ to $\eta = 1.5$. The reason can be found in
the lower longitudinal momentum of photon components with respect to a direct
photon. The next-to-leading order cross section is about 80\% larger than
the leading order cross section, which is compatible to the observation of a
large $k$-factor from the $E_T$-distribution above.

% Plot23

% Plot24

\begin{figure}[p]
 \begin{center}
  {\unitlength1cm
  \begin{picture}(12,8)
   \epsfig{file=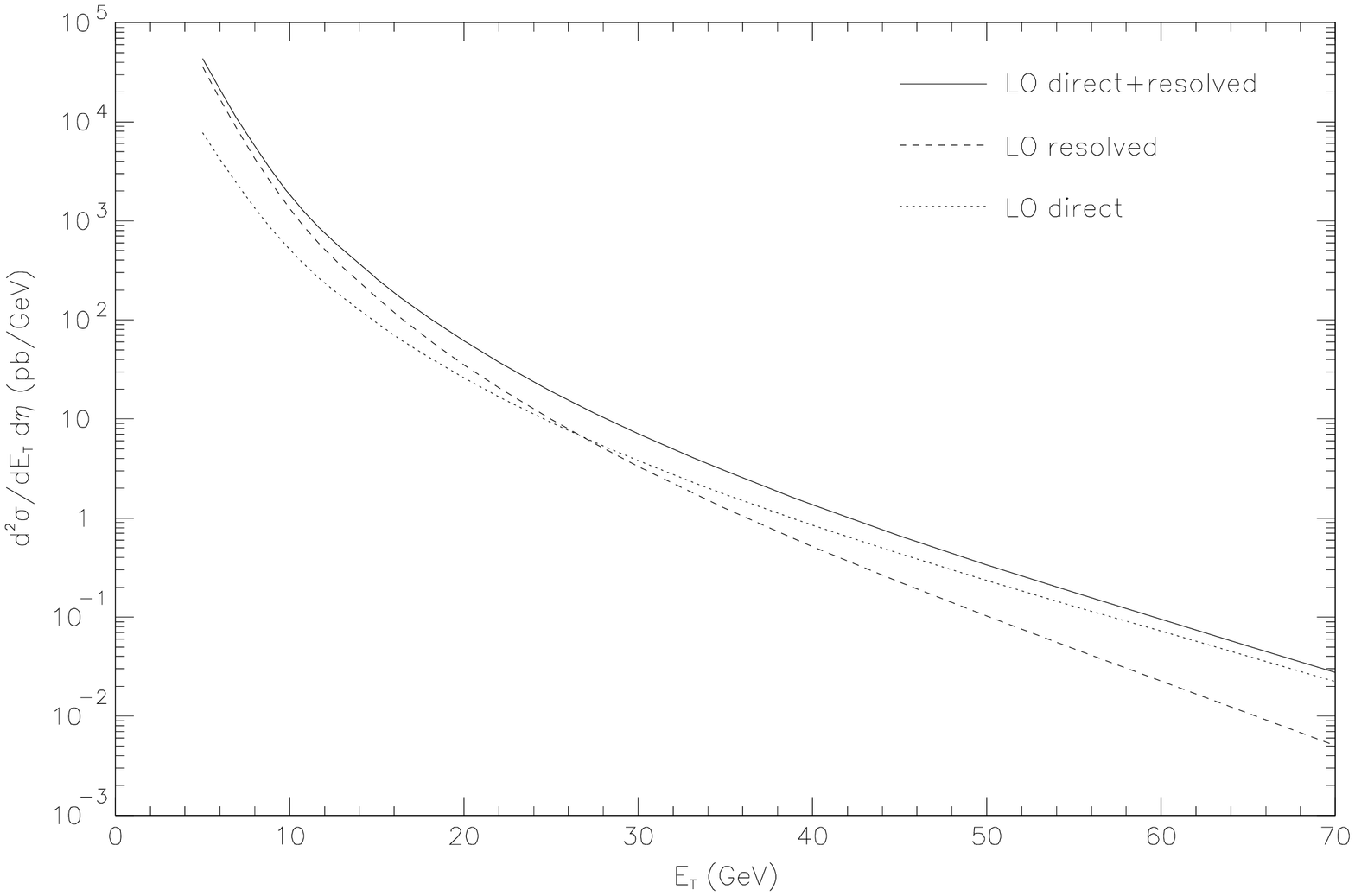,bbllx=520pt,bblly=95pt,bburx=105pt,bbury=710pt,%
           height=12cm,clip=,angle=270}
  \end{picture}}
  \caption[$E_T$-Dependence of Single-Jet Cross Section for Complete
           Photoproduction in LO]
          {\label{plot25}{\it Inclusive single-jet cross section
           $\mbox{d}^2\sigma/\mbox{d}E_T\mbox{d}\eta$ for complete
           photoproduction
           at $\eta = 1$, as a function of $E_T$. The full curve is the sum of
           the LO direct (dotted) and LO resolved (dashed) contributions.}}
 \end{center}
\end{figure}

\begin{figure}[p]
 \begin{center}
  {\unitlength1cm
  \begin{picture}(12,8)
   \epsfig{file=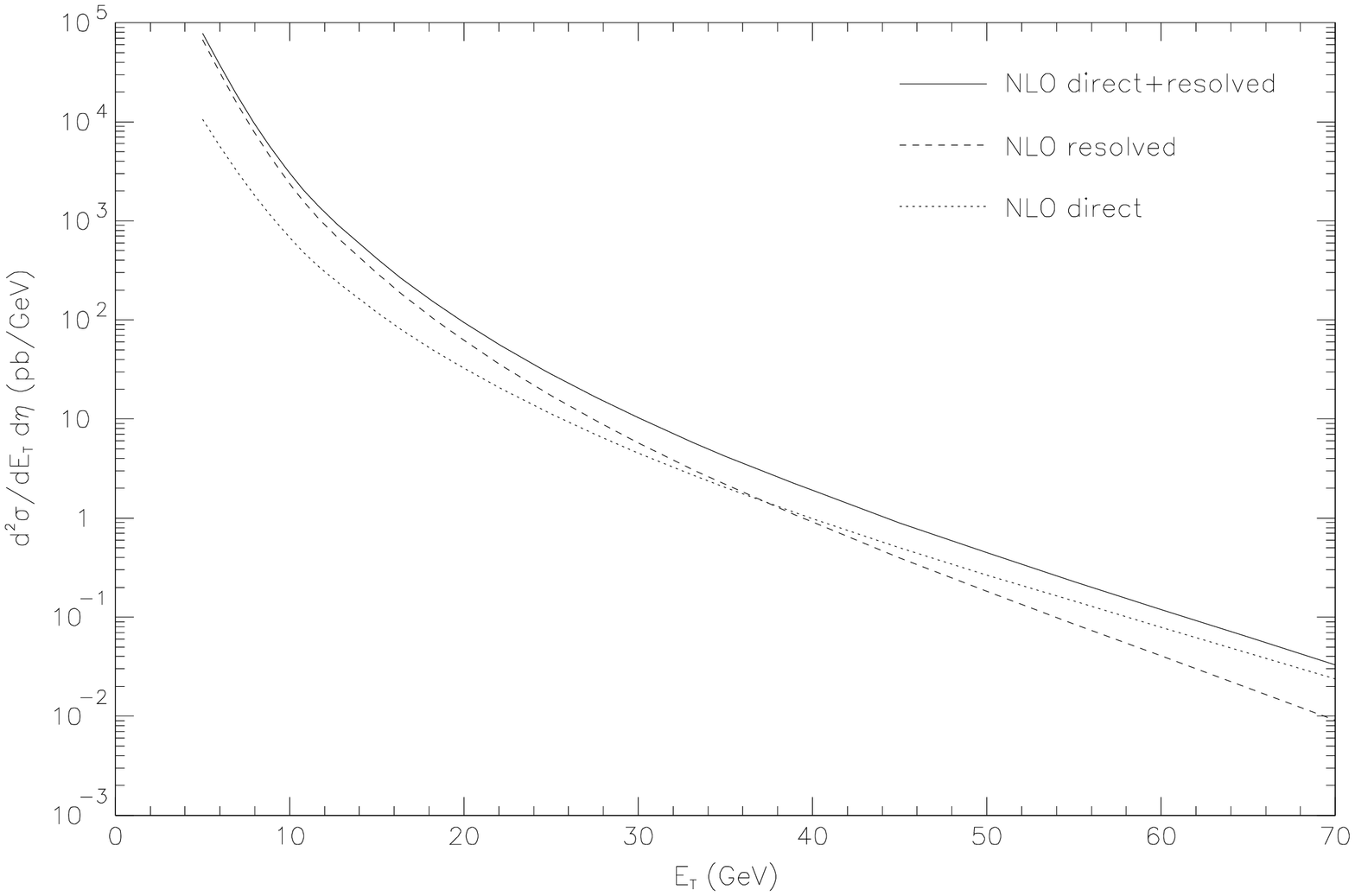,bbllx=520pt,bblly=95pt,bburx=105pt,bbury=710pt,%
           height=12cm,clip=,angle=270}
  \end{picture}}
  \caption[$E_T$-Dependence of Single-Jet Cross Section for Complete
           Photoproduction in NLO]
          {\label{plot26}{\it Inclusive single-jet cross section
           $\mbox{d}^2\sigma/\mbox{d}E_T\mbox{d}\eta$ for complete
           photoproduction
           at $\eta = 1$, as a function of $E_T$. The full curve is the sum of
           the NLO direct (dotted) and NLO resolved (dashed) contributions.}}
 \end{center}
\end{figure}

We now look at the interplay of direct and resolved photoproduction in
inclusive single-jet cross sections. Figures \ref{plot25} and \ref{plot26}
present the leading order and next-to-leading order predictions for complete
photoproduction (full curves), which are the sums of the direct (dotted) and
resolved (dashed) curves already presented above. The rapidity of the observed
jet is fixed at $\eta=1$. Obviously, the resolved photon component is
dominating at low transverse energies due to the photon structure function and
unimportant at large transverse energies, where the perturbative point-like
coupling is large. In leading order, the intersection of the resolved and
direct curves lies near 26~GeV and is shifted towards 37~GeV in next-to-leading
order. The resolved process has larger NLO corrections than the direct
process, so that the region, in which the latter is important, moves to
higher $E_T$.

% Plot25

% Plot26

\begin{figure}[p]
 \begin{center}
  {\unitlength1cm
  \begin{picture}(12,8)
   \epsfig{file=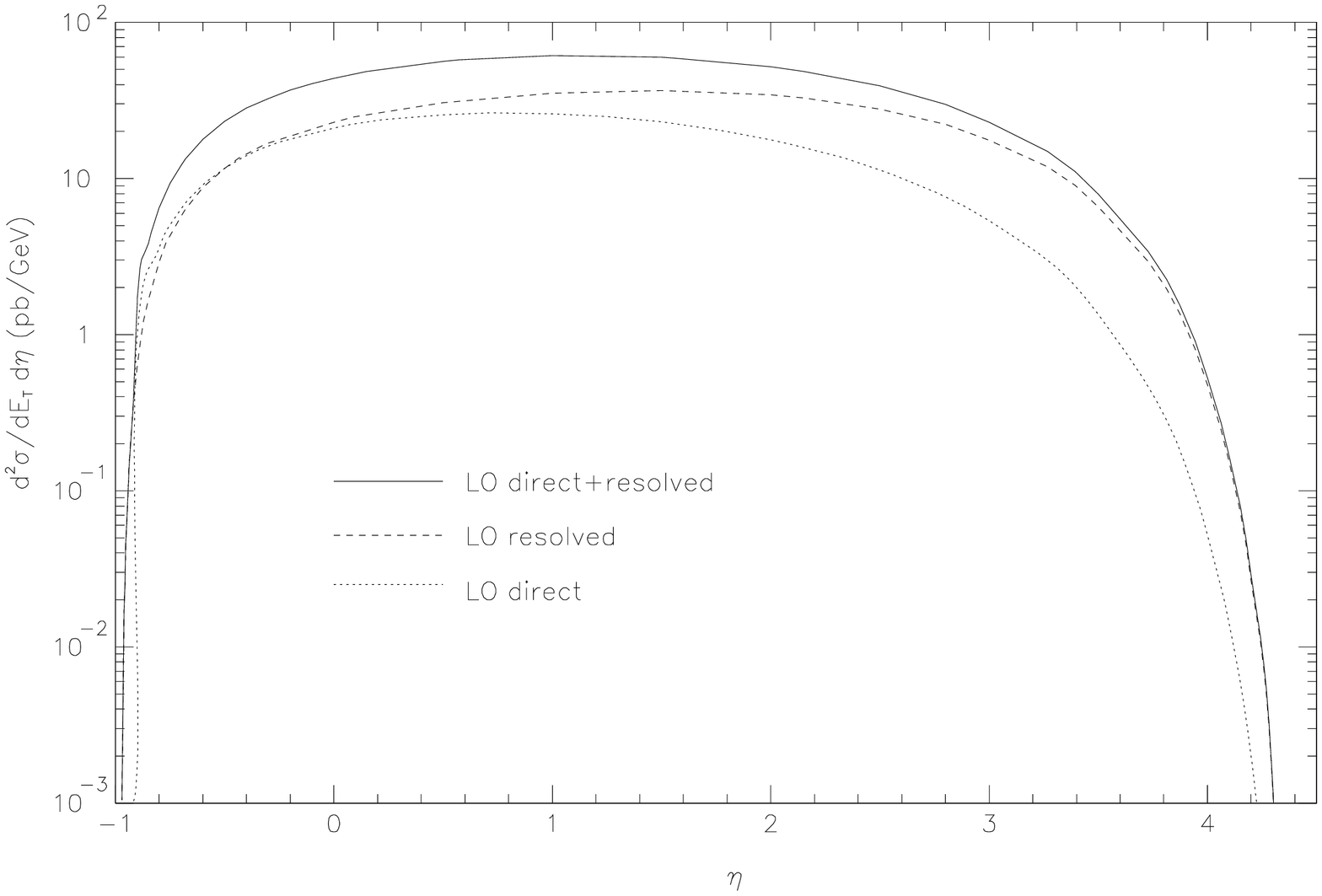,bbllx=520pt,bblly=95pt,bburx=105pt,bbury=710pt,%
           height=12cm,clip=,angle=270}
  \end{picture}}
  \caption[$\eta$-Dependence of Single-Jet Cross Section for Complete
           Photoproduction in LO]
          {\label{plot27}{\it Inclusive single-jet cross section $\mbox{d}^2
           \sigma/\mbox{d}E_T\mbox{d}\eta$ for complete
           photoproduction at
           $E_T=20$~GeV, as a function of $\eta$. The full curve is the sum of
           the LO direct (dotted) and LO resolved (dashed) contributions.}}
 \end{center}
\end{figure}

\begin{figure}[p]
 \begin{center}
  {\unitlength1cm
  \begin{picture}(12,8)
   \epsfig{file=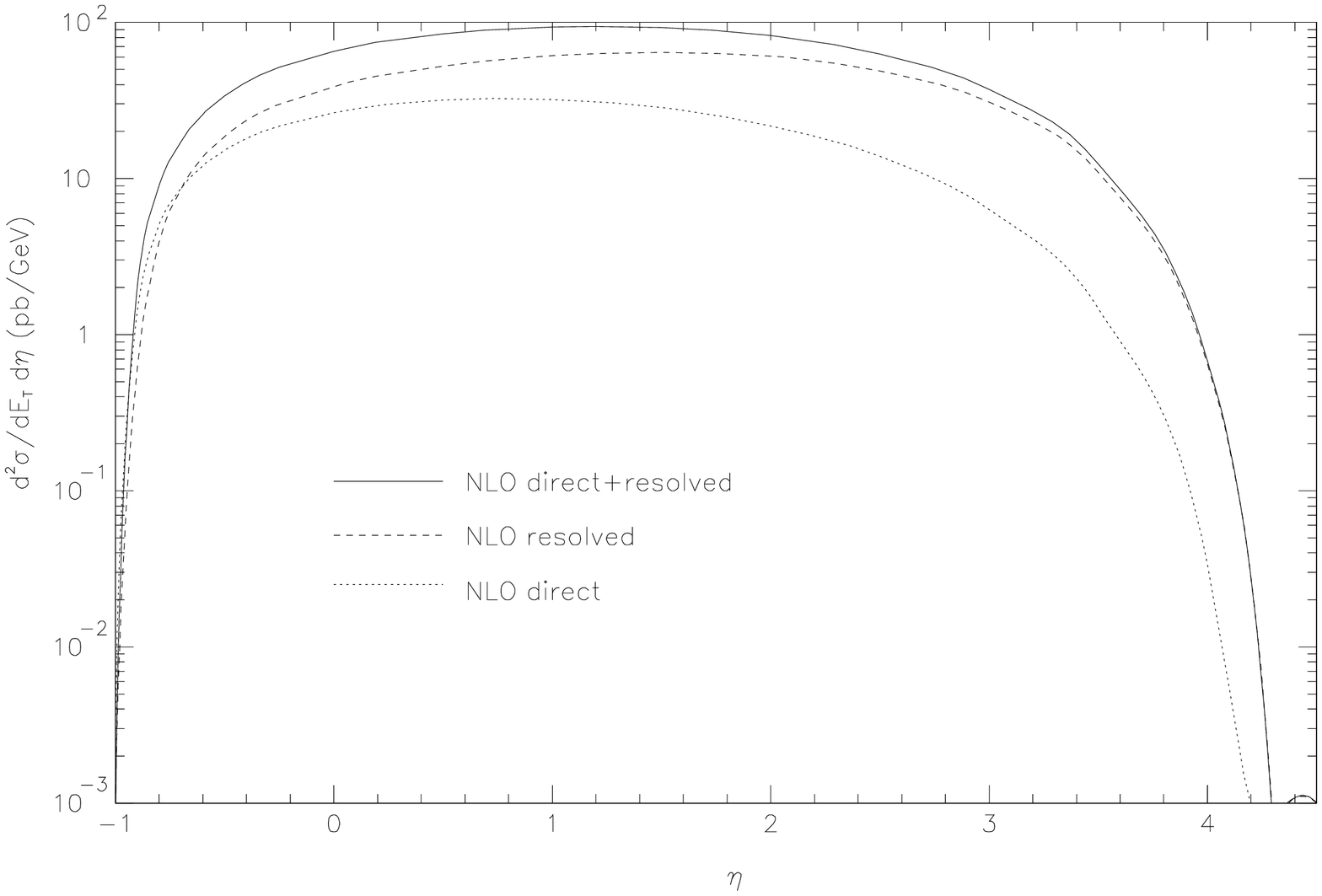,bbllx=520pt,bblly=95pt,bburx=105pt,bbury=710pt,%
           height=12cm,clip=,angle=270}
  \end{picture}}
  \caption[$\eta$-Dependence of Single-Jet Cross Section for Complete
           Photoproduction in NLO]
          {\label{plot28}{\it Inclusive single-jet cross section $\mbox{d}^2
           \sigma/\mbox{d}E_T\mbox{d}\eta$ for complete
           photoproduction at
           $E_T=20$~GeV, as a function of $\eta$. The full curve is the sum of
           the NLO direct (dotted) and NLO resolved (dashed) contributions.}}
 \end{center}
\end{figure}

The rapidity distributions for full photoproduction are shown as full curves
in figures \ref{plot27} and \ref{plot28} for leading and next-to-leading order
at a transverse energy of $E_T=20$~GeV. One expects the direct photon to be
important mostly in the electron direction at small or negative rapidities.
This behavior can be seen in both figures, but only close to the boundary
of phase space below $\eta=0$.

% Plot27

% Plot28

\begin{figure}[p]
 \begin{center}
  {\unitlength1cm
  \begin{picture}(12,8)
   \epsfig{file=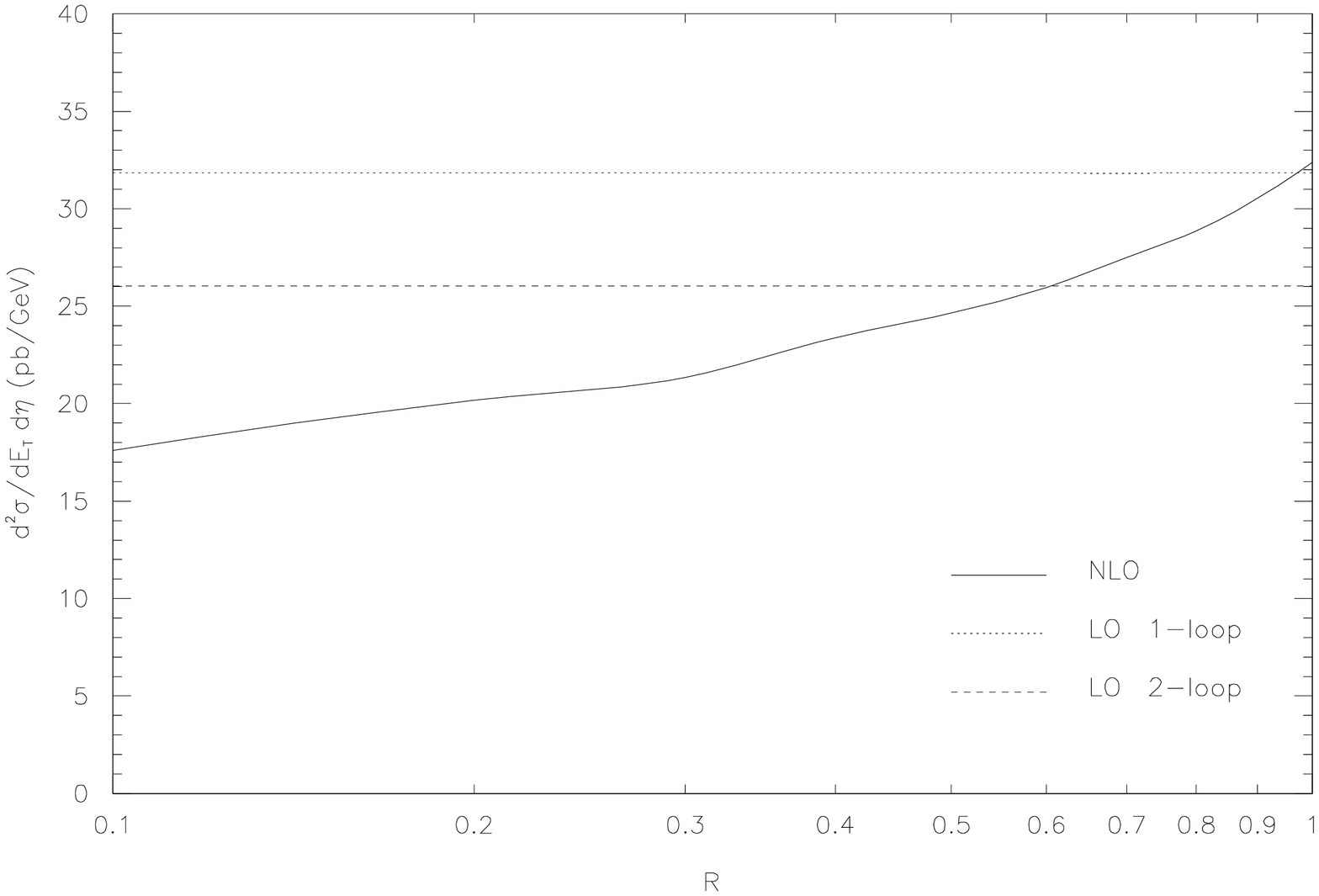,bbllx=520pt,bblly=95pt,bburx=105pt,bbury=710pt,%
           height=12cm,clip=,angle=270}
  \end{picture}}
  \caption[Jet Cone Size Dependence of Single-Jet Cross Section for Direct
           Photons]
          {\label{plot35}{\it Inclusive single-jet cross section $\mbox{d}^2
           \sigma/\mbox{d}E_T\mbox{d}\eta$ for direct photons at
           $E_T=20$~GeV and $\eta=1$, as a function of the jet cone size $R$.
           Only the NLO (full) curve and not the LO curves with one- (dotted)
           or two-loop (dashed) $\alpha_s$ depends on $R$.}}
 \end{center}
\end{figure}

\begin{figure}[p]
 \begin{center}
  {\unitlength1cm
  \begin{picture}(12,8)
   \epsfig{file=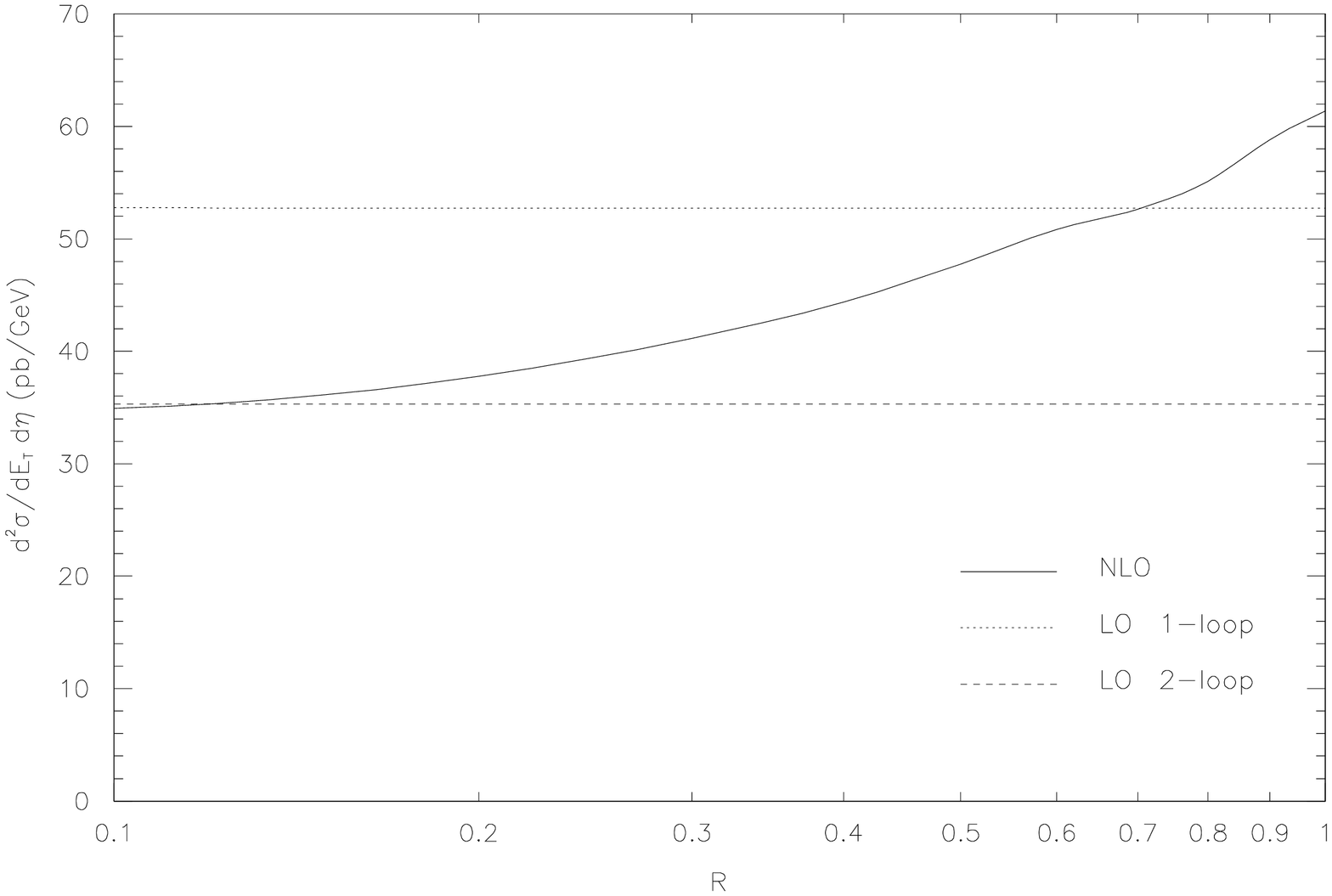,bbllx=520pt,bblly=95pt,bburx=105pt,bbury=710pt,%
           height=12cm,clip=,angle=270}
  \end{picture}}
  \caption[Jet Cone Size Dependence of Single-Jet Cross Section for Resolved
           Photons]
          {\label{plot36}{\it Inclusive single-jet cross section $\mbox{d}^2
           \sigma/\mbox{d}E_T\mbox{d}\eta$ for resolved photons at
           $E_T=20$~GeV and $\eta=1$, as a function of the jet cone size $R$.
           Only the NLO (full) curve and not the LO curves with one- (dotted)
           or two-loop (dashed) $\alpha_s$ depends on $R$.}}
 \end{center}
\end{figure}

In figures \ref{plot35} and \ref{plot36}, we plot the dependence of the
single-jet inclusive cross section on the jet cone size $R$ in the Snowmass
convention \cite{Hut92}. As discussed in section 2.4, a theoretical definition
for jets containing more than one parton is only possible in next-to-leading
order. Therefore only the full curves depend on $R$,
whereas the leading order curves are constant. The dependence has the
functional form of the next-to-leading order result \cite{Sal93,Kra94}
\beq
  \frac{\mbox{d}^2\sigma}{\mbox{d}E_T\mbox{d}\eta}
  = a+b\ln R+cR^2,
\eeq
For the direct photon
contribution in figure \ref{plot35}, the leading and higher order
predictions are equal at $R\simeq 1$ for one-loop $\alpha_s$ (dotted) and at
$R\simeq 0.6$
for two-loop $\alpha_s$ (dashed). For the resolved photon contribution in
figure \ref{plot36}, the situation is different. The stable points lie at
$R\simeq 0.7$ for one-loop (dotted) and $R\simeq 0.12$ for two-loop
$\alpha_s$.

% Plot35

% Plot36

\subsection{Two-Jet Cross Sections}

We now turn to two-jet cross sections, where one does not integrate over the
second rapidity $\eta_2$ or, alternatively, over the momentum fraction $x_a$
of the parton in the electron. The differential cross section
\beq
  \frac{\mbox{d}^3\sigma}{\mbox{d}E_T^2\mbox{d}\eta_1\mbox{d}\eta_2}
  = \sum_b x_a F_{a/e}(x_a,M_a^2) x_b F_{b/p}(x_b,M_b^2)
  \frac{\mbox{d}\sigma}{\mbox{d}t}(ab \rightarrow p_1p_2)
\eeq
then yields the maximum of information possible on the parton distributions
and is better suited to constrain them than the observation of inclusive
single-jets. Since dijet production is a more exclusive process than one-jet
production, the cross sections are smaller and require higher luminosity or
longer running time in the experiments. This is the reason why H1 and ZEUS
have only recently started to analyze dijet data and why we can present here
the first theoretical calculation for complete dijet photoproduction.

It is important to note that only in leading order the transverse energies
of the two observed jets balance $(E_{T_1}=E_{T_2}=E_T)$. In next-to-leading
order inclusive cross sections, there may be a third unobserved jet which
must have full freedom to become infinitely soft. Therefore, the transverse
energies of both jets cannot be observed without spoiling infrared safety.
This is an artifact of fixed order perturbation theory and will go away in
${\cal O} (\alpha\alpha_s^3)$, where one may as well have a fourth jet.
We will calculate similar distributions as in the last section, i.e.~in the
transverse energy {\em of the first jet}, $E_{T_1}$, and in both observable
rapidities $\eta_1$ and $\eta_2$. Yet, the rapidities $\eta_1$ and $\eta_2$
always belong to the two jets with largest transverse energy in the event. 
As in the one-jet case, the direct distributions have already been presented
in a recent paper \cite{x8}, whereas the resolved and complete cross
sections are shown here and partly in \cite{x12}.

\begin{figure}[p]
 \begin{center}
  {\unitlength1cm
  \begin{picture}(12,8)
   \epsfig{file=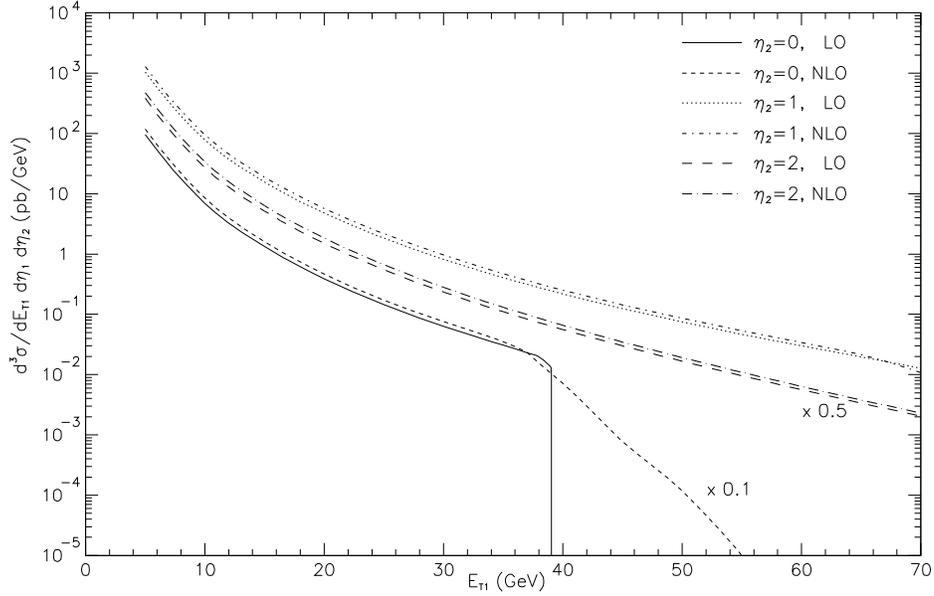,bbllx=520pt,bblly=95pt,bburx=105pt,bbury=710pt,%
           height=12cm,clip=,angle=270}
  \end{picture}}
  \caption[$E_T$-Dependence of Dijet Cross Section for Direct Photoproduction]
          {\label{plot29}{\it Inclusive dijet cross section $\mbox{d}^3
           \sigma/\mbox{d}E_{T_1}\mbox{d}\eta_1\mbox{d}\eta_2$ for direct
           photons as a function of $E_{T_1}$ for $\eta_1=1$ and three values
           of $\eta_2 =0,1,2$. The cross section for $\eta_2=0$ ($\eta_2=2$)
           is multiplied by 0.1 (0.5).}}
 \end{center}
\end{figure}

First, we look at the distributions in the transverse energy $E_{T_1}$ in
figure \ref{plot29} for direct photons. We fix $\eta_1$ at $\eta_1=1$ and
$\eta_2$ at three different values of $\eta_2 = 0,~1,~2$. The curves for
$\eta_2=0$ and $\eta_2=2$ are rescaled by factors of $0.1$ and $0.5$ as
before. We expect the dijet cross sections to be smaller than the single-jet
cross sections in figure \ref{plot21}. Indeed, for $E_T,~E_{T_1}=5$~GeV
and $\eta,~\eta_1,~\eta_2=1$, the cross section drops by almost an
order of magnitude from $10.6$~nb to $1.29$~nb. The $k$-factors in dijet
production are basically the same as in single-jet production. There is,
however, one novel feature: for $\eta_2=0$, the third jet in next-to-leading
order opens up some additional phase space, so that $E_{T_1}$ can go up to
$E_{T_1}< 55$ GeV instead of only 39~GeV in leading order.

\begin{figure}[p]
 \begin{center}
  {\unitlength1cm
  \begin{picture}(12,8)
   \epsfig{file=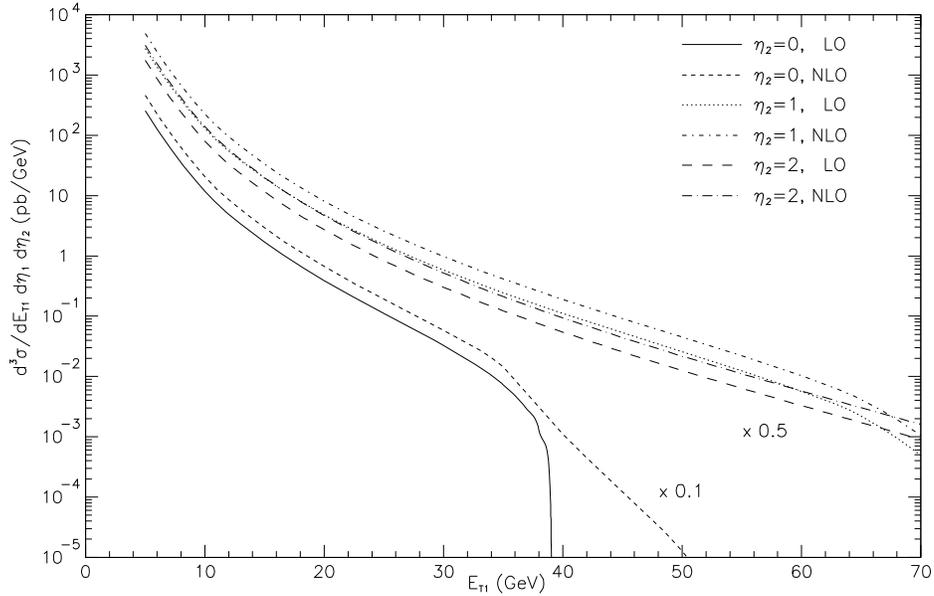,bbllx=520pt,bblly=95pt,bburx=105pt,bbury=710pt,%
           height=12cm,clip=,angle=270}
  \end{picture}}
  \caption[$E_T$-Dependence of Dijet Cross Section for Resolved
           Photoproduction]
          {\label{plot30}{\it Inclusive dijet cross section $\mbox{d}^3
           \sigma/\mbox{d}E_{T_1}\mbox{d}\eta_1\mbox{d}\eta_2$ for resolved
           photons as a function of $E_{T_1}$ for $\eta_1=1$ and three values
           of $\eta_2 =0,1,2$. The cross section for $\eta_2=0$ ($\eta_2=2$)
           is multiplied by 0.1 (0.5).}}
 \end{center}
\end{figure}

A similar behavior is seen for resolved photoproduction in figure \ref{plot30}.
At the lowest value of $E_{T_1}=5$~GeV and back-to-back jets
($\eta_1=\eta_2=1$), the two-jet cross section ($4.91$~nb) is even more than
one order of magnitude smaller than the one-jet cross section ($67.5$~nb). The
ratio of NLO to LO is 1.7 for resolved photons over the whole $E_{T_1}$-range.
The phase space for $\eta_2=0$ increases from
$E_{T_1}=39$~GeV to $E_{T_1}=50$~GeV. This slightly smaller value is due to
the reduced center-of-mass energy in resolved photoproduction, as some of the
energy always goes into the photon remnant.

% Plot29

% Plot30

\begin{figure}[p]
 \begin{center}
  {\unitlength1cm
  \begin{picture}(12,8)
   \epsfig{file=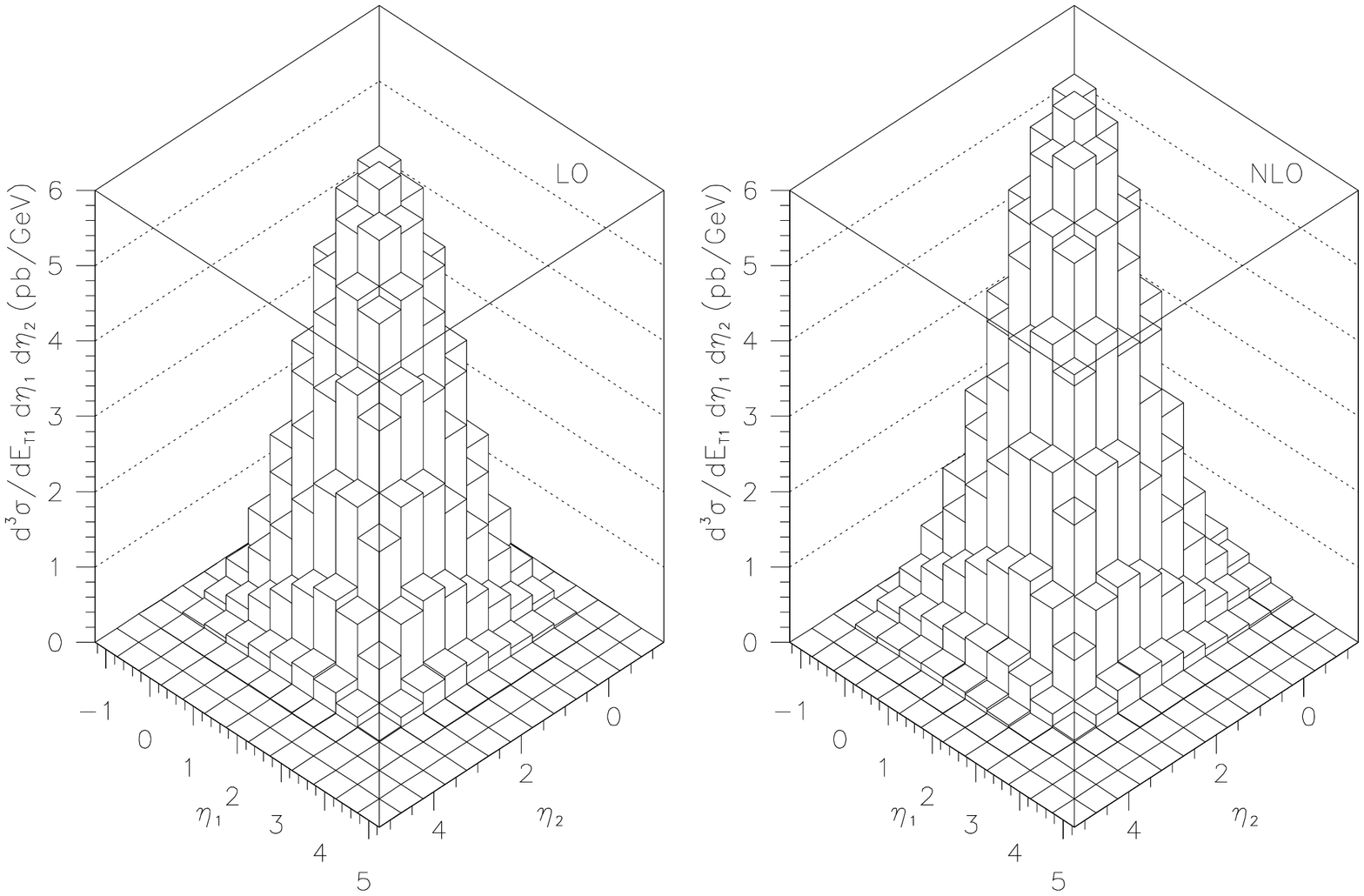,bbllx=520pt,bblly=95pt,bburx=105pt,bbury=710pt,%
           height=12cm,clip=,angle=270}
  \end{picture}}
  \caption[Lego-Plot of Dijet Cross Section for Direct Photoproduction]
          {\label{plot31}{\it Inclusive dijet cross section $\mbox{d}^3
           \sigma/\mbox{d}E_{T_1}\mbox{d}\eta_1\mbox{d}\eta_2$ at $E_{T_1}=
           20$~GeV for direct photons, as a function of $\eta_1$ and $\eta_2$.
           The LO plot (left) is exactly symmetric, the NLO plot (right) only
           approximately.}}
 \end{center}
\end{figure}

\begin{figure}[p]
 \begin{center}
  {\unitlength1cm
  \begin{picture}(12,8)
   \epsfig{file=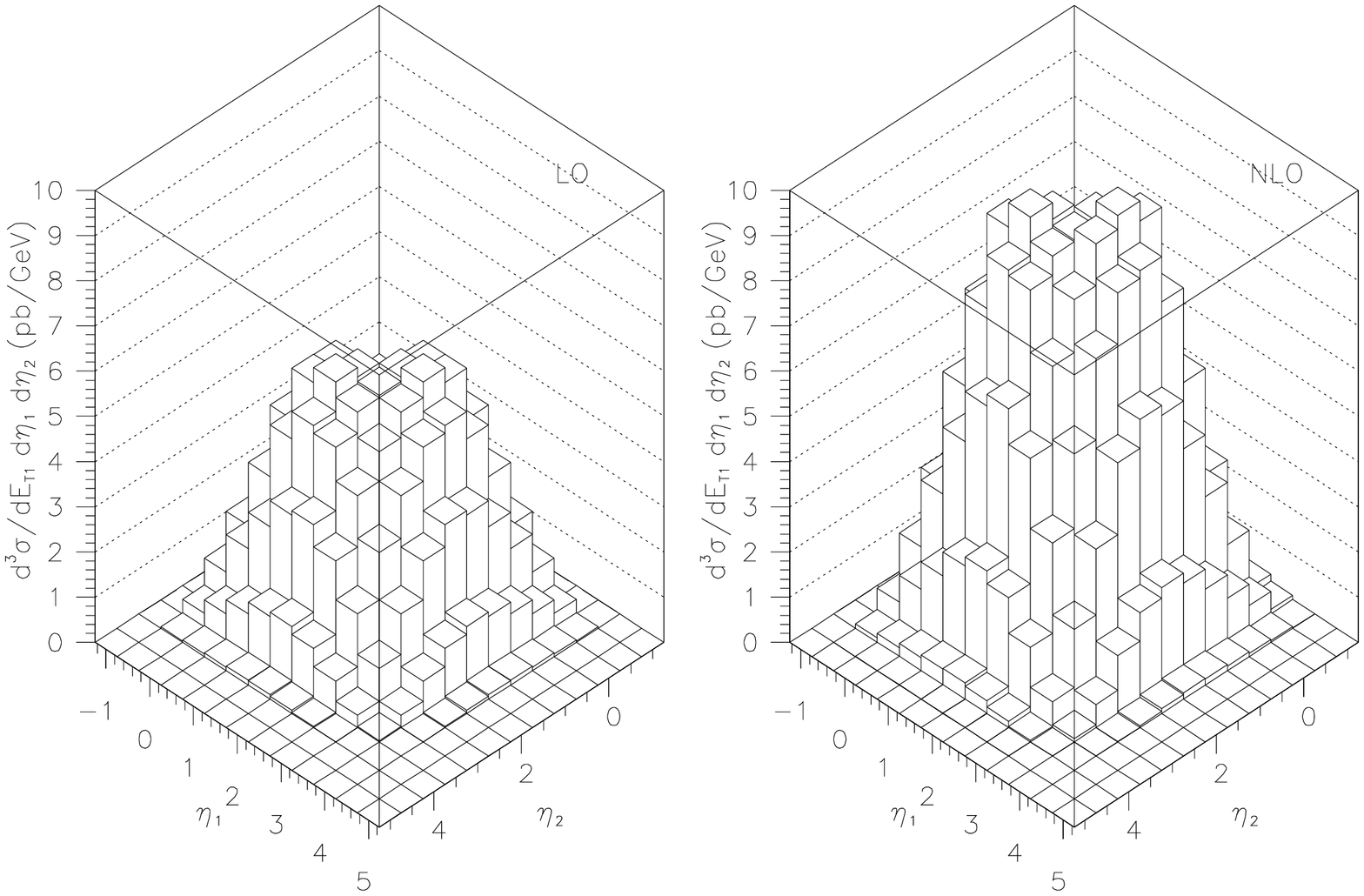,bbllx=520pt,bblly=95pt,bburx=105pt,bbury=710pt,%
           height=12cm,clip=,angle=270}
  \end{picture}}
  \caption[Lego-Plot of Dijet Cross Section for Resolved Photoproduction]
          {\label{plot32}{\it Inclusive dijet cross section $\mbox{d}^3
           \sigma/\mbox{d}E_{T_1}\mbox{d}\eta_1\mbox{d}\eta_2$ at $E_{T_1}=
           20$~GeV for resolved photons, as a function of $\eta_1$ and
           $\eta_2$.
           The LO plot (left) is exactly symmetric, the NLO plot (right) only
           approximately.}}
 \end{center}
\end{figure}

Next, we present the dependence of the cross sections on the two rapidities in
form of the three-dimensional lego-plots \ref{plot31} and \ref{plot32}.
The leading order is always shown on the left side and is completely symmetric
in $\eta_1$ and $\eta_2$. For the next-to-leading order cross sections on the
right hand sides of figures \ref{plot31} and \ref{plot32}, this is no longer
exactly true due to the presence of a ``trigger'' jet with transverse energy
$E_{T_1}$, which is fixed at $E_{T_1}=20$~GeV. The next-to-leading order
lego-plots are only approximately
symmetric. This can best be seen at the bottom of the contour plots, where
at least one of the two observed jets is far off the central region. The
NLO predictions are considerably larger than the LO predictions, especially
in the resolved case.

% Plot31

% Plot32

\begin{figure}[p]
 \begin{center}
  {\unitlength1cm
  \begin{picture}(12,8)
   \epsfig{file=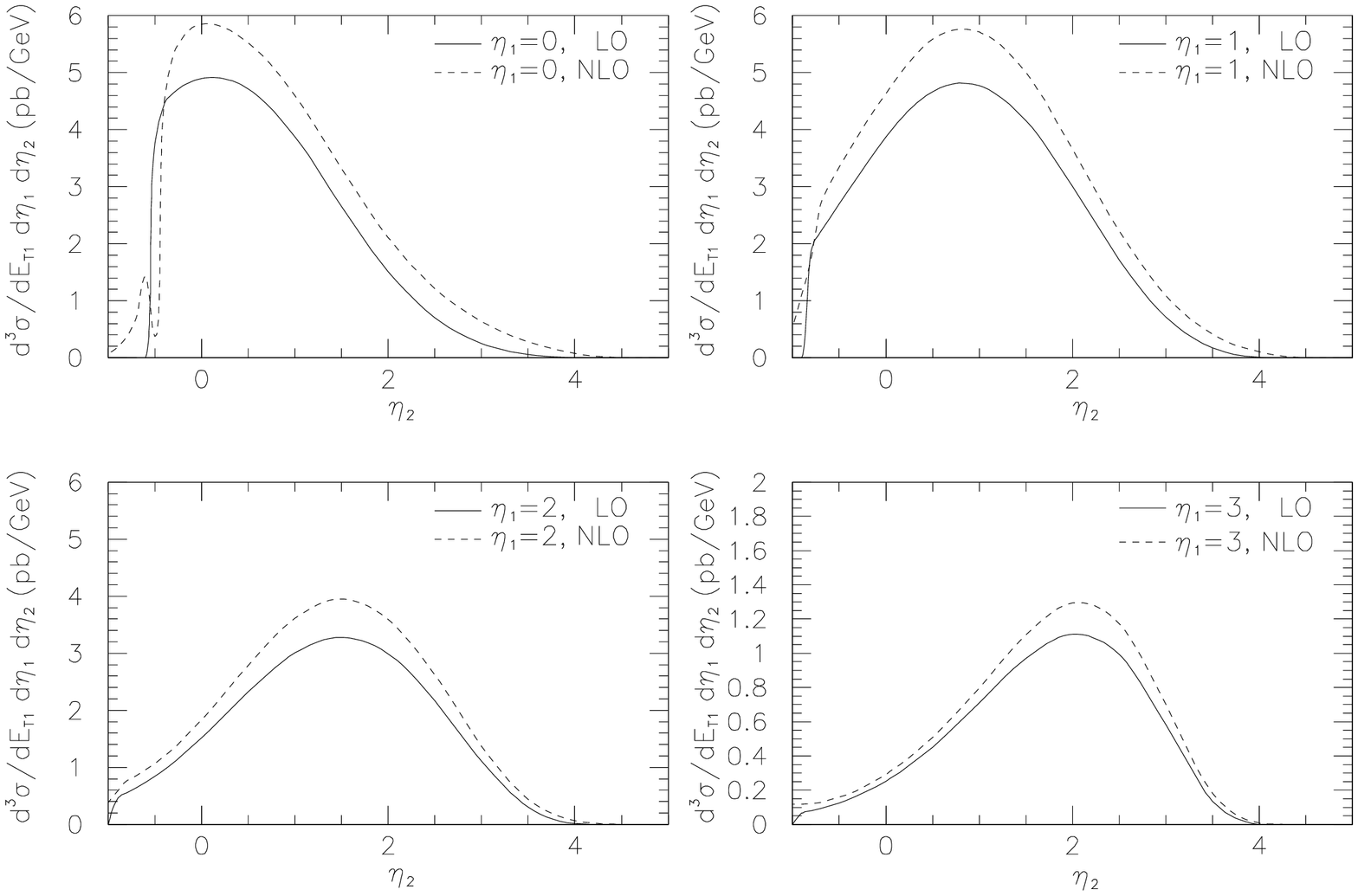,bbllx=520pt,bblly=95pt,bburx=105pt,bbury=710pt,%
           height=12cm,clip=,angle=270}
  \end{picture}}
  \caption[Rapidity Dependence of Dijet Cross Section for Direct
           Photoproduction]
          {\label{plot33}{\it Projections of the LO (full curves) and NLO 
           (dashed curves) triple differential dijet cross section for direct
           photons at $E_{T_1}=20$~GeV and fixed values of $\eta_1=0,~1,~2,$
           and $3$, as a function of $\eta_2$.}}
 \end{center}
\end{figure}

This becomes even clearer when we plot the projections of the lego-plots for
fixed $\eta_1=0,~1,~2$ and $3$. In figure \ref{plot33}, we plot the leading
and next-to-leading order distributions in $\eta_2$ for direct photoproduction.
It is clearly seen that the second jet tends to be back-to-back with the first
jet, since the maximum always occurs at $\eta_2\simeq\eta_1$. However, at
$\eta_1=3$ this is no more permitted by phase space. We obtain the same
$k$-factor of 1.25 in the central regions as in single-jet production.

\begin{figure}[p]
 \begin{center}
  {\unitlength1cm
  \begin{picture}(12,8)
   \epsfig{file=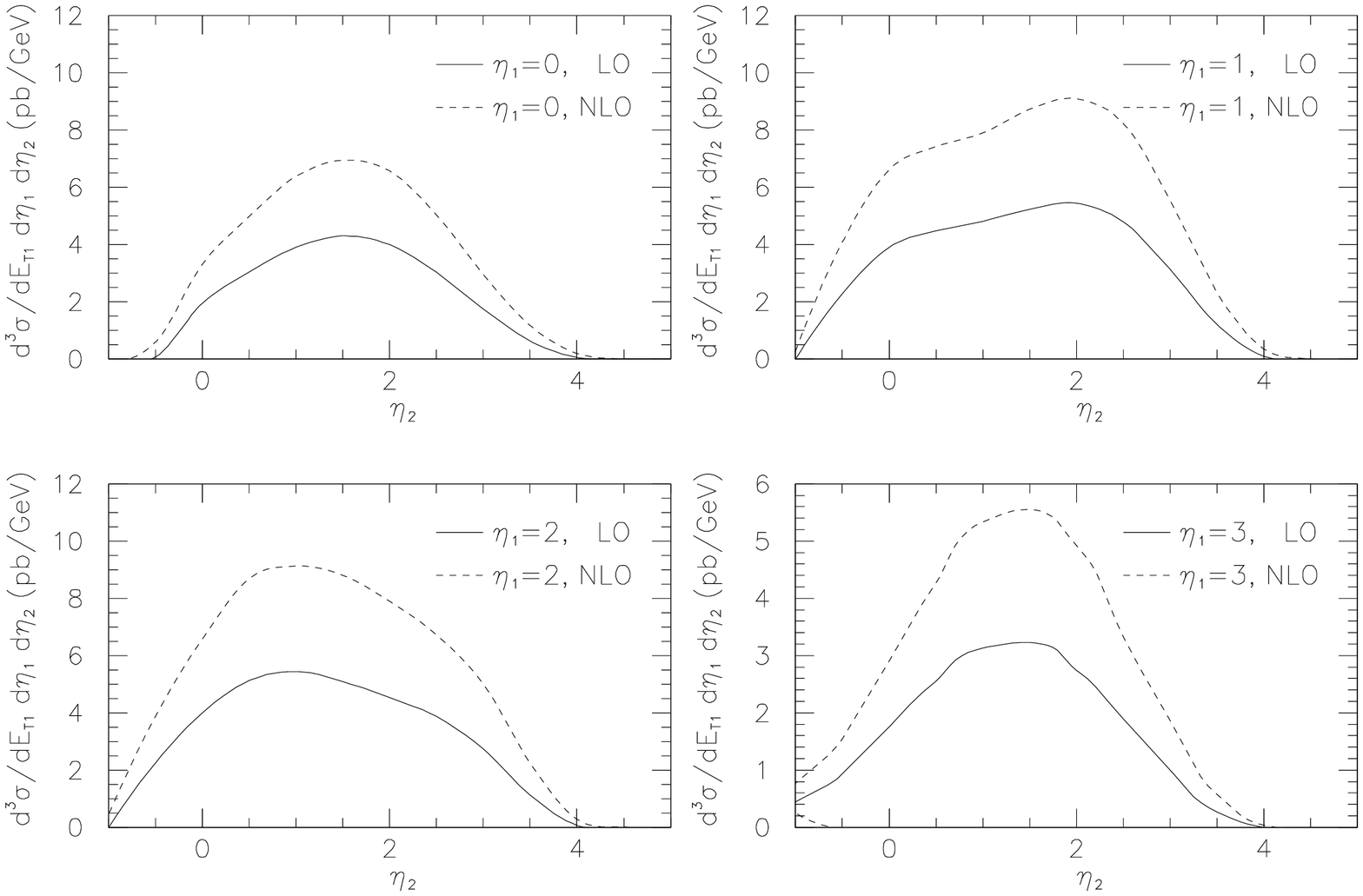,bbllx=520pt,bblly=95pt,bburx=105pt,bbury=710pt,%
           height=12cm,clip=,angle=270}
  \end{picture}}
  \caption[Rapidity Dependence of Dijet Cross Section for Resolved
           Photoproduction]
          {\label{plot34}{\it Projections of the LO (full curves) and NLO 
           (dashed curves) triple differential dijet cross section for resolved
           photons at $E_{T_1}=20$~GeV and fixed values of $\eta_1=0,~1,~2,$
           and $3$, as a function of $\eta_2$.}}
 \end{center}
\end{figure}

The $\eta_2$-distributions for resolved photons in figure \ref{plot34} are
considerably broader than those in figure \ref{plot33} due to the smearing
of the hard cross sections with the distribution function of partons in the
photon. The maxima of the plots are also not so much dominated by kinematics
but more by the quark and gluon structure of the photon in different $x$
regimes. They do not lie at $\eta_2=\eta_1$ any more.
Therefore, dijet rapidity distributions are best suited to
constrain the photon structure. We will come back to this point in the
next section 5.5, when we compare similar plots to data from ZEUS. The
$k$-factors range from 1.65 in the central regions to more than 3 in the
proton forward direction. The shapes of the distributions are very similar
in LO and in NLO. The absolute values make, however, an important
difference.

% Plot33

% Plot34

\begin{figure}[p]
 \begin{center}
  {\unitlength1cm
  \begin{picture}(12,8)
   \epsfig{file=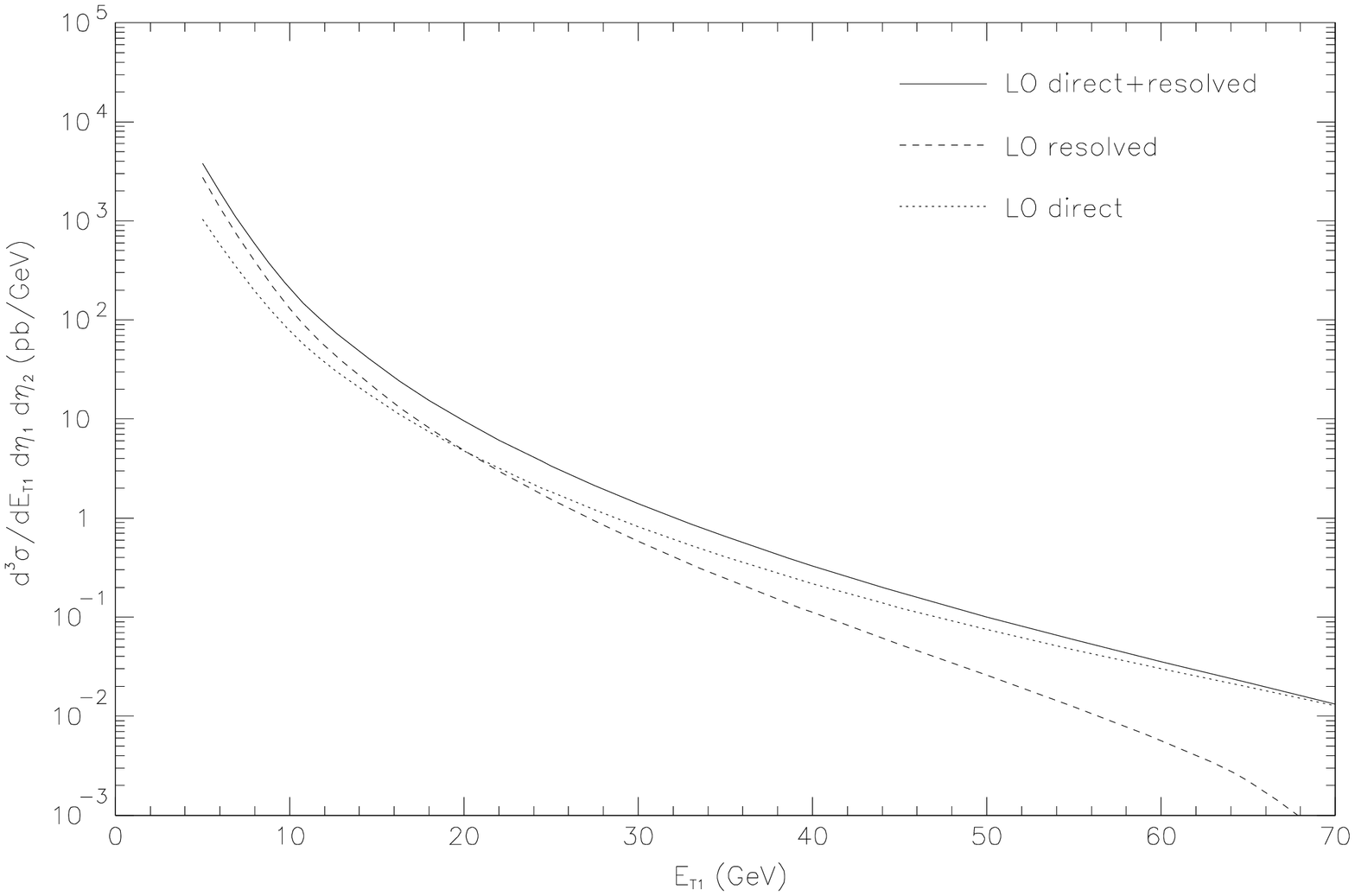,bbllx=520pt,bblly=95pt,bburx=105pt,bbury=710pt,%
           height=12cm,clip=,angle=270}
  \end{picture}}
  \caption[$E_T$-Dependence of Dijet Cross Section for Complete
           Photoproduction in LO]
          {\label{plot37}{\it Inclusive dijet cross section $\mbox{d}^3\sigma
           /\mbox{d}E_{T_1}\mbox{d}\eta_1\mbox{d}\eta_2$ for full
           photoproduction at $\eta_1=\eta_2=1$ as a function of $E_{T_1}$.
           The full curve is the sum of the LO direct (dotted) and LO resolved
           (dashed) contributions.}}
 \end{center}
\end{figure}

\begin{figure}[p]
 \begin{center}
  {\unitlength1cm
  \begin{picture}(12,8)
   \epsfig{file=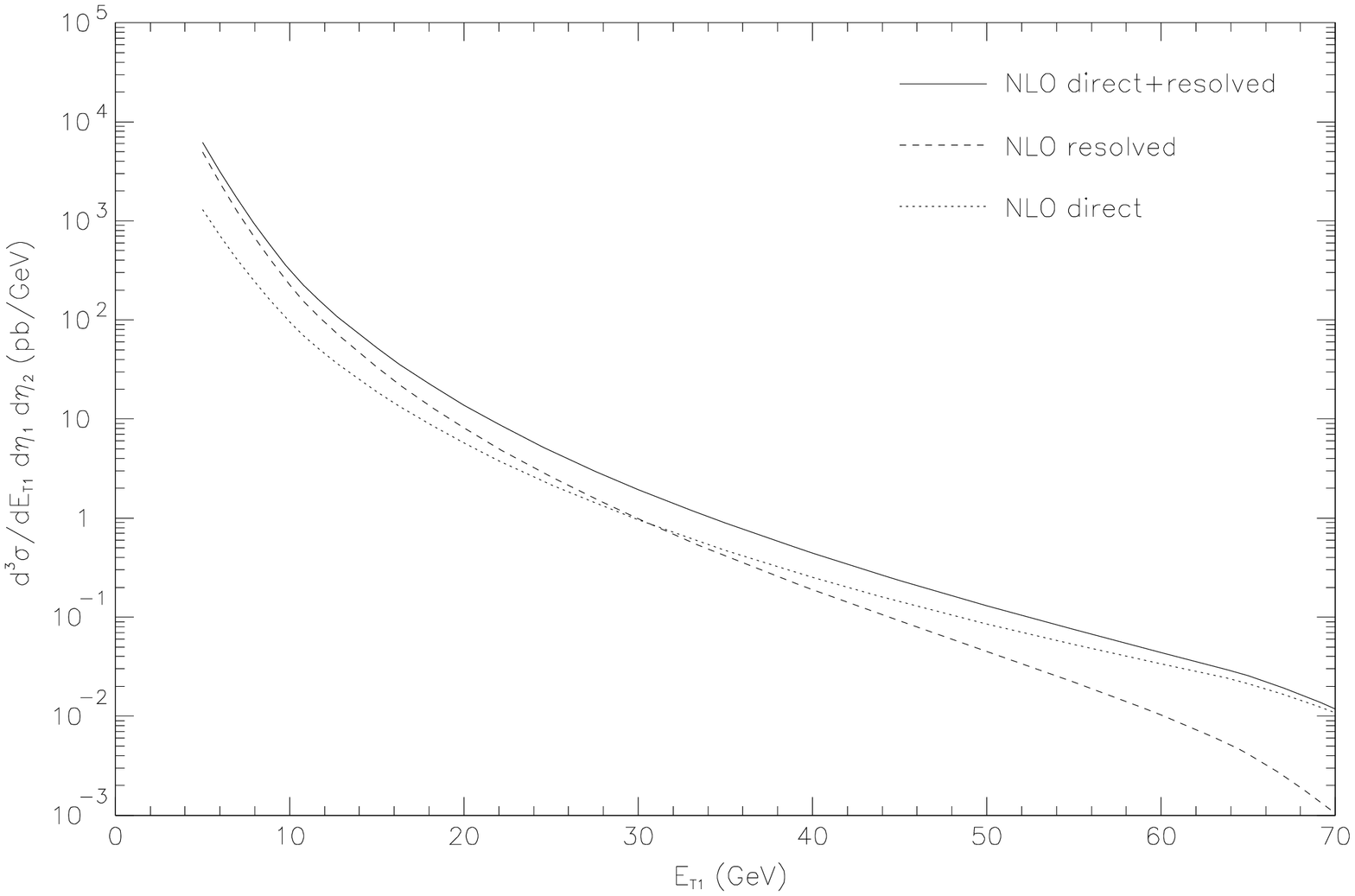,bbllx=520pt,bblly=95pt,bburx=105pt,bbury=710pt,%
           height=12cm,clip=,angle=270}
  \end{picture}}
  \caption[$E_T$-Dependence of Dijet Cross Section for Complete
           Photoproduction in NLO]
          {\label{plot38}{\it Inclusive dijet cross section $\mbox{d}^3\sigma
           /\mbox{d}E_{T_1}\mbox{d}\eta_1\mbox{d}\eta_2$ for full
           photoproduction at $\eta_1=\eta_2=1$ as a function of $E_{T_1}$.
           The full curve is the sum of the NLO direct (dotted) and NLO
           resolved (dashed) contributions.}}
 \end{center}
\end{figure}

Direct and resolved contributions are now added to give physical, complete
photoproduction results. The dijet cross section is first plotted as a function
of the transverse energy $E_{T_1}$. Figure \ref{plot37} gives the LO result,
figure \ref{plot38} the NLO result. Like the direct and resolved cross sections
alone, the full two-jet cross sections are about an order of magnitude smaller
than the one-jet cross sections (see figures \ref{plot25} and \ref{plot26}).
The point where direct and resolved contributions are equally important is
lowered towards $E_{T_1}=20$~GeV in leading order and $E_{T_1}=30$~GeV in
next-to-leading order, so that direct photons are better observed in
dijet production.

% Plot37

% Plot38

\begin{figure}[p]
 \begin{center}
  {\unitlength1cm
  \begin{picture}(12,8)
   \epsfig{file=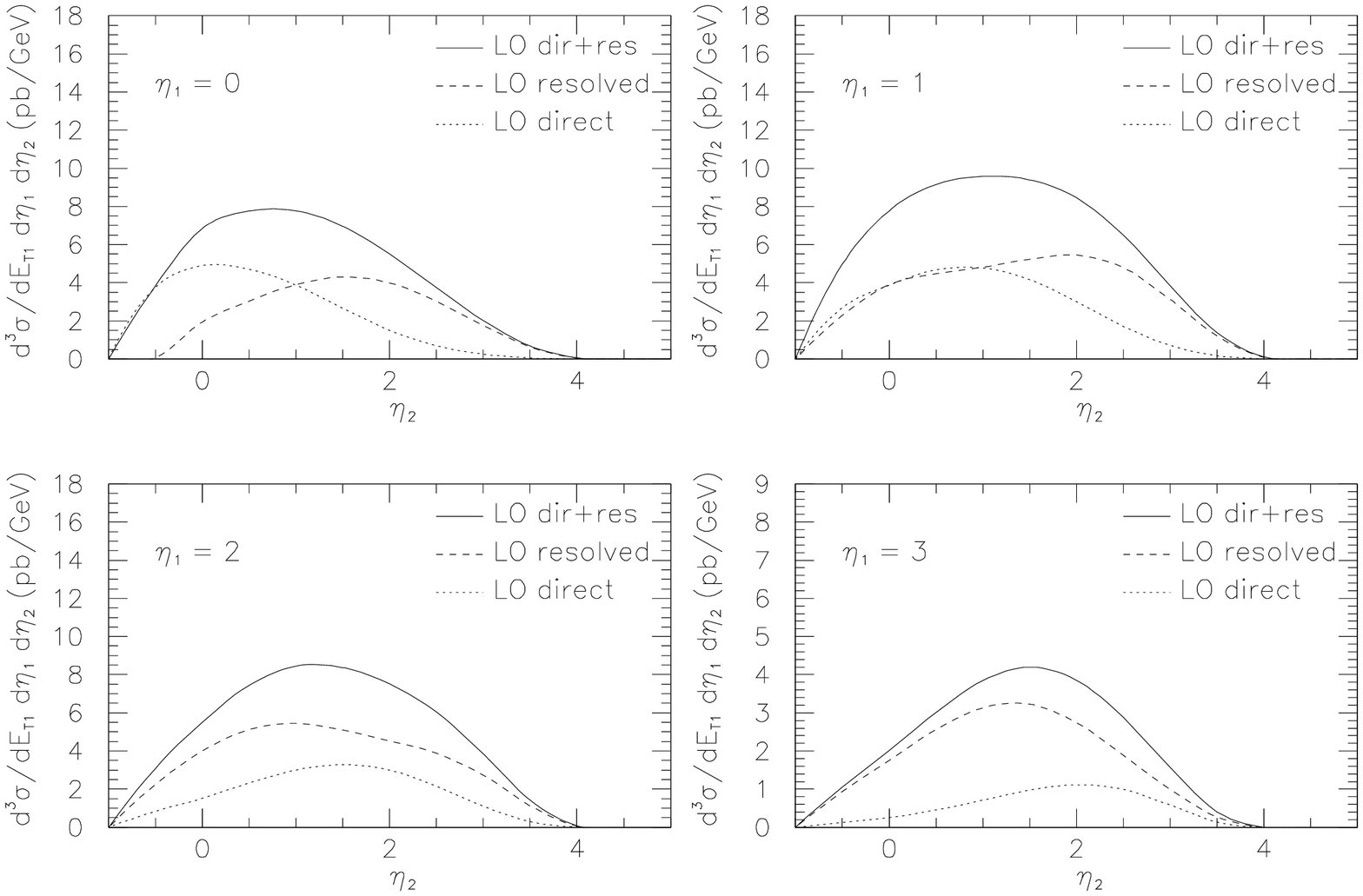,bbllx=520pt,bblly=95pt,bburx=105pt,bbury=710pt,%
           height=12cm,clip=,angle=270}
  \end{picture}}
  \caption[Rapidity Dependence of Dijet Cross Section for Complete
           Photoproduction in LO]
          {\label{plot39}{\it Projections of the complete triple differential
           dijet cross section at $E_{T_1}=20$~GeV and fixed values of
           $\eta_1=0,~1,~2,$ and $3$, as a function of $\eta_2$.
           The full curve is the sum of the LO direct (dotted) and LO
           resolved (dashed) contributions.}}
 \end{center}
\end{figure}

\begin{figure}[p]
 \begin{center}
  {\unitlength1cm
  \begin{picture}(12,8)
   \epsfig{file=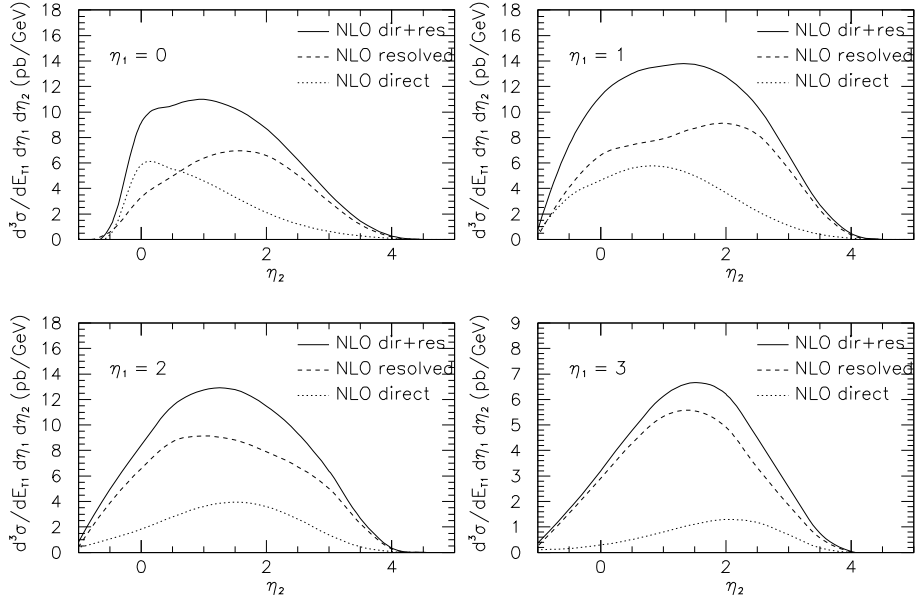,bbllx=520pt,bblly=95pt,bburx=105pt,bbury=710pt,%
           height=12cm,clip=,angle=270}
  \end{picture}}
  \caption[Rapidity Dependence of Dijet Cross Section for Complete
           Photoproduction in NLO]
          {\label{plot40}{\it Projections of the complete triple differential
           dijet cross section at $E_{T_1}=20$~GeV and fixed values of
           $\eta_1=0,~1,~2,$ and $3$, as a function of $\eta_2$.
           The full curve is the sum of the NLO direct (dotted) and NLO
           resolved (dashed) contributions.}}
 \end{center}
\end{figure}

If one plots the complete two-jet cross sections as a function of $\eta_2$,
the different behaviors of direct and resolved photons add up to the full
curves in figures \ref{plot39} and \ref{plot40}. These plots are best
suited to decide in which rapidity regions one can look best for the resolved
photon structure. We have already seen that this will be in situations where
the two jets are not back-to-back, e.g.~for $\eta_1=0$ and positive $\eta_2$
in the upper left plots of figures \ref{plot39} and \ref{plot40}. On the other
hand, the proton structure can best be studied with direct photons, when the
cross section is not folded with another distribution. A possible scenario is
$\eta_1=0$ and negative values of $\eta_2$. This is especially interesting
for the small-$x$ components of the proton like the gluons and the quark sea.
Another interesting observation is that the relative importance of direct
and resolved processes changes dramatically when calculating dijet
photoproduction in next-to-leading order ${\cal O} (\alpha\alpha_s^2)$:
resolved processes are much more important at $E_{T_1}=20$~GeV than one would
have guessed from a leading order estimate.

% Plot39

% Plot40

\subsection{Comparison of Photoproduction Results to H1 and ZEUS Data}
In this section we compare the next-to-leading order calculation to
recent one- and two-jet data from the H1 and ZEUS collaborations at HERA.
Both collaborations have continuously measured various cross sections for
the photoproduction of jets since HERA started running in 1992. With the
increased luminosity in recent years, the data have improved and many
aspects of jet production could be studied. In our comparison we restrict
ourselves to the measurements of one- and two-jet cross sections just
recently published which are based either on 1994 or 1995 data. In particular,
we shall compare with the inclusive single-jet data of 1994 from the ZEUS
collaboration \cite{y1}, with the inclusive dijet data of 1994 from ZEUS
\cite{y2}, with inclusive dijet data of 1995 from ZEUS \cite{y2}, and
with inclusive dijet data of 1994 from the H1 collaboration \cite{y3}.

\begin{figure}[p]
 \begin{center}
  {\unitlength1cm
  \begin{picture}(12,8)
   \epsfig{file=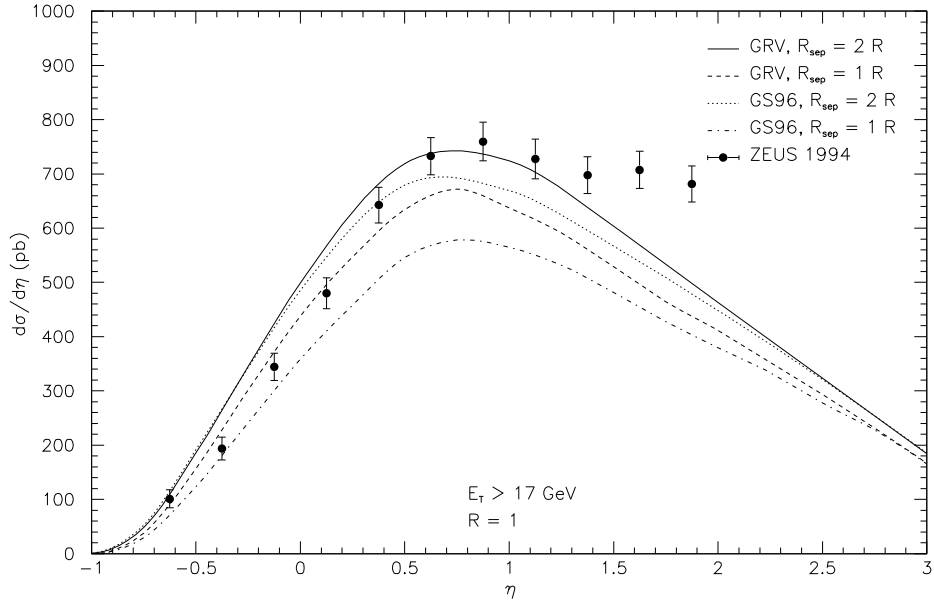,bbllx=520pt,bblly=95pt,bburx=105pt,bbury=710pt,%
           height=12cm,angle=270,clip=}
  \end{picture}}
 \end{center}
 \caption{\label{kkkplot1a17r1}{\it $\eta$ dependence of the inclusive
           single-jet photoproduction cross
           section integrated over $E_T>17$GeV with jet cone size $R=1$.
           We compare our NLO
           prediction with GRV and GS96 photon parton densities and the two
           extreme $R_{\rm sep}$ values to 1994 data
           from ZEUS.}}
\end{figure}

\begin{figure}[p]
 \begin{center}
  {\unitlength1cm
  \begin{picture}(12,8)
   \epsfig{file=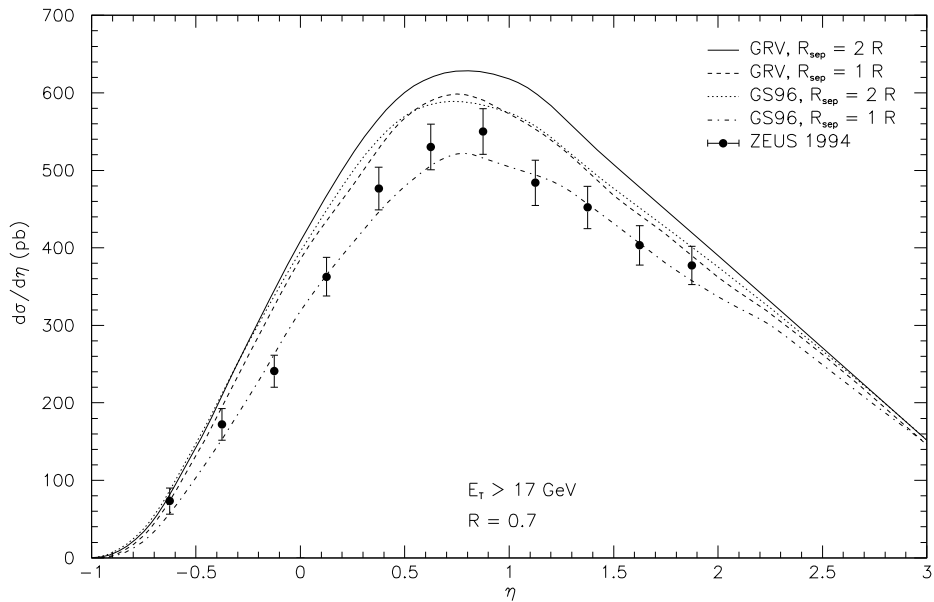,bbllx=520pt,bblly=95pt,bburx=105pt,bbury=710pt,%
           height=12cm,angle=270,clip=}
  \end{picture}}
 \end{center}
 \caption{\label{kkkplot1a17r07}{\it $\eta$ dependence of the inclusive
           single-jet photoproduction cross
           section integrated over $E_T>17$GeV with jet cone size $R=0.7$.
           We compare our NLO
           prediction with GRV and GS96 photon parton densities and the two
           extreme $R_{\rm sep}$ values to 1994 data
           from ZEUS.}}
\end{figure}

\begin{figure}[p]
 \begin{center}
  {\unitlength1cm
  \begin{picture}(12,8)
   \epsfig{file=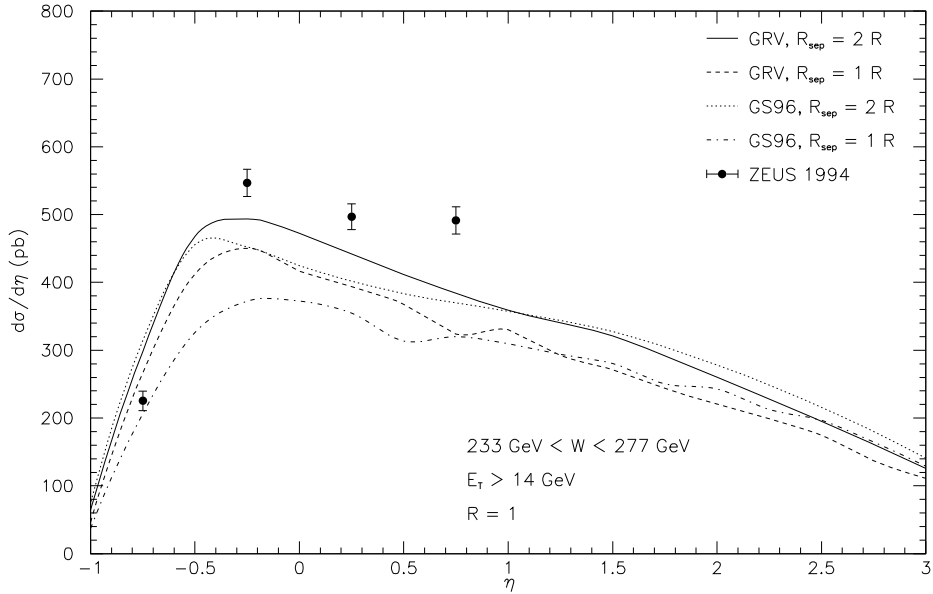,bbllx=520pt,bblly=95pt,bburx=105pt,bbury=710pt,%
           height=12cm,angle=270,clip=}
  \end{picture}}
 \end{center}
 \caption{\label{kkkplot1b06r1}{\it $\eta$ dependence of the inclusive
           single-jet photoproduction cross
           section integrated over 233 GeV $< W <$ 277 GeV with jet cone size
           $R=1$. We compare our NLO
           prediction with GRV and GS96 photon parton densities and the two
           extreme $R_{\rm sep}$ values to 1994 data
           from ZEUS.}}
\end{figure}

\begin{figure}[p]
 \begin{center}
  {\unitlength1cm
  \begin{picture}(12,8)
   \epsfig{file=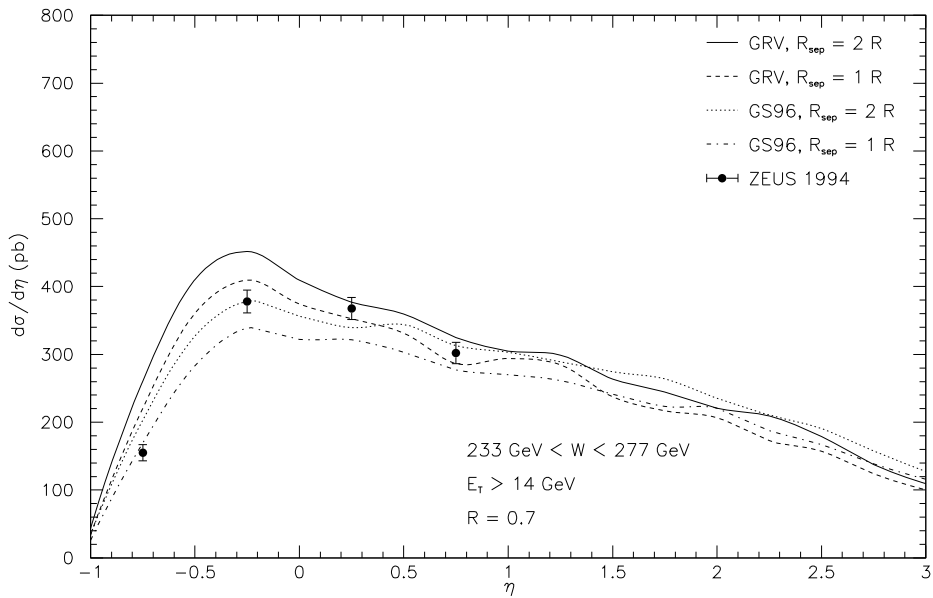,bbllx=520pt,bblly=95pt,bburx=105pt,bbury=710pt,%
           height=12cm,angle=270,clip=}
  \end{picture}}
 \end{center}
 \caption{\label{kkkplot1b06r07}{\it $\eta$ dependence of the inclusive
           single-jet photoproduction cross
           section integrated over 233 GeV $< W <$ 277 GeV with jet cone size
           $R=0.7$. We compare our NLO
           prediction with GRV and GS96 photon parton densities and the two
           extreme $R_{\rm sep}$ values to 1994 data
           from ZEUS.}}
\end{figure}

We start with the single-jet cross section d$^2\sigma$/d$\eta$d$E_T$
integrated over $E_T\geq E_{T_{\min}}$ as a function of $\eta$. This
cross section d$\sigma$/d$\eta$ has been measured in the $\eta$ range
between -1 and 2 and with the $E_T$ thresholds $E_{T_{\min}} = 14,17,21,$ and 25
GeV. The cross section d$\sigma$/d$\eta$ for $E_T>14$ GeV has also been
measured in three different regions of $W$: 134 GeV $< W <$ 190 GeV,
190 GeV $< W <$ 233 GeV, and 233 GeV $< W <$ 277 GeV. The measurements
refer to jets at the hadron level and are performed for two cone radii
in the $\eta-\phi$ plane, $R=0.7$ and $R=1$ using the iterative cone
algorithm PUCELL. The complete data have
already been compared to our next-to-leading order calculations in the
ZEUS publication \cite{y1}. Therefore, we show only a selection for some
specific kinematical ranges and compare them with the data. Our results for
d$\sigma$/d$\eta$ for $E_T>17$ GeV, $R=1$ and $R=0.7$ are shown in
figures \ref{kkkplot1a17r1} and \ref{kkkplot1a17r07},
% KKK-Plot 1a17r1
%
% KKK-Plot 1a17r07
respectively, and are
compared to the ZEUS data. The error bars in figures \ref{kkkplot1a17r1},
\ref{kkkplot1a17r07}, \ref{kkkplot1b06r1}, and \ref{kkkplot1b06r07} only contain
the statistical error. The systematic error and the uncertainty associated
with the absolute energy scale is not included, which adds an additional
30\% error (see \cite{y1}). The theoretical predictions include resolved
and direct processes in NLO. For the proton, the CTEQ4M \cite{Lai96} parton
densities have been used. For the photon distribution, the GRV-HO
\cite{Glu92}, converted to $\overline{\mbox{MS}}$ factorization, and as
an alternative set the recent GS96 \cite{Gor96} parametrizations have been
chosen as input. The renormalization and factorization scales have
been put equal to $E_T$, and $\alpha_s$ was calculated with the
two-loop formula with $\Lambda_{\overline{\mbox{MS}}}^{(4)}=296$ MeV as
used in the proton parton densities. In figures \ref{kkkplot1a17r1} and
\ref{kkkplot1a17r07}, two curves are presented for both photon distribution
sets labeled as $R_{\rm sep}=2 R$ and $R_{\rm sep}=1 R$. They correspond
to two choices of the $R_{\rm sep}$ parameter. Since our calculations
include only up to three partons in the final state, the maximum number of
partons in a single jet is two. Therefore, the overlapping and merging
effects of the experimental jet algorithm cannot be simulated in the
theoretical calculation \cite{y7}. To account for these effects, the
$R_{\rm sep}$ parameter was introduced \cite{y7}. It has the effect
that two partons are not merged into a single jet if their separation
in the $\eta-\phi$ plane is more than $R_{\rm sep}$. Then $R_{\rm sep} = 2 R$
means that no further restriction is introduced and the cone algorithm
is applied in its original form. Experimentally, the two extreme values
of $R_{\rm sep}=2 R$ and $1 R$ correspond to a fixed cone algorithm
(like EUCELL) and to the $k_T$ clustering algorithm (like KTCLUS), whereas
an iterative cone algorithm (like PUCELL) is described by some intermediate
value. In both calculation and data analysis,
the maximal virtuality of the photon is equal to $Q_{\max}^2=4$ GeV$^2$,
and the full $W$ range, which corresponds to 0.20 $< x_a <$ 0.85 in the
EPA formula, is used. Looking at figures \ref{kkkplot1a17r1} and
\ref{kkkplot1a17r07}, we observe that the behavior of the measured cross
sections is different for $R=0.7$ and $R=1$. For $R=1$, the shape of the
cross section is well described for -1 $< \eta <$ 0.5. For higher values
of $\eta$, the data stay almost constant as a function of $\eta$, whereas
the theoretical curves decrease as a function of $\eta$ for both
radii $R=1$ and $R=0.7$. However, when $R=0.7$ is used, the shape and
magnitude of the NLO results agree quite well with the measured differential
cross section in the entire $\eta$ range. This is also the case in the
comparison for the lower $E_T$ threshold, $E_{T_{\min}}=14$ GeV, with
$R=0.7$ shown in \cite{y1}. For the higher $E_T$ thresholds, 21 and 25 GeV,
the NLO predictions give a good description of the measured cross
sections in magnitude and shape for both cone radii $R=0.7$ and $R=1$ 
(see \cite{y1}). In general, the choice $R=0.7$ should be preferred
for the comparison between data and theory. Figure \ref{kkkplot1a17r07}
shows that the predictions with the GS96 parametrization of the
photon parton distributions agree better with the data than the GRV-HO
parametrization which is above the data for both $R_{\rm sep}$
parameter values over the whole $\eta$ range. Concerning the $R_{\rm sep}$
parameter the curve for $R_{\rm sep}=1 R$ is in somewhat better
agreement than the curve for $R_{\rm sep}=2 R$. We have checked that
a value of $R_{\rm sep}=1.4 R$ gives the best agreement. This supports
a recent study of the jet shape function, which depends sensitively
on this parameter \cite{y8}. By comparing with recent measurements
of this jet shape by the ZEUS collaboration \cite{y9}, it was found
that $R_{\rm sep}= 1.5 R$ gives very good agreement with the jet shape
data for PUCELL in the same $\eta$ and $E_T$ range \cite{y8}.

A comparison of d$\sigma$/d$\eta$ for $E_T>14$ GeV in different $W$
regions has also been presented. As an example, we show d$\sigma$/d$\eta$
as a function of $\eta$ for the largest $W$ range: 233 GeV $< W <$ 277 GeV
(corresponding to 0.55 $< x_a <$ 0.85) in figure \ref{kkkplot1b06r1}
($R=1$) and in figure \ref{kkkplot1b06r07} ($R=0.7$). 
% KKK-Plot 1b06r1
%
% KKK-Plot 1b06r07
Whereas the
$R=1$ theoretical cross section (figure \ref{kkkplot1b06r1}) agrees for
low values of $\eta$, it disagrees in the high $\eta$ region with the
data. This disagreement shows up particularly in the high $W$ range \cite{y1}.
The measured differential cross section is again well described by
the NLO calculation for $R=0.7$ (see figure \ref{kkkplot1b06r07}).%
\footnote{The wiggles in the curves in figures \ref{kkkplot1b06r1} and
\ref{kkkplot1b06r07} are due to insufficient numerical accuracy and have
no physical significance.}

The excess of the measured cross section with respect to the calculations
in the high $\eta$ range and for the smaller $E_T$ thresholds for $R=1$
is supposed to be due to the presence of the underlying event in the data
which yields a larger amount of extra energy lying in the jet cone
for $R=1$ but not for the smaller cone $R=0.7$. From figure \ref{kkkplot1a17r1}
it is clear that the excess occurs only in the large $\eta$ range where the resolved
cross section dominates and where additional interactions of the
photon and proton remnants are supposed to occur which are not included
in the NLO calculations. These deviations between NLO theory and the
data at large $\eta$ and smaller $E_T$'s were found earlier \cite{y10} when
the theoretical predictions were compared with the 1993 ZEUS data \cite{x5}.

In conclusion, we can say that the NLO calculations describe reasonably
well the experimental inclusive single-jet cross section for jets defined
with $R=0.7$ in the entire $\eta$ range and for $R=1$, if $E_T$ is large
enough.

Next, we compare the NLO predictions with inclusive dijet cross sections measured
by ZEUS \cite{y2} and H1 \cite{y3}. Inclusive two-jet cross sections depend on one
more variable as compared to the inclusive one-jet cross sections considered above.
Therefore they are supposed to give a much more stringent test of the theoretical
predictions than the inclusive one-jet cross sections. For the comparison with the
data it is essential that in the theoretical calculations the same jet definitions
are introduced as in the experimental analysis. Furthermore it is important that
the theoretical calculations contain the same cuts on the kinematical variables
as for the measured cross sections. First experimental data for inclusive two-jet
production have been published by the ZEUS collaboration in \cite{x5} and
\cite{y13}. The more recent ZEUS analysis based on the 1994 data taking presented
in \cite{Der96a} and recently in \cite{y2} extends the earlier analysis in
\cite{x5} based on 1993 data in several ways. The larger luminosity obtained in 1994
lead to a reduction of the statistical errors as well as allowing for the
measurement of the cross section at higher $E_T$, a region, where uncertainties due
to hadronization of partons into jets and of underlying event effects are reduced
making the comparison with the NLO predictions more meaningful. Furthermore, the ZEUS
collaboration applied three different jet definitions: two variations of the cone
algorithm \cite{Hut92} called ``EUCELL'' and ``PUCELL'', and the $k_T$-cluster
algorithm ``KTCLUS'' as introduced for hadron-hadron collisions \cite{y16}. The
two cone algorithms treat seed finding and jet merging in different ways. Since the
NLO calculations contain only up to three partons in the final state, these
experimental seed finding and jet merging conditions cannot be fully reproduced.
This ambiguity is largely reduced in the $k_T$-cluster algorithm. In the NLO
calculations, the two cone algorithms can be simulated by introducing $R_{\rm sep}$
already considered for the one-jet cross sections. The EUCELL definition corresponds
to $R_{\rm sep}=2 R$, whereas PUCELL is best simulated with $R_{\rm sep}= 1.4 R$ for
the $E_T$ range considered in the experimental analysis (see above) \cite{y8, But96}.
The $k_T$-cluster algorithm for hadron-hadron collisions is identical to using
$R_{\rm sep}=R$. Therefore by introducing the $R_{\rm sep}$ parameter into the NLO
calculations of the two-jet cross sections, all three jet finding definitions used
in the experimental analysis can be accounted for. The ZEUS results of the earlier
analysis \cite{Der96a} have been compared to the NLO predictions in \cite{x12} for
the case of the KTCLUS algorithm and in \cite{x10} for the EUCELL algorithm. These
comparisons were done for the differential cross section d$\sigma$/d$\overline{\eta}$,
where $\overline{\eta}=1/2(\eta_1+\eta_2)$ is the average rapidity of the observed
jets with $E_T$ larger than $E_{T_{\min}}$ for both observed jets. This common cut
on the $E_T$ of both jets causes some theoretical problems as has been noticed
already some time ago \cite{Kla96}.

The new measurements \cite{y2} are for the triple differential cross section
d$^3\sigma$/d$E_T$d$\eta_1$d$\eta_2$ using the $k_T$-cluster or the PUCELL algorithm.
The jet with the highest $E_T$ (leading jet, $E_T=E_{T_1}$) is required to have
$E_T > 14$ GeV and the second highest-$E_T$ jet to have $E_{T_2} > 11$ GeV. This cross
section is symmetrized with respect to $\eta_1$ and $\eta_2$ and therefore double
counted. By this symmetrization, the experimental ambiguity of determining the leading
jet is avoided and the measured cross section corresponds to the calculated cross
section where $E_T$ corresponds to the trigger jet which is not necessarily the leading
jet with the highest $E_T$. In order to have a handle to enhance direct over
resolved photoproduction, one determines also the variable \cite{x5,y21}
\beq
 x_{\gamma}^{\rm OBS} = \frac{\sum_iE_{T_i}e^{-\eta_i}}{2x_aE_e},
 \label{eq51}
\eeq
where the sum runs over the two jets of highest $E_T$ and $x_aE_e$ is the initial
photon energy. $x_{\gamma}^{\rm OBS}$ measures the fraction of the photon energy
that goes into the production of the two hardest jets. The LO direct and resolved
processes populate different regions of $x_{\gamma}^{\rm OBS}$:
$x_{\gamma}^{\rm OBS} = 1$ for the direct process and $x_{\gamma}^{\rm OBS} < 1$
for the resolved process. In NLO, the direct process populates also the region
$x_{\gamma}^{\rm OBS} < 1$. To obtain a measurement of the enriched direct
photoproduction cross section, the cut $x_{\gamma}^{\rm OBS} > 0.75$ is usually
introduced.

\begin{figure}[p]
 \begin{center}
  {\unitlength1cm
  \begin{picture}(14,19)
   \epsfig{file=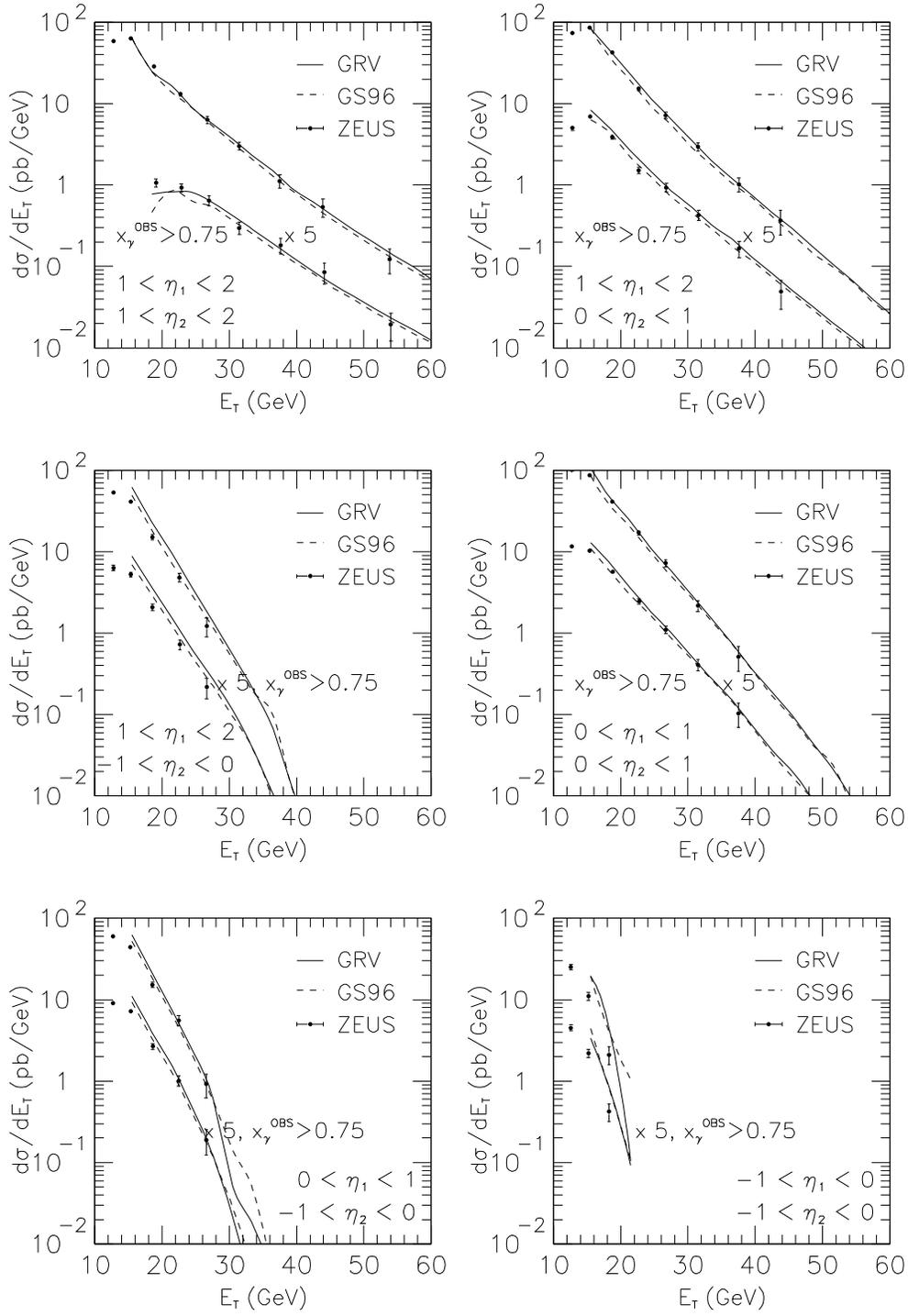,bbllx=70pt,bblly=105pt,bburx=490pt,bbury=715pt,%
           height=19cm,clip=}
  \end{picture}}
 \end{center}
 \caption{\label{kkkplot2a}{\it $E_T$ dependence of the symmetrized dijet
           photoproduction cross
           section integrated over different rapidity bins. We compare our NLO
           prediction with GRV and GS96 photon parton densities and the full
           and upper range of $x_{\gamma}^{\rm OBS}$ to preliminary 1995 data
           from ZEUS.}}
\end{figure}

Figure \ref{kkkplot2a} shows d$\sigma$/d$E_T$ for six independent regions in the
% KKK-Plot 2a
$(\eta_1,\eta_2)$ plane. The upper curves in each plot give the dijet cross section
for the entire $x_{\gamma}^{\rm OBS}$ region, and the lower curves which, to
separate the curves, all have been scaled down by a factor of 5, present the
cross section for the direct $\gamma$ region $x_{\gamma}^{\rm OBS} > 0.75$.
The NLO calculations are performed for $Q_{\max}^2 = 4$ GeV$^2$ and 134 GeV $< W <$
277 GeV using the same parton densities for the proton and the photon as in the
inclusive one-jet calculations. Furthermore we use $R_{\rm sep}= 1.4 R$ with $R=1$
to simulate the PUCELL algorithm. We compare with the corresponding data from ZEUS
\cite{y2} analyzed with the PUCELL algorithm. The agreement between the data
and the theoretical predictions is quite reasonable. Except near the backward
regions (the last two $\eta_1,\eta_2$ regions), the cross sections for the GRV and
GS96 photon densities are very similar. To discriminate between them, the experimental
errors must be reduced. We observe that the cross section for $x_{\gamma}^{\rm OBS} >
0.75$ increases as compared to the full cross section (all $x_{\gamma}^{\rm OBS}$)
as $E_T$ increases in agreement with the prediction from the calculation. This is
the effect of the direct component which shows in general a flatter distribution
with increasing $E_T$ than the resolved cross section \cite{x12,x10}. We
emphasize that the magnitude as well as the shape of the measured cross section
is well reproduced by the calculations except for the first $E_T$ bin and for
the case that both jets are in the region $-1 < \eta_{1,2} < 0$. In this region
(the last plot), the predictions lie above the data. The same cross section has been
calculated also for the $k_T$-cluster algorithm. The results for the GS96 photon
densities have been presented together with the corresponding experimental data
in \cite{y2}. The agreement between data and theory is quite similar. In \cite{y2}
also data for d$\sigma$/d$\eta_2$ for $E_T > 14$ GeV in three regions of $\eta_1$
and the KTCLUS algorithm are compared to our NLO calculations with the GRV-HO and
GS96 parametrizations of the photon densities. Both shape and magnitude of the
cross sections are roughly reproduced by the calculations except for the small
$\eta_1$ region. Comparisons for the same cross section with the PUCELL cone
algorithm have been done, too, but are not shown here \cite{y22}.

The comparison between measurements and calculations of the triple differential
cross section shown so far covers the dependence of this cross section in all three
variables $E_T$, $\eta_1$ and $\eta_2$. Another equivalent set of variables consists
of the dijet invariant mass $M_{JJ}$, the rapidity of the dijet system $y_{JJ}$,
and the scattering angle $\theta^{\ast}$ in the dijet center-of-mass system. The
dijet invariant mass is obtained from the relationship
\beq
 M_{JJ}^2 = 2 E_{T_1} E_{T_2} \le \cosh(\eta_1-\eta_2) - \cos(\phi_1-\phi_2)\re,
\eeq
where $\phi_1$ and $\phi_2$ are the azimuthal angles of the two jets in the HERA
frame. For two jets back-to-back in $\phi$ and with equal $E_T$,
\beq
 M_{JJ} = 2 E_T \cosh\le(\eta_1-\eta_2)/2\re = 2 E_T/\sin\theta^{\ast}
\eeq
and $\cos\theta^{\ast}=\tanh[(\eta_1-\eta_2)/2]$. For events with more than
two jets, the two highest $E_T$ jets are used to calculate $M_{JJ}$.

The distribution in the dijet mass $M_{JJ}$ provides an additional test and is
sensitive to the presence of resonances that decay into two jets. The dijet
cross section as a function of $\cos\theta^{\ast}$ is sensitive to the
parton-parton dynamics in the direct and resolved contributions \cite{Bae89a}.
Direct processes involve quark propagators in the $t$ and $u$ channels leading
to a characteristic angular  dependence proportional to $(1-|\cos\theta^{\ast}|)
^{-1}$. In the case of the resolved process, $t$-channel gluon exchange processes
dominate which lead to an angular dependence proportional to $(1-|\cos\theta^{\ast}|)
^{-2}$. This exchange rises more steeply with increasing $|\cos\theta^{\ast}|$
than in the case of the direct processes. This different behavior in the
angular dependence for resolved and direct processes was observed for $M_{JJ} > 23$
GeV \cite{y13}. We have calculated the NLO cross section d$\sigma$/d$\cos\theta
^{\ast}$ for $x_{\gamma}^{\rm OBS} < 0.75$ (= ``resolved'') and
$x_{\gamma}^{\rm OBS} > 0.75$ (= ``direct'') and have confirmed the different
behavior in the two $x_{\gamma}^{\rm OBS}$ bins as shown in the data (not shown
here).

\begin{figure}[p]
 \begin{center}
  {\unitlength1cm
  \begin{picture}(12,8)
   \epsfig{file=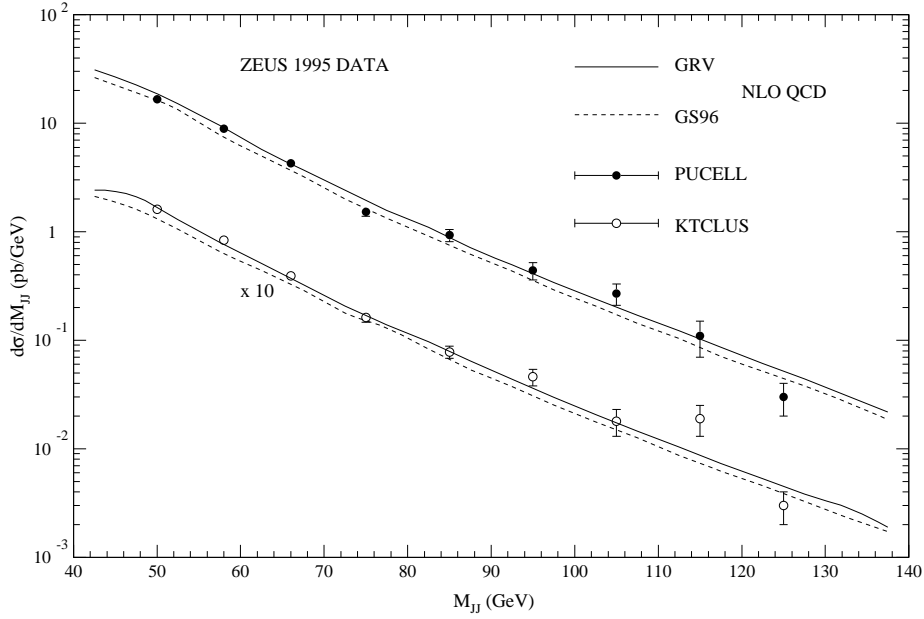,bbllx=520pt,bblly=95pt,bburx=105pt,bbury=710pt,%
           height=12cm,angle=270,clip=}
  \end{picture}}
 \end{center}
 \caption{\label{kkkplot2dnlo}{\it $M_{JJ}$ dependence of the dijet
           photoproduction cross
           section integrated over $|\cos\theta^{\ast}| < 0.85$. We compare our NLO
           prediction with GRV and GS96 photon parton densities to 1995 data from
           ZEUS taken with the PUCELL and KTCLUS jet algorithms.}}
\end{figure}

The cross sections d$\sigma$/d$M_{JJ}$ and d$\sigma$/d$\cos\theta^{\ast}$ have been
measured recently using the sample of dijet events found with the PUCELL cone
and the KTCLUS cluster algorithms. This analysis also includes the 1995 data
which have an even higher statistics than used in the previous analysis \cite{y13}
based on 1994 data. These cross sections have been measured in the kinematic region
$Q^2 < 4$ GeV$^2$, $0.2 < x_a < 0.85$ as used previously. The two jets with highest
$E_T$ are required to have $E_{T_1}, E_{T_2} > 14$ GeV and the rapidities of these
two jets are restricted to $-1 < \eta_1,\eta_2 < 2.5$. The cone radius is $R=1$.
The cross section d$\sigma$/d$M_{JJ}$ has been measured in the $M_{JJ}$ range
between 47 GeV and 120 GeV integrated over $|\cos\theta^{\ast}| < 0.85$. The cross
section d$\sigma$/d$\cos\theta^{\ast}$ has been measured in the interval
$0 < |\cos\theta^{\ast}| < 0.8$ integrated over $M_{JJ} > 47$ GeV. The experimental
results \cite{y2,y25} for d$\sigma$/d$M_{JJ}$ are shown in figure
\ref{kkkplot2dnlo}
% KKK-Plot 2dnlo
separately for the two jet definitions, where the cross
sections for the $k_T$ algorithm have been scaled down by a factor of 10 in order
to separate the data obtained for the two jet definitions. Systematic errors are
available, but only statistical errors are shown here. In figure \ref{kkkplot2dnlo},
two NLO curves are compared to the measurements, the full curve being the result
for the GRV-HO and the dashed curve the result for the GS96 parametrization of the
photon densities. For the lower $M_{JJ}$ values, the GRV density seems to describe
the data better than the GS96 density. However, we have to consider that the data
points have an additional systematic error from the energy scale uncertainty
\cite{y2}, which is also not shown in figure \ref{kkkplot2dnlo}. It is remarkable
that the predictions for both jet algorithms, PUCELL and KTCLUS, agree well with the
data over the full range of $M_{JJ}$ where the cross sections exhibits a fall-off of
almost three orders of magnitude.

\begin{figure}[p]
 \begin{center}
  {\unitlength1cm
  \begin{picture}(12,8)
   \epsfig{file=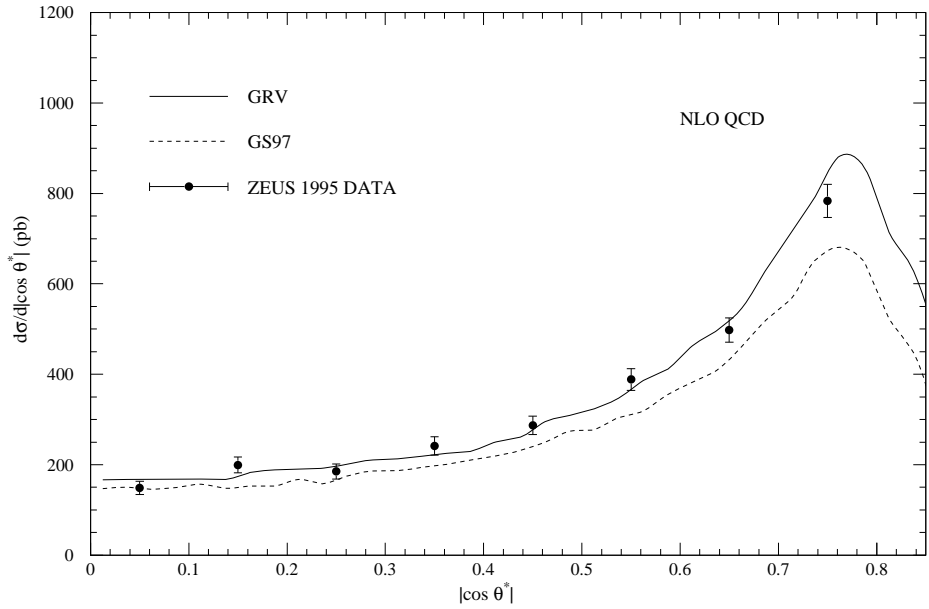,bbllx=520pt,bblly=95pt,bburx=105pt,bbury=710pt,%
           height=12cm,angle=270,clip=}
  \end{picture}}
 \end{center}
 \caption{\label{kkkplot2cnlo}{\it $|\cos\theta^{\ast}|$ dependence of the dijet
           photoproduction cross
           section integrated over $M_{JJ} > 47$GeV. We compare our NLO
           prediction with GRV and GS96 photon parton densities to 1995 data from
           ZEUS using the PUCELL jet algorithm.}}
\end{figure}

The cross section d$\sigma$/d$\cos\theta^{\ast}$ as a function of
$|\cos\theta^{\ast}|$ between 0 and 0.8 is plotted in figure \ref{kkkplot2cnlo}.
% KKK-Plot 2cnlo
Here only the results for the PUCELL algorithm are shown. For the KTCLUS
algorithm, the corresponding cross section is presented in \cite{y2}.
Again two curves are shown, for the GRV and GS96 photon densities, respectively.
The theoretical curves agree reasonably well in magnitude and shape with
the data. If it were not for the systematic error and the energy scale uncertainty,
the comparison of data and theory would lead to a preference of the GRV over the
GS96 density. We conclude, that the NLO calculations account reasonably well
for the shape and the magnitude of the measured d$\sigma$/d$M_{JJ}$ and
d$\sigma$/d$\cos\theta^{\ast}$ as well as the triple differential cross section
d$^3\sigma$/d$E_T$d$\eta_1$d$\eta_2$ as a function of $E_T$ for various bins
in $\eta_1$ and $\eta_2$.

The last comparison between data an our NLO theory concerns the double-differential
inclusive dijet cross section d$^2\sigma$/d$x_{\gamma}^{\rm OBS}$d$E_T$ published
just recently by the H1 collaboration \cite{y3}. This H1 analysis is based on 1994
data. The photoproduction events have been selected with the constraint
$Q_{\max}^2 = 4$ GeV$^2$ and $0.2 < x_a < 0.83$. The jets were constructed with the
cone algorithm with cone size $R=0.7$. The implementation of the cone algorithm
in the H1 analysis is using a fixed cone and is therefore similar to the EUCELL
algorithm used by ZEUS. The rapidities of all jets are restricted to the region
$-0.5 < \eta < 2.5$. In this specific two-jet analysis, $E_T$ is the average
transverse energy of the two jets with the highest $E_T$. The average rapidities
of the two jets were between $0 < (\eta_1+\eta_2)/2 < 2$ their difference 
being $|\eta_1-\eta_2| < 1$, which corresponds to $|\cos\theta^{\ast}| < 0.46$.
These two cuts on $\eta_1$ and $\eta_2$ ensured that the jets are in a region
with good measurements of the hadronic energy in the detector. The transverse
energies of the jets were restricted further to the range
\beq
 \frac{|E_{T_1}-E_{T_2}|}{E_{T_1}+E_{T_2}} < 0.25,
 \label{eq50}
\eeq
and $E_T$ was required to lie above 10 GeV. This cut and the cut (\ref{eq50})
ensure that the transverse energy of both observed jets is above 7.5 GeV (to
avoid underlying event problems) without using the same $E_T$ cut for both
jets, which would cause problems in the NLO calculations \cite{Kla96}.
The observable $x_{\gamma}^{\rm OBS}$ was calculated from the same formula
(\ref{eq51}) as in the ZEUS analysis.

\begin{figure}[p]
 \begin{center}
  {\unitlength1cm
  \begin{picture}(14,19)
   \epsfig{file=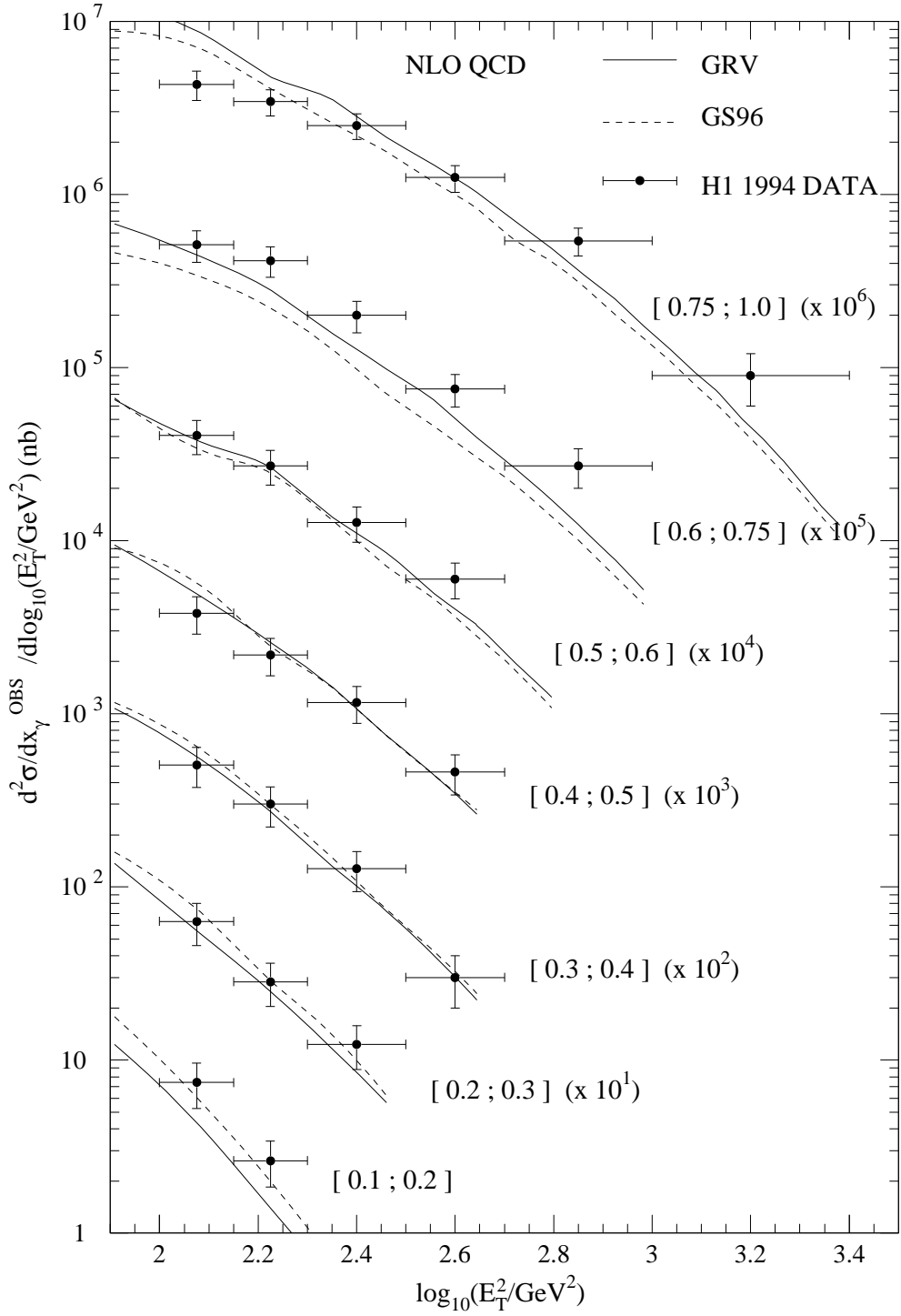,bbllx=70pt,bblly=105pt,bburx=490pt,bbury=715pt,%
           height=19cm,clip=}
  \end{picture}}
 \end{center}
 \caption{\label{kkkplot4a}{\it $E_T$ dependence of the symmetrized dijet
           photoproduction cross
           section integrated over different $x_{\gamma}^{\rm OBS}$ bins.
           We compare our NLO
           prediction with GRV and GS96 photon parton densities and $R_{\rm sep}
           = 2 R$ to 1994 data from H1.}}
\end{figure}

The NLO calculations of the inclusive dijet cross section
d$^2\sigma$/d$x_{\gamma}^{\rm OBS}$d$\log_{10}(E_T^2/$GeV$^2$) is based on the
same parton distributions for the proton and photon, respectively, as used
in the previous sections. The scales are chosen equal to $E_T$ as in the
comparisons above. We have chosen $R_{\rm sep} = 2 R$, which we believe
simulates best the fixed cone algorithm in the H1 analysis. Otherwise
the same cuts on $E_{T_1}, E_{T_2}, \eta_1, \eta_2$ are applied as in the
experimental analysis of the two-jet data. Our predictions are shown in
figure \ref{kkkplot4a},
% KKK-Plot 4a
again for the GRV-HO (full line) and the GS96 (dotted line) photon parton
distributions. The curves are compared to the measured cross sections using
the statistical and systematic errors added in quadrature.
The data and the theoretical cross sections in the
various $x_{\gamma}^{\rm OBS}$ bins between 0.1 and 1 have been multiplied
by factors $10^n (n=0,1,2, ... ,6)$ in order to separate the results for the
seven bins in $x_{\gamma}^{\rm OBS}$. Our calculations give a good description
of the data in magnitude and shape as a function of $E_T$ and $x_{\gamma}^{\rm
OBS}$, except for some deficiencies in the two highest $x_{\gamma}^{\rm OBS}$
ranges. The GRV-HO and the GS96 parton distribution functions each give
a satisfactory description of the measured cross sections with slight
preference for GS96 in the lowest bin $0.1 < x_{\gamma}^{\rm OBS} < 0.2$. It
is remarkable that even for $x_{\gamma}^{\rm OBS} < 0.3$ the theoretical
cross section agrees so well with the data. This region was always a problem
in earlier analyses due to the underlying event problems (see for example the
inclusive single-jet cross sections for $R=1$ of the ZEUS collaboration
discussed above). These problems are apparently avoided in this analysis by
choosing $R=0.7$ and applying sufficiently large $E_T$ cuts. The deviations
of the calculated from the measured cross sections could easily be due to
hadronization effects which are not included in the NLO calculations.

\begin{figure}[p]
 \begin{center}
  {\unitlength1cm
  \begin{picture}(12,8)
   \epsfig{file=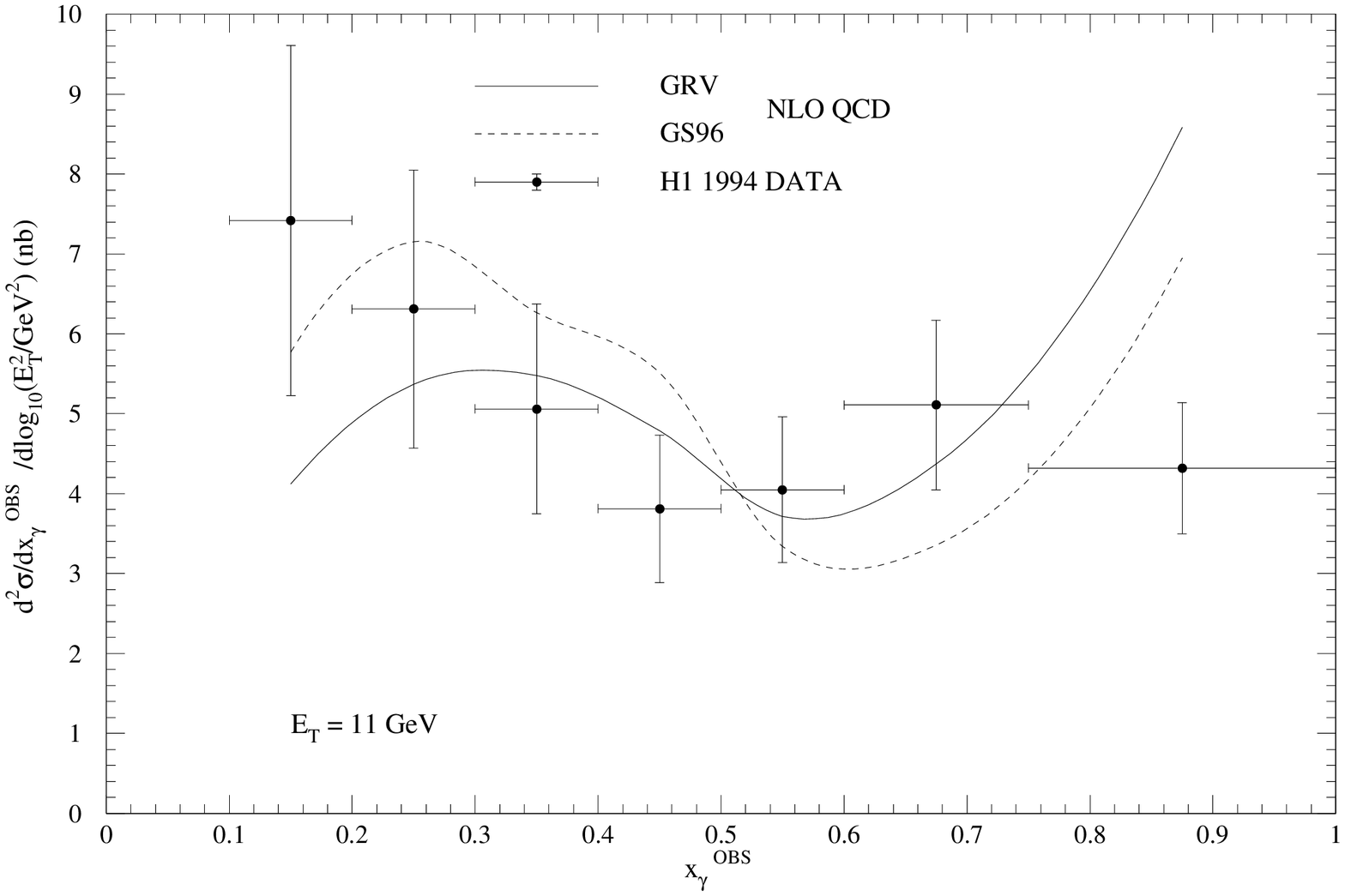,bbllx=520pt,bblly=95pt,bburx=105pt,bbury=710pt,%
           height=12cm,angle=270,clip=}
  \end{picture}}
 \end{center}
 \caption{\label{kkkplot4c}{\it $x_{\gamma}^{\rm OBS}$ dependence of the dijet
           photoproduction cross
           section at $E_T = 11$GeV. We compare our NLO
           prediction with GRV and GS96 photon parton densities to 1994 data from
           H1 using a fixed cone algorithm.}}
\end{figure}

\begin{figure}[p]
 \begin{center}
  {\unitlength1cm
  \begin{picture}(12,8)
   \epsfig{file=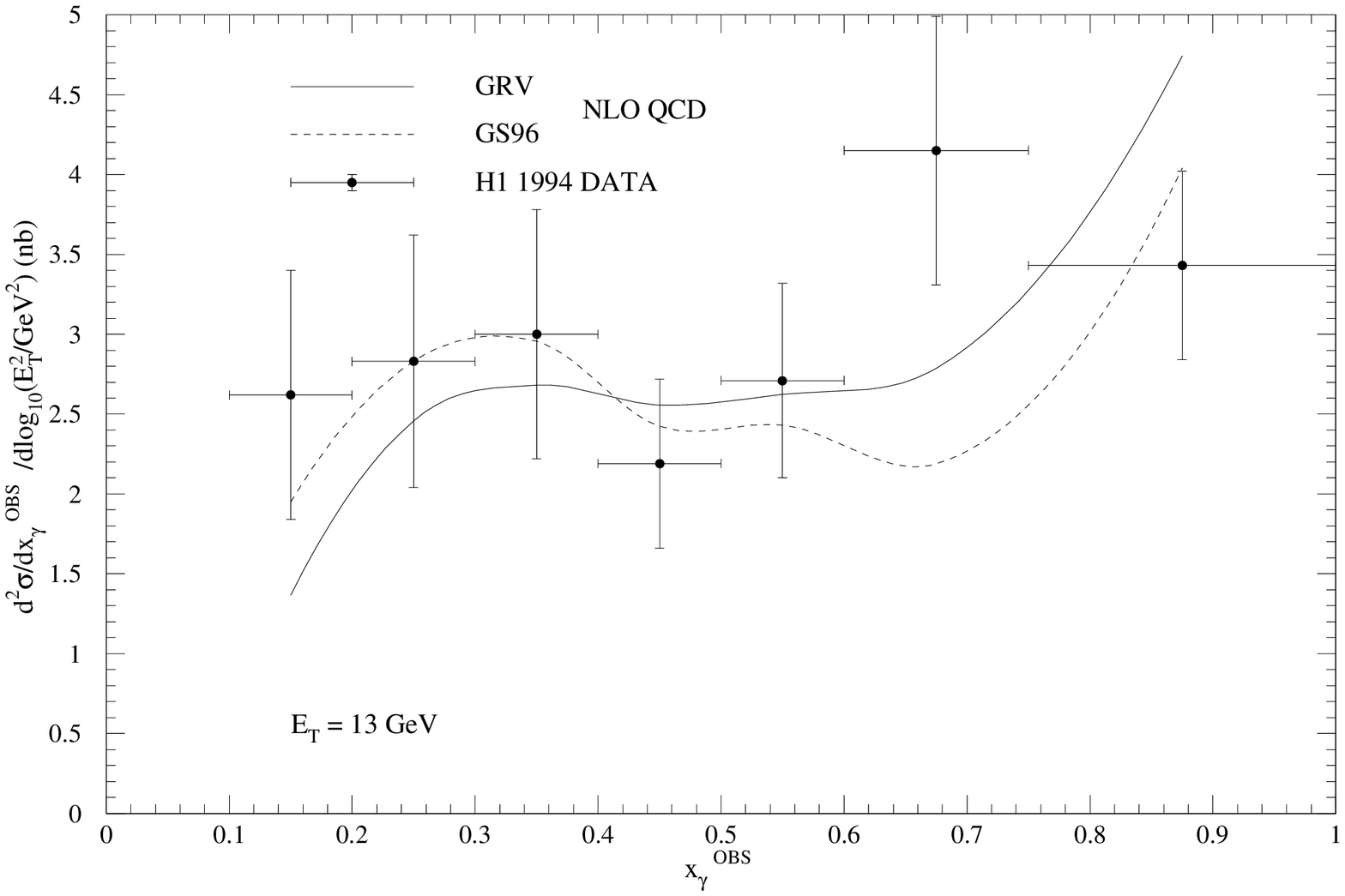,bbllx=520pt,bblly=95pt,bburx=105pt,bbury=710pt,%
           height=12cm,angle=270,clip=}
  \end{picture}}
 \end{center}
 \caption{\label{kkkplot4d}{\it $x_{\gamma}^{\rm OBS}$ dependence of the dijet
           photoproduction cross
           section at $E_T = 13$GeV. We compare our NLO
           prediction with GRV and GS96 photon parton densities to 1994 data from
           H1 using a fixed cone algorithm.}}
\end{figure}

The cross section d$^2\sigma$/d$x_{\gamma}^{\rm OBS}$d$\log_{10}(E_T^2$/GeV$^2$)
as a function of $x_{\gamma}^{\rm OBS}$ for fixed $E_T$ should show more
clearly the dependence on the photon parton distribution sets. Therefore,
we have calculated this cross section for $E_T = 11$ GeV and 13 GeV. The
result is shown in figures \ref{kkkplot4c} and \ref{kkkplot4d} for the GRV-HO
% KKK-Plot 4c
%
% KKK-Plot 4d
and the GS96 parametrizations and compared to the H1 data \cite{y3}, again including
statistical and systematic errors added in quadrature. The agreement with the
data is reasonable for both $E_T$'s. It seems that the hadronic maximum at
$x_{\gamma}^{\rm OBS} \simeq 0.3$ is reduced from $E_T=11$ GeV to $E_T=13$ GeV,
whereas the pointlike/direct peak at $x_{\gamma}^{\rm OBS} > 0.8$ is enhanced
flattening out the valley in between. This might be due to the QCD evolution of the
VMD part of the photon structure function and the increased importance of the
anomalous piece and the direct contribution at larger scales.
For $x_{\gamma}^{\rm OBS} < 0.3$, 
we see a somewhat better agreement for the prediction with GS96 which we
already noticed in connection with the comparison in figure \ref{kkkplot4a}.
Both parton distribution sets give very similar predictions in the intermediate
$x_{\gamma}^{\rm OBS}$ range. The different results in the low and high
$x_{\gamma}^{\rm OBS}$ regions show us clearly how the experimental data must
improve in accuracy before we can discriminate between the two alternative
parametrizations of the photon distributions used in our calculations.

\setcounter{equation}{0}

\section{Numerical Results for Photon-Photon Scattering}
In this section we report on numerical results for inclusive one-jet and two-jet
cross sections in $\gamma\gamma$ collisions. We use the analytical results
that have been calculated in leading order in section 3 and in next-to-leading
order in section 4. The calculation proceeds as in the photoproduction case
except that we now have three contributions: (a) the direct contribution,
where both virtual photons interact directly with the quarks, (b) the
single-resolved contribution, where only one of the photons interacts
directly and the other photon is resolved, and (c) the double-resolved
contribution, where both photons are resolved. It is clear that (b) and
(c) have their analogs in the photoproduction case. For case (a) the
analytical results are given in section 3.4 (Born matrix element), in section
4.1.3 (virtual corrections), and in section 4.2.9 (real corrections).

Three partons appear in the final state of all three contributions when we include
the NLO corrections. Two of these partons are recombined if they obey the
cluster or the jet-cone condition. For the $\gamma\gamma$ process we shall
use only the Snowmass jet algorithm already introduced in section 2.4 with
$R=1$ and $R_{\rm sep}=2$. This algorithm was applied in the analysis
of the TOPAZ, AMY \cite{x1}, and OPAL \cite{x2} data. However,
there is no problem to introduce other jet definitions in the calculations,
as we have done it for the $\gamma p$ case. Before we compare our results
with recent data from the OPAL collaboration \cite{y26} obtained at LEP2,
we shall present in the next section some general results and discuss some
tests of the numerical program.

\subsection{Tests and General Results}
\begin{figure}[h]
 \epsfig{file=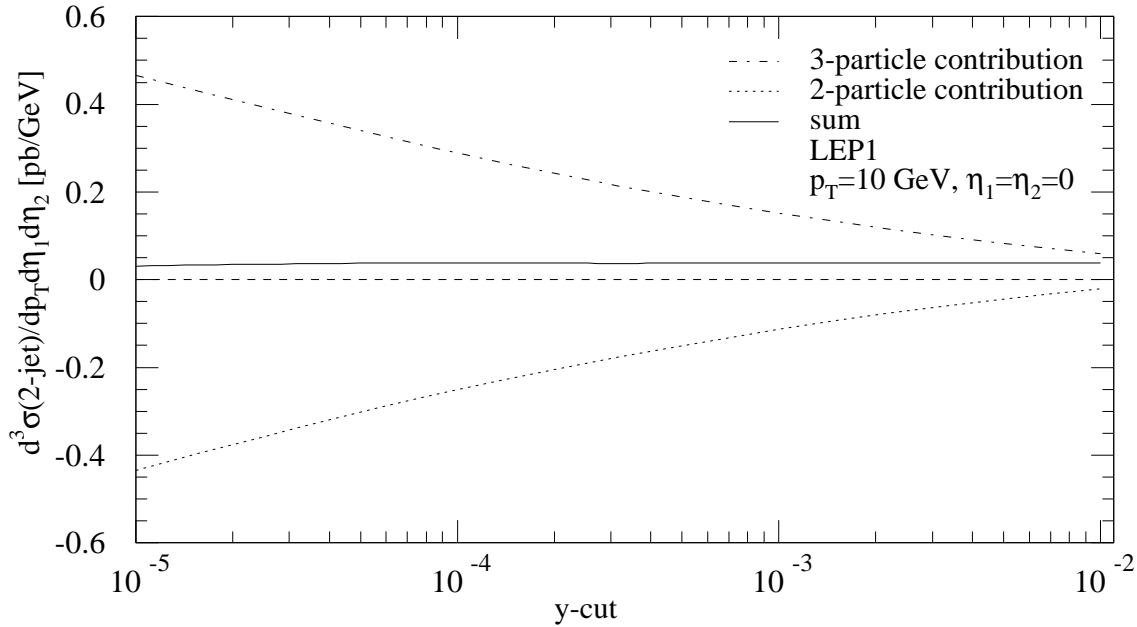}
 \caption{\label{kkkplot7}{\it Dependence of the inclusive dijet
           cross section on the $y$-cut, the boundary between analytical
           and numerical integration.  Only the double-resolved part
           is shown.}}
\end{figure}

Similar to the $\gamma p$ case, we have checked that our results are
independent of the $y$-cut when the analytic results are added to the
numerical results of the $2\rightarrow 3$ processes. This has been studied
for all $2\rightarrow 3$ processes separately. As an example, we show in
figure \ref{kkkplot7} the $y$-dependence of the complete double-resolved
% KKK-Plot 7
two-jet cross section for LEP1 kinematic conditions and at
$\eta_1=\eta_2=0$, $E_T=10$ GeV. As to be expected, the 2-particle
contribution is negative and decreases with decreasing $y$ whereas
the 3-particle contribution is positive and shows the opposite
behavior. The sum of both contributions is independent in the range
$10^{-4} < y < 10^{-2}$. For larger $y$-cuts the independence breaks
down because we neglected terms of order $y$ in the analytic contributions.
The slight decrease for $y < 10^{-4}$ is caused by insufficient accuracy
in the numerical integrations. This could be improved with more CPU time.

\begin{figure}[p]
 \epsfig{file=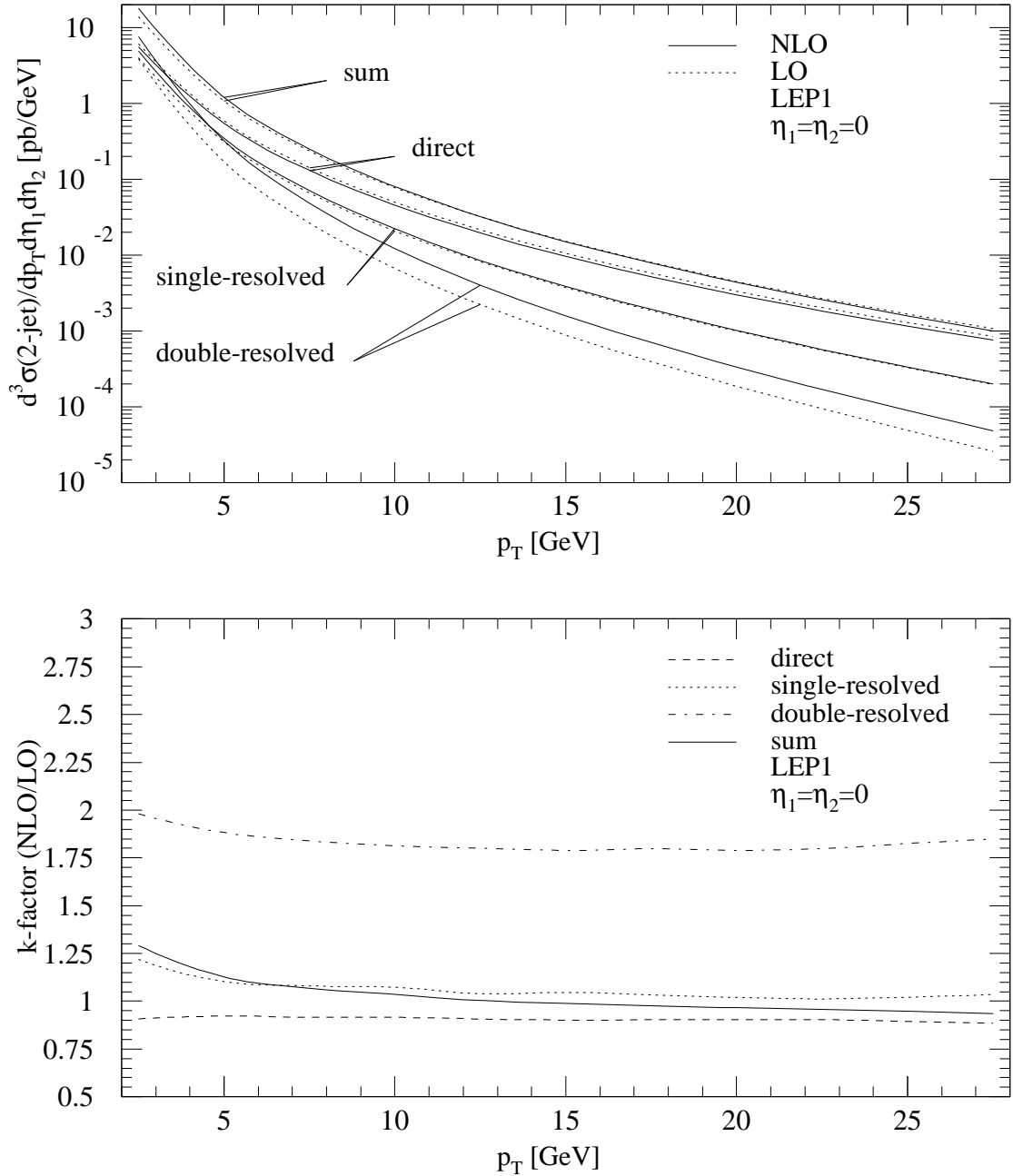}
 \caption{\label{kkkplot8}{\it Direct, single-resolved, and double-resolved
  contributions and their sum in LO and NLO for the inclusive two-jet cross
  section at LEP1. Upper figure: $E_T$-spectrum, lower figure: correspondent
  $k$-factors.}}
\end{figure}

\begin{figure}[p]
 \epsfig{file=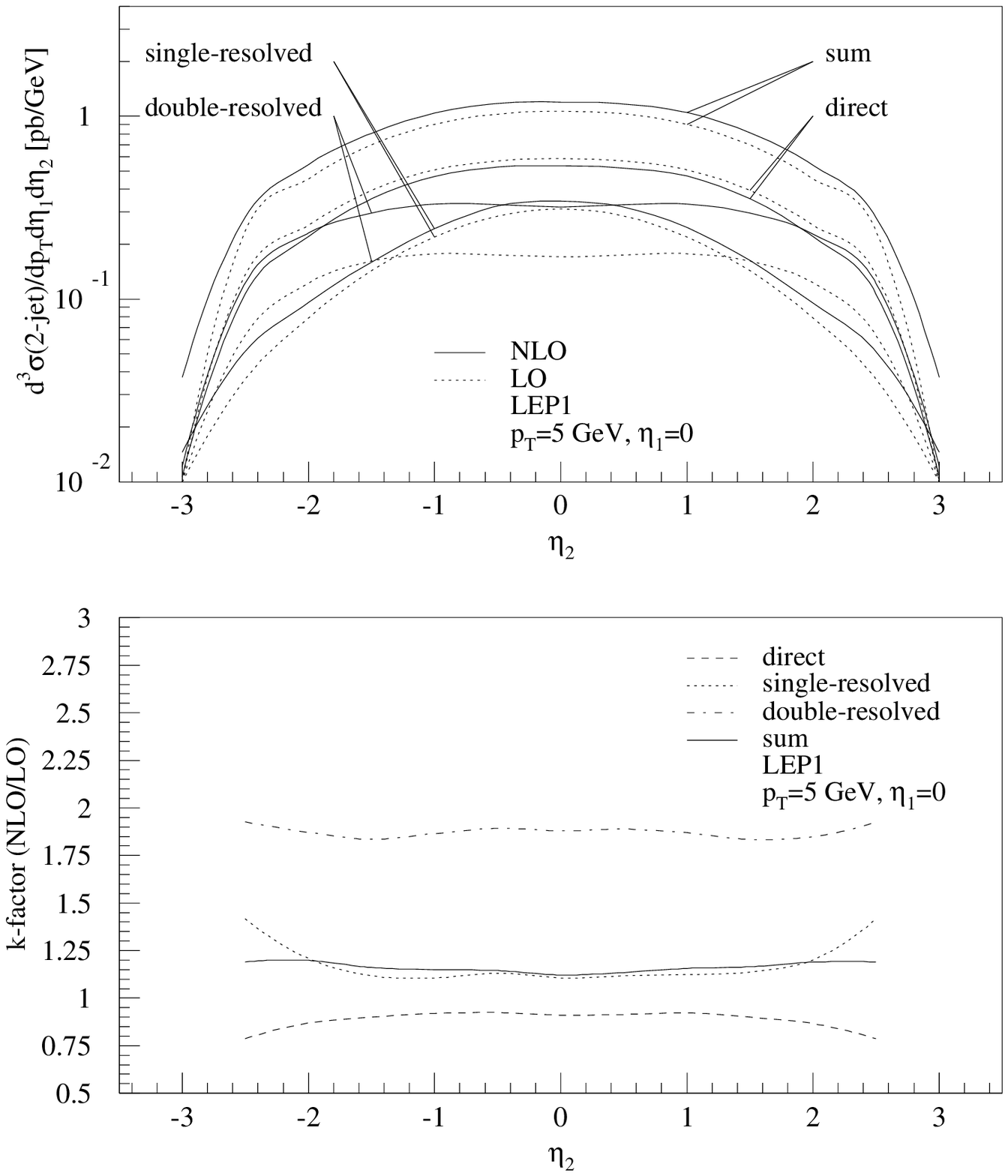}
 \caption{\label{kkkplot9}{\it Direct, single-resolved, and double-resolved
  contributions and their sum in LO and NLO for the inclusive two-jet cross
  section at LEP1. Upper figure: $\eta$-spectrum, lower figure: correspondent
  $k$-factors.}}
\end{figure}

For the results in figures \ref{kkkplot7} to \ref{kkkplot9},
we use the LEP1 center-of-mass energy $\sqrt{s} = 90$ GeV.
The photon spectrum is calculated from the formula
\beq
 F_{\gamma/e}(x) = \frac{\alpha}{2\pi} \le \frac{1+(1-x)^2}{x}
 \ln\frac{E^2\theta_c^2(1-x)^2+m_e^2x^2}{m_e^2x^2}+2(1-x)\lr\frac{m_e^2x}
 {E^2\theta_c^2(1-x)^2+m_e^2x^2}-\frac{1}{x}\rr\re
 \label{eq61}
\eeq
where $E$ is the beam energy and $\theta_c=3^\circ$. The parton distributions
are computed with the NLO parametrization of GRV \cite{Glu92} transformed
to the $\overline{\mbox{MS}}$ subtraction scheme. For all scales we
set $\mu=M_a=M_b=E_T$ and put $N_f=5$. $\Lambda_{\overline{\mbox{MS}}}^{(5)}
= 130$ MeV corresponding to $\Lambda_{\overline{\mbox{MS}}}^{(4)} = 200$ MeV
as fixed in the GRV distributions. Inclusive single-jet cross sections
have been calculated earlier \cite{x7,y27}. In these calculations, the
double-resolved contribution has been obtained with a different method
for canceling infrared and collinear singularities \cite{Sal93} which, however,
can be applied only to the computation of this particular cross section.
These results were compared to TOPAZ and AMY \cite{x1} jet production
data and good agreement was found \cite{y27}.

The fact that a different method for calculation of the double-resolved
contribution to the inclusive one-jet cross section was available was utilized
to test the new results based on the $y$-cut slicing method. Good agreement
between the two independent methods was achieved for all double-resolved
processes ($qq',q\overline{q}',qq,q\overline{q},qg,gg\rightarrow$ one jet)
separately.

In the following figures we show results for d$^3\sigma/$d$E_T$d$\eta_1$d$\eta_2$
as a function of $E_T$ for special $\eta$ values and as function of $\eta_2$
for $E_T=5$ GeV and $\eta_1=0$ using LEP1 kinematics. In figure \ref{kkkplot8}
% KKK-Plot 8
we show the $E_T$ distribution for $\eta_1=\eta_2=0$ in LO and NLO. We plotted
the direct, the single-resolved, and the double-resolved contributions
separately and show the sum of all three contributions. From the $E_T$ spectrum
it is already visible that the direct contribution is reduced by the NLO 
corrections whereas for the resolved contribution the NLO corrections are
positive. The correction in the double-resolved contribution is near 100\%.
This is seen more clearly in the lower part of figure \ref{kkkplot8}, where
we plotted the $k$-factor as a function of $E_T$ for the three contributions and
the sum. We remark that the LO curves are calculated with the two-loop $\alpha_s$
and the same NLO parton distributions of the photon as in the NLO calculation.
For a genuine LO prediction, one would choose the one-loop $\alpha_s$ formula
and LO parton distributions. This would change the $k$-factors. In figure
\ref{kkkplot8}, the $k$-factors show just the influence of the NLO corrections
to the parton-parton scattering cross sections. The upper part of figure
\ref{kkkplot8} also shows how the contributions of the three components (direct,
single-resolved, and double-resolved) sum up to the full inclusive two-jet
cross section. For relatively small $E_T$, this cross section is dominated
by the two resolved components. In this region we have a positive NLO correction
and a $k$-factor larger than 1. In the medium $E_T$ range and for large $E_T$,
the direct component dominates, the net NLO correction is negative, and the
$k$-factor is less than 1. It is clear that the importance of the resolved parts
for larger $E_T$ increases with increasing center-of-mass energy.

In figure \ref{kkkplot9}, we show the $\eta_2$ distribution for $\eta_1=0$ and
% KKK-Plot 9
$E_T=5$ GeV, again for the three components separately and their sum in LO
and in NLO. The $k$-factors corresponding to the upper part of figure \ref{kkkplot9}
are shown in the lower part. The three components show a very distinct behavior
as a function of $\eta_2$. The single-resolved contribution does not have a
plateau as broad as the other two components. So, by measuring in different $\eta_2$
regions, it might be possible to enhance one or two of the three contributions.

\begin{figure}[p]
 \epsfig{file=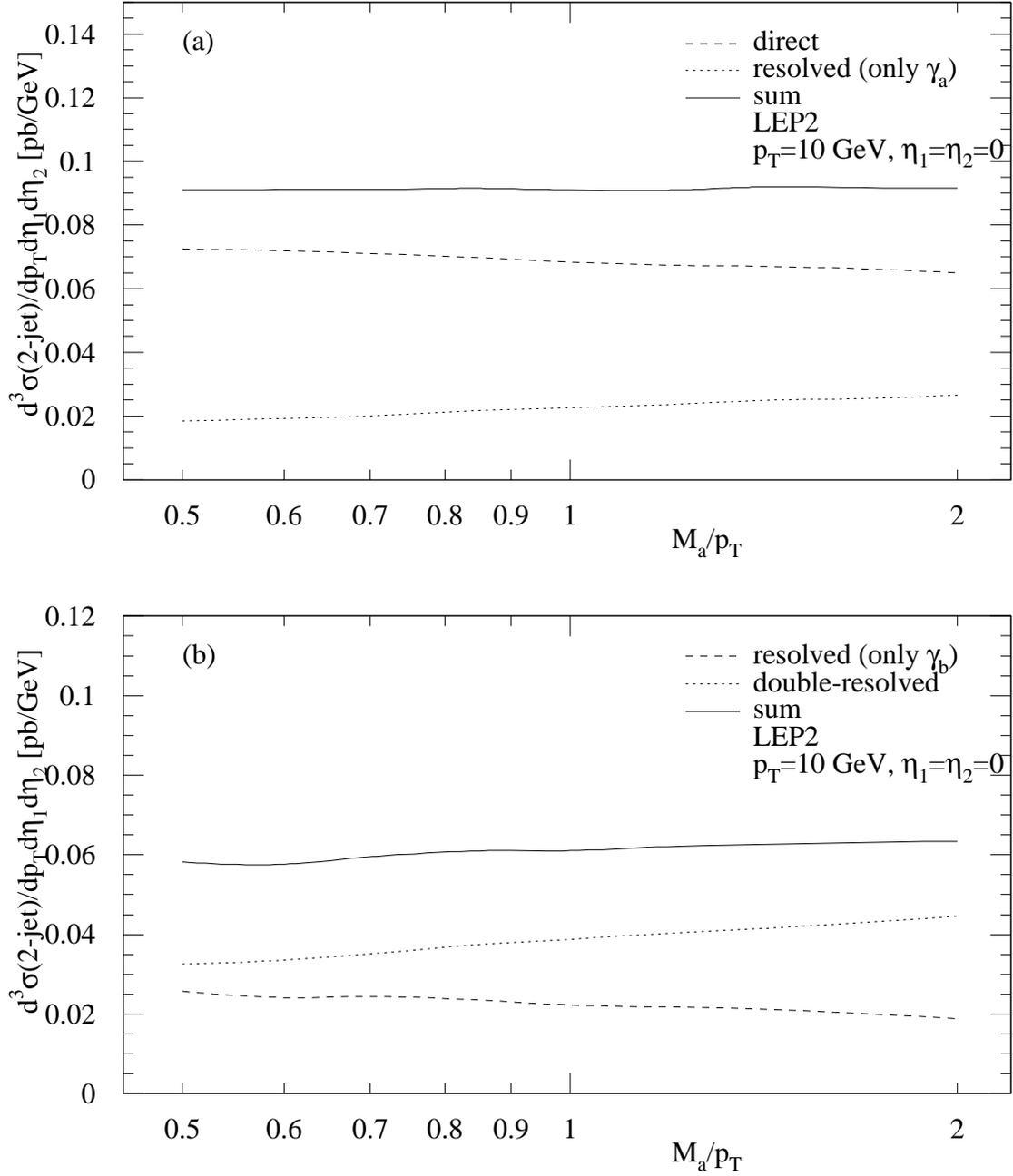}
 \caption{\label{kkkplot10}{\it Dependence of the inclusive two-jet cross section
  on the factorization scale $M_a$ in the photon $\gamma_a$. Upper figure: NLO direct
  and LO single-resolved photon (only $\gamma_a$ resolved), lower figure: NLO
  single-resolved (only $\gamma_b$ resolved) and LO double-resolved photon.}}
\end{figure}

\begin{figure}[p]
 \epsfig{file=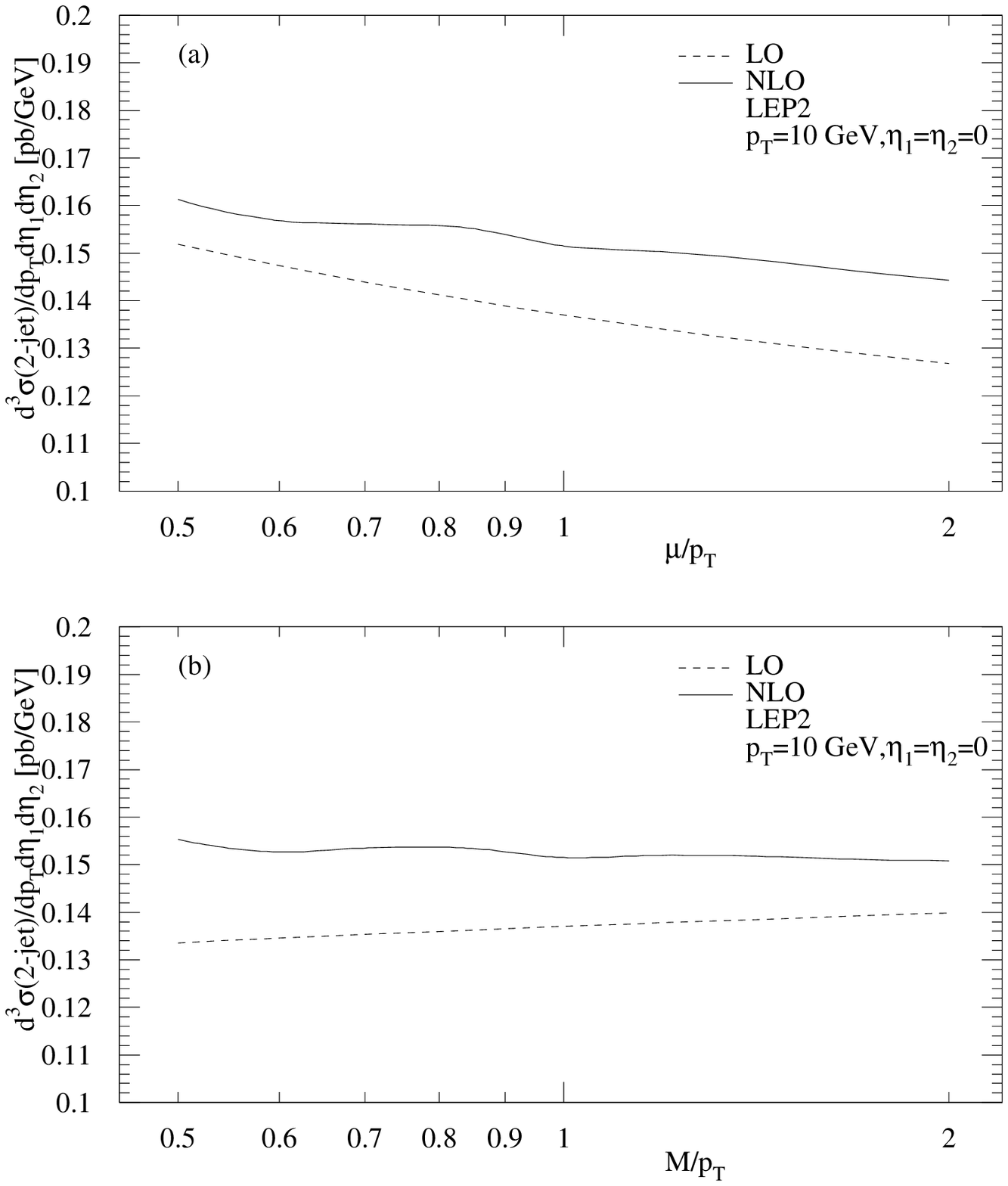}
 \caption{\label{kkkplot11}{\it Dependence of the inclusive two-jet cross section
  on the renormalization scale $\mu$ (a) and on the variation of all scales
  $\mu=M_a=M_b=M$ (b).}}
\end{figure}

An important test of our calculations is the compensation of the factorization
scale dependence between the direct and the single-resolved components and between
the single-resolved and the double-resolved components. That this compensation
works is shown in figures \ref{kkkplot10} a) and b) for the inclusive two-jet
% KKK-Plot 10
cross section at $E_T=10$ GeV and $\eta_1=\eta_2=0$. For this test we applied
LEP2 kinematics, i.e.\ $\sqrt{s}=175$ GeV and $\theta_c=1.72^\circ$ in the photon
spectra. In figure \ref{kkkplot10} a), the NLO direct and the LO single-resolved
cross section, where only the upper photon $\gamma_a$ is resolved, are plotted
as a function of $M_a/E_T$. The dependence of the NLO direct contribution on
$M_a$ is clearly visible. This is opposite to the dependence of the single-resolved
contribution originating from the scale dependence of the parton distributions.
As one can see, the sum is constant as a function of $M_a/E_T$. The same
compensation occurs between the NLO single-resolved and the LO double-resolved
cross section. In the single-resolved cross section, only the lower photon $\gamma_b$
is resolved. Also in this case (figure \ref{kkkplot10} b) the sum of both
contributions is rather independent of $M_a$. Another important topic for the
NLO theory is the question of the overall scale dependence and whether this
is reduced in NLO as compared to the LO cross section. First we show the dependence
on the renormalization scale $\mu$ alone with the factorization scales $M_a=M_b=E_T$
fixed. In figure \ref{kkkplot11} a) it is clearly visible that in NLO the
% KKK-Plot 11
dependence on $\mu$ is reduced as compared to LO when $\mu$ is varied between
$\mu=E_T/2$ and $2E_T$. If all scales $\mu=M_a=M_b=M$ are equal and this common
scale $M$ varies in the same range for the LO and NLO cross section we obtain
the curves in figure \ref{kkkplot11} b). We see that the dependence in LO
is such that the cross section as a function of $M$ increases which is
opposite to the behavior as a function of the renormalization scale $\mu$
in figure \ref{kkkplot11} a). This comes from the dependence on the factorization
scale which is not compensated (see figures \ref{kkkplot10} a) and b)). In NLO
we have the effect that the dependence on the factorization scale is
reduced due to the presence of the NLO corrections, so that in NLO the
cross section as a function of $M$ is fairly constant, i.e.\ it decreases
only slightly with increasing $M/E_T$. In conclusion we can say that the NLO
cross section is nearly independent of the scales and therefore presents a much
more solid prediction than the LO cross section.

\begin{figure}[h]
 \epsfig{file=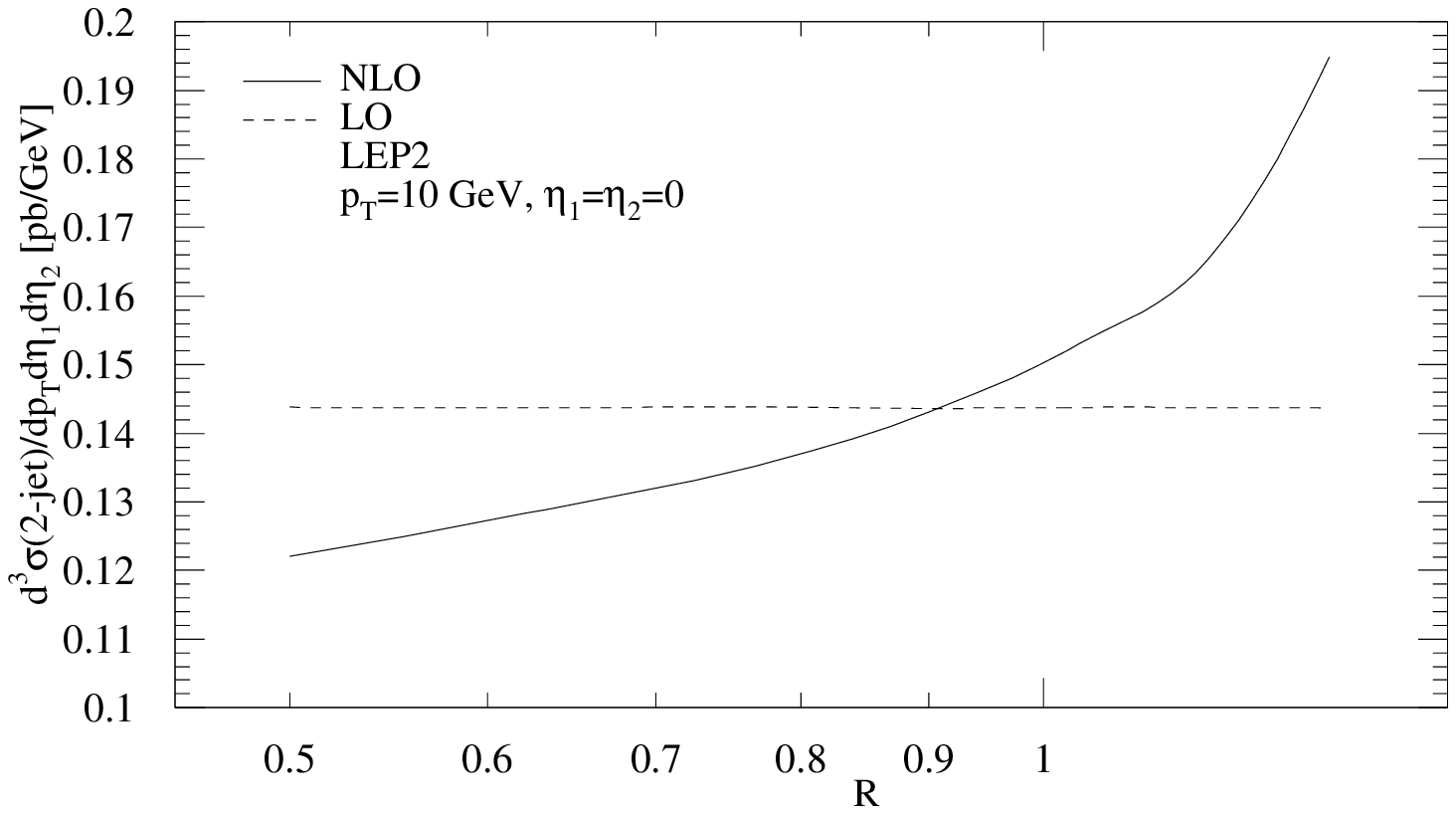}
 \caption{\label{kkkplot12}{\it Dependence of the inclusive two-jet cross section
  on the size of the jet cone $R$ for LO and NLO.}}
\end{figure}

When high statistics data become available, one could test the variation of the jet
cross section with changing cone radius $R$. This has been studied for the dijet
cross section for $E_T=10$ GeV and $\eta_1=\eta_2=0$. The result is displayed
in figure \ref{kkkplot12} where the NLO cross section
% KKK-Plot 12
d$^3\sigma$/d$E_T$d$\eta_1$d$\eta_2$ is shown as a function of $R$ for $R \geq 0.5$.
It increases as a function of $R$ almost like $a+b\ln R$ and somewhat stronger
for $R>1$. With our definition of the LO cross section the NLO result at $R=0.9$
is equal to the LO cross section, so that the NLO corrections stay moderate if $R$
varies between 0.7 and 1.

\subsection{Comparison with Experimental Data}
The first measurements of jet production in $\gamma\gamma$ reactions were
done by the TOPAZ and AMY \cite{x1} collaborations at TRISTAN.
They presented data for the inclusive one- and two-jet cross sections as a
function of $E_T$ in the range 2.5 GeV $< E_T <$ 8 GeV with different
ranges of rapidity in the two experiments. These data were compared to
predictions based on the theoretical work presented here in \cite{x9}.
Good agreement between the data of both experiments and the theoretical
results was achieved for the one- and two-jet cross sections. The
distribution in $E_T$ and also the absolute normalization agreed when
the parton distributions were described by the GRV set. At the energy
of the TRISTAN collider ($\sqrt{s} = 58$ GeV), the main contribution
to the jet cross sections comes from the direct process. However, the data
could not be reproduced by the direct component alone. The resolved
contributions were necessary to obtain agreement over the whole
$E_T$ range.

In the fall of 1995 the LEP ring was operated at the center-of-mass
energy of 133 GeV. During the short run period, the OPAL collaboration 
collected data for jet cross sections in $\gamma\gamma$ collisions.
They measured the inclusive one-jet cross section integrated over $|\eta|<1.0$
and the inclusive two-jet cross section for $|\eta_1|,|\eta_2|<1$ as a function
of $E_T$ in the range 3.0 GeV $< E_T < 14$ GeV \cite{x2}. These data were
compared to our calculated NLO one- and two-jet cross sections in
\cite{x2,x9}, and good agreement between the measurements and the predictions was
found. The calculations were done with $N_f=5$, $\Lambda_{\overline{\mbox{MS}}}^{(5)}
=130$ MeV and the NLO GRV parton distributions. In \cite{x11}, also the dependence of
the two-jet prediction on other parton distribution sets was studied with the
result that the other NLO sets, GS \cite{Gor92} and ACFGP \cite{Aur92}, lead to
almost the same results and the OPAL data could not be used to rule out one of
these sets. In these calculations, the photon spectra were described by
eq.~(\ref{eq61}) with $\theta_c=1.43^\circ$ as in the experimental setup.
The cone radius was $R=1$ and $R_{\rm sep}=2$.

\begin{figure}[h]
 \epsfig{file=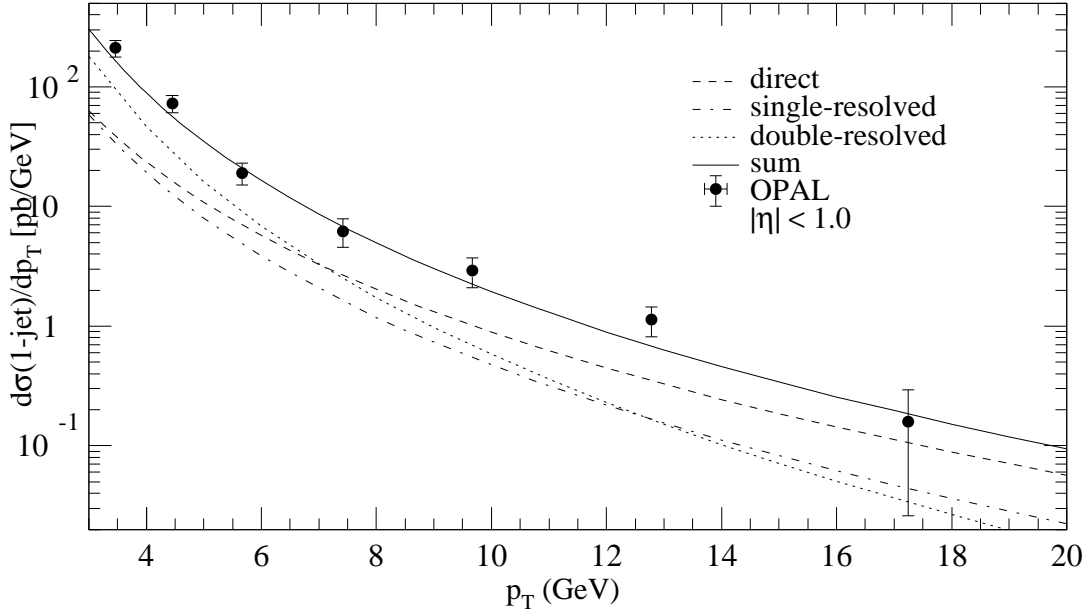}
 \caption{\label{kkkplot13}{\it The inclusive one-jet cross section as a function
  of $E_T$ with $|\eta| < 1$ compared to our NLO calculation. The direct,
  single-resolved, and double-resolved cross sections and the sum (full line)
  are shown separately.}}
\end{figure}

\begin{figure}[h]
 \epsfig{file=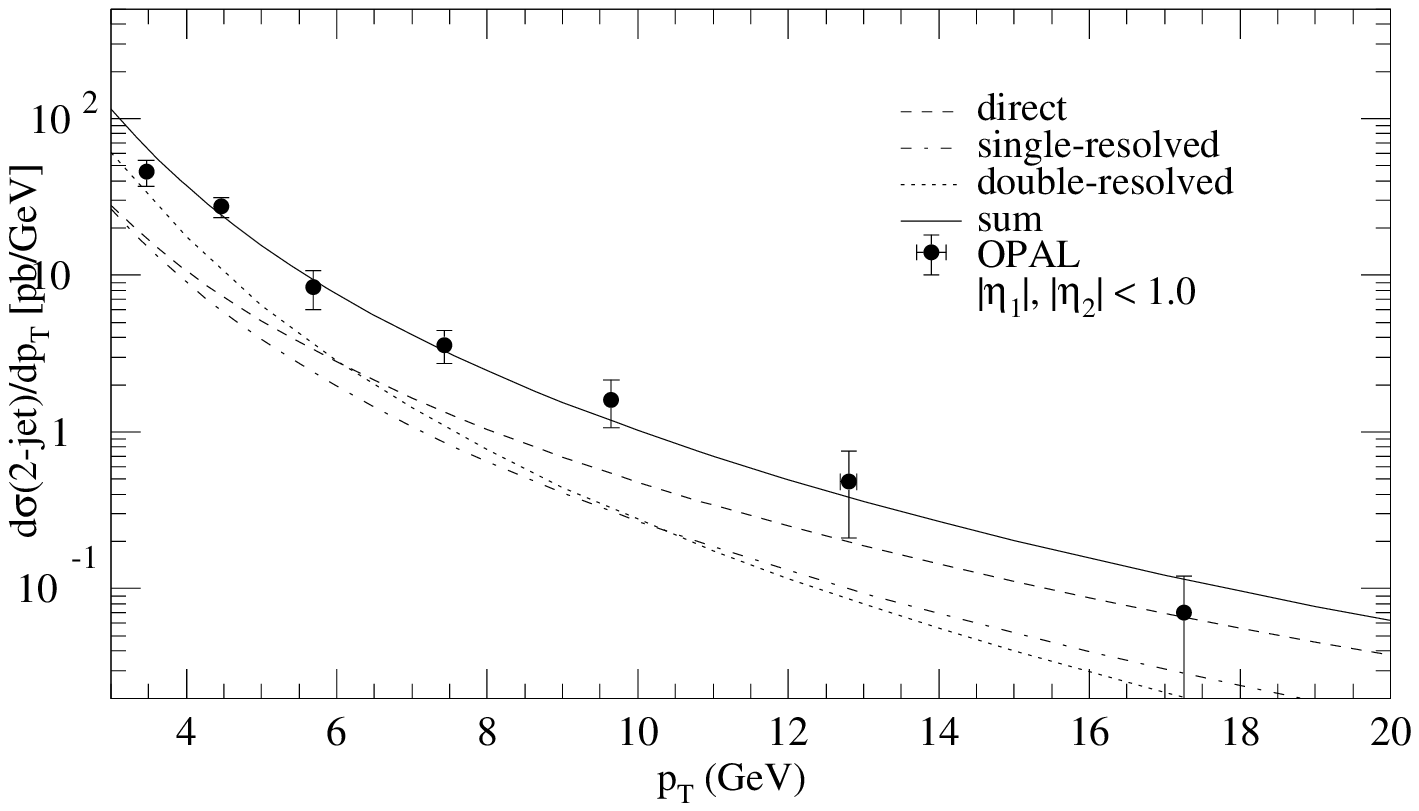}
 \caption{\label{kkkplot14}{\it The inclusive two-jet cross section as a function
  of $E_T$ with $|\eta_1|,|\eta_2| < 1$ compared to our NLO calculation. The direct,
  single-resolved, and double-resolved cross sections and the sum (full line)
  are shown separately.}}
\end{figure}

In the meantime the LEP energy was raised and the OPAL collaboration extended
their measurements to the center-of-mass energies of $\sqrt{s}=161$ GeV and
172 GeV \cite{y26}. They presented one-jet cross sections for $|\eta|<1$ and two-jet
cross sections for $|\eta_1|,|\eta_2|<1$ ($\sqrt{s} = 161$ GeV) and two-jet
cross sections for $|\eta_1|,|\eta_2|<2$ ($\sqrt{s} = 161$ GeV and 172 GeV).
The latter cross section was compared to our predictions already in \cite{y26}.
As an example, we show the experimental results of the OPAL collaboration
from the $\sqrt{s}=161$ GeV run. In figures \ref{kkkplot13} and \ref{kkkplot14}
% KKK-Plot 13
% KKK-Plot 14
the inclusive one- and two-jet cross section data as a function of $E_T$
for jets with $|\eta|< 1$ are compared to the NLO calculations with $R=1$,
$R_{\rm sep}=2$, and the NLO GRV parametrization for the photon. The parameters
were $\theta_c=33$ mrad, $\Lambda_{\overline{\mbox{MS}}}^{(5)} = 130$ MeV, and
$\mu=M_a=M_b=E_T$. The direct, single-resolved, double-resolved cross sections
and their sum (full line) are shown separately. As we can see, the agreement
between data and the predictions is good. The resolved cross sections dominate
in the region $E_T \leq 7$ GeV, whereas at high $E_T$ the direct cross section
dominates. We emphasize that the inclusive two-jet cross section is measured
using events with at least two jets. If an event contains more than two
jets, only the two jets with the highest $E_T$ values are used. This definition
of the inclusive two-jet cross section coincides with what is done in
the theoretical calculation. The good agreement between measured and
calculated cross section is remarkable, since the NLO calculation gives the
cross sections for massless partons, which are combined to jets, whereas
the experimental jet cross sections are measured for jets built of hadrons.
The good agreement then tells us that in $\gamma\gamma$ jet production,
parton-hadron duality is realized to a high degree and no major disturbance
due to underlying event energy is present. This is in contrast to what we
observed for $\gamma p$ jet production at lower $E_T$ where there
was disagreement at the larger rapidity values.

\begin{figure}[h]
 \begin{center}
  {\unitlength1cm
  \begin{picture}(12,8)
   \epsfig{file=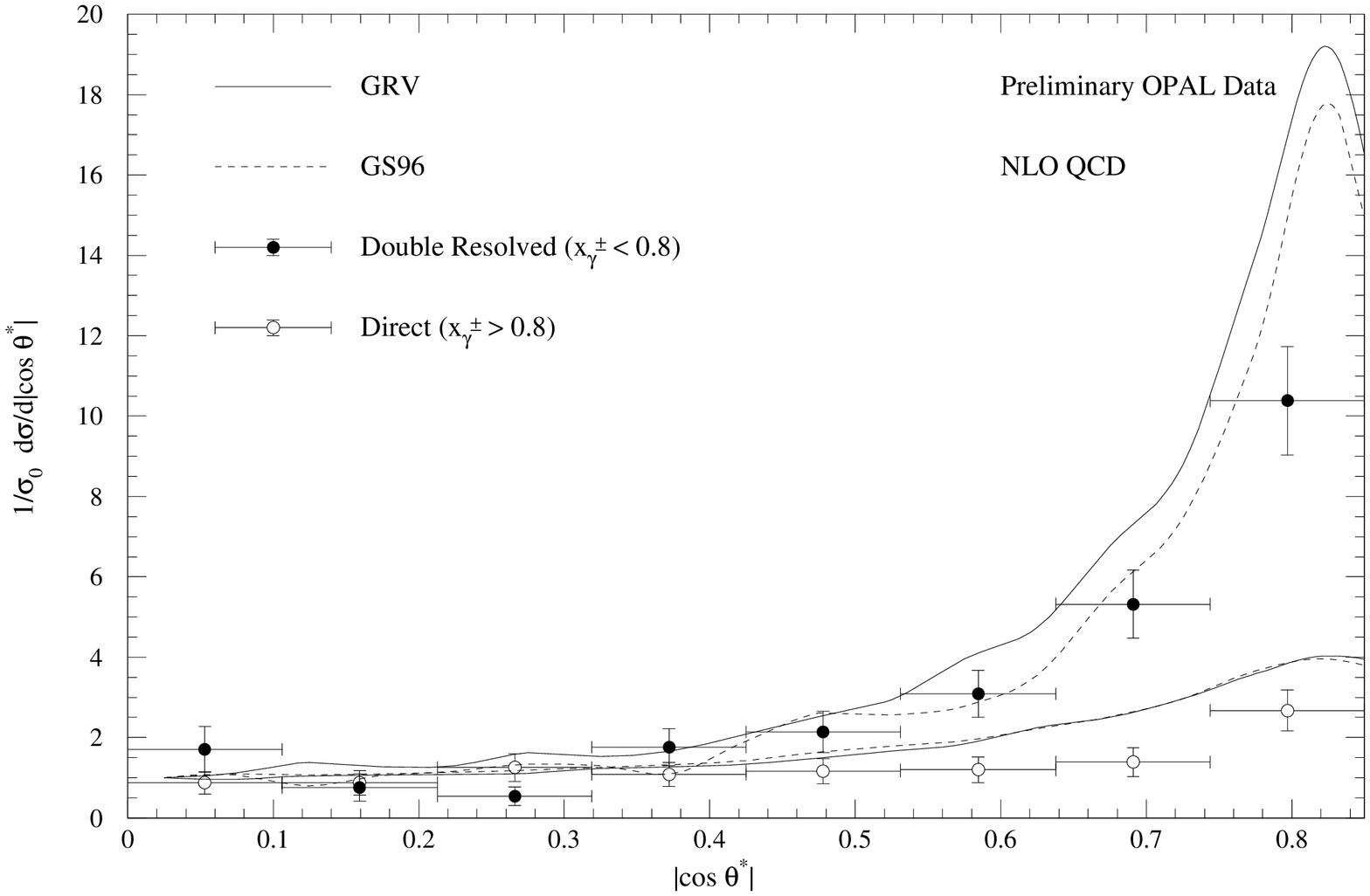,bbllx=520pt,bblly=95pt,bburx=105pt,bbury=710pt,%
           height=12cm,angle=270,clip=}
  \end{picture}}
 \end{center}
 \caption{\label{kkkplot15}{\it $|\cos\theta^\ast|$ dependence of the dijet
           $\gamma\gamma$ cross section for $M_{JJ} > 12$GeV, $|\eta_1|,|\eta_2| < 1$
           and $|\overline{\eta}| < 1$. We compare our NLO
           prediction with GRV and GS96 photon parton densities for direct-enhanced
           and double-resolved enhanced regions to preliminary data from OPAL.}}
\end{figure}

As a last point we confront the recently published $|\cos\theta^\ast|$
distribution of the OPAL collaboration with the theoretical predictions.
These data were taken for two $x_{\gamma}^{\rm OBS}$ intervals to separate
direct and resolved dominated contributions similar as it was done for jet
production in $ep$ scattering (see section 5). The selection cuts were
$x_{\gamma}^{\pm} > 0.8$ (direct-dominated) and $x_{\gamma}^{\pm} < 0.8$
(resolved-dominated), where
\bea
 x_{\gamma}^+ &=& \frac{\sum_{i=1}^2 E_{T_i} e^{-\eta_i}}{2x_aE_{e^+}}\\
 x_{\gamma}^- &=& \frac{\sum_{i=1}^2 E_{T_i} e^{ \eta_i}}{2x_bE_{e^-}}
\eea
are in LO equal to the fractions of the photon energies of the upper and
lower vertex entering the hard parton-parton scattering process. Thus, in LO
the direct process has $x_{\gamma}^+=x_{\gamma}^-=1$, whereas the double-resolved
process occurs only for $x_{\gamma}^+,x_{\gamma}^- < 1$. The direct cross
section data for $|\cos\theta^\ast|$ between 0 and 0.85 for the two event
classes $x_{\gamma}^{\pm}>0.8$ and $x_{\gamma}^{\pm}<0.8$ are exhibited in
figure \ref{kkkplot15}. The data points are normalized to have an
% KKK-Plot 15
average value of 1 in the first three bins and are plotted at the center
of the bins. The data with $x_{\gamma}^{\pm}> 0.8$ show a small rise
with increasing $|\cos\theta^\ast|$, whereas the data for $x_{\gamma}^{\pm}< 0.8$
show a much stronger rise in $|\cos\theta^\ast|$, similar to the findings
in $\gamma p$ scattering, where we pointed out how this qualitative behavior
is related to the exchanges of the LO cross sections. The theoretical curves
for the cross section in the two $x_{\gamma}^{\pm}$ ranges are also shown using
the two sets for the parton distribution functions, NLO GRV and NLO GS96 \cite{Gor96}.
In these calculations, $|\eta_1|,|\eta_2| < 1$, $|\overline{\eta}| < 1$, where
$\overline{\eta} = \frac{1}{2} (\eta_1+\eta_2)$, and the invariant two-jet
mass $M_{JJ} > 12 $ GeV, as in the experimental selection \cite{y26}.
$M_{JJ}$ is calculated from the two jets with highest $E_T$. All other
parameters are as stated previously. The theoretical curves in figure
\ref{kkkplot15} are normalized at $|\cos\theta^\ast|=0$. We observe
that the shape of the $|\cos\theta^\ast|$ distributions in the two
$x_{\gamma}^{\pm}$ bins is nicely reproduced in comparison with the data.
This means that the qualitative behavior known from LO arguments is still
present in NLO and agrees with the experimental data. On the quantitative
side, the theoretical cross sections seem to increase stronger towards
$|\cos\theta^\ast| = 1$ than the data, both the direct and the
double-resolved event sample, indicate.
However, a new OPAL analysis shows that the quantitative
agreement there is much better for the resolved sample, if
the $k_T$-cluster algorithm is used instead of the
cone algorithm. We remark that the data
analysis is still going on, i.e.\ the experimental data are still preliminary
and we have to wait for the final analysis before further conclusions
can be drawn.

\setcounter{equation}{0}

\section{Summary and Outlook}

In this work we have presented a complete
next-to-leading order calculation of direct and resolved photoproduction of
one and two jets in $\gamma p$ and $\gamma\gamma$ collisions.
Photon-proton and photon-photon scattering were considered simultaneously
since the two processes are intimately related to each other.
The results are of great importance not only as a test of
quantum chromodynamics (QCD), but also for the measurement of the proton and
photon parton densities currently performed at HERA and LEP.

First, we embedded the perturbatively calculable photon-parton scattering in
the experimentally observable electron-proton scattering process, using the
Weizs\"acker-Williams approximation and universal structure functions for the
proton and the photon. For these structure functions, we chose recent
next-to-leading order parametrizations from the CTEQ and GRV collaborations.
Special emphasis was given to experimental and theoretical ambiguities in the
Snowmass jet definition.

The hard photon-parton and photon-photon
scattering cross section was calculated in leading and
in next-to-leading order. The analytical calculation included the tree-level
Born matrix elements, the virtual corrections with one internal loop, and the
real
corrections with the radiation of a third particle in the initial or final
states. We integrated the latter over singular regions of phase space
up to an invariant mass cut $y$. This was done in $d$ dimensions in order to
regularize the soft and collinear divergencies. All infrared singularities
canceled, as they must according to the Kinoshita-Lee-Nauenberg theorem. The
ultraviolet 
poles in the virtual corrections and the collinear poles in the initial state
corrections were absorbed into the Lagrangian and the structure functions,
respectively.

The cross sections proved to be independent of the technical $y$-cut and less
dependent on the renormalization and factorization scales, when added to the
regular three-body contributions and integrated numerically. Excellent
agreement was found in inclusive single-jet predictions with the existing
programs of B\"odeker and Salesch. We extensively studied the direct,
resolved, and complete (in the case of $\gamma p$) and the direct, single-resolved,
double-resolved, and complete (in the case of $\gamma\gamma$)
one- and two-jet distributions in the transverse
energies and rapidities of the observed jets as well as the dependence on the
jet cone size $R$. Finally, we compared similar distributions to inclusive and
dijet data from H1 and ZEUS and OPAL and found good agreement in all cases.

It is possible to extend the work presented here in a number of different ways.
First, the formalism presented can easily be extended to the calculation of
$\gamma p$ and $\gamma\gamma$ collisions, where one of the photons has
a virtuality larger than zero, although still small compared to $E_T$
\cite{y34}. Second, other observables than those considered in this
work will allow to test the theory further and make it easier to isolate
the resolved contributions to obtain information on the parton distributions
in the photon for various scales. Third, our program can be used to predict
single and dijet cross sections in proton-proton and proton-antiproton
scattering at the TEVATRON at Fermilab in Chicago or at the designated LHC at
CERN in Geneva.

\setcounter{equation}{0}

\begin{appendix}
\renewcommand{\theequation}{\mbox{\Alph{section}.\arabic{equation}}}
\section{Phase Space Integrals for Final State Singularities}

This appendix contains the formul{\ae} needed to integrate the real
$2\rightarrow 3$ matrix elements in sections 4.2.2 and 4.2.3 over phase space
regions with final state singularities. To this aim, we have defined in
section 4.2.1 a measure
\beq
  \int \mbox{d}\mu_F = \int\limits_0^{y_F} \mbox{d}z'
  z'^{-\eps} \left( 1+\frac{z's}{t} \right) ^{-\eps}
  \int\limits_0^1 \frac{\mbox{d}b}{N_b} b^{-\eps} \left( 1-b \right)
  ^{-\eps}
  \int\limits_0^\pi \frac{\mbox{d}\phi}{N_\phi} \sin ^{-2\eps}\phi,
\eeq
which contains all variables associated with the phase space of the
unobserved subsystem $\overline{p}_{1}=p_1+p_3$. The integral over the
azimuthal angle $\phi$ is trivial, as the matrix elements are independent of
$\phi$.

We give the results for the four generic types of integrals in the following:
\bea
  f_1(a) & = & \int \mbox{d}\mu_F\frac{1}{z'}\frac{a}{z'+ab} 
              =\int \mbox{d}\mu_F\frac{1}{z'}\frac{a}{z'+a(1-b)} \nonumber \\
         & = & \frac{1}{2\eps^2}+\frac{1}{\eps}\left(-1-\frac{1}{2}\ln a\right)
               +\ln a-\frac{1}{2}\ln^2\frac{y_F}{a}+\frac{1}{4}
               \ln^2 a-\li\left( -\frac{y_F}{a} \right)-\frac{\pi^2}{6}+
               {\cal O}(\eps), \\
         &   & \nonumber \\
  f_2(a) & = & \int \mbox{d}\mu_F\frac{1}{z'}(1-b)(1-\eps)\nonumber \\
         & = & - \frac{1}{2\eps}+\frac{1}{2} + \frac{1}{2}\ln y_F +{\cal O}
                 (\eps), \\
         &   & \nonumber \\
  f_3(a) & = & \int \mbox{d}\mu_F\frac{1}{z'} \nonumber \\
         & = & - \frac{1}{\eps}+\ln y_F +{\cal O}(\eps), \\
         &   & \nonumber \\
  f_4(a) & = & \int \mbox{d}\mu_F\frac{1}{z'}(1-b+b^2) \nonumber \\
         & = & - \frac{5}{6\eps}+\frac{5}{6}\ln y_F -\frac{1}{18}+{\cal O}
               (\eps).
\eea
Terms of order $\eps$ have been omitted since they vanish in the limit
$d\rightarrow 4$.
\setcounter{equation}{0}

\section{Phase Space Integrals for Initial State Singularities}

In this appendix, we calculate the formul{\ae} needed to integrate the real
$2\rightarrow 3$ matrix elements in sections 4.2.5 through 4.2.8 over phase
space regions with initial state singularities. The integration measure was
defined in section 4.2.4 and is given by
\beq
  \int \mbox{d}\mu_I = \int\limits_0^{y_I} \mbox{d}z''
  z''^{-\eps} \int\limits_{X_a}^1 \frac{\mbox{d}z_a}{z_a}
  \lr\frac{z_a}{1-z_a}\rr^{\eps} \int\limits_0^\pi \frac{\mbox{d}\phi}
  {N_\phi}\sin^{-2\eps}\phi
  \frac{\Gamma(1-2\eps)}{\Gamma^2(1-\eps)}.
\eeq
The integration variables are associated with the phase space of the
unobserved particle $p_3$. The integral over the azimuthal angle $\phi$ is
again trivial, as the matrix elements do not depend on $\phi$. The longitudinal
momentum fraction $z_a$ is integrated over numerically, since the cross 
sections still have to be folded with the parton densities. These are contained
in the functions $g(z_a)$ below.

We give the results for the four generic types of integrals in the following:
\bea
  i_1(a) & = & \int \mbox{d}\mu_I g(z_a) \frac{1}{z''}\frac{a}{z''+a(1-z_a)}
               \nonumber \\
         & = & \int\limits_{X_a}^1\frac{\mbox{d}z_a}{z_a}g(z_a)
               \le-\ede-\ede\frac{z_a}{(1-z_a)_+}+\lr\frac{\ln\lr a\lr
               \frac{1-z_a}{z_a}\rr^2\rr}{1-z_a}\rr_+-\ln\lr a\lr\frac{1-z_a}
               {z_a}\rr^2\rr\rp\nonumber\\
         &   & \hspace{2.25cm}\lp-\frac{z_a}{1-z_a}\ln\lr 1+\frac{a(1-z_a)}
               {y_Iz_a}\rr+\ln\lr y_I\lr\frac{1-z_a}{z_a}\rr\rr\re\nonumber \\
         &   & +\hspace{1.05cm}g(1)\le\frac{1}{2\eps^2}-\frac{1}{2\eps}\ln a
               +\frac{1}{4}\ln^2a+\frac{\pi^2}{2}\re +{\cal O}(\eps),\\
         &   & \nonumber \\
  i_2(a) & = & \int \mbox{d}\mu_I g(z_a) \frac{1}{z''}\nonumber \\
         & = & \int\limits_{X_a}^1\frac{\mbox{d}z_a}{z_a}g(z_a)
               \le-\ede+\ln\lr y_I\frac{1-z_a}{z_a}\rr\re+{\cal O}(\eps),\\
         &   & \nonumber \\
  i_3(a) & = & \int \mbox{d}\mu_I g(z_a) \frac{1}{z''}(1-z_a)(1-\eps)
               \nonumber \\
         & = & \int\limits_{X_a}^1\frac{\mbox{d}z_a}{z_a}g(z_a)
               \le-\ede (1-z_a)+(1-z_a)+(1-z_a)\ln\lr y_I\frac{1-z_a}{z_a}\rr
               \re+{\cal O}(\eps),\\
         &   & \nonumber \\
  i_4(a) & = & \int \mbox{d}\mu_I g(z_a) \frac{1}{z''}\lr\frac{z_a^2-2z_a+2
               -\eps z_a^2}{z_a}\rr\frac{1}{1-\eps}\nonumber \\
         & = & \int\limits_{X_a}^1\frac{\mbox{d}z_a}{z_a}g(z_a)
               \le-\ede\frac{z_a^2-2z_a+2}{z_a}+\frac{z_a^2-2z_a+2}{z_a}
               \ln\lr y_I\frac{1-z_a}{z_a}\rr-2\frac{1-z_a}{z_a}
               \re+{\cal O}(\eps).
\eea
Terms of order $\eps$ have again been omitted.

\end{appendix}

\end{document}